\documentclass[aps,pra,twocolumn,superscriptaddress]{revtex4-1}

\usepackage{amsmath}
\usepackage{amssymb}
\usepackage{bm}
\usepackage{url}
\usepackage[dvipdfmx]{graphicx}
\usepackage{color}

\makeatletter
\def\l@subsubsection#1#2{}
\makeatother

\begin{document}

\title{Topological superconductors: a review}

\author{Masatoshi Sato}
\email{msato@yukawa.kyoto-u.ac.jp}
\affiliation{Yukawa Institute for Theoretical Physics, Kyoto University, 
Kyoto 606-8502, Japan}

\author{Yoichi Ando}
\email{ando@ph2.uni-koeln.de}
\affiliation{Physics Institute II, University of Cologne,
Z\"ulpicher Str. 77, 50937 Cologne, Germany}

\begin{abstract}

This review elaborates pedagogically on the fundamental concept, basic theory, expected properties, and materials realizations of topological superconductors. The relation between topological superconductivity and Majorana fermions are explained, and the difference between dispersive Majorana fermions and a localized Majorana zero mode is emphasized. A variety of routes to topological superconductivity are explained with an emphasis on the roles of spin-orbit coupling. Present experimental situations and possible signatures of topological superconductivity are summarized with an emphasis on intrinsic topological superconductors. 

\end{abstract}


\maketitle

\tableofcontents

\section{Introduction}

Topology is a mathematical concept to classify shapes. 
When two shapes can be continuously deformed into each other (like the shapes of a doughnut and a teacup), they belong to the same topological class. Similarly, when a quantum mechanical wavefunction is adiabatically connected to a different wavefunction, they may be called topologically identical. In the case of quantum many-body systems, simple combinations of atomic wavefunctions are considered to be trivial, and hence any condensed matter system whose wavefunction is adiabatically connected to the atomic limit is topologically trivial. However, when the wavefunction is adiabatically distinct from the atomic limit, such a system may be called topological. The first quantum mechanical system which was recognized to be topological was the quantum Hall system; in this case, the adiabatic continuity can be mathematically judged by an integer number called Chern number \cite{TKNN,Kohmoto85}, which will be explained in detail in the next section. In general, topological classifications are done by finding an integer number called {\it topological invariant} which characterize the topology; the Chern number is a prominent example of the topological invariant in quantum-mechanical systems.

Although the nontrivial topology in the quantum Hall system was
recognized in as early as 1982 \cite{TKNN}, topological quantum systems gained wide interest after the theoretical discovery in 2005 of the quantum spin Hall insulator phase in a two-dimensional (2D) time-reversal-invariant insulator \cite{Kane-Mele1}, whose topology is characterized by a $Z_2$ topological invariant \cite{Kane-Mele2}. This topology was extended to three-dimensional (3D) systems \cite{Moore-Balents,Fu-Kane-Mele,Roy2009}, leading to the birth of the term {\it topological insulators} \cite{Moore-Balents}. The historical perspective of topological insulators are concisely summarized in a recent review \cite{AndoReview}. 

It was soon recognized that topological classifications are possible in principle for various quantum many-body systems having a gap in the energy-band spectrum to protect the occupied states. As a result, systematic topological classifications of not only insulators but also superconductors based on the symmetry properties were conducted \cite{Schnyder2008,Kitaev2009}. Such theories elucidated how the symmetry and dimensionality of a gapped system dictate the types of topological invariants; for example, a time-reversal-breaking superconductor can be topological in 2D but not in 3D. In general, topological superconductors are adiabatically distinct from the Bose-Einstein condensate of Cooper pairs, which obviously forms a trivial superconducting state, like the atomic limit of an insulator.

Interestingly, before the above-mentioned developments regarding the topological quantum systems took place, nontrivial topology in superfluid helium 3 (He-3) was discussed by Volovik \cite{VolovikHe3}. Also, topologically nontrivial superconductors were discussed in 2000 in a 2D model by Read and Green \cite{Read-Green} and in a one-dimensional (1D) model by Kitaev \cite{Kitaev2001}. Both models consider spinless, time-reversal-breaking $p$-wave pairing for the superconducting state, which gives rise to a non-Abelian Majorana zero mode in the vortex core in the 2D case or at the edge in the 1D case. In 2003, Sato proposed a 2D model to realize a similar non-Abelian Majorana zero mode even in the case of $s$-wave pairing \cite{Sato2003}. These models were motivated to realize topologically-protected quantum computations that are free from decoherence. It is noteworthy that those models of topological superconductors preceded the recent developments of topological quantum systems, but concrete ideas to realize such topological superconductors emerged only after the discovery of topological insulators, which provide a convenient platform for ``spinless'' superconductivity.

As one can see in this history, the interest in topological superconductors are strongly tied to the Majorana fermions that are exotic in that particles are their own antiparticles \cite{Alicea2012,Beenakker2013,Sato2016}. In condensed matter systems, when an quasiparticle is a superposition of electron and hole excitations and its creation operator $\gamma^{\dagger}$ becomes identical to the annihilation operator $\gamma$, such a particle can be identified as a Majorana fermion. In the Read-Green model, the Bogoliubov quasiparticles in the bulk become dispersive Majorana fermions and the bound state formed in the vortex core becomes a Majorana zero mode. The former is interesting as a new type of itinerant quasiparticles, while the latter is useful as a qubit for topological quantum computation.

In the following, we first present the basics of topological quantum systems and superconductivity to prepare the readers to the theoretical discussion in this topic. We then elaborate on the theory of topological superconductors and Majorana fermions. The materials search for topological superconductors are summarized and the novel properties to look for in the materials realizations are discussed.

\section{Concept of topology in quantum mechanics}

We will first familiarize the readers with the concept of topology in
quantum-mechanical wavefunctions, by elaborating on a simple example.

\subsection{Berry phase, Berry connection, and gauge field}

When a Hamiltonian depends on a certain set of parameters, its eigenstates are defined
in the corresponding parameter space.
The wavefunctions may have a ``twist'' in such a parameter space, i.e., they may have a
topologically nontrivial structure. 
The Berry phase typically characterizes such a nontrivial topology in
the parameter space. 

In particular, the Berry phase in the momentum space plays
an important role in the topology of insulators and superconductors. 
Let us consider a band insulator described by a Bloch Hamiltonian
${\cal H}({\bm k})$ with the crystal momentum ${\bm k}$.
The eigenstates are given by the solutions of the Bloch equation,
\begin{eqnarray}
H({\bm k})|u_n({\bm k})\rangle=E_n({\bm k})|u_n({\bm k})\rangle, 
\label{eq2:eigen}
\end{eqnarray}
so that they contain ${\bm k}$ as a parameter of the wavefunctions.
The Berry connection $\bm{\mathcal A}^{(n)}({\bm k})$ defined by
\begin{eqnarray}
\bm{\mathcal A}^{(n)}({\bm k})=i\langle u_n({\bm k})|{\bm \partial}_{{\bm
k}}u_n({\bm k})\rangle 
\label{eq2:Berryconnection}
\end{eqnarray}
measures the rate of change in the wavefunction $|u_n({\bm k})\rangle$ in the
momentum space, and therefore it vanishes when $|u_n({\bm k})\rangle$ does not change with ${\bm k}$.

To consider the topology based on $\bm{\mathcal A}^{(n)}({\bm k})$, we need to take
into account the ambiguity in the solutions of Eq. (\ref{eq2:eigen}); 
this Bloch equation does not fix the phase factor of the solution, and hence there remains a gauge degree of freedom, 
\begin{eqnarray}
|u_n({\bm k})\rangle \rightarrow e^{i\phi_n({\bm k})}|u_n({\bm k})\rangle.
\end{eqnarray}
This leads to a gauge transformation in $\bm{\mathcal A}^{(n)}({\bm k})$ as
\begin{eqnarray}
\bm{\mathcal A}^{(n)}({\bm k})\rightarrow \bm{\mathcal A}^{(n)}({\bm k})
-{\bm \partial}_{\bm k}\phi_n({\bm k}).
\end{eqnarray}
Now we note that any physical quantity should be gauge invariant.
A gauge-invariant quantity to be constructed from $\bm{\mathcal A}^{(n)}({\bm k})$ is the 
``field strength'' of the Berry connection,
\begin{eqnarray}
{\cal F}^{(n)}_{ij}({\bm k})=\partial_{k_i}{\cal A}^{(n)}_{k_j}({\bm k})
-\partial_{k_j}{\cal A}^{(n)}_{k_i}({\bm k}),
\end{eqnarray}
which has a physical meaning. As we will show in the next subsection, the integration of the field
strength over the whole Brillouin zone defines a topological invariant
named Chern number.

Another gauge invariant quantity constructed from $\bm{\mathcal A}^{(n)}({\bm k})$ is the Berry phase,
which is given as its line integral along a closed path $C$ in the
momentum space. 
We first take a gauge with which $\bm{\mathcal A}^{(n)}({\bm k})$ is non-singular on $C$ (which is always possible),
and then calculate the line integral along $C$,
\begin{eqnarray}
\oint_C d{\bm k}\cdot\bm{\mathcal A}^{(n)}({\bm k}). 
\label{eq2:integral}
\end{eqnarray}
Since various choices are possible for the gauge taken above, there remains a freedom for 
a non-singular gauge transformation, with $e^{i\phi_n({\bm k})}$ a unique function on $C$. 
Such a gauge transformation changes the line integral Eq. (\ref{eq2:integral}) as
\begin{eqnarray}
&&\oint_C d{\bm k}\cdot\bm{\mathcal A}^{(n)}({\bm k}) 
\nonumber\\
&&\rightarrow
\oint_C d{\bm k}\cdot\bm{\mathcal A}^{(n)}({\bm k})
-\oint_C d{\bm k}\cdot{\bm \partial}_{\bm k}\phi_n({\bm k}).
\end{eqnarray} 
The second line integral in the right hand side is easily evaluated: Due to the uniqueness of 
$e^{i\phi_n({\bm k})}$ on $C$, this term must give $2\pi N$ with an integer $N$. 
Consequently, one can see that
\begin{eqnarray}
\exp\left[
i\oint_C d{\bm k}\cdot\bm{\mathcal A}^{(n)}({\bm k}) 
\right] 
\end{eqnarray}
is gauge-invariant. The phase factor of this gauge-invariant quantity is the Berry
phase, which, in general, can take any real value and change
continuously, and thus is not a topological invariant. 
Nevertheless, when a certain symmetry is imposed and a suitable $C$ is considered, the Berry phase 
can be quantized and give a topological invariant, as we will show later.




\subsection{Chern number}
\label{sec2:Chern}

Shortly before Berry discovered his famous phase in quantum
mechanics \cite{Berry84}, 
Thouless, Kohmoto, Nightingale and den Nijs introduced a topological
number (TKNN integer) to characterize the quantum Hall states \cite{TKNN}.
Soon after these works, Simon pointed out a mathematical relation between the
two quantities \cite{Simon83} and Kohmoto derived the TKNN integer in the form of
the (first) Chern number \cite{Kohmoto85}.
It is now generally recognized that the Chern number is one of the most
fundamental topological numbers in topological phases of matter.

For a 2D system, the Chern number of the $n$-th
band is defined by using the field strength of the Berry connection as
\begin{eqnarray}
Ch^{(n)}_1=\frac{1}{2\pi}\int_{\rm 2dBZ} dk_x dk_y {\cal F}^{(n)}_{xy}({\bm k}),
\label{eq2:Ch1}
\end{eqnarray}
where the integration is performed on the 2D Brillouin zone.
This integration is facilitated by using the Stokes' theorem:
Due to the periodicity of the Brillouin zone, the integral vanishes if
the Berry connection $\bm{\mathcal A}^{(n)}({\bm k})$ has no singularity 
over the Brillouin zone.
On the other hand, if $\bm{\mathcal A}^{(n)}({\bm k})$ has a singularity
at ${\bm k}_0$, one performs the gauge transformation
\begin{eqnarray}
\bm{\mathcal A}^{(n)'}({\bm k}) 
=\bm{\mathcal A}^{(n)}({\bm k})-{\bm \partial}_{\bm k}\phi_n({\bm k})
\end{eqnarray}
in the region $R$ including ${\bm k}_0$, so that
$\bm{\mathcal A}^{(n)'}({\bm k})$ has no singularity in $R$. 
From the Stokes' theorem, the integral in Eq. (\ref{eq2:Ch1}) becomes
the line integral of the gauge function,
\begin{eqnarray}
Ch^{(n)}_1= \frac{1}{2\pi}\int_{\partial R} d{\bm k}\cdot {\bm \partial}_{\bm
k}\phi_n({\bm k}),
\end{eqnarray}
where $\partial R$ is the boundary of $R$.
Because $e^{i\phi_n({\bm k})}$ is a unique function on $\partial R$, 
this $Ch^{(n)}_1$ must be an integer.
For a 2D insulator, the total Chern number of the occupied bands
\begin{eqnarray}
Ch=\sum_{E_n<E_{\rm F}}Ch^{(n)} 
\label{eq3:totalCh}
\end{eqnarray} 
is of particular importance, because it is directly related to the Hall conductance through
\begin{eqnarray}
\sigma_{xy}=-\frac{e^2}{h}Ch. 
\end{eqnarray}

Under time-reversal, the total Berry connection of the occupied bands and
its field strength transform as
\begin{eqnarray}
\sum_{E_n<E_{\rm F}}
\bm{\mathcal A}^{(n)}({\bm k})
\rightarrow 
\sum_{E_n<E_{\rm F}}
\bm{\mathcal A}^{(n)}(-{\bm k}),
\nonumber\\
\sum_{E_n<E_{\rm F}}
{\mathcal F}_{ij}^{(n)}({\bm k})
\rightarrow 
-\sum_{E_n<E_{\rm F}}{\mathcal
F}_{ij}^{(n)}(-{\bm k}).
\end{eqnarray}
Therefore, if the system retains time-reversal symmetry, 
the total Chern number of the occupied bands satisfies
\begin{eqnarray}
Ch&=&
\frac{1}{2\pi}\int_{\rm 2d BZ} dk_x dk_y\sum_{E_n<E_{\rm F}}{\mathcal
F}_{xy}({\bm k}) 
\nonumber\\
&=&
-\frac{1}{2\pi}\int_{\rm 2d BZ} dk_x dk_y \sum_{E_n<E_{\rm F}}{\mathcal
F}_{xy}(-{\bm k}) 
\nonumber\\
&=&-Ch,
\end{eqnarray}
which leads to $Ch=0$.
This means that time-reversal breaking is necessary for realizing a quantum
Hall state with a nonzero $Ch$.

\subsection{Role of symmetry}
\label{sec2:symmetry}

The symmetry of the Hamiltonian plays a crucial role in determining the topology of the occupied states. 
Actually, if no symmetry is assumed, the only possible topological phase in up to
three dimensions is the quantum Hall state with a nonzero Chern number \cite{Avron1983}.
To see this, let us consider a general Hamiltonian $H({\bm k})$.
The Hamiltonian is diagonalized by using a unitary matrix $U({\bm k})$ as
\begin{eqnarray}
U^{\dagger}({\bm k})H({\bm k})U({\bm k})=
\left(
\begin{array}{ccc}
E_1({\bm k}) & & \\
& \ddots & \\
& & E_n({\bm k}) \\
\end{array}
\right). 
\end{eqnarray} 
We suppose that the Hamiltonian describes a band insulator with 
a band gap at the Fermi energy $E_{\rm F}$. For convenience, the bands are numbered from the 
highest energy, with the first $m$ bands to be empty; in other words, we assume $E_i({\bm k})>E_{\rm F}$
for $i=1,\dots, m$, and $E_i({\bm k})<E_{\rm F}$ for $i=m+1, \dots, n$.

To examine the topology of this system, 
we perform an adiabatic deformation of the Hamiltonian.
If no symmetry is imposed on $H({\bm k})$, any deformation is allowed
as long as it keeps a band gap. In particular, one can make the energies
of all the occupied (empty) states to become $-1$ (1) without closing the gap. 
Such a deformation does not change the topological property of occupied bands.

After this deformation, we have
\begin{eqnarray}
U^{\dagger}({\bm k})H({\bm k})U({\bm k})=
\left(
\begin{array}{cc}
{\bm 1}_{m\times m} & \\
& -{\bm 1}_{(n-m)\times (n-m)} \\
\end{array}
\right),
\nonumber\\
\label{eq4:flatHamiltonian}
\end{eqnarray} 
where ${\bm 1}_{m\times m}$ and ${\bm 1}_{(n-m)\times (n-m)}$ are
the $m\times m$ and $(n-m)\times(n-m)$ unit matrices, respectively.
Therefore, the information of the topology is encoded
in the unitary matrix $U({\bm k})$.
It should be noted here that $U({\bm k})$ itself has a gauge redundancy.
Indeed, using a $m\times m$ unitary matrix $U_{m\times m}({\bm k})$ and 
another $(n-m)\times (n-m)$ unitary matrix $U_{(n-m)\times(n-m)}({\bm k})$, 
one can construct the gauge transformation 
\begin{eqnarray}
U({\bm k})\rightarrow U({\bm k})
\left(
\begin{array}{cc}
U_{m\times m}({\bm k}) & \\
& U_{(n-m)\times (n-m)}({\bm k})
\end{array}
\right),
\label{eq4:gaugetr}
\end{eqnarray}
which does not change $H({\bm k})$ in Eq. (\ref{eq4:flatHamiltonian}).
Consequently, $U({\bm k})$ is not merely an element of the group of $n\times n$
unitary matrices, ${\rm U}(n)$, but it should be regarded as an
element of its coset space
\begin{eqnarray}
{\cal M}=\frac{{\rm U}(n)}{{\rm U}(m)\times {\rm U}(n-m)}, 
\end{eqnarray}
where the equivalence relation of the coset space is obtained on the
basis of the gauge redundancy expressed in Eq. (\ref{eq4:gaugetr}).
As a result, the unitary matrix $U({\bm k})$ defines a map from the
Brillouin zone to the coset
space ${\cal M}$.

Such a map can be topologically classified by the homotopy theory.
In the present case, the homotopy group is given by 
\begin{eqnarray}
\pi_d({\cal M})=
\left\{
\begin{array}{cl}
0, & d=1,3 \\
{\bm Z}, & d=2 
\end{array}
\right. .
\end{eqnarray}
This means that a topologically non-trivial structure exists only in $d=2$
dimensions for $d\le 3$.
The integer topological number of $\pi_2({\cal M})$ is given by the
Chern number defined in Eq. (\ref{eq3:totalCh}).

\subsection{Time-reversal symmetry and ${\bm Z}_2$ index}

The above result implies that some additional symmetry is needed to obtain
topological phases other than the quantum Hall states.
Indeed, in the presence of time-reversal symmetry, one can obtain
new topological phases, i.e. the quantum spin Hall phase in 2D \cite{Kane-Mele1, Kane-Mele2, Bernevig2006}
and the topological insulator phase in 3D \cite{Moore-Balents,Fu-Kane-Mele, Roy2009}. 

A key structure of these new topological phases is the Kramers degeneracy
due to time-reversal symmetry.
The time-reversal operator ${\cal T}$ obeys
${\cal T}^2=-1$ for spin-$\frac{1}{2}$ electrons, 
and the anti-unitarity of ${\cal T}$ yields 
\begin{eqnarray}
\langle {\cal T} u | {\cal T} v\rangle=\langle v| u \rangle, 
\end{eqnarray}
for any states $|u\rangle$ and $|v\rangle$.
Therefore, a state $|u \rangle$ and its
time-reversal partner ${\cal T}|u\rangle$ is orthogonal, i.e. $\langle
u|{\cal T}u\rangle=0$, because of the relation
\begin{eqnarray}
\langle u|{\cal T}u\rangle=\langle {\cal T}^2 u|{\cal T}u\rangle=-\langle u|{\cal T} u\rangle. 
\end{eqnarray}
In time-reversal-invariant systems, $|u\rangle$ and ${\cal T} |u\rangle$
have the same energy, and thus the above result implies that any energy
eigenstate has two-fold degeneracy, which is called Kramers degeneracy.

For the Bloch Hamiltonian $H({\bm k})$ in Eq. (\ref{eq2:eigen}), time-reversal
symmetry leads to
\begin{eqnarray}
{\cal T} H({\bm k}){\cal T}^{-1}=H(-{\bm k}). 
\end{eqnarray} 
Thus, if we consider an energy eigenstate $|u_n({\bm k})\rangle$ of
$H({\bm k})$, its Kramers partner ${\cal T}|u_n({\bm k})\rangle$
is an eigenstate of $H(-{\bm k})$, which is also written as
$|u_{n'}(-{\bm k})\rangle$ with a different eigenstate $|u_{n'}({\bm
k})\rangle$ of $H({\bm k})$.
For later convenience, 
we relabel the band indices of the Kramers pair $(|u_n({\bm k})\rangle,
|u_{n'}({\bm k})\rangle)$ to $(|u_n^{\rm I}({\bm k})\rangle, |u^{\rm
II}_{n}({\bm k})\rangle)$.
Considering the phase ambiguity of the eigenstates, they satisfy
\begin{eqnarray}
|u_n^{\rm II}({\bm k})\rangle=e^{i\varphi_n({\bm k})}{\cal T}|u^{\rm I}_n(-{\bm
k})\rangle, 
\end{eqnarray}
where $\varphi_n({\bm k})$ signifies the gauge degree of freedom.

As discovered first by Kane and Mele \cite{Kane-Mele2}, 
the Kramers degeneracy makes it possible to introduce a new topological
number called ${\bm Z}_2$ index in 2D.
Originally, the ${\bm Z}_2$ index was provided in the form of a
Pfaffian, but there are several different ways to define the same ${\bm Z}_2$
index \cite{Moore-Balents, Roy2009, Qi2008}.
Here, we define it as a variant of the so-called {\it spin Chern number} \cite{Sheng2006}.

For a 2D system with a fixed spin orientation, say $S_z$,
one can introduce the spin Chern number. In the diagonal basis of $S_z$, 
the Hamiltonian is written as 
\begin{eqnarray}
H({\bm k})=\left(
\begin{array}{cc}
H_{\uparrow}({\bm k}) & 0\\
0 & H_{\downarrow}({\bm k})
\end{array}
\right) ,
\end{eqnarray}
where $H_{\uparrow}({\bm k})$ ($H_{\downarrow}({\bm k})$) is
the Hamiltonian in the $S_z=1/2$ ($S_z=-1/2$) sector.
With this separation of $H({\bm k})$, the spin Chern number $Ch_{\uparrow}$ ($Ch_{\downarrow}$) is
defined as the Chern number of $H_{\uparrow}({\bm k})$
($H_{\downarrow}({\bm k})$). 
Since the spin flips under time-reversal, each spin sector does not
have time-reversal invariance. Hence, the spin Chern number can be nonzero even for a time-reversal-symmetric
system, although the total Chern number always vanishes (i.e. $Ch_{\uparrow}+Ch_{\downarrow}=0$) due 
to time-reversal symmetry.

In general, the spin-orbit coupling breaks spin conservation, 
and thus the spin Chern number is not well-defined in a spin-orbit coupled system. 
However, in the presence of time-reversal symmetry, 
one can derive an analogous topological number even for a spin-orbit coupled system in 2D. 
Instead of the spin eigensectors, 
we use Kramers pairs $(|u_n^{\rm I}({\bm k})\rangle, |u_n^{\rm II}({\bm
k})\rangle)$ to divide the Hilbert space into two subspaces, and then
introduce the Chern numbers in those subspaces,
\begin{eqnarray} 
&&Ch_{\rm I}=\frac{1}{2\pi}\int_{\rm 2dBZ} dk_xdk_y
{\cal F}^{{\rm I}(-)}_{xy}({\bm k}), 
\nonumber\\
&&Ch_{\rm II}=\frac{1}{2\pi}\int_{\rm 2dBZ} dk_xdk_y
{\cal F}^{{\rm II}(-)}_{xy}({\bm k}), 
\end{eqnarray}
where ${\cal F}^{{\rm I}(-)}_{xy}({\bm k})$ 
and ${\cal F}^{{\rm II}(-)}_{xy}({\bm k})$ 
are the field strengths of the
Berry curvatures $\bm{\mathcal A}^{{{\rm I}(-)}}({\bm k})$ 
and $\bm{\mathcal A}^{{{\rm II}(-)}}({\bm k})$, namely, 
\begin{eqnarray}
&&\bm{\mathcal A}^{{{\rm I}(-)}}({\bm k})=i\sum_{E_n<E_{\rm F}}
\langle u_n^{\rm I}({\bm k})|\partial_{\bm k} u_n^{\rm I}({\bm k})\rangle, 
\nonumber\\
&&\bm{\mathcal A}^{{{\rm II}(-)}}({\bm k})=i\sum_{E_n<E_{\rm F}}
\langle u_n^{\rm II}({\bm k})|\partial_{\bm k} u_n^{\rm II}({\bm k})
\rangle. 
\end{eqnarray}
These Chern numbers share the same property as the spin Chern numbers.
They take integer numbers, and $Ch_{\rm I}+Ch_{\rm II}=0$ due to time-reversal
symmetry. 
However, $Ch_{\rm I}$ and $Ch_{\rm II}$ themselves are not really well-defined,
since there is the following ambiguity: In Kramers
pairs, the superscripts ${\rm I}$ and ${\rm II}$ do not have any physical
meaning, and hence they can be exchanged; this process changes the sign
of $Ch_{\rm I}$ ($Ch_{\rm II}$) as $Ch_{\rm I}\rightarrow Ch_{\rm
II}=-Ch_{\rm I}$ ($Ch_{\rm II}\rightarrow Ch_{\rm I}=-Ch_{\rm II}$), 
and thus the Chern numbers $Ch_{\rm I}$ and $Ch_{\rm II}$ cannot be unique. 
%
Nevertheless, this difficulty can be resolved by considering the {\it parities} of
the Chern numbers, 
$(-1)^{Ch_{\rm I}}$ and $(-1)^{Ch_{\rm II}}$.
Because of the constraint $Ch_{\rm I}=-Ch_{\rm II}$, these parities are
always the same, making them robust against the exchange of the superscripts.
As a result, we have a well-defined ${\bm Z}_2$ index 
\begin{eqnarray}
(-1)^{\nu_{\rm 2d}}\equiv (-1)^{Ch_{\rm I}}=(-1)^{Ch_{\rm II}}. 
\label{eq2:2dZ2}
\end{eqnarray}
For a spin-preserving time-reversal-invariant system, 
the ${\bm Z}_2$ index coincides with the parity of the spin Chern
number. 
An insulator with $(-1)^{\nu_{\rm 2d}}=-1$ is topologically distinct from
an ordinary insulator and such a system is called a quantum spin Hall insulator. 

\begin{figure}[tb]
\centering
\includegraphics[width=0.8\columnwidth]{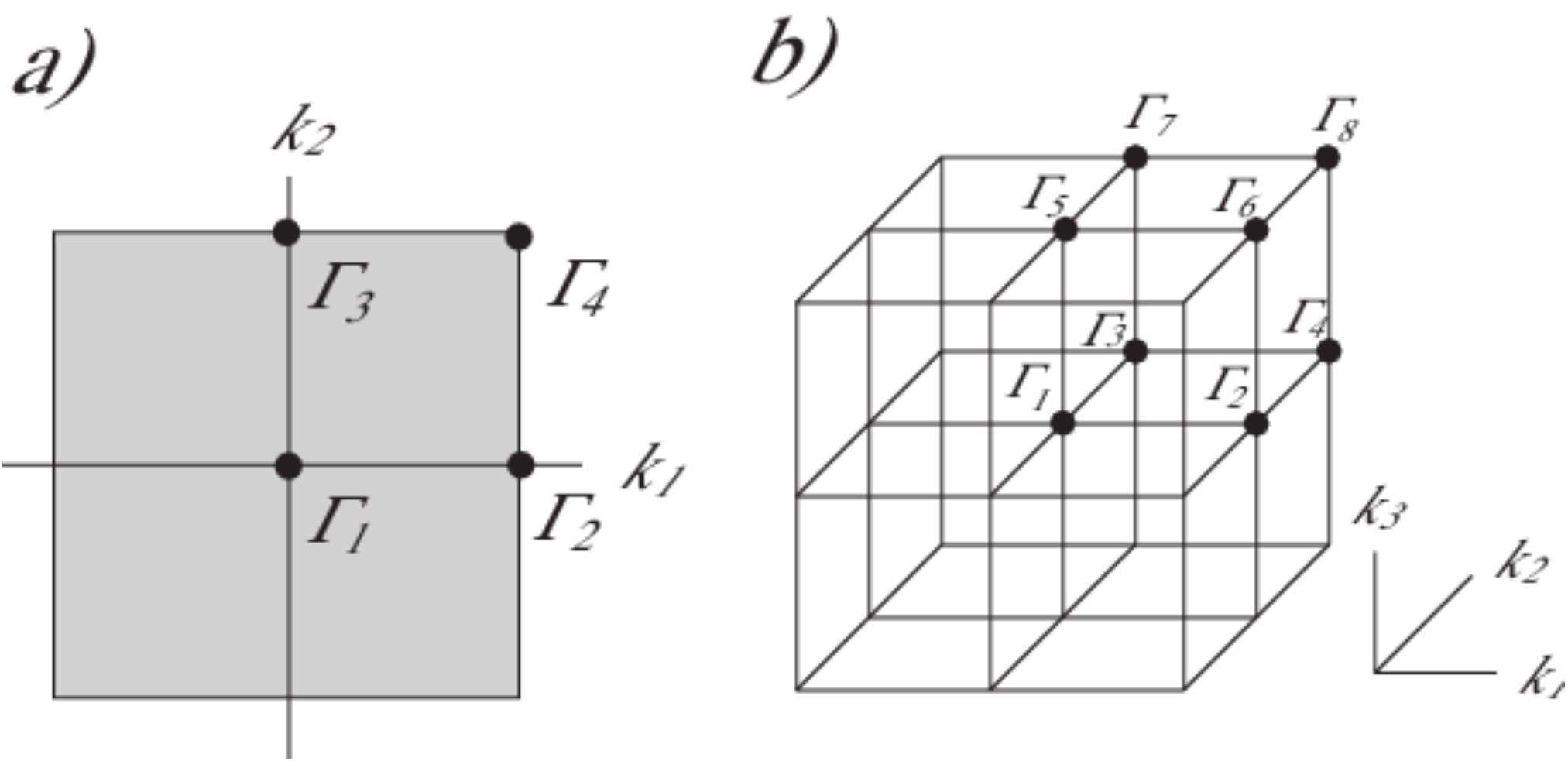}
\caption
{Time-reversal-invariant momenta $\Gamma_i$ $(i=1,2,\dots)$ in the
Brillouin zone. (a) 2D case and (b) 3D case. 
} 
\label{fig:bz}
\end{figure}

Using the time-reversal invariance, one can also introduce ${\bm
Z}_2$ indices in three dimensions \cite{Moore-Balents}. 
We consider the time-reversal-invariant
momenta $\Gamma_i$ in the 3D Brillouin zone, which satisfy
$-\Gamma_i=\Gamma_i+{\bm G}$ for a reciprocal lattice vector ${\bm G}$.
There are eight such $\Gamma_i$'s in 3D, which can be indexed
by using three integers $n_a$ ($a$ =1, 2, 3) taking the value of 0 or 1: 
\begin{eqnarray}
\Gamma_{i=(n_1,n_2,n_3)}=\frac{1}{2}
\left(n_1 {\bm b}_1+n_2{\bm b}_2+n_3{\bm b}_3\right), 
\end{eqnarray}
where ${\bm b}_i$ are the primitive reciprocal lattice vectors
(see Fig. \ref{fig:bz}).
One can choose four time-reversal-invariant momenta by fixing one of 
the three $n_a$'s. For instance, by fixing
$n_1$ as $n_1=0$, one obtains four time-reversal-invariant momenta 
$\Gamma_{(0,n_2,n_3)}$ with $n_2=0,1$ and $n_3=0,1$.
These four time-reversal-invariant momenta can be used for specifying a
distinct time-reversal-invariant plane:
A time-reversal-invariant plane $\Sigma^a_{n_a}$ is defined as a plane in
the 3D Brillouin zone which includes those four time-reversal-invariant
momenta obtained by fixing $n_a$. 
One can easily see that there are six distinct time-reversal-invariant planes in the
3D Brillouin zone.

For each time-reversal-invariant plane $\Sigma^a_{n_a}$, one can introduce
the ${\bm Z}_2$ index $(-1)^{Ch_{\rm I}(\Sigma_{n_a}^a)}$.
A total of six ${\bm Z}_2$ indices can be introduced in this manner, 
but there exist the following constraints \cite{Moore-Balents}: 
\begin{eqnarray}
(-1)^{Ch_{\rm I}(\Sigma_{1}^1)-Ch_{\rm I}(\Sigma_{0}^1)} 
&=&(-1)^{Ch_{\rm I}(\Sigma_{1}^2)-Ch_{\rm I}(\Sigma_{0}^2)} 
\nonumber\\
&=&(-1)^{Ch_{\rm I}(\Sigma_{1}^3)-Ch_{\rm I}(\Sigma_{0}^3)}.
\end{eqnarray}
Hence, only the following four ${\bm Z}_2$ indices, 
\begin{eqnarray}
&&(-1)^{\nu_{{\rm 2d},a}}\equiv
(-1)^{Ch_{\rm I}(\Sigma_1^a)}, \quad (a=1,2,3)
\nonumber\\
&&(-1)^{\nu_{\rm 3d}}\equiv
(-1)^{Ch_{\rm I}(\Sigma_{1}^1)-Ch_{\rm I}(\Sigma_{0}^1)}, 
\label{eq2:3dZ2}
\end{eqnarray}
are independent \cite{Moore-Balents}.
The first three ${\bm Z}_2$ indices can be nontrivial (i.e. $-1$)
even when the system is quasi-2D.
Indeed, $(-1)^{\nu_{{\rm 2d},a}}$ becomes $-1$ even for a
$k_a$-independent Hamiltonian if $(-1)^{Ch_{\rm I}(\Sigma_0^a)}=-1$.
On the other hand, the last ${\bm Z}_2$ index $(-1)^{\nu_{\rm 3d}}$ is
intrinsic to 3D, and hence this is called 3D ${\bm Z}_2$ index. 
An insulator having the nontrivial 3D ${\bm Z}_2$ index 
is called a {\it strong topological insulator}.

In the case of centrosymmetric band insulators, 
there is a useful formula for evaluating the ${\bm Z}_2$ indices \cite{Fu-Kane2007}.
The inversion operator $P$ has the property $P^2=1$ and inversion symmetry leads to 
\begin{eqnarray}
PH({\bm k})P^{-1}=H(-{\bm k}).
\end{eqnarray}
Hence, an energy eigenstate of $H({\bm k})$ at ${\bm k}=\Gamma_i$
is simultaneously an eigenstate of $P$.
Because $[{\cal T}, P]=0$, a Kramers pair of the energy eigenstates has a common
eigenvalue $\xi_i=\pm 1$ of $P$, namely,
\begin{eqnarray}
&&P |u_n^{\rm I}(\Gamma_i)\rangle=\xi_i |u_n^{\rm I}(\Gamma_i)\rangle, 
\nonumber\\
&&P |u_n^{\rm II}(\Gamma_i)\rangle=\xi_i |u_n^{\rm II}(\Gamma_i)\rangle.
\end{eqnarray}
By using $\xi_i$, the ${\bm Z}_2$ index in 2D is evaluated as 
\begin{eqnarray}
(-1)^{\nu_{\rm 2d}}=\prod_{i}\xi_i, 
\end{eqnarray} 
where the product is taken for all inequivalent $\Gamma_i$'s in 2D. 
In the case of 3D, we have
\begin{eqnarray}
(-1)^{\nu_{{\rm 2d},a}}=\prod_{n_a=1;n_{b\neq a}}\xi_{i=(n_1n_2n_3)} 
\end{eqnarray} 
and 
\begin{eqnarray}
(-1)^{\nu_{\rm 3d}}=\prod_i\xi_i \,, 
\label{eq2:3dparity_formula}
\end{eqnarray}
where the product in the latter equation is taken for all eight $\Gamma_i$'s in the 3D
Brillouin zone.

\subsection{Topological materials}

Here, we briefly summarize the recently discovered topological materials,
i.e. topological insulators, Weyl semimetals, and Dirac
semimetals. As we shall discuss later, upon carrier doping these materials may show 
topological superconductivity without the necessity of strong electron correlations.

\subsubsection{Topological insulator}
\label{sec2:TI}

The spin-orbit coupling is
essentially important in topological insulators.
Distinct from ordinary insulators, topological insulators have gapless
surface Dirac fermions with a spin-momentum-locked energy dispersion
\cite{Hasan10, Qi11, AndoReview}.
Such a nontrivial electronic structure stems from a strong spin-orbit coupling.
For instance, consider a prototypical topological insulator, Bi$_2$Se$_3$.
The bulk band near the Fermi energy consists of $p_z$ orbitals of two
inequivalent Se atoms in the unit cell, and the Hamiltonian is 
written as \cite{Zhang2009}
\begin{eqnarray}
H_{\rm TI}({\bm k})=(m_0-m_1{\bm k}^2)\sigma_x+v_zk_z\sigma_y
+v\sigma_z(k_x s_y-k_y s_x).
\nonumber\\ 
\label{eq2:HamiltonianTI}
\end{eqnarray} 
Here, $\sigma_i$ denotes the Pauli matrix in the orbital space where
the eigenstates of $\sigma_z$ (i.e. $\sigma_z=\pm 1$) correspond to 
two $p_z$ orbitals, and $s_i$ is the Pauli matrix in the spin space.
Reflecting the crystal structure of Bi$_2$Se$_3$, $H_{\rm TI}({\bm k})$
is invariant under the inversion $P$, which exchanges the $p_z$ orbitals:
\begin{eqnarray}
PH_{\rm TI}(-{\bm k})P^{-1}=H_{\rm TI}({\bm k}), \quad P=\sigma_x. 
\label{eq2:inversion_BiSe}
\end{eqnarray} 
As shown in the previous subsection, inversion symmetry leads to a simple 
criterion for topological insulators \cite{Fu-Kane2007}.
For the continuum Hamiltonian Eq. (\ref{eq2:HamiltonianTI}), there are
only two time-reversal-invariant momenta at ${\bm k}={\bm 0}$ and ${\bm k}=\infty$, 
so the summation in Eq. (\ref{eq2:3dparity_formula}) is taken only for
these two points.
At these time-reversal-invariant momenta, the Hamiltonian is simplified to
\begin{eqnarray}
H_{\rm TI}({\bm k})=
\left\{
\begin{array}{cl}
m_0 P, &\mbox{for ${\bm k}={\bm 0}$}\\
-m_1 P, &\mbox{for ${\bm k}=\infty$}
\end{array}
\right.,
\end{eqnarray}
so that the parity $\xi$ of the occupied states at these momenta are given by
\begin{eqnarray}
\xi= 
\left\{
\begin{array}{cl}
-{\rm sgn}(m_0), &\mbox{for ${\bm k}={\bm 0}$}\\
{\rm sgn}(m_1), &\mbox{for ${\bm k}=\infty$}
\end{array}
\right..
\end{eqnarray}
Therefore, the 3D ${\bm Z}_2$ index of a topological insulator is evaluated as
\begin{eqnarray}
(-1)^{\nu_{\rm 3d}}=-{\rm sgn}\left(m_0m_1\right). 
\end{eqnarray}
From this formula, one can see that when $m_0$ and $m_1$ have the same sign, 
the system is a topological insulator.

\begin{figure}[tb]
\centering
\includegraphics[width=0.6\columnwidth]{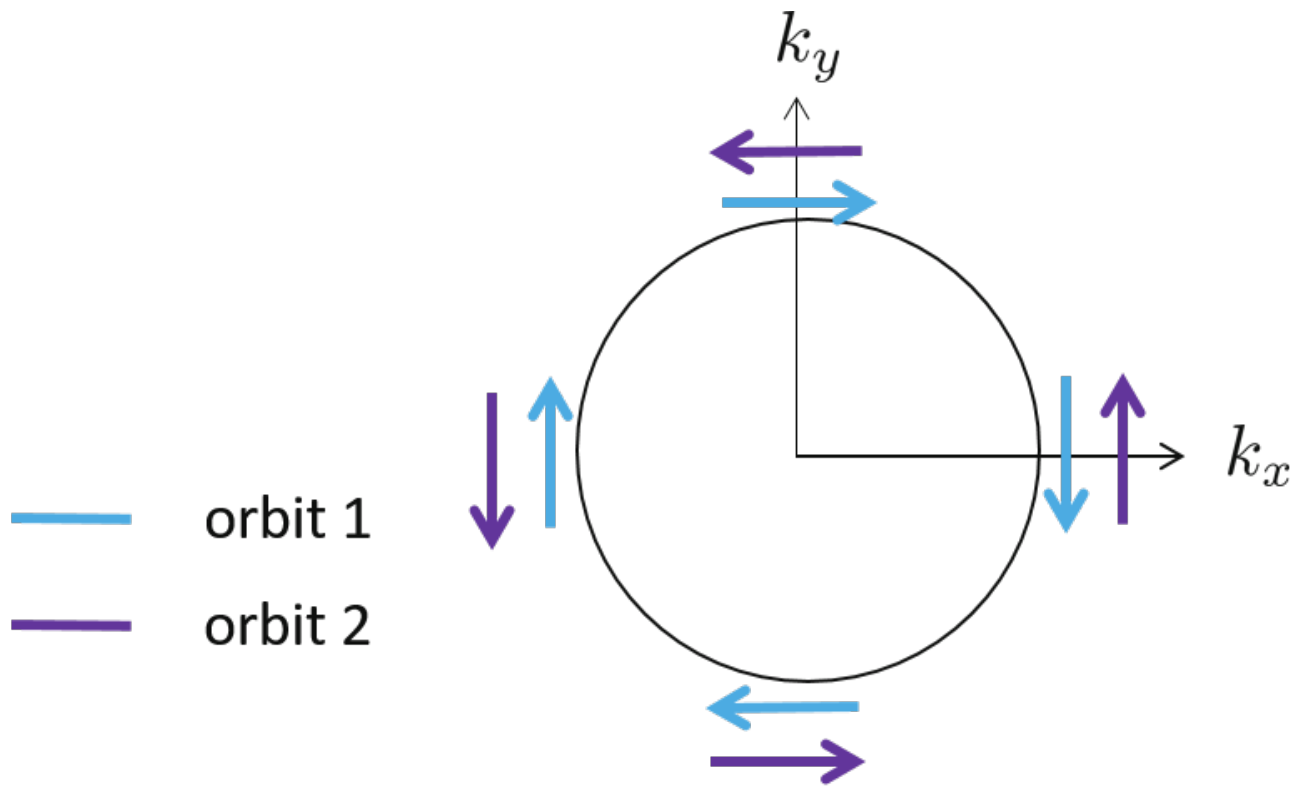}
\caption
{Spin texture on the bulk Fermi surface of a topological insulator due to the
term $v\sigma_z(k_xs_y-k_ys_x)$ in $H_{\rm TI}({\bm k})$. 
} 
\label{fig:SCTI}
\end{figure}

Indeed, one can confirm the existence of surface Dirac fermions in the
following way: 
Consider a surface at $z=0$ and assume that the topological insulator extends to
the positive $z$-direction. The wavefunction at the surface is given by 
\begin{eqnarray}
\psi_{k_x, k_y}(z)=\left(e^{-\kappa_-z}-e^{-\kappa_+z}\right)
\left(
\begin{array}{c}
0 \\
1
\end{array}
\right)_{\sigma}
\otimes u_s(k_x, k_y), 
\nonumber\\
\end{eqnarray} 
with 
\begin{eqnarray}
\kappa_{\pm}=-\frac{v_z}{2m_1}\pm\sqrt{\left(\frac{v_z}{2m_1}\right)^2+k_x^2+k_y^2-\frac{m_0}{m_1}}. 
\end{eqnarray} 
By substituting $\psi_{k_x, k_y}(z)$ into the Schr\"{o}dinger equation 
\begin{eqnarray}
H_{\rm TI}(k_x, k_y, -i\partial_z)\psi_{k_x, k_y}(z)=E\psi_{k_x, k_y}(z) ,
\end{eqnarray}
one can obtain the Dirac equation 
\begin{eqnarray}
-v(k_x s_y-k_y s_x)u_s(k_x, k_y)=Eu_s(k_x, k_y). 
\end{eqnarray}
Here it should be noted that the spin-momentum locking in the surface
states originates from the spin-orbit-coupling term
$v\sigma_z(k_xs_y-k_ys_x)$ in $H_{\rm TI}({\bm k})$ (see Fig. \ref{fig:SCTI}).

\subsubsection{Weyl semimetal}

Weyl semimetals are 3D materials that support bulk
gapless excitations described by the $2\times 2$ Weyl Hamiltonian
\begin{eqnarray}
H(\bm k)=v({\bm k}-{\bm k}_0)\cdot{\bm s}, 
\end{eqnarray}
where ${\bm s}$ is the Pauli matrices of (pseudo) spin
\cite{Nielsen-Ninomiya1983, Murakami2007, Wan2011, Burkov-Balents2011}.
The spin-momentum-locked nature of the Hamiltonian implies a strong 
spin-orbit coupling in the system.

A band structure described by the Weyl Hamiltonian is obtained as a
result of band crossing between a pair of spin-non-degenerate bands. 
Hence, Weyl semimetals can be realized only when time-reversal or
inversion symmetry is broken. Otherwise, the bands are spin-degenerate at
each momentum due to the Kramers theorem, and an isolated Weyl node cannot
be obtained.
Weyl semimetals have the following characteristics:
\begin{enumerate}
\item Each Weyl point has a non-zero Chern number
calculated on a sphere surrounding that point \cite{Nielsen-Ninomiya1983}. 
\item The summation of the Chern numbers for all the Weyl points vanishes 
(Nielsen-Ninomiya theorem \cite{Nielsen-Ninomiya1981, Nielsen-Ninomiya1981b}).
\item They host gapless metallic surface states forming a Fermi
arc in the surface Brillouin zone \cite{Wan2011}.
\end{enumerate} 
The surface Fermi arc in a Weyl semimetal is terminated at the points in the surface Brillouin zone onto which a pair of Weyl points are projected.
This is because the surface Fermi arc originates from the Weyl points; namely,
the non-zero Chern numbers associated with the Weyl points give rise to the gapless 
surface states which connect the two singularities in the momentum space.

\subsubsection{Dirac semimetal}
\label{sec2:Dirac}

Finally, we consider Dirac semimetals, which are 3D semimetals or metals that support 
spin-degenerate gapless bulk nodes \cite{Young2012, Wang2012, Wang2013, Yang2014, Novak2015}. 
Like Weyl semimetals, a gapless node in a Dirac semimetal is
realized as a crossing point of two bands, but the difference from Weyl
semimetals lies in the fact that Dirac semimetals have both time-reversal and inversion
symmetries.
In such a case, the Kramers theorem dictates that each energy band has two-fold
spin degeneracy, and thus the band crossing point is four-fold degenerate. 
Therefore, a gapless node in a Dirac semimetal is described by a
$4\times 4$ matrix Hamiltonian like in the case of a Dirac fermion.

As an example, let us consider Cd$_3$As$_2$
\cite{Wang2013, Neupane2014,Borisenko2014,Liu2014,Yi2014}. In this Dirac semimetal, 
the Dirac points for gapless excitations appear at two points in the momentum space, $k_z=\pm k_0$ on the
$k_z$ axis. 
Near these points, the Hamiltonian to describe the gapless dispersion is given by 
\begin{eqnarray}
H_{\pm}({\bm k})=v_x(\pm k_z-k_0)\sigma_zs_0+v(k_x\sigma_x s_z-k_y\sigma_ys_0) ,
\nonumber\\
\label{eq6:Dirac}
\end{eqnarray} 
where $\sigma_{\mu}$ and $s_{\mu}$ are the Pauli matrices 
in the orbital and spin spaces, respectively.
The time-reversal operator ${\cal T}$ and the inversion operator $P$ are written
as ${\cal T}=i\sigma_0s_yK$ ($\sigma_0$ is the $2 \times 2$ identity matrix) and $P=\sigma_zs_0$, and the Hamiltonian transforms as
\begin{eqnarray}
{\cal T}H_{\pm}(-{\bm k}){\cal T}^{-1}=H_{\mp}({\bm k}),
\quad
PH_{\pm}(-{\bm k})P^{-1}=H_{\mp}({\bm k}).
\nonumber\\
\end{eqnarray}
Hence, the system as a whole is invariant under ${\cal T}$ and $P$.
If we introduce the Gamma matrices $\gamma_{i=x,y,z}$ as
\begin{eqnarray}
\gamma_x=\sigma_x s_z, \quad \gamma_y=-\sigma_ys_0, \quad
\gamma_z=-\sigma_z s_0, 
\end{eqnarray}
it follows that the relation $\{\gamma_i, \gamma_j\}=2\delta_{i,j}$ holds and
the Hamiltonian $H_{\pm}({\bm k})$ can be rewritten in the form of the
Dirac Hamiltonian,
\begin{eqnarray}
H_{\pm}({\bm k})=vk_x\gamma_x+vk_y\gamma_y+v_z(\pm k_z-k_0)\gamma_z. 
\end{eqnarray} 

For Dirac semimetals, neither time-reversal symmetry nor inversion
symmetry ensures the stability of gapless nodes. 
For instance, one can add 
the term $\Delta H({\bm k})=mk_z\sigma_x s_x$,
which preserves both time-reversal and inversion symmetries, to the above $H_{\pm}({\bm k})$. 
The resulting energy dispersion is
\begin{eqnarray}
E_{\pm}({\bm k})=\pm \sqrt{v_z^2(k_z\mp k_0)^2+m^2k_z^2+v^2(k_x^2+k_y^2)}, 
\nonumber\\
\end{eqnarray}
which is gapped. Therefore, there must be an additional symmetry to prohibit such a
gap-opening term.
In general, energy bands of electrons in a solid are formed by electron
orbitals of the constituent atoms.
If one chooses a symmetry of the crystal and consider a momentum
subspace which is invariant under that symmetry, each band has a
quantum number associated with the symmetry of the electron orbital.
When two bands having different quantum numbers intersect, they belong to different 
subspaces and hence no mixing is allowed, protecting the crossing point to remain gapless.
In the case of Cd$_3$As$_2$, four-fold rotation ($C_4$) symmetry around the
$z$-axis forbids the gap-opening mass term \cite{Yang2014}. To see this, we note that 
the $C_4$ operation transforms $H_{\pm}({\bm k})$ as
\begin{eqnarray}
C_4 H_{\pm}(k_y, -k_x, k_z)C_4^{-1}=H_{\pm}({\bm k}). 
\end{eqnarray}
In the $C_4$-invariant momentum subspace at $k_x=k_y=0$, $C_4$ and
$H_{\pm}(0,0,k_z)$ commute, and hence the energy band has an
eigenvalue of $C_4$ as a good quantum number.
In Cd$_3$As$_2$, the $\sigma_z=1$ band and the $\sigma_z=-1$ band have
different eigenvalues of $C_4$. Indeed, on the $k_z$-axis, the
Hamiltonian Eq. (\ref{eq6:Dirac}) becomes 
\begin{eqnarray}
H_{\pm}({\bm k})=v_z(\pm k_z-k_0)\sigma_z s_0,
\end{eqnarray}
which includes no off-diagonal band-mixing term.
Moreover, the gap-opening term $\Delta H({\bm k})=mk_z \sigma_x s_0$ 
mentioned above is not allowed by the $C_4$ symmetry.

\section{Basics of superconductivity}

\subsection{Bogoliubov quasiparticles and particle-hole symmetry}
\label{sec:3_1}

We first consider a single-band description of a superconductor.
With an appropriate pairing interaction in the momentum space, a general
single-band effective Hamiltonian is given by 
\begin{eqnarray}
&&{\cal H}=
\sum_{{\bm k}, s_1, s_2}
\varepsilon_{s_1 s_2}({\bm k})c^{\dagger}_{{\bm k}s_1}c_{{\bm k}s_2} 
\nonumber\\ 
&&\hspace{3ex}
+\frac{1}{2}\sum_{{\bm k},{\bm k}',s_1,s_2,s_3,s_4}V_{s_1s_2s_3s_4}({\bm
k}, {\bm k}')c^{\dagger}_{-{\bm k}s_1}
c^{\dagger}_{{\bm k}s_2}
c_{{\bm k}'s_3}c_{-{\bm k}'s_4},
\nonumber\\
\label{eq3:Hamiltonian}
\end{eqnarray}
where $c_{{\bm k}s}$ ($c^{\dagger}_{{\bm k}s}$) is the annihilation
(creation) operator of the electron with
momentum ${\bm k}$ and spin $s$, and $\varepsilon_{s_1 s_2}({\bm k})$ is
the band Hamiltonian to give the momentum- and spin-dependent 
band energy measured from the chemical potential; this can be viewed as
a $2 \times 2$ matrix operator in the spin basis. Its spin dependence comes
from the spin-orbit coupling.
The anticommutation relation between $c_{{\bm k}s}$ and
$c^{\dagger}_{{\bm k}s}$ implies the following relation for the 
pairing interaction $V_{s_1 s_2 s_3 s_4}({\bm k}, {\bm k}')$:
\begin{eqnarray}
V_{s_1 s_2 s_3 s_4}({\bm k}, {\bm k}')&=& 
-V_{s_2 s_1 s_3 s_4}(-{\bm k}, {\bm k}')
\nonumber\\
&=& 
-V_{s_1 s_2 s_4 s_3}({\bm k}, -{\bm k}')
\nonumber\\
&=& 
V_{s_4 s_3 s_2 s_1}({\bm k}', {\bm k}). 
\end{eqnarray}
In a superconducting state, Cooper pairs form so that the two-body operator
$c_{{\bm k}s}c_{-{\bm k}s'}$ has a nonzero expectation value $\langle
c_{{\bm k}s}c_{-{\bm k}s'}\rangle$. 
By defining a {\it pair potential} 
\begin{eqnarray}
\Delta_{ss'}({\bm k})=-\sum_{{\bm k}',s_3,s_4}V_{s'ss_3 s_4}({\bm k},
{\bm k}')\langle c_{{\bm k}'s_3}c_{-{\bm k}'s_4}\rangle,
\end{eqnarray}
one can decouple the quadratic term and simplify the Hamiltonian within the mean-field approximation to
\begin{eqnarray}
&&{\cal H}=
\sum_{{\bm k}, s_1, s_2}
\varepsilon_{s_1 s_2}({\bm k})c^{\dagger}_{{\bm k}s_1}c_{{\bm k}s_2} 
\nonumber\\
&&\hspace{3ex}
+\frac{1}{2}\sum_{{\bm k},s_1,s_2}
\left[
\Delta_{s_1 s_2}({\bm k})
c^{\dagger}_{{\bm k}s_1}c^{\dagger}_{-{\bm k}s_2}
+\mbox{h.c.}
\right].
\end{eqnarray}
It is convenient to rewrite this into a $4\times 4$ matrix form 
\begin{eqnarray}
{\cal H}=\frac{1}{2}\sum_{{\bm k} s_1s_2}
\left(c^{\dagger}_{{\bm k}s_1}, c_{-{\bm k}s_1}\right)
{\cal H}_{4\times 4}({\bm k})
\left(
\begin{array}{c}
c_{{\bm k} s_2} \\
c^{\dagger}_{-{\bm k} s_2}
\end{array}
\right)
\label{eq3:Hamiltonian2}
\end{eqnarray}
with
\begin{eqnarray}
{\cal H}_{4\times 4}({\bm k})=
\left(
\begin{array}{cc}
\varepsilon_{s_1 s_2}({\bm k}) & \Delta_{s_1 s_2}({\bm k})\\
\Delta_{s_1 s_2}^{\dagger}({\bm k}) & -\varepsilon^t_{s_1 s_2}(-{\bm k})
\end{array}
\right),
\label{eq3:Hamiltonian_matrix}
\end{eqnarray}
where the spin indices make $\varepsilon_{s_1 s_2}({\bm k})$ {\it etc.} 
to be $2\times 2$ matrices. Also note that $\left(c^{\dagger}_{{\bm k}s_1}, c_{-{\bm k}s_1}\right)$ 
is actually a four-component vector $\left(c^{\dagger}_{{\bm k}\uparrow}, c^{\dagger}_{{\bm k}\downarrow}, 
c_{-{\bm k}\uparrow}, c_{-{\bm k}\downarrow}\right)$.
In the above, we have neglected constant terms since they merely shift
the ground state energy.

The matrix form of the Hamiltonian Eq. (\ref{eq3:Hamiltonian_matrix}) has a
particular symmetry called {\it particle-hole symmetry}. To see this, we define an
anti-unitary operator 
\begin{eqnarray}
{\cal C}=\left(
\begin{array}{cc}
0 & {\bm 1}_{2\times 2}\\
{\bm 1}_{2\times 2} & 0
\end{array}
\right)K, 
\end{eqnarray}
with the complex conjugation operator $K$ and 
the $2\times 2$ unit matrix ${\bm 1}_{2\times 2}$.
This operator exchanges the particle sector with the hole sector, and 
it transforms ${\cal H}_{4\times 4}({\bm k})$ as
\begin{eqnarray}
{\cal C}{\cal H}_{4\times 4}({\bm k}){\cal C}^{-1}
=-{\cal H}_{4\times 4}(-{\bm k}),
\label{eq3:particle_hole}
\end{eqnarray}
which is the mathematical expression of the particle-hole symmetry.
Here, one should note the relation 
\begin{eqnarray}
\Delta_{s_1 s_2}({\bm k})=-\Delta_{s_2 s_1}(-{\bm k}) 
\end{eqnarray}
coming from the Fermi statistics of $c_{{\bm k},s}$.
The particle-hole symmetry is associated with the redundancy in the
Hamiltonian. That is, the components of the vector $(c^{\dagger}_{{\bm k}s}, c_{-{\bm k}s})$ 
in Eq. (\ref{eq3:Hamiltonian2}) are related by complex conjugate, and hence they 
are not independent; accordingly, the matrix
components of ${\cal H}_{4\times 4}({\bm k})$ are not independent of each other, and 
the particle-hole symmetry expresses this mutual dependence of the components. 

The effective Hamiltonian Eq. (\ref{eq3:Hamiltonian2}) is diagonalized
by the eigenvectors $(u_s({\bm k}), v^*_s(-{\bm k}))^t$ of the eigen equation
\begin{eqnarray}
{\cal H}_{4\times 4}({\bm k})
\left(
\begin{array}{c}
u_s({\bm k}) \\
v_s^*(-{\bm k})
\end{array}
\right)
=E({\bm k})
\left(
\begin{array}{c}
u_s({\bm k}) \\
v_s^*(-{\bm k})
\end{array}
\right).
\label{eq3:eigen}
\end{eqnarray}
The particle-hole symmetry imposes a relation between the solutions of 
Eq. (\ref{eq3:eigen}); namely, using Eq. (\ref{eq3:particle_hole}), one can rewrite
Eq. (\ref{eq3:eigen}) into
\begin{eqnarray}
{\cal H}_{4\times 4}({\bm k}){\cal C}
\left(
\begin{array}{c}
u_s(-{\bm k}) \\
v_s^*({\bm k})
\end{array}
\right)
=-E(-{\bm k})
{\cal C}
\left(
\begin{array}{c}
u_s(-{\bm k}) \\
v_s^*({\bm k})
\end{array}
\right),
\end{eqnarray} 
which implies that the eigenvalues $E({\bm k})$ and $-E(-{\bm k})$ come
in pairs.
One can thus write the set of four eigenvalues of the $4\times 4$ matrix ${\cal H}_{4\times 4}({\bm k})$ as
$(E_1({\bm k}), E_2({\bm k}), -E_1(-{\bm k}), -E_2(-{\bm
k}))$ with $E_i({\bm k})\ge 0$.
Expressing the eigenvector for $E_i({\bm k})$ $(i=1,2)$ as $(u_s^{(i)}({\bm
k}), v_s^{(i)*}(-{\bm k}))$, we can diagonalize ${\cal H}_{4\times 4}({\bm k})$ as
\begin{eqnarray}
&&U^{\dagger}({\bm k})
{\cal H}_{4\times 4}({\bm k})
U({\bm k})
\nonumber\\
&&=\left(
\begin{array}{cccc}
E_1({\bm k}) & & & \\
& E_2({\bm k}) & &\\
& & -E_1(-{\bm k})& \\
& & & -E_2(-{\bm k})
\end{array}
\right), 
\nonumber\\
\label{eq3:diagonal}
\end{eqnarray} 
where the unitary matrix $U({\bm k})$ is given by
\begin{eqnarray}
U({\bm k})=\left(
\begin{array}{cc}
u_s^{(i)}({\bm k}) & v_s^{(i)}({\bm k})\\
v_s^{(i)*}(-{\bm k}) & u_s^{(i)*}(-{\bm k})\\
\end{array}
\right). 
\end{eqnarray}
Here we used the relation ${\cal C}(u_s^{(i)}(-{\bm k}), v_s^{(i)*}({\bm
k}))^t=(v_s^{(i)}({\bm k}), u_s^{(i)*}(-{\bm k}))$. 
Without loss of generality, one can assume $E_i({\bm k})\ge 0$ for any
weak-pairing superconducting states. A ``weak Cooper pair'' means that the energy
scale of the pair potential $\Delta({\bm k})$ is much smaller than that of
$\varepsilon_{s_1 s_2}({\bm k})$ in Eq. (\ref{eq3:Hamiltonian}). Thus, the pair potential
can be neglected except near the Fermi surface where $\varepsilon_{s_1 s_2}({\bm k})$ vanishes. 
At the Fermi energy, the pair potential opens a gap and separates the positive energy
branch from the negative one.
Reflecting the spin degrees of freedom, there are two positive
energy branches $E_1({\bm k})$ and $E_2({\bm k})$.
Substituting the expression Eq. (\ref{eq3:diagonal}) for ${\cal H}_{4\times 4}({\bm k})$ into
Eq. (\ref{eq3:Hamiltonian2}), one obtains
\begin{eqnarray}
{\cal H}=\sum_{{\bm k} i}E_i({\bm k})\alpha_{{\bm k}i}^{\dagger}
\alpha_{{\bm k}i},
\label{eq3:Hamiltonian3}
\end{eqnarray}
with
\begin{eqnarray}
\alpha_{{\bm k}i}=\sum_s \left(
u_s^{(i)*}({\bm k})c_{{\bm k}s}+v_s^{(i)}(-{\bm k})c^{\dagger}_{-{\bm k}s} 
\right).
\end{eqnarray}
The operator $\alpha_{{\bm k}i}$ satisfies the anti-commutation relation
\begin{eqnarray}
&&\{\alpha^{\dagger}_{{\bm k}i}, 
\alpha_{{\bm k}'j}\}=\delta_{ij}\delta_{{\bm k}, {\bm k}'},
\nonumber\\
&&\{\alpha_{{\bm k}i}, \alpha_{{\bm k}' j}\}= 
\{\alpha^{\dagger}_{{\bm k}i}, \alpha^{\dagger}_{{\bm k}' j}\}=0, 
\end{eqnarray} 
and it describes thermal excitations of quasiparticles called {\it Bogoliubov quasiparticles}
with energy $E_i({\bm k})$. 
The ground state of a superconductor is annihilated by
$\alpha_{{\bm k}i}$, 
\begin{eqnarray}
\alpha_{{\bm k}i}|0\rangle =0,
\end{eqnarray}
which implies that the negative energy states are fully occupied as in an
insulating state.

The above properties also hold for multi-orbital (i.e. multi-band) superconductors.
In a general multi-orbital system, electrons have the orbital index
$\sigma=1,\dots,N$ as
well the spin index $s=\uparrow, \downarrow$. Expressing these internal
indices with $\alpha \equiv (s, \sigma)$,
the Hamiltonian is written as
\begin{eqnarray}
{\cal H}=\frac{1}{2}\sum_{{\bm k},\alpha,\beta} 
\left(c^{\dagger}_{{\bm k}\alpha}, c_{-{\bm k}\alpha}\right)
{\cal H}({\bm k})
\left(
\begin{array}{c}
c_{{\bm k}\beta}\\
c^{\dagger}_{-{\bm k}\beta}
\end{array}
\right)
\end{eqnarray}
with the following $4N\times 4N$ Hamiltonian
\begin{eqnarray}
{\cal H}({\bm k})=
\left(
\begin{array}{cc}
{\cal E}_{\alpha\beta}({\bm k}) & \Delta_{\alpha\beta}({\bm k}) \\
\Delta^{\dagger}_{\alpha\beta}({\bm k}) & -{\cal E}^t_{\alpha\beta}(-{\bm k}).
\end{array}
\right),
\label{eq3:general_BdG}
\end{eqnarray}
where ${\cal E}({\bm k})=H({\bm k})-\mu$ is the normal-state Bloch Hamiltonian
$H({\bm k})$ measured relative to the chemical potential $\mu$. 
The anti-commutation relation of $c_{{\bm k},\alpha}$ implies 
\begin{eqnarray}
\Delta_{\alpha\beta}({\bm k})=-\Delta_{\beta\alpha}(-{\bm k}), 
\label{eq3:antisymmetric_condition}
\end{eqnarray} 
and the BdG Hamiltonian ${\cal H}({\bm k})$ has the particle-hole
symmetry
\begin{eqnarray}
{\cal C}{\cal H}({\bm k}){\cal C}^{-1}=-{\cal H}(-{\bm k}), 
\end{eqnarray}
with 
\begin{eqnarray}
{\cal C}=\left(
\begin{array}{cc}
0 & {\bm 1}_{2N\times 2N} \\
{\bm 1}_{2N\times 2N} & 0
\end{array}
\right)K, 
\end{eqnarray}
in a similar manner as the single band case.
The spectrum of the Bogoliubov quasiparticles in the superconducting state is determined by
\begin{eqnarray}
{\cal H}({\bm k})|u_n({\bm k})\rangle=E_n({\bm k})|u_n({\bm k})\rangle, 
\label{eq3:BdGmulti}
\end{eqnarray}
with a $4N$-component wavefunction $|u_n({\bm k})\rangle$. 
Because of the particle-hole symmetry, positive- and negative-energy states
come in pairs, and the negative-energy states are fully occupied in
the superconducting ground state. 

\subsection{Pairing symmetry}
\label{sec3:pairing_symmetry}

The pair potentials are often classified based on their spin angular momentum.
In the simple single-band case discussed above, spin is the only internal degrees
of freedom for electrons. 
Since a Cooper pair is formed by pairing two spin-$1/2$ electrons, the
spin angular momentum of a pair potential is either 0 (spin-singlet) or 1 (spin-triplet).
A spin-singlet pair potential is antisymmetric in the spin space, 
which is written as
\begin{eqnarray}
\Delta_{ss'}({\bm k})=i\psi({\bm k})[s_y]_{ss'}.
\label{eq3:spin_singlet}
\end{eqnarray}
On the other hand, a spin-triplet pair potential is symmetric in the spin space, and it 
can be expressed as
\begin{eqnarray}
\Delta_{ss'}({\bm k})=i{\bm d}({\bm k})\cdot[{\bm s}s_y]_{ss'}
\label{eq3:spin_triplet}
\end{eqnarray}
by using a vector ${\bm d}({\bm k})$ called ``${\bm d}$-vector''.
Since $\Delta({\bm k})=-\Delta^t(-{\bm k})$, $\psi({\bm
k})$ (${\bm d}({\bm k})$) is an even (odd) function of ${\bm k}$:
\begin{eqnarray}
\psi({\bm k})=\psi(-{\bm k}), \quad
{\bm d}({\bm k})=-{\bm d}(-{\bm k}).
\label{eq3:parity}
\end{eqnarray}
A spin-singlet pair potential has the total angular momentum $l$ of $0, 2, 4, \dots$, 
while a spin-triplet one has $l=1,3,5,\dots$. In analogy with atomic orbitals, 
Cooper pairs having $l=0,1,2,3$ are called $s$-wave, $p$-wave, $d$-wave, and $f$-wave,
respectively.

Whereas the above classification is convenient and has been widely used
in the literature, it is not well-defined in a strict sense, because
the spin and the angular momentum are not good quantum numbers in the
presence of the spin-orbit coupling and/or the crystal field. 
In particular, in heavy-fermion or topological materials, where
unconventional superconductivity can be expected, these two effects cannot be neglected.
In such cases, the pair potential should instead be classified by using crystal
symmetry \cite{Sigrist-Ueda1991}.

Among the various crystal symmetries, special attention is often paid to inversion symmetry.
Any pair potential in a centrosymmetric material has a definite parity under inversion. 
When a pair potential is even (odd) under inversion, i.e.
\begin{eqnarray}
P\Delta({\bm k})P^t=\Delta(-{\bm k}), 
\quad
(P\Delta({\bm k})P^t=-\Delta(-{\bm k})), 
\end{eqnarray}
with the inversion operator $P$, it is called an even-parity
(odd-parity) pair potential. 
Here it should be noted that there is a simple correspondence between the spin
classification and the parity classification in the single-band case. 
When there is only one band, $P$ is just an identity operator (i.e. $P=1$), and hence
Eq. (\ref{eq3:parity}) means that a spin-singlet (spin-triplet) pair potential is even-parity
(odd-parity). 
In multi-orbital cases, however, no such simple correspondence exists.

Inversion symmetry is broken in noncentrosymmetric systems.
In this case, parity mixing occurs in the pair potential \cite{Frigeri2004}. 
In particular, spin-singlet and spin-triplet pair potentials can coexist even 
in a single-band model:
\begin{eqnarray}
\Delta({\bm k})=i\psi({\bm k})s_y+i{\bm d}({\bm k})\cdot{\bm s}s_y. 
\end{eqnarray}
This is the essential characteristic of noncentrosymmetric superconductors \cite{Noncentro}.

\subsection{Bogoliubov-de Gennes equation}

Equations (\ref{eq3:eigen}) and (\ref{eq3:BdGmulti}) determine the
quasiparticle spectrum in the bulk superconducting state. Here we generalize
these equations to systems with boundaries or defects. 
As we will show later, the quasiparticle states localized at 
boundaries or defects play important roles in topological superconductors.
Below, we only discuss the single-band case for simplicity. The
generalization to a multi-orbital case is straightforward. 

First, to consider a system with a boundary, we replace
the momentum ${\bm k}$ in the Hamiltonian Eq. (\ref{eq3:Hamiltonian_matrix})
with its operator form $-i{\bm \partial_{\bm x}}$. 
This changes the eigen equation (\ref{eq3:eigen}) into a
differential equation
\begin{eqnarray}
{\cal H}_{4\times 4}(-i{\bm \partial}_{\bm x})
\left(
\begin{array}{c}
\tilde{u}_s({\bm x})\\
\tilde{v}_s^*({\bm x})
\end{array}
\right)
=E 
\left(
\begin{array}{c}
\tilde{u}_s({\bm x})\\
\tilde{v}_s^*({\bm x})
\end{array}
\right),
\label{eq3:eigen2}
\end{eqnarray}
with $\tilde{u}_s({\bm x})=(1/\sqrt{V})
\sum_{\bm k}e^{i{\bm k}\cdot{\bm x}}u_s({\bm
k})$ and
$\tilde{v}_s({\bm x})=(1/\sqrt{V})
\sum_{\bm k}e^{i{\bm k}\cdot{\bm x}}v_s({\bm k})$.
($V$ is the system volume.)

There are several ways to introduce a boundary.
The simplest way is to impose a boundary condition on
the differential equation (\ref{eq3:eigen2}).
For instance, to consider a system extending to the positive $x$-direction
with the boundary at $x=0$, 
we place the boundary condition 
\begin{eqnarray}
\left(
\begin{array}{c}
\tilde{u}_s({\bm x})\\
\tilde{v}_s^*({\bm x})
\end{array}
\right)=0 
\label{eq3:boundary}
\end{eqnarray}
at $x=0$. At the same time, a physical wavefunction $(\tilde{u}_s({\bm
x}), \tilde{v}_s^*({\bm x}))^t$ should not diverge at anywhere with $x>0$.
The solution of Eq. (\ref{eq3:eigen2}) under these two requirements
determines the quasiparticle spectrum of the system with the boundary at $x=0$. 

We can modify the differential equation (\ref{eq3:eigen2})
so as to mimic a real system more closely.
Instead of imposing the boundary condition (\ref{eq3:boundary}), one
can add to the kinetic energy $\varepsilon_{s_1 s_2}(-i{\bm \partial}_{\bm x})$ 
a confining potential $V({\bm x})$ which blows up at $x=0$. 
In a similar way, one can take into account various boundary effects such as a surface potential
or a modulation of the gap function near the boundary.

Alternatively, one can use a lattice model to consider a system with a
boundary. Discretization of the space ${\bm x}$ and the derivative ${\bm
\partial_{\bm x}}$ in Eq. (\ref{eq3:eigen2}) into sites ${\bm m}$ and ${\bm n}$ on a lattice
leads to a tight-binding equation
\begin{eqnarray}
\sum_{\bm n}{\cal H}_{4\times 4}({\bm m},{\bm n}) 
\left(
\begin{array}{c}
\tilde{u}_s({\bm n})\\
\tilde{v}^*_s({\bm n})
\end{array}
\right)
=E
\left(
\begin{array}{c}
\tilde{u}_s({\bm m})\\
\tilde{v}^*_s({\bm m})
\end{array}
\right).
\end{eqnarray}
By restricting ${\bm m}$ and ${\bm n}$ to a finite region, 
one can obtain the equation with a boundary, which also determines the
quasiparticle spectrum in a finite region.

The influence of a defect can be taken into account in Eq. (\ref{eq3:eigen2}).
When a defect is present, quasiparticles at different positions 
feel different values of the gap function. Thus, $\Delta_{s_1 s_2}(-i{\bm \partial}_{\bm x})$ in
Eq. (\ref{eq3:eigen2}) becomes position dependent and is replaced by $\Delta_{s_1 s_2}(-i{\bm
\partial}_{\bm x},{\bm x})$:
\begin{eqnarray}
&&
\left(
\begin{array}{cc}
\varepsilon_{s_1 s_2}(-i{\bm \partial}_{\bm x}) 
& \Delta_{s_1
s_2}(-i{\bm \partial}_{\bm x}, {\bm x})\\
\Delta^{\dagger}_{s_1
s_2}(-i{\bm \partial}_{\bm x}, {\bm x})
&-\varepsilon^t_{s_1 s_2}(i{\bm \partial}_{\bm x}) 
\end{array}
\right)
\left(
\begin{array}{cc}
\tilde{u}_{s_2}({\bm x}) \\
\tilde{v}^*_{s_2}({\bm x}) 
\end{array}
\right)
\nonumber\\
&&=E
\left(
\begin{array}{cc}
\tilde{u}_{s_2}({\bm x}) \\
\tilde{v}^*_{s_2}({\bm x}) 
\end{array}
\right).
\label{eq3:eigen3}
\end{eqnarray}
For instance, let us consider a vortex located at the origin in a 2D $s$-wave superconductor.
In this case, $\Delta_{s_1 s_2}(-i{\bm \partial}_{\bm x}, {\bm x})$ is given by
\begin{eqnarray}
\Delta_{s_1 s_2}(-i{\bm \partial}_{\bm x}, {\bm x})=i\psi(\rho)e^{i\phi}[s_y]_{s_1 s_2}, 
\end{eqnarray}
with ${\bm x}=(\rho \cos\phi, \rho\sin\phi)$ and the Pauli matrix $s_\mu$
in the spin space.
Here $\psi(\rho)$ is the pairing amplitude which vanishes at the origin ($\psi(0)=0$) 
and approaches a constant value at $\rho=\infty$ ($\psi(\infty)=\psi_0$). 
For other defects, explicit forms of $\Delta_{s_1 s_2}(-i{\bm \partial}_{\bm x}, {\bm
x})$ can be provided in a similar manner, although they can be more complicated. 
Equation (\ref{eq3:eigen3}) determines the quasiparticle spectrum in the
presence of a defect.

Equations (\ref{eq3:eigen}), (\ref{eq3:eigen2}) and (\ref{eq3:eigen3})
are called Bogoliubov-de Gennes (BdG) equation and determine the 
quasiparticle spectrum in a superconductor under
various conditions.

\subsection{Andreev bound states}

Near a boundary or a vortex core, Cooper pairs collapse. 
As a result, spatially-localized in-gap states can be formed in a superconductor.
For instance, a vortex in an $s$-wave superconductor is known to host mini-gap
states bound in the vortex core \cite{Caroli1964}.
Such bound states in a superconductor are generally called {\it Andreev bound states}.
They are very sensitive to the phase of the superconducting
wavefunctions, and thus provide useful information on the pairing
symmetry \cite{Kashiwaya00,Lofwander2001}.

Whereas Andreev bound states are in general mini-gap states, 
for a class of superconductors they can be a gapless mode or, more
drastically, they can be zero-energy (mid-gap) states \cite{Hara1986,Hu1994,Kopnin-Salomaa1991}.
As an example, let us consider a 2D $d_{xy}$-wave
superconductor where $\varepsilon_{s_1s_2}({\bm k})$ and
$\Delta_{s1s_2}({\bm k})$ in Eq. (\ref{eq3:Hamiltonian_matrix}) are given by
\begin{eqnarray}
\varepsilon_{s_1s_2}({\bm k})=\left(\frac{\hbar^2{\bm k}^2}{2m}-\mu\right)
\delta_{s_1s_2},
\nonumber\\
\Delta_{s_1s_2}({\bm k})=\Delta_0 \frac{k_x k_y}{k_{\rm F}^2}i[s_y]_{s_1s_2}. 
\end{eqnarray}
Here $\Delta_0$ is a positive constant, $\mu$ $(>0)$ is the chemical
potential, and $k_{\rm F}=\sqrt{2m\mu}/\hbar$ is the Fermi momentum. 
In this case, the BdG Hamiltonian reduces to
\begin{eqnarray}
{\cal H}=\sum_{\bm k}
(c^{\dagger}_{{\bm k}\uparrow}, c_{-{\bm k}\downarrow})
{\cal H}_{2\times 2}({\bm k}) 
\left(
\begin{array}{c}
c_{{\bm k}\uparrow}\\
c_{-{\bm k}\downarrow}^{\dagger}
\end{array}
\right)
\end{eqnarray}
where the $2\times 2$ matrix Hamiltonian is given by
\begin{eqnarray}
{\cal H}_{2\times 2}({\bm k})=
\left(
\begin{array}{cc}
\epsilon({\bm k}) & \psi({\bm k}) \\
\psi({\bm k}) & -\epsilon({\bm k})
\end{array}
\right) 
\end{eqnarray}
with 
\begin{eqnarray}
\epsilon({\bm k})=\left(\frac{\hbar^2{\bm k}^2}{2m}-\mu\right),
\quad
\psi({\bm k})=\Delta_0 \frac{k_x k_y}{k_{\rm F}^2}. 
\end{eqnarray}
Now we solve the BdG equation for a semi-infinite $d_{xy}$-wave
superconductor at $x \geq 0$. Replacing $k_x$ with $-i\partial_{x}$,
we have the BdG equation for a zero-energy state,
\begin{eqnarray}
{\cal H}_{2 \times 2}(-i\partial_x, k_y)|u_0(x)\rangle=0. 
\end{eqnarray}
For $|k_y|<k_{\rm F}$, 
the above equation with the boundary condition
$|u(x=0)\rangle=0$ has a solution \cite{Hu1994}
\begin{eqnarray}
|u_0(x)\rangle=C
\left(
\begin{array}{c}
1 \\
-i{\rm sgn} k_y
\end{array}
\right)e^{ik_y y}\sin \left( \sqrt{k_{\rm F}^2-k_y^2}x \right)e^{-x/\xi}
\nonumber\\
\end{eqnarray}
for weak pairing (i.e. $m\Delta_0/\hbar^2k_{\rm F}^2\ll 1$).
Here, $C$ is a normalization constant and $\xi^{-1}\equiv m\Delta_0
k_y/\hbar^2k_{\rm F}^2$.
The zero-energy state is localized at the boundary, which strongly affects the
transport properties through the edge \cite{Tanaka-Kashiwaya1995}. 

In the above example, the Andreev bound state has a flat
dispersion with zero energy. However, there are a variety of gapless Andreev bound
states in other unconventional superconductors, depending on the symmetry of the system.
For example, in the time-reversal-breaking $^3$He-A
phase in 2D, there appear gapless chiral edge states with a linear dispersion, 
while in the time-reversal-invariant $^3$He-B phase in 3D, gapless helical
surface states with a linear dispersion show up \cite{VolovikHe3,Hara1986}. 



\section{Theory of topological superconductors}

\subsection{General definition}
\label{sec:general}

As discussed in Sec. \ref{sec:3_1}, all negative
energy states of the BdG Hamiltonian are fully-occupied in a superconducting state. 
Thus, like in insulators, one can define topological numbers for the occupied states.
Depending on the dimension and symmetries of the system, various
topological numbers can be introduced. 
In the broadest sense, a superconductor is regarded topological 
if any of such topological numbers is nonzero.
Unconventional superconductors often have nodal superconducting gaps,
where the nodes themselves have topological numbers. 
From the above definition, such unconventional superconductors are 
topological, and they may be called weak topological superconductors. 

In a narrower sense, a fully-opened gap is required besides a nonzero topological number for
topological superconductivity. 
In this case, no gapless bulk excitation exists and the system may be called a strong topological
superconductor. Similar to the case of a quantum Hall state, the transport properties
of such a system at low temperature are purely determined by topologically-protected gapless
excitations localized at a boundary or at a topological defect.

\subsection{Particle-hole symmetry and topological superconductor}

In Sec. \ref{sec2:symmetry}, we have argued that general Hamiltonians do
not have any non-trivial topological structure in one and three-dimensions.
Nevertheless, as we shall discuss below, topological superconductors are
possible in both one and three dimensions, as well as in two-dimensions.
The key to understanding this result is the symmetry specific to superconductors. 
As was shown in Sec. \ref{sec:3_1}, superconductors are generically 
characterized by particle-hole symmetry.
This symmetry enables us to introduce topological numbers
other than the Chern number.

To see this, let us consider a 1D spinless superconductor,
\begin{eqnarray}
{\cal H}=\frac{1}{2}\sum_{k}\left(c_k^{\dagger}, c_{-k}\right){\cal
H}'_{2\times 2}(k)
\left(
\begin{array}{c}
c_k\\
c_{-k}^{\dagger}
\end{array}
\right) 
\end{eqnarray}
with
\begin{eqnarray}
{\cal H}'_{2\times 2}(k)=\left(
\begin{array}{cc}
\varepsilon(k) & d(k)\\
d^*(k) & -\varepsilon(-k)
\end{array}
\right).
\label{eq4:2x2}
\end{eqnarray}
Here, due to the Fermi statistics (i.e. the anti-commutation relation) of $c_k$, the term 
$d(k)$ must be an odd function of $k$, which makes
${\cal H}'_{2\times 2}(k)$ to have the particle-hole symmetry, ${\cal
C}{\cal H}'_{2\times 2}(k){\cal
C}^{-1}=-{\cal H}'_{2\times 2}(-k)$. 
Below, we additionally assume that $\varepsilon(k)$ is an even function of $k$, since
otherwise fully-gapped superconductivity is not realized in general.
For simplicity, the lattice constant is taken to be 1, so that the Hamiltonian is $2\pi$-periodic, ${\cal
H}'_{2\times 2}(k)={\cal H}'_{2\times 2}(k+2\pi)$.

Let us first reproduce the argument in Sec. \ref{sec2:symmetry}:
Using a $2\times 2$ unitary matrix $U(k)$, 
one can diagonalize ${\cal H}_{2\times 2}(k)$ in Eq. (\ref{eq4:2x2}) as
\begin{eqnarray}
U^{\dagger}(k){\cal H}(k)U(k)=
\left(
\begin{array}{cc}
E(k) & 0\\
0 & -E(k)
\end{array}
\right),
\end{eqnarray}
where $E(k)=\sqrt{\varepsilon^2(k)+|d(k)|^2}>0$.
Since $U(k)$ in the above equation is not unique but has a redundancy of the gauge
degree of freedom, we can generalize it to
\begin{eqnarray}
U(k)\rightarrow U(k)
\left(
\begin{array}{cc}
e^{i\theta_1(k)} & \\
& e^{i\theta_2(k)}
\end{array}
\right)
\end{eqnarray}
with ${\rm U(1)}$ rotations $e^{i\theta_i(k)}$ $(i=1,2)$.
This $U(k)$ is not merely an element of the unitary group ${\rm
U}(2)$, but it should be
regarded as an element of the coset space 
\begin{eqnarray}
{\cal M}=\frac{{\rm U}(2)}{{\rm U}(1)\times {\rm U}(1)},
\end{eqnarray}
which is equivalent to a 2D sphere, i.e. ${\cal M}=S^2$ \cite{Soliton}.
Since the 1D Brillouin zone is equivalent to a 1D circle $S^1$, 
the image of the Brillouin zone by the map $U(k)$ is a circle $S^1$ 
on the sphere $S^2$ of ${\cal M}$. 
The circle $S^1$ can smoothly shrink to a point on the sphere $S^2$, 
and thus no topological constraint exists on this map, which reproduces
the homotopy result $\pi_1({\cal M})=0$ in Sec. \ref{sec2:symmetry}.

The above argument, however, is not really correct due to the particle-hole
symmetry. To see this, we specify the 2D sphere $S^2$ in a different manner.
By using the Pauli matrices ${\bm \tau}$ in the Nambu space, 
the Hamiltonian ${\cal H}(k)$ can be written as
\begin{eqnarray}
{\cal H}(k)=E(k){\bm x}(k)\cdot{\bm \tau}
\end{eqnarray}
with ${\bm x}(k)=(x_1(k), x_2(k), x_3(k))$ where
\begin{eqnarray}
x_1(k)=\frac{{\rm Re}\,d(k)}{E(k)},
\quad
x_2(k)=-\frac{{\rm Im}\,d(k)}{E(k)},
\quad
x_3(k)=\frac{\varepsilon(k)}{E(k)}.
\nonumber\\
\end{eqnarray}
Since ${\bm x}^2(k)=1$, the space spanned by the vector ${\bm x}(k)$ is the
2D sphere $S^2$. The image of the Brillouin zone by the map
${\bm x}(k)$ provides a circle $S^1$ on $S^2$.

Now one can see that the particle-hole symmetry imposes an additional
constraint on the circle $S^1$.
Due to the particle-hole symmetry, $d(k)$ satisfies $d(k)=-d(-k)$, and thus
$d(k)=0$ at $k=0$ and $\pi$.
This imposes a constraint on the $S^1$ image of the Brillouin zone that it 
must pass a
pole of the 2D sphere, i.e. $x_1(k)=x_2(k)=0$, at $k=0$ and
$k=\pi$.
This constraint allows us to tell topologically distinct circles: 
If $x_3(0)$ and $x_3(\pi)$ have the same sign, $S^1$ passes the same
pole at $k=0$ and $\pi$, and thus $S^1$ can smoothly shrink to a
point; on the other hand, if $x_3(0)$ and $x_3(\pi)$ have opposite signs, 
$S^1$ passes different poles at $k=0$ and $\pi$, which means that 
$S^1$ cannot shrink to a point. 
Therefore, $S^1$ in the former (latter) case is topologically trivial
(nontrivial). 
Since the sign difference between $x_3(0)$ and $x_3(\pi)$ is specified by
${\rm sgn}[\varepsilon(0)\varepsilon(\pi)]$, the corresponding
topological number $\nu_{\rm 1d}$ is a ${\bm Z}_2$ index \cite{Kitaev2001}:
\begin{eqnarray}
(-1)^{\nu_{\rm 1d}}={\rm sgn}\left[
\varepsilon(0)\varepsilon(\pi)
\right]. 
\label{eq4:z2simple}
\end{eqnarray}
For a tight-binding model with $\varepsilon(k)=-t\cos k-\mu$ ($t>0$),
the above ${\bm Z}_2$ index is evaluated as
\begin{eqnarray}
(-1)^{\nu_{\rm 1d}}=-{\rm sgn}\left[
(t+\mu)(t-\mu) 
\right],
\end{eqnarray}
and thus it is nontrivial when $-t<\mu<t$.
When $\mu=t$ ($\mu=-t$), $E(k)$ vanishes at $k=\pi$ ($k=0$), so that the gap
in the spectrum closes and the topological phase transition occurs.

In the above example, we considered the simplest $2\times 2$ matrix
Hamiltonian, but a similar result can be obtained for a general 1D
superconductor, for which the ${\bm Z}_2$ index is given in terms of 
the Berry phase \cite{Qi2008} as follows:
Suppose that $|u_n(k)\rangle$ is a solution of 
$H(k)|u_n(k)\rangle=E_n(k)|u_n(k)\rangle$ with a positive energy
$E_n(k)>0$. The particle-hole symmetry dictates that 
$C|u_n(-k)\rangle$ is also a solution of the same equation with 
the energy $E_n(-k)<0$. When we assign a positive (negative) integer 
$n$ to such $|u_n(k)\rangle$ which gives a positive
(negative) energy eigenvalue, the particle-hole symmetry allows us to set
\begin{eqnarray}
|u_{-n}(k)\rangle=C|u_n(-k)\rangle. 
\end{eqnarray}
To calculate the ${\bm Z}_2$ index, we introduce the gauge fields 
\begin{eqnarray}
{\mathcal A}^{(+)}(k)=i\sum_{n>0}\langle u_n(k)|\partial_k
u_n(k)\rangle,
\nonumber\\
{\mathcal A}^{(-)}(k)=i\sum_{n<0}\langle u_n(k)|\partial_k
u_n(k)\rangle,
\label{eq4:gauge}
\end{eqnarray}
and their sum ${\mathcal A}(k)={\mathcal A}^{(+)}(k)+{\mathcal
A}^{(-)}(k)$. Using the particle-hole symmetry, one obtains the following
relation
\begin{eqnarray}
{\mathcal A}^{(+)}(k)={\mathcal A}^{(-)}(-k). 
\end{eqnarray}
Writing the $m$-th component of the eigenvector $|u_n(k)\rangle$
as $u_{mn}(k)$, ${\mathcal A}(k)$ can be recast into
\begin{eqnarray}
{\mathcal A}(k)&=&i\sum_{n}\langle u_n(k)|\partial_k u_n(k)\rangle 
\nonumber\\
&=&
i \sum_{nm}u^{\dagger}_{nm}(k)\partial_k u_{mn}(k)
\nonumber\\
&=& i \,{\rm tr}[U^{\dagger}(k)\partial_k U(k)],
\end{eqnarray}
where $U_{mn}(k)\equiv u_{mn}(k)$. The normalization condition of $|u_n(k)\rangle$
implies that $U(k)$ is a unitary matrix, and 
the derivative in the last equation is rewritten as
\begin{eqnarray}
{\rm tr}[U^{\dagger}(k)\partial_k U(k)]
&=&\partial_k \left[\ln {\rm det}U(k) \right]
\nonumber\\
&=& i\partial_k \theta(k),
\end{eqnarray}
where $\theta(k)$ is the phase angle of ${\rm det}U(k)$.
Meanwhile, the Berry phase obtained from
${\mathcal A}^{(-)}(k)$ along the 1D Brillouin zone 
\begin{eqnarray}
\gamma=\int_{-\pi}^{\pi}dk{\mathcal A}^{(-)}(k) 
\label{eq4:Berry}
\end{eqnarray}
is evaluated as
\begin{eqnarray}
\gamma&=&\frac{1}{2}\int_{-\pi}^{\pi}dk {\mathcal A}(k)
\nonumber\\
&=&\frac{i}{2}\int_{-\pi}^{\pi}dk 
\partial_k\left[\ln {\rm det} U(k)\right] 
\nonumber\\
&=& -\frac{1}{2}\left[\theta(\pi)-\theta(-\pi)\right].
\end{eqnarray}
Since ${\rm det}U(k)$ is $2\pi$-periodic in $k$, we have 
$\theta(-\pi)-\theta(\pi)=2\pi N$ with an integer $N$.
Consequently, the Berry phase is quantized as 
$e^{i\gamma}=e^{i\pi N}=\pm 1$, which defines the 1D ${\bm Z}_2$
index $(-1)^{\nu_{\rm 1d}}$ as
\begin{eqnarray}
(-1)^{\nu_{\rm 1d}}=e^{i\gamma}. 
\label{eq4:1dz2}
\end{eqnarray}
When $(-1)^{\nu_{\rm 1d}}=-1$, the system is topologically nontrivial.
This ${\bm Z}_2$ index can be shown to coincide with that in
Eq. (\ref{eq4:z2simple}) for the $2\times 2$ BdG Hamiltonian
Eq. (\ref{eq4:2x2}).

When a superconductor preserves time-reversal symmetry, there is
another constraint. 
The time-reversal operation flips the electron spin,
which results in the two-fold Kramers degeneracy in the energy eigenstates.
As in the case of topological insulators,
the Kramers degeneracy plays a crucial role in a time-reversal-invariant
superconductor.
For instance, the two-fold degeneracy makes the
1D ${\bm Z}_2$ index in Eq. (\ref{eq4:1dz2}) trivial.
However, using time-reversal symmetry, 
one can define another nontrivial 1D ${\bm Z}_2$ index instead of 
Eq. (\ref{eq4:1dz2}).
To see this, we consider Kramers pairs $(|u^{\rm I}_n(k)\rangle,
|u^{\rm II}_n(k)\rangle)$ and define the gauge fields for the occupied states
\begin{eqnarray}
{\mathcal A}^{{\rm I}(-)}(k)&=&i\sum_{n<0}\langle u_n^{\rm
I}(k)|\partial_k u_n^{\rm I}(k)\rangle,
\nonumber\\ 
{\mathcal A}^{{\rm II}(-)}(k)&=&i\sum_{n<0}\langle u_n^{\rm
II}(k)|\partial_k u_n^{\rm II}(k)\rangle.
\end{eqnarray}
The Kramers pairs are related by time-reversal ${\cal T}$ as
\begin{eqnarray}
|u_n^{\rm II}(k)\rangle=e^{i\varphi_n(k)}{\cal T}|u_n^{\rm I}(-k)\rangle 
\end{eqnarray}
with the gauge degree of freedom $\varphi_n(k)$, and this leads to the relation
\begin{eqnarray}
{\mathcal A}^{{\rm II}(-)}(k)= {\mathcal A}^{{\rm I}(-)}(-k)
-\sum_{n<0}\partial_k\varphi_n(k).
\end{eqnarray}
By using the fact that the 1D ${\bm Z}_2$ index in Eq. (\ref{eq4:1dz2}) 
is trivial in the present situation, one can derive
\begin{eqnarray}
\int_{-\pi}^{\pi} dk \left[{\mathcal A}^{{\rm I}(-)}(k)
+{\mathcal A}^{{\rm II}(-)}(k) 
\right] =2\pi M
\end{eqnarray}
with an integer $M$.
The Berry phase obtained from ${\mathcal A}^{{\rm I}(-)}(k)$ along the
1D Brillouin zone is evaluated as
\begin{eqnarray}
\gamma^{\rm I}&=&\int_{-\pi}^{\pi}dk{\mathcal A}^{{\rm I}(-)}(k)
\nonumber\\
&=&\frac{1}{2}\int_{-\pi}^{\pi}dk 
\left[{\mathcal A}^{{\rm I}(-)}(k)+{\mathcal A}^{{\rm II}(-)}(-k)
\right] \, (\mbox{mod.}\, 2\pi)
\nonumber\\
&=&\pi M \, (\mbox{mod.}\, 2\pi).
\end{eqnarray}
The similar Berry phase $\gamma^{\rm II}$ for 
${\mathcal A}^{{\rm II}(-)}(k)$ also satisfies
\begin{eqnarray}
\gamma^{\rm II}=\pi M \, (\mbox{mod.}\, 2\pi).
\end{eqnarray}
Therefore, a time-reversal-invariant 1D ${\bm Z}_2$ index can
be defined as
\begin{eqnarray}
(-1)^{\nu_{\rm 1d}^{\cal T}}=e^{i\gamma^{\rm I}}= e^{i\gamma^{\rm II}}, 
\label{eq4:1dTRIz2}
\end{eqnarray}
which is quantized to $\pm 1$.
The system is topologically nontrivial (trivial) when $(-1)^{\nu_{\rm
1d}^{\cal T}}=-1$ ($(-1)^{\nu_{\rm 1d}^{\cal T}}=1$).

In 2D, the particle-hole symmetry does not introduce a new topological structure, 
although this symmetry always pairs up the positive and negative energy states.
Like the quantum Hall states or the quantum spin Hall states, one can define the 
Chern number $Ch$ or the 2D ${\bm Z}_2$ index $(-1)^{\nu_{\rm 2d}}$ in the absence
or presence of time-reversal symmetry, respectively
\cite{Schnyder2008, Sato-Fujimoto2009,Tanaka-Yokoyama2009}.
The only change from Eqs. (\ref{eq3:totalCh}) and (\ref{eq2:2dZ2}) is that the occupied states
of the Bloch Hamiltonian $H({\bm k})$ is replaced by the negative energy
states of the BdG Hamiltonian ${\cal H}({\bm k})$.


The particle-hole symmetry plays an essential role again in 3D, and it 
allows us to introduce a new topological number.
In the presence of time-reversal symmetry, the 3D ${\bm Z}_2$ index 
$(-1)^{\nu_{\rm 3d}}$ in Eq. (\ref{eq2:3dZ2}) can
be defined in the same manner as in topological insulators,
but the coexistence of the particle-hole symmetry makes it possible to
introduce a more refined integer topological number.
The BdG Hamiltonian in such a case has combined symmetry of 
particle hole ${\cal C}$ and time-reversal ${\cal T}$ as 
\begin{eqnarray}
\Gamma {\cal H}({\bm k})\Gamma^{-1}=-{\cal H}({\bm k}),
\quad \Gamma=i{\cal C}{\cal T}. 
\label{eq4:chiral}
\end{eqnarray}
which is called {\it chiral symmetry}.
Using the chiral operator $\Gamma$, the 3D winding number $w_{\rm 3d}$ is defined as \cite{Grinevich1988, Schnyder2008}
\begin{align}
&w_{\rm 3d}
\nonumber\\
=&\frac{1}{48\pi^2}\int_{\rm BZ} d^3k
\epsilon^{ijl}{\rm tr}\left[\Gamma 
{\cal H}^{-1}(\partial_{k_i}{\cal H}) 
{\cal H}^{-1}(\partial_{k_j}{\cal H}) 
{\cal H}^{-1}(\partial_{k_l}{\cal H}) 
\right].
\nonumber\\
\label{eq4:3dW}
\end{align}
One can find that the parity of $w_{\rm 3d}$ is the same as the
3D ${\bm Z}_2$ index of topological insulators. 

The presence of time-reversal and/or particle-hole (charge conjugation) symmetry
is robust against (non-magnetic) disorder.
Based on these general symmetries, 
Schnyder {\it et al.} constructed a table of topological numbers for fully-gapped 
insulators and superconductors in various dimensions \cite{Schnyder2008}.
In Table \ref{table1}, we show a part of the topological table relevant to
superconductors, summarizing the topological numbers discussed
in this subsection.

\begin{table}[t]
\begin{center}
\begin{tabular}{|c|cc|ccc|}
\hline
AZ class & TRS & PHS & 1d & 2d & 3d \\
\hline
TRB SCs (class D) & - & +1 & ${\bm Z}_2^{(\rm CS)}$ & ${\bm
Z}^{(\rm Ch)}$ & 0 \\
TRI SCs (class DIII) & $-1$ &+1 & ${\bm Z}_2^{\rm (CS_{\rm T})}$
& ${\bm Z}_2^{(\rm KM)}$ &
${\bm Z}^{(\rm W)}$ \\
\hline
\end{tabular}
\caption{Topological table for time-reversal-breaking (TRB)
and time-reversal-invariant (TRI) superconductors (SCs). In terms of the
Altland-Zirnbauer classification, they belong to class D and class
DIII, respectively \cite{Schnyder2008}. ${\bm Z}_2^{(\rm CS)}$ and
${\bm Z}_2^{(\rm CS_{\rm T})}$ are given in Eqs. (\ref{eq4:1dz2}) and (\ref{eq4:1dTRIz2}),
respectively. ${\bm Z}^{(\rm Ch)}$ and ${\bm Z}_2^{(\rm KM)}$ indicate
the first Chern number (TKNN integer) and the Kane-Mele's ${\bm Z}_2$ index,
respectively. ${\bm Z}^{(\rm W)}$ corresponds to the 3D winding number in
Eq. (\ref{eq4:3dW}). In the middle block, $+1$ and $-1$ indicate that ${\cal C}^2=1$ and ${\cal
T}^2=-1$, respectively.
}
\label{table1}
\end{center} 
\end{table}

\subsection{Topological boundary and defect states}
\label{sec:tbs}

A characteristic feature of topological superconductors is the
existence of gapless boundary states.
As explained above, a topological superconductor hosts a bulk nonzero
topological number. When it is interfaced with a topologically
trivial state such as vacuum, there arises a mismatch of topology,
which cannot be resolved without having a singularity at the boundary. 
The singularity is physically realized as gapless boundary states
\cite{Tanaka-Sato-Nagaosa2012}.

Depending on the symmetries and dimensions of the system, 
there exist a variety of topological numbers defining topological
superconductors; correspondingly, we have a variety of gapless
boundary states consistent with the symmetries and dimensions. 
The correspondence between a topological number and its peculiar 
boundary states is called bulk-boundary correspondence
\cite{Hatsugai1993, VolovikHe3, Schnyder2008, Teo-Kane2010,
Sato-Tanaka-Yada-Yokoyama2011, Essin2011}.
The bulk-boundary correspondence makes it possible to detect a definite
fingerprint of topological superconductivity. 

A topological superconductor may also support zero-energy states localized on
topological defects.
A topological defect can be considered as a sort of a boundary
of the system, and thus zero-energy states appear for the same reason as the
gapless boundary states
\cite{Teo-Kane2010,Teo2013,Ueno2013, Benalcazar2014,Shiozaki2014}.
For example, let us consider a vortex in a 2D spinless chiral
$p$-wave superconductor. The vortex can be considered as a small hole in the
superconductor; since $|Ch|=1$ in the spinless chiral $p$-wave
superconductor, the boundary of the small hole supports a gapless boundary
state, which eventually becomes a zero-energy state (called ``zero mode'') when localized in the
vortex core \cite{Kopnin-Salomaa1991,Read-Green}. 
Vortices in a variety of topological superconductors may support zero
modes in a similar manner \cite{Tsutsumi2013, Chang2014,Lee2016,Tsutsumi2015}.

The gapless/zero-energy states at the boundary or defects are topologically 
classified in a uniform manner in the mathematical framework of the
K-theory \cite{Teo-Kane2010, Shiozaki2014}. 
The classification of these gapless modes reduces to the classification
of a semiclassical Hamiltonian 
\begin{eqnarray}
{\cal H}({\bm k}, {\bm r}) ,
\end{eqnarray}
where ${\bm k}$ is the momentum in the $d$-dimensional Brillouin zone
and ${\bm r}$ is the coordinate of the $D$-dimensional sphere $S^D$
surrounding a defect.
Using the K-theory, one can show that the possible topological number for
${\cal H}({\bm k}, {\bm r})$ depends only on $\delta=d-D$.
This result also implies that a defect can be considered as a boundary:
Since $D$ is the defect codimension,
a topological defect surrounded by $S^D$ in $d$-dimensions defines a 
$\delta-1$ dimensional submanifold. 
(For instance, a line defect in three dimensions has $\delta=2$ ($d=3$,
$D=1$), and thus it defines a one-dimensional submanifold.)
Therefore, the defect submanifold can be considered as a boundary of
a $\delta$-dimensional insulator/superconductor \cite{Read-Green, Shiozaki2014}. 
Consequently, the classification of the gapless modes on defects reduces to that of the
$\delta$-dimensional insulators/superconductors \cite{Teo-Kane2010}.

\subsection{Topological crystalline superconductor}

Superconductors may have space-group symmetry according to their crystal structures.
Like the case of topological crystalline insulators \cite{Fu2011},
such a material-based symmetry can give rise to a nontrivial topology in
the superconducting state \cite{Mizushima2012, Teo2013, Ueno2013,
Zhang-Kane-Mele2013, Chiu2013, Morimoto2013, Benalcazar2014,
Tsutsumi2013, Shiozaki2014, 
Shiozaki-Sato-Gomi2016}. 

To see this, let us consider mirror reflection $M_{xy}$ with respect
to the $xy$ plane. When the base material has mirror-reflection
symmetry, the Hamiltonian in the normal state ${\cal E}({\bm k})$ obeys
\begin{eqnarray}
M_{xy}{\cal E}(k_x, k_y, -k_z)M_{xy}^{-1}
={\cal E}({\bm k}). 
\end{eqnarray}
The superconducting state keeps the mirror-reflection symmetry if the
pair potential $\Delta({\bm k})$ does not break it spontaneously.
When the pair potential is invariant under mirror reflection,
\begin{eqnarray}
M_{xy}\Delta(k_x, k_y -k_z)M_{xy}^t =\Delta({\bm k}), 
\end{eqnarray}
the BdG Hamiltonian also retains the mirror-reflection symmetry,
which can be expressed as 
\begin{eqnarray}
\tilde{M}_{xy}
{\cal H}(k_x, k_y, -k_z) 
\tilde{M}_{xy}^{-1}=
{\cal H}({\bm k}), 
\end{eqnarray}
where $\tilde{M}_{xy}$ is the mirror-reflection operator acting on
the Nambu space,
\begin{eqnarray}
\tilde{M}_{xy}=
\left(
\begin{array}{cc}
M_{xy} & 0\\
0 & M_{xy}^* 
\end{array}
\right). 
\label{eq4:mirror1}
\end{eqnarray}
It should be noted here that the above scenario is not the only way to keep the 
mirror-reflection symmetry in the superconducting state.
Even when the pair potential changes sign under mirror reflection, 
\begin{eqnarray}
M_{xy}\Delta(k_x, k_y, -k_z)M_{xy}^t =-\Delta({\bm k}), 
\end{eqnarray}
the superconducting state can still support the mirror-reflection symmetry,
and the key to this scenario is the U(1) electromagnetic gauge symmetry.
When a Cooper pair forms in the above pair potential, both the original mirror
reflection symmetry and the U(1) gauge symmetry are spontaneously broken, 
but their combination can be preserved. 
This is because the BdG Hamiltonian has the following mirror reflection symmetry,
\begin{eqnarray}
\tilde{M}'_{xy}
{\cal H}(k_x, k_y, -k_z) 
\tilde{M}_{xy}^{'-1}=
{\cal H}({\bm k}), 
\end{eqnarray}
with
\begin{eqnarray}
\tilde{M}'_{xy}=
\left(
\begin{array}{cc}
M_{xy} & 0\\
0 & -M_{xy}^* 
\end{array}
\right). 
\label{eq4:mirror2}
\end{eqnarray}
In general, when a crystal-symmetry operation changes the phase of the 
pair potential, it can be made to be retained by combining the crystal-symmetry operation 
with a U(1)-gauge-symmetry operation.
For mirror reflection, only a sign change of the pair potential is allowed since ${\cal
M}_{xy}^2=-1$, and hence there are two possible realizations of the
mirror reflection in the BdG Hamiltonian, which correspond to
Eqs. (\ref{eq4:mirror1}) and (\ref{eq4:mirror2}).

Now we show that mirror-reflection symmetry provides novel topological structures
through the introduction of a new topological number called mirror Chern number \cite{Teo-Fu-Kane2008}. 
On the mirror-invariant plane $k_z=0$ in the Brillouin zone, the BdG
Hamiltonian commutes with the mirror reflection operator, 
\begin{eqnarray}
[{\cal H}(k_x, k_y, 0), \tilde{M}_{xy}]=0. 
\end{eqnarray}
Therefore, an eigenstate of the BdG Hamiltonian on the mirror-invariant
plane can be simultaneously an eigenstate of the mirror operator. 
The mirror Chern number is defined as the Chern number of each mirror
eigensector. 
In contrast to the ordinary Chern number, the mirror Chern number can be
nonzero even when the system preserves time-reversal symmetry, because the
individual mirror subsector is not time-reversal invariant.
Furthermore, the mirror Chern number provides a more
detailed topological structure than the Kane-Mele's ${\bm Z}_2$ index.
For example, in the case of insulators, SnTe hosts gapless
surface states ensured by a nonzero mirror Chern number, while it
is trivial in the ${\bm Z}_2$ topology \cite{HsiehTCI2012, Tanaka2012}.

The two different mirror-reflection operations for the BdG Hamiltonian discussed
above leads to different properties of the topological states in superconductors \cite{Ueno2013}.
While the mirror reflection-operator $\tilde{M}_{xy}$ in
Eq. (\ref{eq4:mirror1}) commutes with the charge-conjugation operator ${\cal C}$
\begin{eqnarray}
\left[{\cal C}, \tilde{M}_{xy}\right]=0, 
\label{eq4:CM+}
\end{eqnarray}
the other mirror operator $\tilde{M}_{xy}'$ in Eq. (\ref{eq4:mirror2})
anti-commutes
\begin{eqnarray}
\left\{{\cal C}, \tilde{M}'_{xy}\right\}=0. 
\label{eq4:CM-}
\end{eqnarray}
This difference gives rise to different realizations of the particle-hole symmetry in
mirror subsectors. 
For the commuting case, the mirror subsector with the mirror eigenvalue $\tilde{M}_{xy}=i$ 
is interchanged with the subsector with the opposite eigenvalue 
$\tilde{M}_{xy}=-i$ when one applies ${\cal C}$.
Thus, each mirror subsector does not have its own particle-hole
symmetry, although the system as a whole does.
This means that the mirror subsector belongs to the same topological
class as a quantum Hall state.
On the other hand, in the anti-commuting case, the operator
${\cal C}$ maps each mirror eigensector onto itself, and hence
the mirror subsector has its own particle-hole symmetry.
Moreover, the mirror subsector in the anti-commuting case 
effectively realizes a spinless system, since electrons with an
opposite $z$-component of the spin have an opposite eigenvalue of ${M}_{xy}$; 
as a result, each subsector is topologically equivalent to a spinless superconductor, 
and it may support Majorana fermions of a spinless superconductor, which is discussed in
the next section. 


\subsection{Topology of nodal superconductors}
\label{sec4:e}

Unconventional superconductors often host some nodes in the bulk 
superconducting gap.
A traditional approach to the nodal structure is based on the irreducible
representation of gap functions under space-group
symmetry \cite{Sigrist-Ueda1991}. However, it has been recognized that nodes
themselves can have their own topological numbers.
Such topological arguments have revealed new aspects of nodal
superconductors.

From the viewpoint of group theory, stable nodes in superconductors are 
classified into two: those protected by
crystal symmetry, and those unprotected by crystal symmetry. 
For both classes of nodes, one can define topological numbers. 

Let us first consider nodes that are not protected by crystal symmetry. 
Such nodes may appear at any position on the Fermi surface, and
thus, in principle, they can move freely in response to perturbations to the system.
In the group theoretical approach, these nodes are called accidental nodes.
It should be noted, however, that they are not necessarily unstable. 
They can be stable due to their intrinsic topological numbers.

The topological numbers for accidental nodes are defined on
momentum-space submanifolds enclosing the nodes: 
A time-reversal-breaking superconductor may host an accidental point
node, whose topological number is given by the 
Chern number on a sphere enclosing the point node.
The point node is a superconducting analogue of a Weyl node in a Weyl semimetal. 
For this reason, a superconductor with an accidental point node is
dubbed {\it Weyl superconductor} \cite{Meng2012}. 
In contrast to a Weyl semimetal, which may keep time-reversal symmetry
when it breaks inversion symmetry, 
a Weyl superconductor must always break time-reversal symmetry.
This difference originates from the intrinsic symmetry of
superconductors, i.e. the particle-hole symmetry.
When a superconductor preserves time-reversal symmetry, it must also host 
chiral symmetry (which is the combination of particle-hole and time-reversal symmetries)
defined in Eq. (\ref{eq4:chiral}), from which one can show that the Chern number always vanishes.
Hence, breaking of time-reversal symmetry is necessary for
realizing a point node with a nonzero Chern number.
The $^3$He A-phase in 3D supports such topologically-protected accidental point
nodes at poles on the Fermi surface \cite{VolovikHe3}.

A superconductor may also host an accidental line node with a non-zero
topological number in the presence of time-reversal symmetry \cite{Sato2006, Beri2010}. 
Such a topological number is defined on a circle $C$ enclosing the
line node. Using the chiral symmetry [Eq. (\ref{eq4:chiral})], the
topological number is introduced as the 1D winding
number \cite{Sato-Tanaka-Yada-Yokoyama2011, Schnyder-Ryu2011},
\begin{eqnarray}
w_{\rm 1d}=\frac{i}{4\pi}\oint_{C} d{\bm k} \cdot
{\rm tr}\left[\Gamma {\cal H}^{-1}({\bm k})
{\bm \partial}_{\bm k}{\cal H}({\bm k}) \right] .
\label{eq4:winding}
\end{eqnarray}
The line nodes in high-$T_{\rm c}$ cuprates and noncentrosymmetric
superconductors have non-zero values of
the 1D winding number \cite{Sato2006, Beri2010, Mizuno2010, Yada2011,
Sato-Tanaka-Yada-Yokoyama2011, Schnyder-Ryu2011, Schnyder2012, Ikegaya2016}. 

Accidental nodes in superconductors are classified by the presence or
absence of time-reversal and/or inversion symmetries. 
The topological classification of accidental nodes in
superconductors is given in Ref. \cite{Kobayashi2014}.
A similar classification is also discussed in Ref. \cite{Zhao2016}.
In the former theory, the topological version of the Blount's theorem
\cite{Blount} was employed. 
These classifications indicate that in addition to point and line nodes,
accidental area nodes with a ${\bm Z}_2$ topological
index are possible in even-parity, time-reversal-breaking superconductors
\cite{Kobayashi2014, Zhao2016}. 
A possible realization of such area nodes in heavy-fermion
superconductors was discussed recently \cite{Agterberg2016}. 

The second class of gap nodes, which are protected by crystal symmetry, 
appear at high-symmetry lines or planes in the Brillouin zone.
Both symmorphic symmetry (like mirror reflection) and non-symmorphic
symmetry (such as glide) give rise to stable line nodes on the Brillouin-zone 
boundary \cite{Norman1995, Micklitz2009, Nomoto2016}.
Using crystal symmetry as well as particle-hole and/or
time-reversal symmetries, node-protecting topological numbers can be
introduced for such nodes \cite{Kobayashi2014,
Yang-Pan2014,Chiu2014, Kobayashi2015, Kobayashi2016, Micklitz2016}. 

For both classes of nodes, the relevant topological numbers are not
defined globally, but are defined only in restricted regions in the
momentum space. Hence, they are considered to be ``weak'' topological indices. 
Such weak indices often lead to gapless surface states with a flat dispersion.
For instance, the non-zero Chern number of Weyl superconductors gives rise to
a superconducting analogue of the surface Fermi arc having a flat dispersion 
\cite{VolovikHe3, Mizushima2016, Meng2012, Silaev2012, Goswami2013,Fischer2014,Goswami2015, Bo2015}.
Also, the 1D winding number [Eq. (\ref{eq4:winding})] of line-nodal
superconductors is responsible for zero-energy surface Andreev bound states 
with a flat dispersion \cite{Ryu2002, Mizuno2010, Yada2011,
Sato-Tanaka-Yada-Yokoyama2011, Schnyder-Ryu2011, Brydon2011,
Schnyder2012}. 

For superconductors having the second class of nodes, one can 
introduce ``strong'' indices, despite the fact that they are gapless in the bulk
\cite{Sato-Fujimoto2010, Sasaki2011}. This is because
the nodes in this class can be easily gapped out by a local perturbation
breaking the relevant crystal symmetry. By opening a gap with such a 
perturbation, one can define the bulk topological numbers listed in Table \ref{table1}.
However, 
for an integer-valued topological number such as the Chern number in
2D Class D superconductors or the
winding number in 3D Class DIII superconductors, 
the obtained values depend on details of the perturbation.
Therefore, these integer topological numbers are not uniquely defined
in this manner. Nevertheless, it has been shown that this perturbation procedure 
provide a unique {\it parity} of the integer topological numbers, irrespective of the 
details of the perturbation. As a result, for nodal superconductors of the second class, 
one can rigorously define mod-2 topological numbers in terms of their parity.
When the mod-2 Chern number (mod-2 winding number) is non-trivial, 
there exists an odd number of chiral edge states (surface
helical Majorana fermions) on the boundary. 
Concrete examples of such mod-2 gapless topological superconductors have
been given in Refs. \cite{Sato-Fujimoto2010, Sasaki2011}.
Other models to yield gapless topological phases have also been discussed
recently \cite{Yuval2015, Yuval2015b}.

\section{Majorana fermions}

The emergence of Majorana fermions is the most prominent characteristic of topological superconductors.
Here we present phenomenological properties of Majorana fermions.

\subsection{Concept of Majorana fermions}

When Dirac introduced the Dirac equation to describe the relativistic motion of
electrons, he found that it
also predicts an antiparticle of an electron, i.e. positron. The
antiparticle has the same mass and the same spin as the electron, but has an
opposite charge.
Whereas the Dirac equation was derived mathematically by
demanding a compatibility with the special relativity in quantum mechanics, 
the existence of positron was verified in a cosmic ray soon after the prediction. 
The discovery of antiparticle was one of the great success stories in the marriage
of relativity and quantum theory. 

While an electron is different from a positron due to the opposite
charge, a neutral particle can be identical to
its antiparticle.
Indeed, in 1937, Majorana found that the Dirac equation can describe a
particle which is identical to its antiparticle \cite{Majorana37}. 
The Dirac equation reads
\begin{eqnarray}
i\frac{\hbar}{c}\partial_t\psi({\bm x},t)=
\left[-i\hbar{\bm \alpha}\cdot{\bm \partial}_{\bm x}+\beta mc \right]\psi({\bm x},t),
\label{eq5:Dirac}
\end{eqnarray}
where $m$ is the mass of a particle, and 
${\bm \alpha}$ and $\beta$ are $4\times 4$ matrices obeying the
anti-commutation relations,
\begin{eqnarray}
\{
\alpha_i,\alpha_j
\}=2\delta_{ij},
\quad 
\{
\alpha_i,\beta
\}=0,
\quad
\beta^2=1.
\label{eq5:alphabeta}
\end{eqnarray}
The following ${\bm \alpha}$ and $\beta$ are often used,
\begin{eqnarray}
\alpha^{\rm D}_i=
\left(
\begin{array}{cc}
0 & \sigma_i \\
\sigma_i & 0
\end{array}
\right),
\quad 
\beta^{\rm D} 
=
\left(
\begin{array}{cc}
1 & 0\\
0 & -1
\end{array}
\right),
\end{eqnarray}
but this is not the only choice. Any ${\bm \alpha}$ and $\beta$
satisfying Eq. (\ref{eq5:alphabeta}) describe a relativistic fermion
particle, so one can take the following ${\bm \alpha}$
and $\beta$, 
\begin{eqnarray}
&&\alpha^{\rm M}_1=
\left(
\begin{array}{cc}
0 & \sigma_1 \\
\sigma_1 & 0
\end{array}
\right),
\quad 
\alpha^{\rm M}_{2}
=
\left(
\begin{array}{cc}
0 & \sigma_3\\
\sigma_3 & 0
\end{array}
\right),
\nonumber\\
&&\alpha^{\rm M}_3=
\left(
\begin{array}{cc}
1 & 0 \\
0 & -1
\end{array}
\right),
\quad 
\beta^{\rm M}
=
\left(
\begin{array}{cc}
0 & \sigma_2\\
\sigma_2 & 0
\end{array}
\right),
\end{eqnarray}
which satisfy the same anti-commutation relations as Eq. (\ref{eq5:alphabeta}). 
In this basis, the complex conjugate of Eq. (\ref{eq5:Dirac}) reads,
\begin{eqnarray}
i\frac{\hbar}{c}\partial_t\psi^*({\bm x},t)=
\left[-i\hbar{\bm \alpha}\cdot{\bm \partial}_{\bm x}+\beta mc \right]\psi^*({\bm x},t),
\end{eqnarray}
and thus $\psi$ and $\psi^{*}$ satisfy the same Dirac equation.
This means that the reality condition $\psi=\psi^*$ can be imposed without contradiction. 
Since $\psi^*$ corresponds to an antiparticle of $\psi$, the obtained
real field $\psi$ describes a particle identical to its antiparticle.
This self-conjugate Dirac particle is today called Majorana fermion.

\subsection{Majorana fermions in particle physics and in condensed matter}

The neutrino in particle physics is a promising candidate of Majorana fermion. 
First, it is the only neutral fermion obeying the Dirac equation in the standard
model of particle physics. Also, a mass term is allowed for Majorana
fermions, and this so-called Majorana mass term naturally explains why
neutrino is extremely light in comparison with other fermions such as electrons.
However, the verification of the self-conjugate property of neutrino is very
difficult, since neutrino rarely interacts with other particles. 
No direct experimental verification has been reported up to now.

If supersymmetry relating boson and fermion is discovered, the
superpartners of gauge bosons should be Majorana fermions.
Gauge bosons such as photons are described by real vector fields, so 
they are identical to their antiparticles.
Therefore, their partner fermions called gauginos should also have the
same property; namely, they are Majorana fermions.
Supersymmetry is expected to be realized in high-energy particle
physics and it predicts many undiscovered particles, but no such partner
particle has been discovered so far.

While Majorana originally introduced his fermion to describe an
elementary particle, it was recently recognized that condensed matter
systems may also support Majorana fermions \cite{Wilczek2009}. 
In condensed matter, the constituent fermions are
electrons. Since electron has a negative charge, it 
cannot be a Majorana fermion, as discussed above.
Nevertheless, Majorana fermions may exist as emergent collective
excitations of electrons. 
Note that the emergent Majorana fermions are 
distinct from the original Majorana fermions in that they do not 
keep the true Lorentz invariance of the Dirac equation, since
they do not move with the speed of light.
Nevertheless, under a proper rescaling of length and time, the emergent
Majorana fermions also obey the Dirac equation. 
Such emergent Majorana fermions appear in boundaries of topological
superconductors or in a class of spin-liquid systems.

\subsection{Relationship between topological superconductivity and
Majorana fermions} 

To be a Majorana fermion in a condensed matter systems,
a collective excitation should satisfy the following two conditions:
The first is that it obeys the Dirac equation. 
An ordinary electron in condensed matter physics has a parabolic energy
dispersion obeying the non-relativistic Schr\"{o}dinger equation, so it
does not satisfy this condition. However, if the system supports a
gapless fermionic excitation near a band crossing point, the low-energy
Hamiltonian of the excitation has a matrix form with band indices and
describes a linear dispersion.
Therefore, the motion of the excitation naturally obeys the
massless Dirac equation. 
The second condition is that the excitation should be its own
antiparticle, which is essential for being a Majorana fermion.

These two conditions are naturally met in topological superconductors.
The first condition is satisfied due to the topological nature.
From the bulk-boundary correspondence, topological superconductors
support gapless excitations on the boundaries, and those excitations are described by the Dirac
equation. The second condition is satisfied by virtue of the fact that, as discussed in 
Sec. \ref{sec:3_1}, the electron and hole excitations
are superimposed in the superconducting state so that they become indistinguishable.
This makes a superconductor to obtain particle-hole symmetry, 
with which the topological gapless boundary excitations become Majorana fermions. 

\subsection{Majorana zero mode and Non-Abelian statistics}

Suppose that there exists a zero mode $\gamma_0$ localized in a vortex core. 
For instance, the vortex in a 2D chiral $p$-wave superconductor supports
such a zero mode, as discussed in Sec. \ref{sec:tbs}. 
The zero mode is self-conjugate, so that it satisfies the
so-called Majorana condition,
\begin{eqnarray}
\gamma_0^{\dagger}=\gamma_0. 
\label{eq5:MCzero}
\end{eqnarray}
This equation implies that a single zero mode $\gamma_0$ cannot define
a creation or an annihilation operator.
Indeed,
if one regards $\gamma_0^{\dagger}$ as a creation
operator of the zero mode, Eq. (\ref{eq5:MCzero}) requires that 
the annihilation operator $\gamma_0$ must be the same, which leads to a contradiction. 
Therefore, $\gamma_0^{\dagger}$ cannot be considered as a creation operator.

This difficulty can be resolved by considering a pair of vortices.
Let us consider vortices 1 and 2, and express their Majorana zero
modes as $\gamma_0^{(1)}$ and $\gamma_0^{(2)}$, which
obey $\{\gamma_0^{(i)}, \gamma_0^{(j)}\}=2\delta_{ij}$ with a suitable
normalization of $\gamma_0$. 
When one defines the operators $c_{12}^{\dagger}$ and $c_{12}$ as
\begin{eqnarray}
c_{12}^{\dagger}=\frac{\gamma_0^{(1)}+i\gamma_0^{(2)}}{2},
\quad
c_{12}=\frac{\gamma_0^{(1)}-i\gamma_0^{(2)}}{2},
\label{eq5:creation}
\end{eqnarray}
they satisfy the conventional anticommutation relation
\begin{eqnarray}
\{c_{12}^{\dagger}, c_{12}\}=1. 
\end{eqnarray}
Therefore, without contradiction, one can identify $c_{12}^{\dagger}$
and $c_{12}$ as a creation and an annihilation operator, respectively. 

It should be noted here that a pair of separated vortices are
necessary for defining the creation and annihilation operators.
This nonlocality leads to a nonlocal quantum correlation between the
vortices, which results in a drastic change in their statistical nature
\cite{Read-Green, Ivanov2001}.

\begin{figure}[tb]
\centering
\includegraphics[width=0.6\columnwidth]{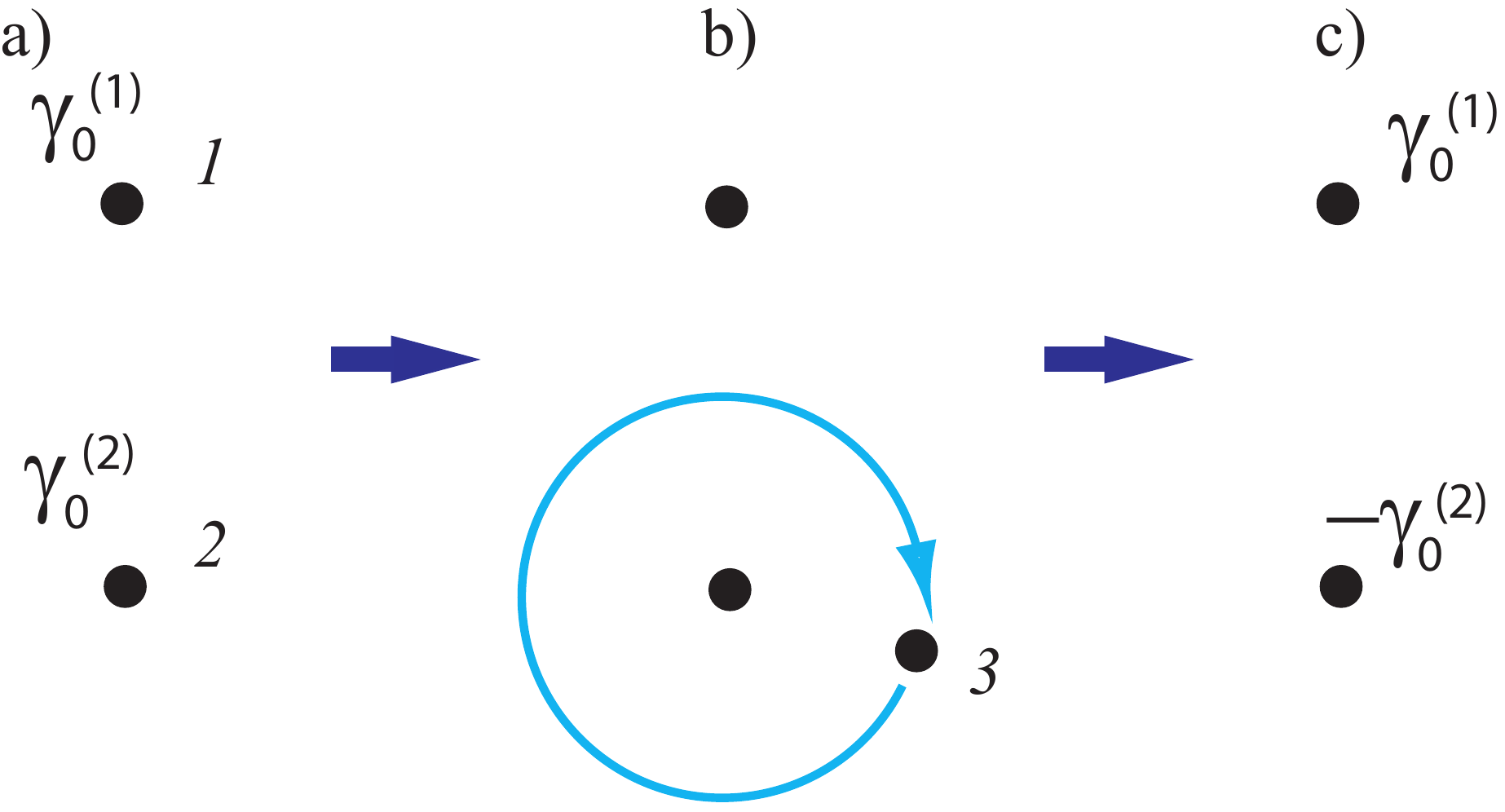}
\caption
{A braiding process of vortices containing a Majorana zero mode. 
} 
\label{fig5:braiding}
\end{figure}

To examine the statistics of the vortices, let us consider the process
illustrated in Fig. \ref{fig5:braiding}.
Initially, only vortices 1 and 2 exist as in Fig. \ref{fig5:braiding}(a). Later,
another vortex 3 comes and it moves around vortex 2
[see Fig. \ref{fig5:braiding}(b)]. Finally, vortex 3 goes far way, and
vortices 1 and 2 remain in the same initial configuration.
In terms of the number of vortex exchange, the process in Fig. \ref{fig5:braiding}(b) is equivalent to a
process of exchanging vortex 1 and vortex 2 twice. Hence, if the vortices obey the conventional bose
or fermi statistics, the finial state remains the same as the initial one. However, 
in a 2D chiral $p$-wave superconductor, due to the existence of the Majorana zero mode,
the wavefunction of the final state is different from that of the initial state, as we show below.

To see this, let us first examine what happens in a quantum state when a
particle
moves around a vortex. 
A vortex in a superconductor contains a flux quantum $\Phi_0=hc/2e$; therefore, 
when an electron moves around a vortex, the state acquires the phase factor
\begin{eqnarray}
e^{-ie\Phi_0/\hbar c}=-1 
\end{eqnarray}
due to the Aharonov-Bohm effect.
In the same manner, when a hole moves around a vortex, 
the state obtains the same phase factor
\begin{eqnarray}
e^{ie\Phi_0/\hbar c}=-1. 
\end{eqnarray}
Consequently, a quasiparticle in the superconductor, which is 
a superposition of an electron and a hole, acquires the same factor $-1$.
In particular, a Majorana zero mode $\gamma_0$ changes its phase as
\begin{eqnarray}
\gamma_0\rightarrow -\gamma_0 
\end{eqnarray}
after moving around a vortex.

Now we suppose that the initial state $|i\rangle$ in the vortex
configuration of Fig. \ref{fig5:braiding}(a) is given
by the state $|0\rangle$ annihilated by $c_{12}$ in Eq. (\ref{eq5:creation}):
\begin{eqnarray}
|i\rangle=|0\rangle, \quad c_{12}|0\rangle=0 .
\label{eq5:initial}
\end{eqnarray}
If one considers the process in Fig. \ref{fig5:braiding}(b), the
Majorana zero mode localized in vortex 3 changes sign as discussed above. 
At the same time, $\gamma_0^{(2)}$ also changes sign, 
\begin{eqnarray}
\gamma_0^{(2)}\rightarrow -\gamma_0^{(2)}, 
\label{eq4:gamma2}
\end{eqnarray}
because vortex 2
goes around vortex 3 in the coordinate system moving with vortex 3.
Therefore, in the process of Fig. \ref{fig5:braiding}(b), the operator
changes as 
\begin{eqnarray}
c_{12}=\frac{\gamma_0^{(1)}-i\gamma_0^{(2)}}{2} 
\rightarrow
c_{12}^{\dagger}=\frac{\gamma_0^{(1)}+i\gamma_0^{(2)}}{2}, 
\end{eqnarray} 
so the final state $|f\rangle$ in Fig. \ref{fig5:braiding}(c) satisfies
\begin{eqnarray}
c_{12}^{\dagger}|f\rangle=0, 
\end{eqnarray}
instead of Eq. (\ref{eq5:initial}).
In other words, the final state $|f\rangle$ become a state annihilated
by $c_{12}^{\dagger}$, $|1\rangle\equiv c_{12}^{\dagger}|0\rangle$,
which is orthogonal to the initial state $|i\rangle$,
\begin{eqnarray}
\langle i|f\rangle=0. 
\end{eqnarray}
This means that vortices are neither bosons nor fermions, since the
braiding to provide a new state should be represented in a matrix form.
The statistics expressed in a matrix form is called non-Abelian statistics, and
particles obeying non-Abelian statistics are dubbed non-Abelian anyons.

In general, as discussed by Ivanov \cite{Ivanov2001}, an exchange of the Majorana zero modes
$\gamma_0^{(i)}$ and $\gamma_0^{(j)}$ is represented by the unitary
transformation 
\begin{eqnarray}
U_{ij}=\exp\left(-\frac{\pi}{4}\gamma_0^{(i)}\gamma_0^{(j)}\right), 
\label{eq4:braid_U}
\end{eqnarray}
from which one obtains
\begin{eqnarray}
U_{ij}\gamma_0^{(i)}U^{\dagger}_{ij}=\gamma_0^{(j)},
\quad 
U_{ij}\gamma_0^{(j)}U^{\dagger}_{ij}=-\gamma_0^{(i)}.
\end{eqnarray}
In the process of Fig. \ref{fig5:braiding}, where $\gamma_0^{(2)}$ and
$\gamma_0^{(3)}$ are exchanged twice, one has
\begin{eqnarray}
(U_{23})^2\gamma_0^{(2)}(U^{\dagger}_{23})^2=-\gamma_0^{(2)}, 
\nonumber\\
(U_{23})^2\gamma_0^{(3)}(U^{\dagger}_{23})^2=-\gamma_0^{(3)}, 
\end{eqnarray}
which reproduces Eq. (\ref{eq4:gamma2}).
$U_{23}$ and $U_{12}$ do not commute with each other
\begin{eqnarray}
U_{12}U_{23}\neq U_{23}U_{12}, 
\end{eqnarray}
and thus the exchange process is actually non-Abelian.

In the above example, we assumed that the initial state $|i\rangle$ is
given by $|0\rangle$, but one can also consider the case with
$|i\rangle=|1\rangle\equiv c_{12}^{\dagger}|0\rangle$ in a similar manner. 
In this case, after the process in Fig. \ref{fig5:braiding}, the state
becomes $|0\rangle$.
After all, the process in Fig. \ref{fig5:braiding} exchanges $|0\rangle$
and $|1\rangle$, which is given by the Pauli matrix $\sigma_x$,
\begin{eqnarray}
\left(
\begin{array}{c}
|0\rangle\\
|1\rangle
\end{array}
\right) 
\rightarrow
\left(
\begin{array}{c}
|1\rangle\\
|0\rangle
\end{array}
\right) 
=
\sigma_x
\left(
\begin{array}{c}
|0\rangle\\
|1\rangle
\end{array}
\right).
\label{eq5:not}
\end{eqnarray}

An isolated Majorana zero mode also appears at each end point of a
1D topological superconductor \cite{Kitaev2001}. One can
exchange such Majorana zero modes in a
network of 1D topological superconductors \cite{Alicea2011,
Sau2011, Kotetes2013, Sau2010b, Halperin2012, Hyart2013, Kraus2013,
Liu2013, Chiu2015, Amorim2015}. 
Again, the exchange operators of those Majorana zero modes are given by $U_{ij}$
in Eq. (\ref{eq4:braid_U}), and the Majorana zero mode at the end point is also a
non-Abelian anyon.

The non-Abelian anyon is expected to have an application in quantum
computing \cite{Kitaev2003, Nayak2008}, because the states 
$|0\rangle$ and $|1\rangle$ defined by a pair of 
Majorana zero modes work as a nonlocal qubit.
In general, if there are $2N$ Majorana zero modes, one can define $N$
independent creation operators, which define $N$ qubits.
Equation (\ref{eq5:not}) implies that the process in
Fig. \ref{fig5:braiding} gives a NOT gate for the qubits.

\section{Routes to topological superconductivity}
\label{sec:route}

Recently, much efforts have been paid to find ways to realize topological superconductivity.
Here we summarize possible routes to topological superconductivity. 

\subsection{Odd-parity superconductors}
\label{sec:odd_parity}

A promising ground for topological superconductivity is the 
spin-triplet (or more precisely, odd-parity) pairing state.
For instance, 2D spinless chiral $p$-wave
superconductors have a nonzero bulk Chern number, $Ch=1$ (or
$Ch=-1$), supporting a gapless chiral Majorana edge mode on the
boundary \cite{Read-Green}. Their vortices host a single Majorana zero mode
in the core, which obeys the non-Abelian statistics \cite{Read-Green}. 
Also, the spin-triplet $^3$He-B phase is known to be a 3D
topological superfluid hosting gapless helical Majorana fermions on its
surface \cite{Hara1986, Chung09, Nagato2009,Shindou10}.

In general, to clarify topological superconductivity, one needs to either 
evaluate topological numbers or
examine the boundary states. However, in the case of
odd-parity superconductors, one can judge the topological nature just 
from the information of the Fermi surface, independently
of the details of gap functions. 
This result follows from the following theorem \cite{Sato2009, Sato2010, Fu-Berg2010}.

\begin{itemize}
\item An odd-parity superconductor is a topological superconductor if the Fermi surface 
encloses an odd number of time-reversal-invariant momenta in the Brillouin zone. 
\item The number of Fermi surfaces are counted as follows:
For a time-reversal-invariant superconductor, the spin-degeneracy of the Fermi surfaces 
is neglected. Each spin-degenerate Fermi surface is counted as a single Fermi surface.
On the other hand, for a time-reversal-breaking superconductor, a spin-degenerate 
Fermi surface is counted as a pair of Fermi surfaces. 
\end{itemize}

Here the time-reversal-invariant momentum ${\bm k}=\Gamma_i$ is defined as a
momentum satisfying $\Gamma_i=-\Gamma_i+{\bm G}$ with a
reciprocal lattice vector ${\bm G}$. The $\Gamma_i$'s in two and three dimensions 
are illustrated in Fig. \ref{fig:bz}.

This theorem is a consequence of the relations between the parity of
topological numbers and the sign of the eigenvalues of the normal-state Hamiltonian \cite{Sato2010}. 
For time-reversal-breaking superconductors, the relations in $d$-dimensions are 
\begin{eqnarray}
&&(-1)^{\nu_{\rm 1D}}=\prod_{\alpha, i=1,2}{\rm sgn}\,
\varepsilon_{\alpha}(\Gamma_i), \quad (\mbox{for $d$=1}),
\nonumber\\
&&(-1)^{Ch}=\prod_{\alpha, i=1,2,3,4}{\rm sgn}\,
\varepsilon_{\alpha}(\Gamma_i), \quad (\mbox{for $d$=2}),
\label{eq5:criterion}
\end{eqnarray}
where $\varepsilon_{\alpha}({\bm k})$ is the $\alpha$-th
eigenvalue of the normal-state Hamiltonian ${\cal E}({\bm k})$ in Eq. (\ref{eq3:general_BdG}), and 
the product in the right hand side with the running variable $(\alpha, i)$ is taken for all eigenvalues of
${\cal E}({\bm k})$ and all time-reversal-invariant momenta in each dimension.
There are similar relations for time-reversal-invariant superconductors,
\begin{eqnarray}
&&(-1)^{\nu_{\rm 1D}^{\cal T}}=\prod_{\alpha, i=1,2}{\rm sgn}\,
\varepsilon^{\rm I}_{\alpha}(\Gamma_i), \quad (\mbox{for $d$=1}),
\nonumber\\
&&(-1)^{Ch}=\prod_{\alpha, i=1,2,3,4}{\rm sgn}\,
\varepsilon^{\rm I}_{\alpha}(\Gamma_i), (\mbox{for $d$=2}),
\nonumber\\
&&(-1)^{\nu_{\rm 3dw}}=\prod_{\alpha, i=1,\dots,8}{\rm sgn}\,
\varepsilon^{\rm I}_{\alpha}(\Gamma_i), (\mbox{for $d$=3}), 
\label{eq5:criterion2}
\end{eqnarray}
where the product of the eigenvalues of ${\cal E}({\bm k})$ is taken only
for one of each Kramers pair $(\varepsilon^{\rm I}_{\alpha}({\bm k}), 
\varepsilon^{\rm II}_{\alpha}({\bm k}))$. 

Since the Fermi surface of the $\alpha$-th band is a surface on which the
momentum ${\bm k}$ satisfies $\varepsilon_{\alpha}({\bm k})=0$ 
(or $\varepsilon^{\rm I}_{\alpha}({\bm k})=0$), 
when the Fermi surface encloses $\Gamma_i$, 
$\varepsilon_{\alpha}(\Gamma_i)$ (or $\varepsilon^{\rm I}_{\alpha}(\Gamma_i)$) is negative.
Therefore, if the Fermi surface encloses an odd number of time-reversal-invariant 
momenta in the Brillouin zone, the right hand side of
Eq. (\ref{eq5:criterion}) or ({\ref{eq5:criterion2}) becomes negative.
As a result, the corresponding topological number in the left hand side 
becomes non-zero, which implies the realization of topological superconductivity. 

Since a topological phase transition may occur only when the gap of the
system closes, the above theorem implies that there arises a gap closing
when the number of the Fermi surfaces enclosing $\Gamma_i$ changes.
This situation can be confirmed directly in a single-band superconductor
described by the $4\times 4$ Hamiltonian Eq. (\ref{eq3:Hamiltonian_matrix}).
As shown in Sec. \ref{sec3:pairing_symmetry}, 
an odd-parity pair potential in a single-band superconductor is
spin-triplet,
$\Delta({\bm k})=i\Delta_0{\bm d}({\bm k})\cdot{\bm s}s_y$, 
which vanishes at $\Gamma_i$. Thus, at $\Gamma_i$, the Hamiltonian becomes
\begin{eqnarray}
{\cal H}_{4\times 4}(\Gamma_i)=
\left(
\begin{array}{cc}
\varepsilon(\Gamma_i) & 0\\
0 & -\varepsilon^t(\Gamma_i)
\end{array}
\right),
\end{eqnarray}
and hence the quasiparticle spectrum reduces to that of 
the normal-state Hamiltonian $\varepsilon(\Gamma_i)$. 
Therefore, when the number of the Fermi surfaces enclosing $\Gamma_i$
changes, a gap closing occurs at $\Gamma_i$. 

To appreciate the above theorem, let us consider a Fermi surface surrounding the $\Gamma$ point in
Fig. \ref{fig:fermi}(a). If the normal state preserves both time-reversal and inversion
symmetries, the Fermi surface is spin-degenerate due to the Kramers
theorem. From the above theorem, one can judge that any time-reversal-invariant 
odd-parity superconductivity realized on this Fermi surface is topological, irrespective of the 
details of the pair potential.
Indeed, assuming the gap function of the $^3$He-B phase, one can confirm
the existence of a gapless surface state as evidence for topological
superconductivity \cite{Sato2009}.
Furthermore, even for the nodal pair potential of the planar phase of $^3$He, one obtains
a topologically protected gapless mode.
The latter property is ensured by the mod-2 winding number discussed in Sec. \ref{sec4:e}.

Next, we consider a quasi-2D Fermi surface illustrated in Fig. \ref{fig:fermi}(b). 
In this case, the Fermi surface encloses two time-reversal-invariant
momenta, and hence the theorem tells us nothing about the topology in 3D.
However, one can argue for {\it weak} topological superconductivity defined in
lower dimensions. If one considers the 2D Brillouin zones defined by $k_z=0$
and $k_z=\pi$ in Fig. \ref{fig:fermi}(b), the Fermi surface encloses a single
time-reversal-invariant momentum in each 2D Brillouin zone. 
Thus, if a time-reversal-invariant odd-party pairing is realized on the
Fermi surface, the theorem tells us that the system is at least a weak
topological superconductor. 

\begin{figure}[tb]
\centering
\includegraphics[width=0.8\columnwidth]{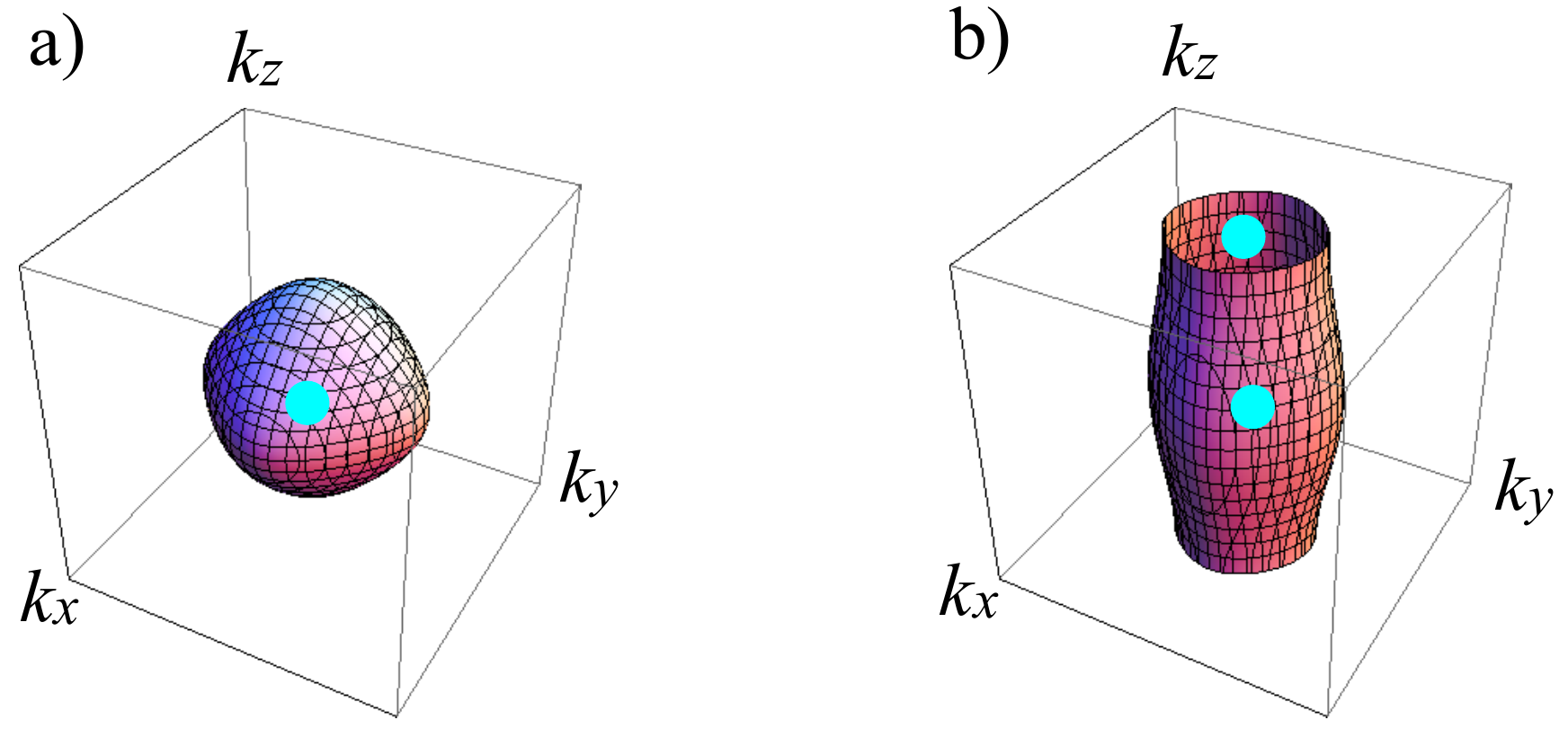}
\caption
{Time-reversal-invariant momenta enclosed by a Fermi surface. (a) 3D
Fermi surface. (b) Quasi-2D Fermi surface. Blue dots mark the time-reversal-invariant momenta.
} 
\label{fig:fermi}
\end{figure}

When an odd-parity Cooper pair breaks time-reversal symmetry, a spin-degenerate Fermi
surface is counted as two different Fermi surfaces in the theorem. 
Since the Fermi surfaces in ordinary superconductors are spin-degenerate, they always surround 
the time-reversal invariant momenta even times, and the above theorem alone is not useful for 
judging the topological nature. In many cases, however, one can examine the topology 
by employing additional symmetry considerations. 
For instance, when the spin-orbit coupling is so small that a superconductor has 
an approximate uniaxial spin-rotation symmetry with respect to, say, the $z$-axis,
one can use it for examining the topology. In this case, upon evaluating a topological number, 
one can completely neglect the spin-orbit coupling which breaks uniaxial spin-rotation symmetry, 
because such a small perturbation does not change the topological nature.
After neglecting the spin-orbit coupling, the BdG Hamiltonian commutes with the spin operator $S_z$, 
so that it becomes block-diagonal in the eigen basis of $S_z$.
As a result, the spin-degenerate Fermi surface is decomposed into a
nondegenerate Fermi surface in each spin sector. For the Fermi surface in
Fig. \ref{fig:fermi}(a), each spin sector has a single Fermi surface enclosing the
$\Gamma$ point, and one can identify topological superconductivity. 
Actually, the spin-Chern number on the $k_z=0$ plane can be evaluated as
\begin{eqnarray}
(-1)^{Ch_{\uparrow}}=\prod_{\alpha,i}\varepsilon^{\uparrow}
_{\alpha}(\Gamma_i), 
\quad
(-1)^{Ch_{\downarrow}}=\prod_{\alpha,i}\varepsilon^{\downarrow}
_{\alpha}(\Gamma_i), 
\label{eq4:spin}
\end{eqnarray}
where $\varepsilon_{\alpha}^{\uparrow}$ ($\varepsilon_{\alpha}^{\downarrow}$) 
is the $\alpha$-th eigenvalue of the normal-state Hamiltonian in the $S_z=1$
($S_z=-1$) sector, and the products with respect to $i$ are taken for all four
time-reversal-invariant momenta on the $k_z=0$ plane.

\begin{figure}[tb]
\centering
\includegraphics[width=0.45\columnwidth]{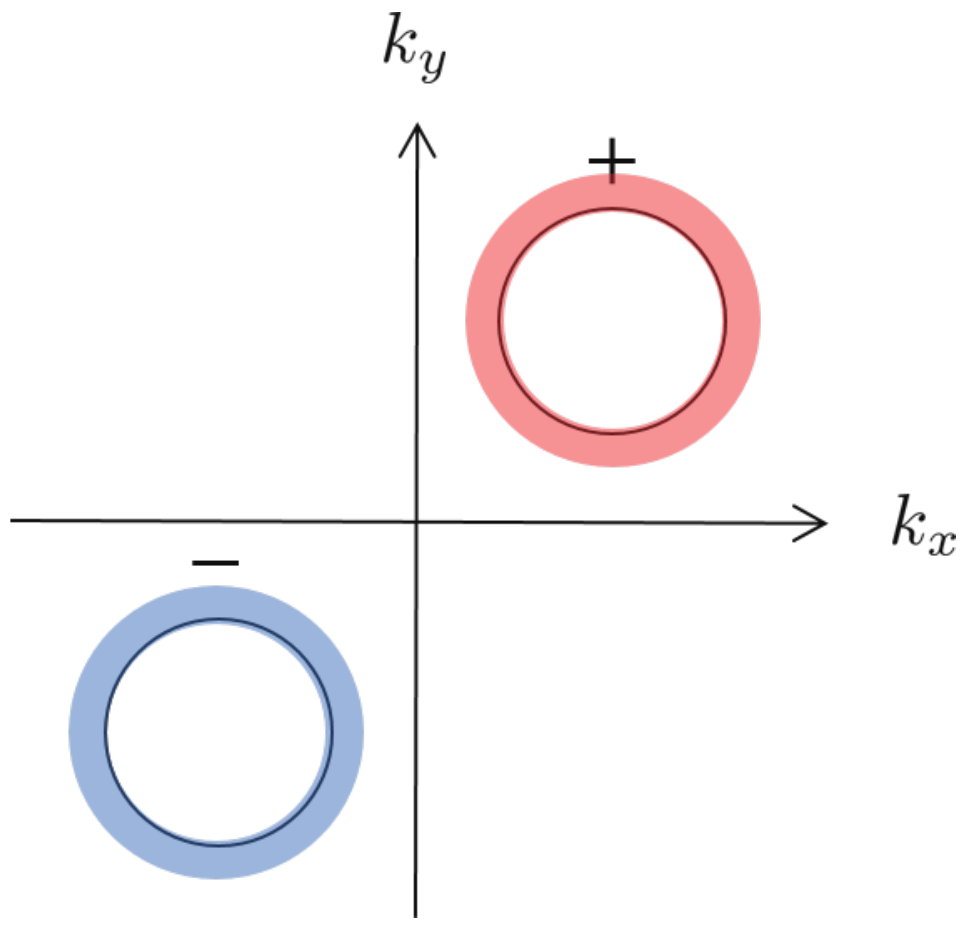}
\caption
{A topologically trivial odd-parity superconductor. Each Fermi surface
realizes an $s$-wave pairing state.
} 
\label{fig:spm}
\end{figure}

For a superconductor with a strong spin-orbit coupling, one can use
mirror-reflection symmetry instead of spin-rotation symmetry. On a mirror-invariant plane in the momentum space, 
the mirror symmetric BdG Hamiltonian commutes with the mirror operator, 
so that it becomes block-diagonal in the eigen basis of the mirror operator.
In such a situation, the mirror Chern number can be defined for each mirror eigensector.
Since the spin degeneracy of the Fermi surface is now decomposed into two
different mirror eigensectors, one can use the Fermi surface criterion to
evaluate the mirror Chern number: By replacing the spin Chern number and the spin sector with the
mirror Chern number and the mirror eigensector in Eq. (\ref{eq4:spin}), 
one obtains the corresponding formulas for the mirror Chern numbers on the $k_z=0$ plane. 
For the Fermi surface in Fig. \ref{fig:fermi}(a), there is only a single Fermi surface enclosing the
$\Gamma$ point in each mirror eigensector on the mirror invariant plane at $k_z=0$, and thus 
the mirror Chern numbers are found to be nonzero. 

The above examples imply that odd-parity superconductors are
intrinsically topological. Unless no Fermi surface encloses $\Gamma_i$ or each
Fermi surface reduces to an $s$-wave pairing state as illustrated in
Fig. \ref{fig:spm}, an odd-parity pairing state leads to topological
superconductivity.

\subsection{Superconducting topological materials}

Traditionally, spin-triplet or odd-parity superconductivity has been
explored in strongly correlated electron systems \cite{Sigrist-Ueda1991}.
However, it has been discussed recently that systems with a strong
spin-orbit coupling tend to realize odd-parity pairing states.
Here we show the mechanism of odd-parity pairing due to spin-orbit
coupling. 

\subsubsection{Topological insulators}
\label{sec:STM_TI}

\begin{figure}[b]
\centering
\includegraphics[width=0.98\columnwidth]{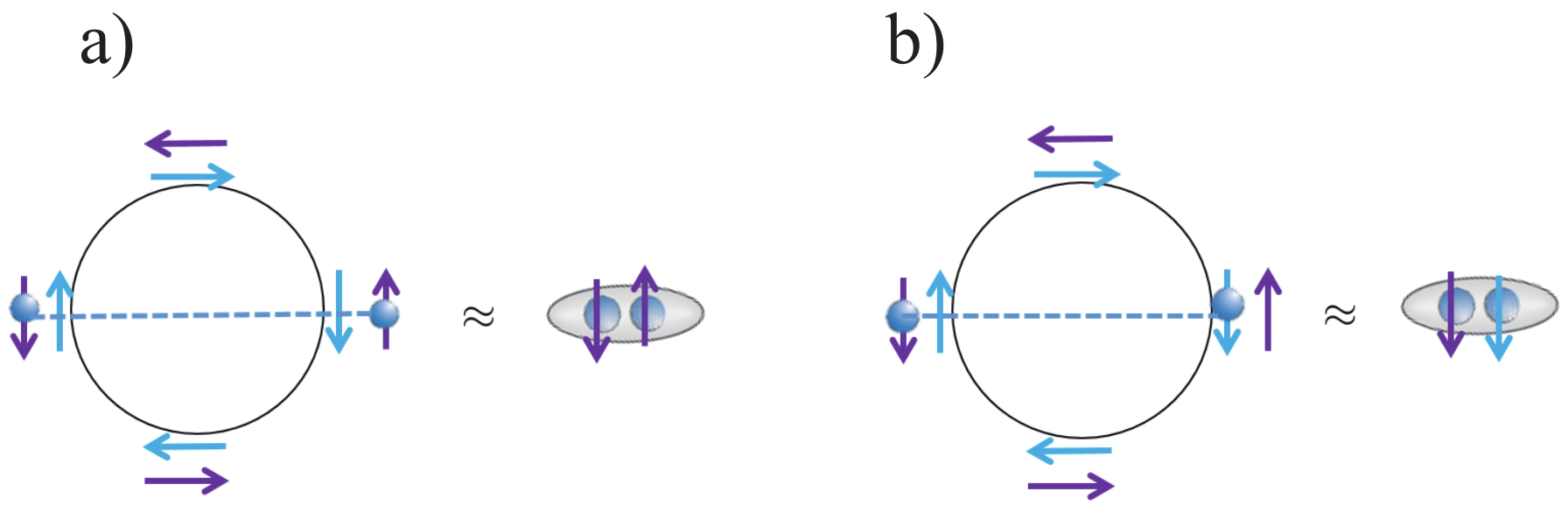}
\caption
{Possible pairing symmetries of superconducting topological insulators.
(a) Intra-orbital pairing. (b) Inter-orbital pairing. Electrons connected
by the dotted lines form Cooper pairs.} 
\label{fig:SCTI2}
\end{figure}

\begin{table*}[tb]
\begin{center}
\begin{tabular}{lcccl}
\hline
Pair potential & Representation & Parity & Spin& Energy gap \\
\hline
$\Delta_1=i\Delta_0 s_y$ & $A_{1g}$ & Even & \,Singlet \,\, & Isotropic full gap \\ 
$\Delta_2=i\Delta_0 \sigma_y s_z s_y$ & $A_{1u}$ & Odd & \,Triplet \,\, & Anisotropic full gap \\ 
$\Delta_3=i\Delta_0 \sigma_z s_y$ & $A_{2u}$ & Odd & \,Singlet \,\, & Point nodes at poles, orbital triplet \\ 
$\Delta_{4}=i\Delta_0 \sigma_y s_x s_y$ & $E$ & Odd & \,Triplet \,\, & Point nodes or gap minima on the equator 
\\\hline
\end{tabular}
\caption{Possible pair potentials in carrier-doped Bi$_2$Se$_3$
 \cite{Fu-Berg2010}. The pair potentials are momentum-independent and satisfy
 Eq. (\ref{eq3:antisymmetric_condition}). Even (odd) parity means 
 $P\Delta_iP^t=\Delta_i$ ($P\Delta_iP^t=-\Delta_i$) with inversion
 operator $P$ defined in Eq. (\ref{eq2:inversion_BiSe}).
}
\label{table4:pair_potential}
\end{center} 
\end{table*}

In Sec. \ref{sec2:TI}, we have considered the prototypical topological
insulator Bi$_2$Se$_3$ having the bulk Hamiltonian 
\begin{eqnarray}
H_{\rm TI}({\bm k})=(m_0-m_1{\bm
k}^2)\sigma_x+v_zk_z\sigma_y+v\sigma_z(k_xs_y-k_ys_x).
\nonumber\\ 
\label{eq4:TIHamiltonian}
\end{eqnarray}
Let us now consider possible superconducting states of this material
\cite{Fu-Berg2010}.
Since an insulator cannot superconduct, we should first dope carriers and make the system a metal.
Correspondingly, we have a non-zero chemical potential $\mu$ in ${\cal E}({\bm k})$,
\begin{eqnarray}
H_{\rm TI}({\bm k})\rightarrow {\cal E}({\bm k})=H_{\rm TI}({\bm k})-\mu,
\label{eq4:Bi2Se3}
\end{eqnarray}
which enters the BdG Hamiltonian Eq. (\ref{eq3:general_BdG}),
\begin{eqnarray}
{\cal H}({\bm k})=\left(
\begin{array}{cc}
{\cal E}({\bm k}) & \Delta({\bm k})\\
\Delta^{\dagger}({\bm k}) & -{\cal E}^t(-\bm k)
\end{array}
\right). 
\end{eqnarray}
The doping gives rise to a Fermi surface surrounding the $\Gamma$ point.
Note that the strong spin-orbit coupling in topological insulators
induce a spin texture on the Fermi surface, as shown in Fig. \ref{fig:SCTI}
for $k_z=0$ plane. 
The spin-orbit coupling term $v\sigma_z(k_xs_y-k_ys_x)$ 
is similar to the Rashba term, and hence it results in a helical spin structure.
However, since the system as a whole preserves inversion symmetry, 
the two orbitals (specified by $\sigma_z$) have opposite helicity. 
As we will see below, this structure makes it possible to host odd-parity
pairing states, without relying on spin-mediated pairing interactions. 

If the attractive interaction is the strongest between
electrons in the same orbital, Cooper pairs are formed
between electrons having antiparallel spins, as illustrated in
Fig. \ref{fig:SCTI2}(a); this means that an ordinary spin-singlet $s$-wave 
pairing state is realized. 
On the other hand, if the attractive interaction primarily acts between electrons in different
orbitals, Cooper pairs will have a parallel spin structure as shown in Fig. \ref{fig:SCTI2}(b); 
in this case a spin-triplet pairing state is realized, although the pairing interaction itself is 
independent of the spin. 
For the carrier-doped Bi$_2$Se$_3$, one can reasonably assume that
the attractive interaction is momentum-independent, since there is no 
strong electron correlations. Under this assumption, the possible pair potentials are 
classified into four types of irreducible representations of the $D_{3d}$ point group relevant to Bi$_2$Se$_3$. 
The matrix forms of the possible pairings $\Delta_1$, $\Delta_2$, $\Delta_3$, and
$\Delta_4$, as well as their properties, are summarized in Table \ref{table4:pair_potential}. 
Among them, spin-triplet Cooper pairs are realized in $\Delta_2$ and $\Delta_4$,
both of which are odd-parity pairings. 

Early microscopic calculations based on the simple model Hamiltonian Eq.
(\ref{eq4:TIHamiltonian}) supported the pair potential $\Delta_2$
\cite{Fu-Berg2010,Brydon2014} rather than $\Delta_4$, but later the possibility of
$\Delta_4$ was also pointed out \cite{Fu2014}.
Whereas the model Hamiltonian Eq. (\ref{eq4:TIHamiltonian}) is fully uniaxial symmetric
along the $z$-axis, the actual crystal structure of Bi$_2$Se$_3$ is
invariant only under its subgroup of three-fold rotation. This effect can be
taken into account by adding the so-called warping term
\begin{eqnarray}
H_{\rm warp}({\bm k})=-i\lambda(k_+^3-k_-^3)\sigma_z s_z,
\end{eqnarray}
where $k_{\pm}=k_x\pm ik_y$.
This term makes the spin texture of Fig. \ref{fig:SCTI2} slightly tilted
in the $k_z$ direction, stabilizing the $\Delta_4$ pair potential in the
phase diagram. 
The pair potential $\Delta_4$ belongs to the $E$ representation of
$D_{3d}$, which spontaneously breaks three-fold rotation
symmetry along the $z$-axis in a manner similar to a nematic order;
therefore, the $\Delta_4$ state is called a {\it nematic superconductor}.

For superconducting topological insulators, the bulk electric properties
\cite{Hashimoto2013, Hashimoto2014, Nagai2012}
as well as the surface tunneling spectra \cite{Hao-Lee2011, Sasaki2011,
Yamakage2012, Hsieh2012,Takami2014,Mizushima2014} have been theoretically calculated.
As discussed in detail in Sec. \ref{sec:CBS}, recent experiments confirmed that the carrier-doped
Bi$_2$Se$_3$ is a nematic topological superconductor. 

Similar odd-parity superconducting states were also proposed for
interacting two-layer Rashba systems \cite{Nakosai2012}.
Because the Rashba coupling in each layers is given by
$\sigma_z(k_xs_y-k_ys_x)$ with the layer index $\sigma_z=\pm 1$, which has the
same form as the spin-orbit coupling term in Eq. (\ref{eq4:TIHamiltonian}), 
interlayer Cooper pairs naturally lead to odd-parity superconductivity. 
Such interacting two-layer Rashba systems may be fabricated in heterostructures of semiconductors and oxides.

\subsubsection{Weyl semimetals}

It is possible that carrier-doped Weyl semimetals show superconductivity at low
temperature \cite{Cho2012,Bednik2015,Zhou2016}. 
As already mentioned, either time-reversal or inversion symmetry must be broken 
to realize Weyl semimetals. 
At the same time, however, at least one of them must be preserved to realize
a bulk superconducting state --- if both are broken, electrons with momenta ${\bm k}$ and $-{\bm k}$
are not at the same energy, and thus they cannot form a Cooper pair.
Therefore, in the following we assume that either time-reversal or inversion symmetry is preserved.

\vspace{2mm}
{\bf Time-reversal-breaking Weyl semimetals --}
We first consider time-reversal-breaking Weyl semimetals which preserve
inversion symmetry. 
For a finite chemical potential, a Weyl node is described by
\begin{eqnarray}
H_{\rm W}({\bm k})=v({\bm k}-{\bm k}_0)\cdot {\bm s}-\mu. 
\label{eq6:Weyl}
\end{eqnarray}
By performing space inversion on Eq. (\ref{eq6:Weyl}), 
we obtain an anti-Weyl node described by
\begin{eqnarray}
H_{\bar{\rm W}}({\bm k})=-v({\bm k}+{\bm k}_0)\cdot {\bm s}-\mu. 
\label{eq6:antiWeyl}
\end{eqnarray}
The anti-Weyl node has an opposite Chern number as the Weyl
node described by Eq. (\ref{eq6:Weyl}), so the total Chern number is zero.
Thus, they form a minimal set of Weyl points satisfying the Nilsen-Ninomiya theorem. 
Expressing the Weyl (anti-Weyl) node with $\sigma_z=1$
($\sigma_z=-1$), 
the sum of the Hamiltonians Eqs. (\ref{eq6:Weyl}) and (\ref{eq6:antiWeyl}) becomes
\begin{eqnarray}
H_{{\rm W}+\bar{\rm W}}({\bm k})=v({\bm k}\sigma_z-{\bm k}_0\sigma_0)\cdot {\bm s}
-\mu\sigma_0, 
\label{eq6:WaW}
\end{eqnarray}
which transforms under inversion $P$ as
\begin{eqnarray}
PH_{{\rm W}+\bar{\rm W}}(-{\bm k})P^{-1}=H_{{\rm W}+\bar{\rm W}}({\bm
k}),
\quad 
P=\sigma_x. 
\end{eqnarray}

\begin{figure}[b]
\centering
\includegraphics[width=0.45\columnwidth]{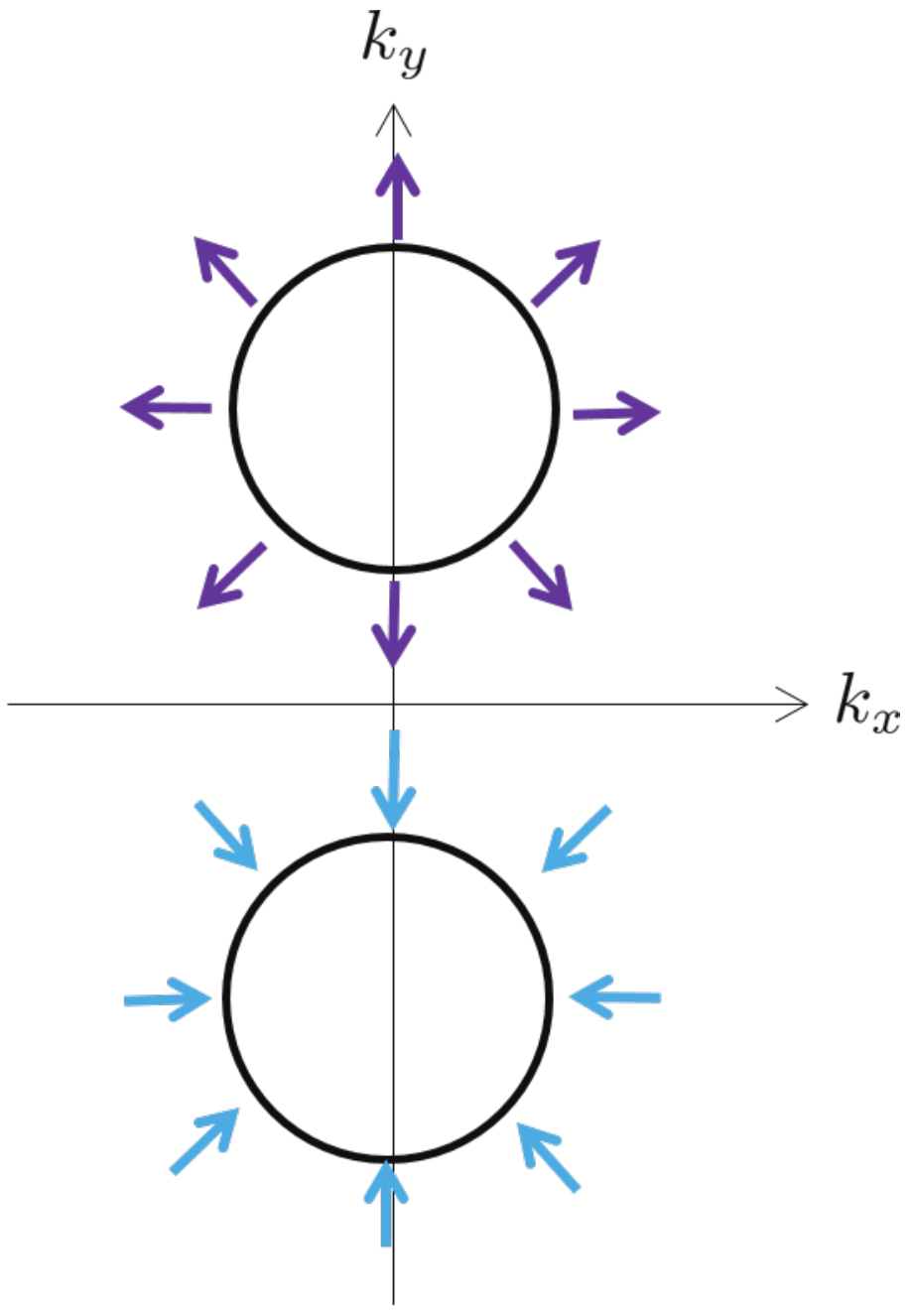}
\caption
{Spin texture on the Fermi surface of a doped Weyl semimetal.
} 
\label{fig:Weyl}
\end{figure}

Now we examine what kind of superconducting state naturally arises in the above system.
First, it should be noted that a spin texture appears on the Fermi
surface due to the spin-orbit coupling.
In Fig. \ref{fig:Weyl}, we illustrate the spin texture obtained from
Eq. (\ref{eq6:WaW}). 
From this structure, we find the following important feature.
\begin{description}
\item[I]\mbox{}\\ 
If the total momentum of a Cooper pair is zero, the Coopers
pair must be formed between electrons with parallel spins. In
particular, the Cooper pairs cannot be in a spin-singlet $s$-wave pairing state. 
\end{description}

To better understand the situation, let us consider the simplest gap functions
with a constant amplitude $\Delta_0$.
From the Fermi statistics, the gap functions must satisfy $\Delta({\bm
k})=-\Delta^t(-{\bm k})$, and there are six such possibilities: 
\begin{eqnarray}
&&\Delta_0 \sigma_0is_y,\quad \Delta_0 \sigma_xis_y, \quad \Delta_0 \sigma_zis_y 
\label{eq6:gapfunction-s}
\end{eqnarray}
and
\begin{eqnarray}
&&
\Delta_0 \sigma_yis_xs_y, \quad
\Delta_0 \sigma_yis_ys_y, \quad 
\Delta_0 \sigma_yis_zs_y.
\label{eq6:gapfunction}
\end{eqnarray}
It is useful to compare these six gap functions 
with those in Eqs. (\ref{eq3:spin_singlet}) and (\ref{eq3:spin_triplet}).
The first three gap functions in Eq. (\ref{eq6:gapfunction-s}) are of the 
Eq. (\ref{eq3:spin_singlet}) type 
and correspond to spin-singlet pair potentials, while the three in 
Eq. (\ref{eq6:gapfunction}) are of the Eq. (\ref{eq3:spin_triplet}) 
type and correspond to spin-triplet pair potentials; also, those three transform
with $P$ as
\begin{eqnarray}
P\Delta({\bm k})P^t=-\Delta(-{\bm k}), 
\end{eqnarray}
and hence their parity is odd. 
The ${\bm d}$-vector of the three spin-triplet gap functions in 
Eq. (\ref{eq6:gapfunction}) point to the $x$, $y$, and $z$-directions, respectively.
Since the Cooper pairs cannot be spin-singlet in doped Weyl semimetals as already 
mentioned, only those in Eq. (\ref{eq6:gapfunction}) can be realized, 

Now let us examine the spin structures of the three spin-triplet gap functions in Eq. (\ref{eq6:gapfunction}).
In general, electrons forming spin-triplet Cooper pairs take an anti-parallel
spin configuration in the direction along the ${\bm d}$-vector, while they take
a parallel spin configuration in the direction normal to the ${\bm
d}$-vector. However, the spin texture of a Weyl semimetal 
does not allow Cooper pairs to take such spin configurations everywhere on the Fermi surface.
For instance, in the case of $\Delta_0is_xs_y$ gap function, electrons forming Cooper
pairs have anti-parallel spins in the $x$-direction and
parallel spins in the $y$- and $z$-directions. On the other hand, the
spin texture of Weyl semimetals only allows parallel spin
configurations. Therefore, $\Delta_0is_xs_y$ is not consistent with the
spin texture on the $k_x$-axis, implying the existence of point nodes on
the $k_x$-axis. In a similar manner, one can see that the other two
spin-triplet gap functions $\Delta_0is_ys_y$ and $\Delta_0is_zs_y$
also have point nodes on the $y$- and $z$-axes, respectively.
This property is summarized as follows.
\begin{description}
\item[I']\mbox{}\\ 
As long as the Cooper pairs of a time-reversal-breaking Weyl semimetal do
not have any finite momentum to break the
translational symmetry, the superconducting state must realize
an odd-parity, spin-triplet gap function with point nodes. 
\end{description} 

One might wonder if the above two properties are specific to the simple
model considered here; in this regard, they actually represent general properties derived
from the topological nature of Weyl semimetals. 
For example, the fact that an $s$-wave pairing cannot be realized in a
time-reversal-breaking Weyl semimetal is a natural consequence of
the following property of superconductors:
\begin{description}
\item[II]\mbox{}\\
It is impossible to form an even-parity Cooper pair on any
spin-nondegenerate Fermi surface if the Cooper pair keeps the
(lattice) translational symmetry 
\end{description}
This property can be derived by using a topological argument combining inversion
and particle-hole symmetries. 
Consider an inversion-symmetric system 
\begin{eqnarray}
P{\cal E}(-{\bm k})P^{-1}={\cal E}({\bm k}) 
\end{eqnarray} 
and its BdG Hamiltonian having the (lattice) translation symmetry
\begin{eqnarray}
{\cal H}({\bm k})=\left(
\begin{array}{cc}
{\cal E}({\bm k}) & \Delta({\bm k})\\
\Delta^{\dagger}({\bm k}) & -{\cal E}^t(-{\bm k})
\end{array}
\right). 
\end{eqnarray}
If the gap function $\Delta({\bm k})$ is even under inversion, 
\begin{eqnarray}
P\Delta(-{\bm k})P^t=\Delta({\bm k}), 
\end{eqnarray}
then the inversion operator for the BdG Hamiltonian is defined as
\begin{eqnarray}
\tilde{P}=\left(
\begin{array}{cc}
P & 0 \\
0 & P^*
\end{array}
\right), 
\end{eqnarray}
and the inversion symmetry reads $\tilde{P} {\cal H}(-{\bm k}) \tilde{P}^{-1}={\cal H}({\bm k})$.
The particle-hole symmetry is expressed by using the charge conjugation operator ${\cal C}$ as
\begin{eqnarray}
&&{\cal C}{\cal H}({\bm k}){\cal C}^{-1}=-{\cal H}({\bm k}),
\nonumber\\
&&{\cal C}=
\left(
\begin{array}{cc}
0 & 1 \\
1 & 0 
\end{array}
\right)K=\tau_x K, 
\end{eqnarray}
where $K$ is the complex conjugate operator. Combining these two symmetries, we obtain
\begin{eqnarray}
(\tilde{P}{\cal C})
{\cal H}({\bm k}) 
(\tilde{P}{\cal C})^{-1}={\cal H}({\bm k}).
\label{eq6:0dclassDeven}
\end{eqnarray}
Note that there is an identity $(\tilde{P}{\cal C})^2=1$, because 
$[\tilde{P}, {\cal C}]=0$ and $\tilde{P}^2={\cal C}^2=1$.
One can interpret Eq. (\ref{eq6:0dclassDeven}) to mean that the BdG Hamiltonian 
${\cal H}({\bm k})$ has zero-dimensional (0D) class D particle-hole
symmetry at each momentum ${\bm k}$. 
Hence, one can define the 0D ${\bm Z}_2$ index for each momentum.
This can be done in the following steps: First, we note that $P=P^{\dagger}$, since
$P^2=1$ and $P$ is unitary. Second, we use Eq. (\ref{eq6:0dclassDeven}) to
prove that the matrix ${\cal H}({\bm k})\tilde{P}\tau_x$ is antisymmetric, i.e.
\begin{eqnarray}
(\tilde{P}\tau_x)^{\dagger}{\cal H}^t({\bm k})=-{\cal H}({\bm
k})\tilde{P}\tau_x, 
\end{eqnarray}
which allows us to define the Pfaffian of ${\cal H}({\bm k})\tilde{P}\tau_x$. 
Finally, the 0D ${\bm Z}_2$ invariant is defined as
\begin{eqnarray}
(-1)^{\nu_{\rm 0d}({\bm k})}={\rm sgn}
\left[
\frac
{{\rm Pf}({\cal H}({\bm k})\tilde{P}\tau_x)}
{{\rm Pf}({\cal H}({\bm k}_0)\tilde{P}\tau_x)}
\right], 
\end{eqnarray}
where ${\bm k}_0$ is a reference momentum which can be taken to be the
$\Gamma$ point, for instance.

Now we evaluate the ${\bm Z}_2$ index $(-1)^{\nu_{\rm 0D}({\bm k})}$ 
at ${\bm k}$ other than on the Fermi surface. 
Since the superconducting gap is usually much smaller
than the Fermi energy, one can neglect the gap function $\Delta({\bm k})$
except on the Fermi surface, and thus the 0D ${\bm Z}_2$ index is
evaluated as
\begin{eqnarray}
(-1)^{\nu_{\rm 0d}({\bm k})}&=& 
{\rm sgn}
\left[
\frac
{{\rm det}({\cal E}({\bm k})P)}
{{\rm det}({\cal E}({\bm k}_0)P)}
\right]
\nonumber\\
&=&{\rm sgn}
\left[
\frac
{{\rm det}({\cal E}({\bm k}))}
{{\rm det}({\cal E}({\bm k}_0))}
\right].
\end{eqnarray}
Physically, ${\rm det}\,{\cal E}({\bm k})$ is a product of all band
energies (measured from  
the chemical potential) 
at ${\bm k}$. 
Since we are considering a 
spin-nondegenerate system, upon crossing the Fermi surface the energy of
only one band changes from negative to positive, and hence ${\rm
det}\,{\cal E}({\bm k})$ 
must change sign across the Fermi surface. 
Therefore, the ${\bm Z}_2$ index $(-1)^{\nu_{\rm 0d}({\bm k})}$ is
different between inside and outside the Fermi surface.
Since the topological number changes only when the gap of the system
closes, this means that there must be a gapless surface separating inside and
outside the Fermi surface. In other words, 
the Fermi surface must remain gapless, and thus no superconducting state is
realized for any even-parity gap function. When applied to Weyl semimetals
which always have spin-nondegenerate Fermi surfaces, the above result dictates 
that an even-parity superconducting state cannot be realized in Weyl semimetals.

It should be noted here that the above constraint does not apply to
an odd-parity gap function $\Delta({\bm k})$, which transforms non-trivially 
under inversion,
\begin{eqnarray}
P\Delta(-{\bm k})P^t=-\Delta({\bm k}). 
\label{eq6:oddparity}
\end{eqnarray}
However, combining $P$ with a U(1) gauge rotation, 
one can eliminate the minus sign in the right hand side of
Eq. (\ref{eq6:oddparity}). This means that the inversion symmetry is effectively 
restored under the combination of the original inversion and a U(1) gauge rotation.
The BdG Hamiltonian is invariant under this modified inversion $\tilde{P}'$, i.e.
$\tilde{P}'{\cal H}(-{\bm k})\tilde{P}^{'-1}={\cal H}({\bm k})$, with $\tilde{P}'$ given by
\begin{eqnarray}
\tilde{P}'= 
\left(
\begin{array}{cc}
P & 0 \\
0 & -P^*
\end{array}
\right).
\end{eqnarray}
Therefore, in a manner similar to the case of an even-parity gap function, one obtains 
\begin{eqnarray}
(\tilde{P}'{\cal C})
{\cal H}({\bm k}) 
(\tilde{P}'{\cal C})^{-1}={\cal H}({\bm k}) 
\label{eq6:0dclassCodd}
\end{eqnarray}
by combining the modified inversion with particle-hole symmetry. 
Nevertheless, the present situation is different from the even-parity case,
in that the operator $\tilde{P}'$ anticommutes with the charge
conjugation operator ${\cal C}$, i.e. $\{\tilde{P}', {\cal C}\}=0$.
As a result, we have $(\tilde{P}'{\cal C})^2=-1$.
The combination of this identity with Eq. (\ref{eq6:0dclassCodd}) implies that 
the system has the 0D class C symmetry at each ${\bm k}$, not class D. In class C, no
topological number is defined in zero dimension. Indeed,
Eq. (\ref{eq6:0dclassCodd}) is equivalent to 
\begin{eqnarray}
(\tilde{P}'\tau_x)^t {\cal H}^t({\bm k})
= {\cal H}({\bm k})(\tilde{P}'\tau_x), 
\end{eqnarray}
so that ${\cal H}({\bm k})(\tilde{P}'\tau_x)$ is symmetric. Hence, one cannot 
define the Pfaffian of ${\cal H}({\bm k})(\tilde{P}'\tau_x)$ nor the ${\bm Z}_2$ index.
Since there is no topological number to protect the system from gap opening in this case, 
an odd-parity superconducting gap is allowed by symmetry.

As for the generality of the conclusions we obtained for the simple model, 
the following property can also be derived by using a topological argument.
\begin{description}
\item[III]\mbox{}\\ 
If a superconducting state preserving translational symmetry is realized in an inversion-symmetric 
Weyl semimetal, it must have an odd-parity gap function with point nodes. 
Also, the superconducting state hosts gapless Andreev bound states forming Fermi arcs. 
\end{description}

We first show that the Fermi arcs of the Weyl semimetals remain
gapless in the superconducting state \cite{Bo2015}.
In the case of a time-reversal-breaking Weyl semimetal, inversion symmetry
must be preserved for Cooper pairs to from. However, inversion symmetry is explicitly broken at the surface
where the Fermi arcs reside. Hence, on a general surface which does not have an accidental two-fold 
rotation symmetry, an electronic state in the Fermi arc does not have a partner state with opposite 
momentum to form a Cooper pair at the same energy (see Fig. \ref{fig:fermiarc}). This means that the
constituent electrons of the Fermi arc cannot form Cooper pairs with zero total momentum, 
and therefore the Fermi arc remains gapless even in the superconducting state. The redundancy in 
the BdG Hamiltonian for electron and hole sectors leads to the appearance of a pair of gapless Fermi arcs 
of Bogoliubov quasiparticles on the surface. 

\begin{figure}[tb]
\centering
\includegraphics[width=0.55\columnwidth]{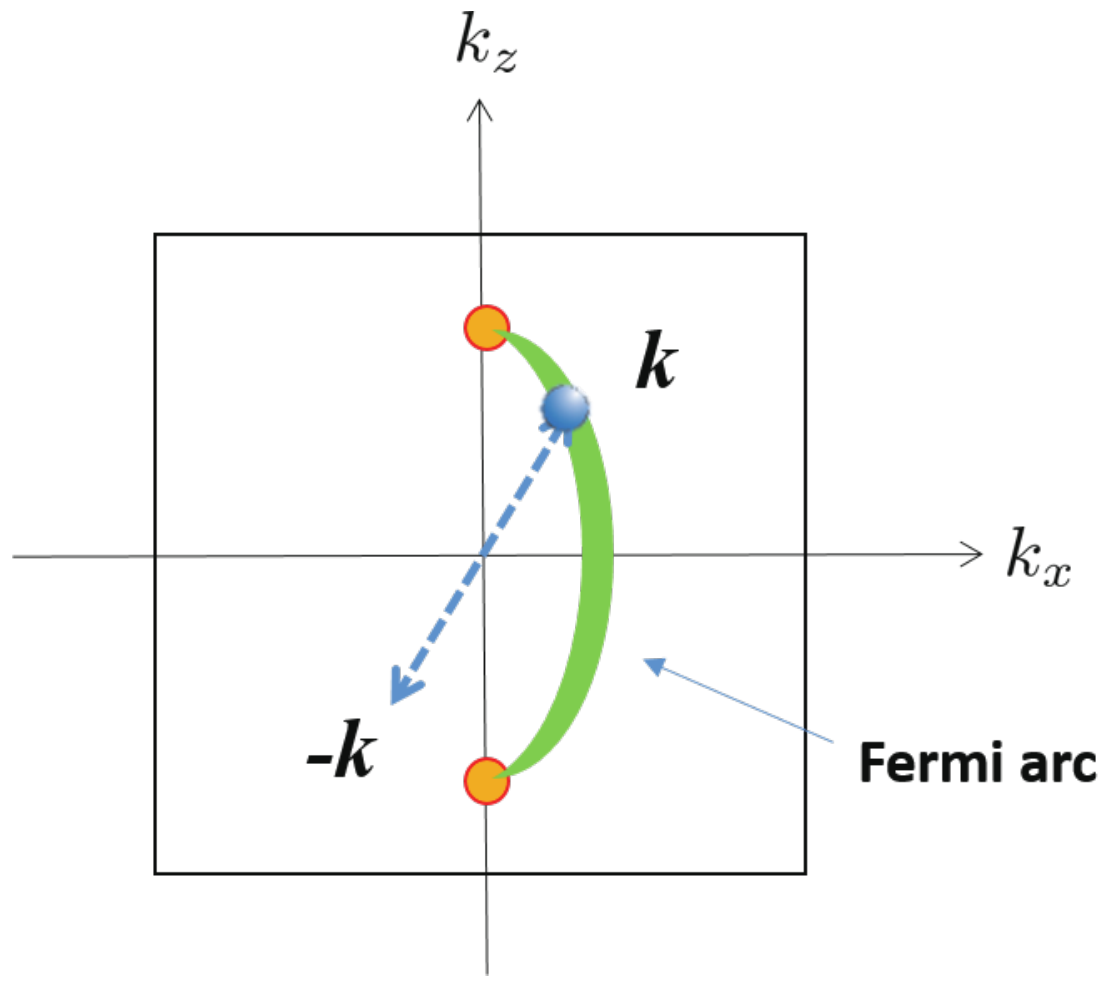}
\caption
{A surface Fermi arc in an inversion-symmetric Weyl semimetal.
There is no partner electron to form a Cooper pair. 
} 
\label{fig:fermiarc}
\end{figure}

The same conclusion can be obtained by calculating the Chern number in the
superconducting state \cite{Bo2015}.
In a Weyl semimetal illustrated in Fig. \ref{fig:Chern}, one can evaluate the Chern
number on two different planes $S_1$ and $S_2$ in the momentum space ($S_1$ and 
$S_2$ should not intersect the Fermi surfaces around the Weyl points).
As already mentioned, except on the Fermi surfaces, one can
neglect the gap function without changing the topological numbers; therefore, the
Chern number on $S_i$ ($i=1,2$) can be evaluated for the following Hamiltonian
\begin{eqnarray}
\left(
\begin{array}{cc}
{\cal E}({\bm k}) & 0 \\
0 & -{\cal E}^t(-{\bm k})
\end{array}
\right). 
\end{eqnarray} 
Since ${\cal E}({\bm k})$ gives the Chern number of the original Weyl
semimetal and the hole part ${\cal E}^t(-{\bm k})$ gives the same
Chern number as its electron counterpart, the Chern number on $S_i$ for the BdG
Hamiltonian is twice the original Chern number which is non-zero for either $i$ = 1 or 2.
Such a topological character of the bulk superconducting state guarantees the existence of 
gapless surface states consisting of Fermi arcs of electrons and holes. 

\begin{figure}[tb]
\centering
\includegraphics[width=0.55\columnwidth]{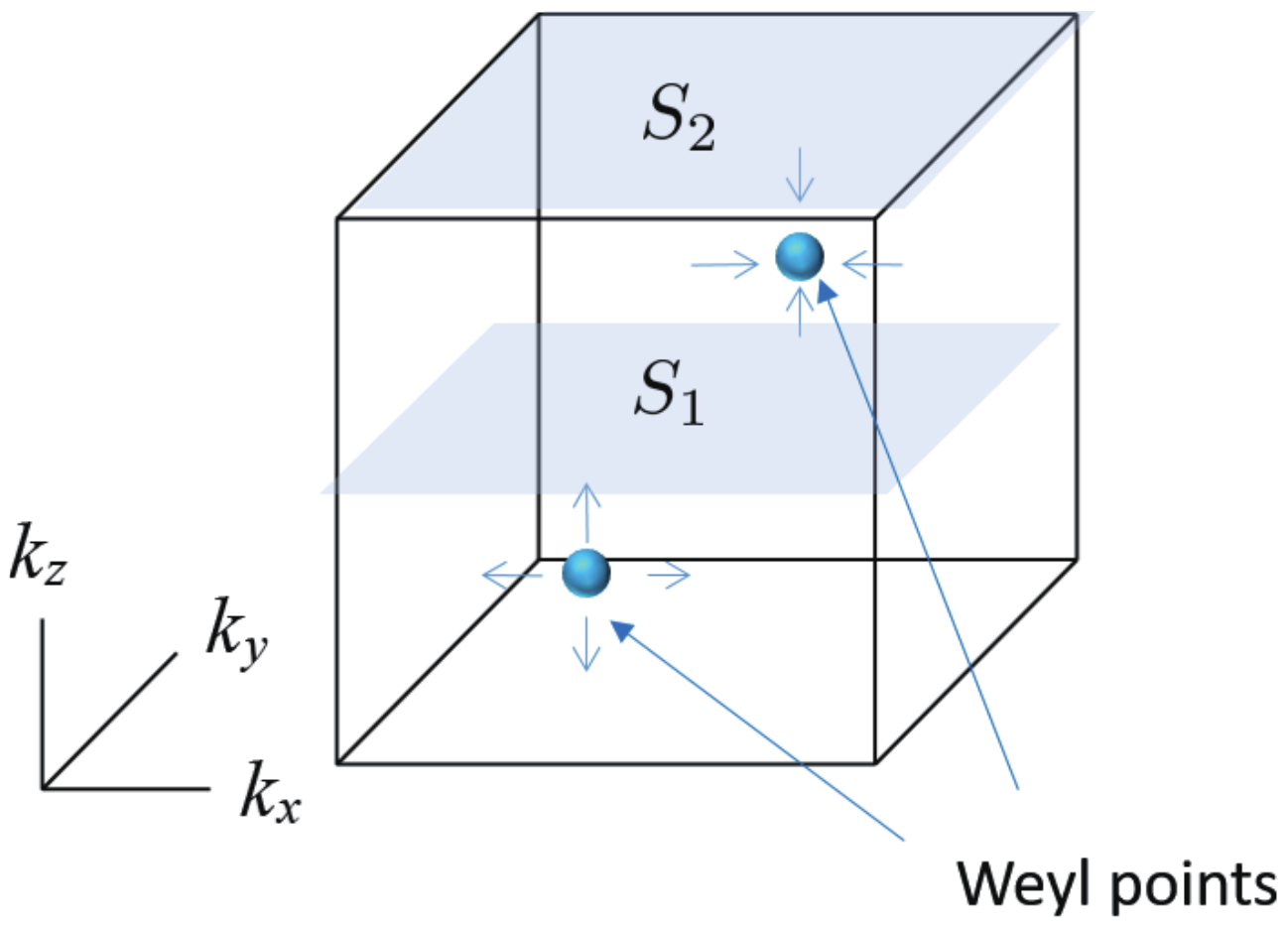}
\caption
{Weyl points in the Brillouin zone of a Weyl semimetal and the planes on which the Chern
number is evaluated.
} 
\label{fig:Chern}
\end{figure}

The analysis of the Chern number also gives us an insight into why the
superconducting state has point nodes \cite{Murakami-Nagaosa2003,
Bo2015, Li-Haldane2015}. 
When a Weyl point is located in-between the planes $S_1$ and $S_2$ as illustrated in 
Fig. \ref{fig:Chern}, the Chern number calculated on $S_1$ is different from that 
on $S_2$, and the difference is given by the Chern number associated with the Weyl point.
Therefore, the Chern number of the BdG Hamiltonian is also different between
$S_1$ and $S_2$, which means that there must be gapless points in the region
between $S_1$ and $S_2$, resulting in point nodes in the superconducting state. 
This argument also tells us that the number of point nodes in the superconducting state 
coincides with twice the Chern number associated with the Weyl point. 

\vspace{2mm}
{\bf Time-reversal-invariant Weyl semimetals --} 
Now we consider superconducting states in doped time-reversal-invariant
Weyl semimetals, which must break inversion symmetry. 
In contrast to inversion-symmetric Weyl semimetals, here we find no special reason
to expect unconventional superconducting states.

This difference comes from the fact that spin transforms differently under
inversion and time reversal. 
On one hand, spin does not change under inversion, and hence a Cooper pair formed 
between an electron and its inversion partner must take parallel spins; 
this means that a spin-singlet $s$-wave pairing does not occur in the inversion-symmetric 
case, as we have already seen.
On the other hand, spin flips under time reversal, and hence a Cooper pair formed 
between an electron and its time-reversal partner takes anti-parallel spins, which 
allows for conventional $s$-wave pairing. 

It is prudent to mention that if the pairing interaction is spin-dependent, one may have spin-triplet
Cooper pairs or other unconventional superconducting states even in the
time-reversal-invariant case. However, such a spin-dependent pairing interaction has been
known only for heavy-fermion systems with strong electron correlations.
Therefore, it is fair to say that time-reversal-invariant Weyl semimetals may become an
unconventional superconductor only when electron correlations are strong.

\subsubsection{Dirac semimetals}

Dirac semimetals are also an interesting platform for topological superconductivity
\cite{Kobayashi-Sato2015, Hashimoto2016}.
Upon carrier doping, Dirac semimetals host Fermi surfaces surrounding Dirac points.
As discussed in Sec. \ref{sec2:Dirac}, a gapless Dirac dispersion in a Dirac semimetal 
originates from bands (or orbitals) with different quantum numbers associated with a 
certain crystal symmetry.
While no orbital mixing is permitted in the symmetry-invariant momentum
subspace (which includes the Dirac point), the Fermi surface at a finite doping does 
not coincide with such an invariant subspace, and hence an orbital mixing is allowed on the Fermi surface.
As a result, Cooper pairing between electrons with different quantum numbers is possible. 
In general, such an inter-orbital Cooper pair is unconventional.

As an example, we consider Cd$_3$As$_2$ mentioned in Sec. \ref{sec2:Dirac}. 
In the model Hamiltonian for this material, Eq. (\ref{eq6:Dirac}), there are
orbital mixing terms proportional to $\sigma_x$ or $\sigma_y$, except on the
$k_z$-axis. Therefore, the Fermi surfaces around the Dirac points naturally
exhibit an orbital mixing, and Cooper pairing between different orbitals are allowed. 
Since the inversion operator is given by $P=\sigma_z$, such orbital-mixed
Cooper pairs should have odd parity. Indeed, it was shown theoretically 
\cite{Kobayashi-Sato2015, Hashimoto2016} that an odd-parity pairing 
can be realized in Cd$_3$As$_2$ depending on the parameters of the 
attractive interactions given in the form
\begin{eqnarray}
H_{\rm int}({\bm x})=-U\left[n_s^2({\bm x})+n_p^2({\bm
x})\right]-2Vn_s({\bm x})n_p({\bm x}), 
\nonumber\\
\end{eqnarray}
where $U$ and $V$ are the intra- and inter-orbital interactions, respectively, and 
$n_\alpha$ ($\alpha=s, p$) is the density operator for the $\alpha$-orbital. 
The obtained phase diagram is shown Fig. \ref{fig4:Dirac}.

\begin{figure}[tb]
\centering
\includegraphics[width=0.5\columnwidth]{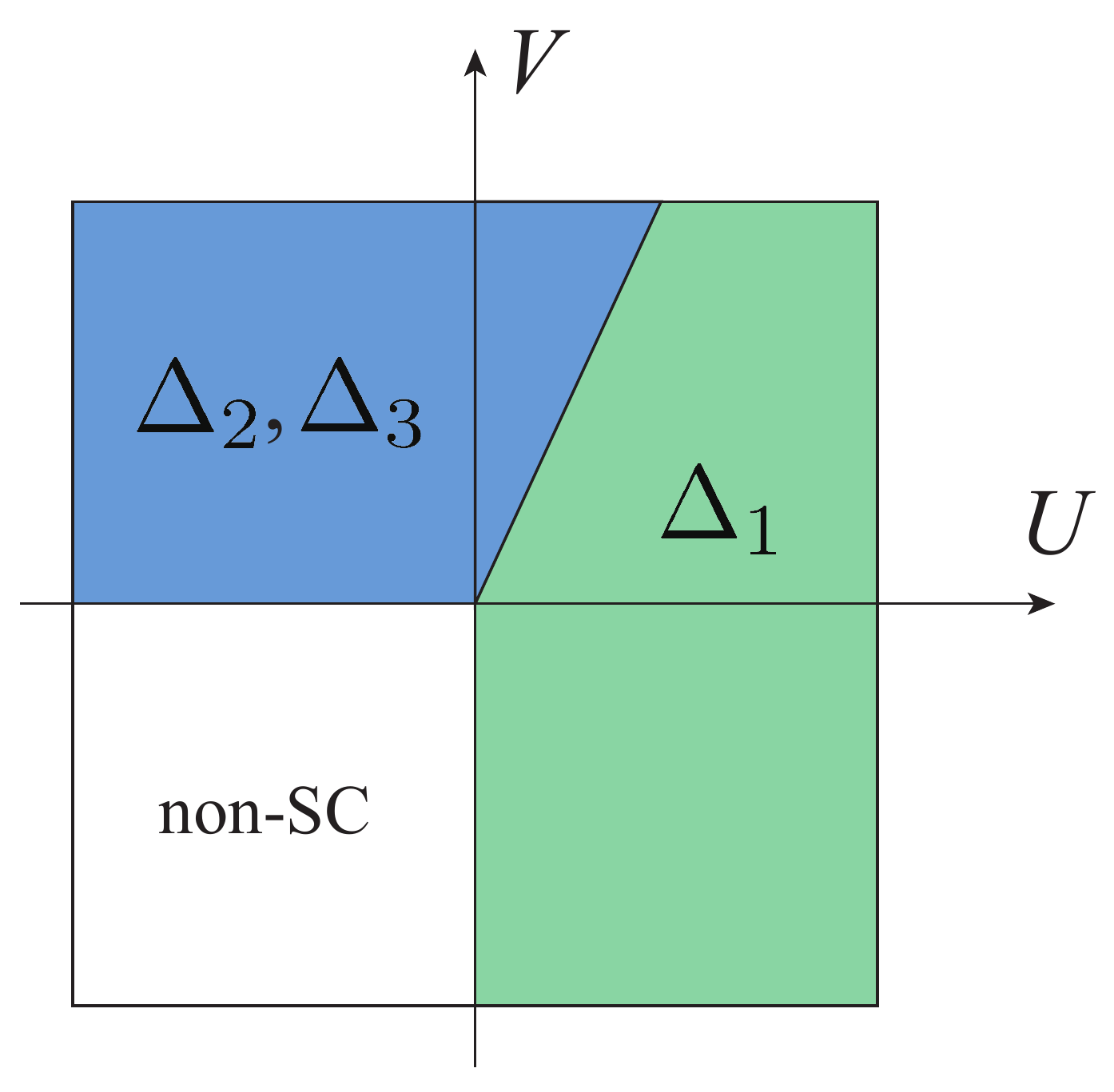}
\caption
{Phase diagram of the superconducting Dirac semimetal Cd$_3$As$_2$. 
$\Delta_1$ is a conventional $s$-wave pairing, while
$\Delta_2$ and $\Delta_3$ are unconventional odd-parity pairings.
For definitions of $\Delta_i$ ($i=1,2,3$), see Ref. \cite{Hashimoto2016}. 
Adapted from Ref. \cite{Hashimoto2016}; copyright (2014) by the American
Physical Society.} 
\label{fig4:Dirac}
\end{figure}

Since the Fermi surfaces in Cd$_3$As$_2$ do not enclose any
time-reversal-invariant momentum, one cannot apply the criterion in
Sec. \ref{sec:odd_parity} to identify topological superconductivity.
Nevertheless, the odd-parity superconducting state in this material is at
least a topological crystalline superconductor. This can be seen by considering
a combination of two-fold rotation (which is twice the $C_4$ rotation responsible for
stabilizing the Dirac points) and inversion, which is equivalent to mirror reflection
with respect to the $xy$-plane; since Cd$_3$As$_2$ is both $C_4$ and inversion 
symmetric, it also has a mirror symmetry and one can define a mirror Chern number. 
It has been shown that its odd-parity superconducting state has a non-zero mirror 
Chern number and hence is topological \cite{Kobayashi-Sato2015, Hashimoto2016}.

A recent experiment found that Cd$_3$As$_2$ shows 
a superconducting transition under pressure \cite{He2015}.
Also, two experimental groups reported that even in ambient pressure, the region
beneath a pressurized point-contact tip develops a superconducting gap with a 
zero-bias peak in the conductance spectrum \cite{Aggarwal2016, Wang2016}. 
Correspondingly, theoretical analyses showed \cite{Hashimoto2016} 
that the odd-parity superconducting states expected in Cd$_3$As$_2$ 
would present a peak structure in the surface density of state.

\subsection{Spin-singlet pairing with spin-orbit coupling}


\subsubsection{$s$-wave pairing in 2D Dirac fermions}
\label{sec:sato2003}

While an odd-parity pairing state is guaranteed to be topological 
when the Fermi surfaces satisfy the condition in Sec. \ref{sec:odd_parity},
most superconducting materials realize even-parity $s$-wave pairing states.
Since ordinary $s$-wave superconductors do not host any gapless Andreev
boundary state, it was believed that they cannot be topological.
Consequently, it was also believed that non-Abelian anyons, which may appear in
topological superconductors, cannot be realized in an $s$-wave pairing state. 

In particular, it is usually considered that time-reversal-symmetry breaking is necessary
for non-Abelian anyons; when there are two non-Abelian anyons, their clockwise exchange gives a
final state which is different from that of the time-reversed counterclockwise exchange, and this explicitly
shows that time-reversal symmetry is broken. 
Indeed, all platforms for non-Abelian anyons known in the early stage of the field, such as the 
$\nu=5/2$ fractional quantum Hall state \cite{Moore-Read1991} or the spinless chiral $p$-wave superconductor, 
break time-reversal symmetry. 
In contrast, an $s$-wave pairing state preserves time-reversal symmetry, which seems to disqualify
it as a host for non-Abelian anyons. 



Nevertheless, it was theoretically discovered in 2003 that 
an $s$-wave pairing state can support non-Abelian anyon excitations
if the pairing is realized in 2D Dirac fermions \cite{Sato2003}. 
To understand its mechanism, let us consider the BdG Hamiltonian for 2D Dirac fermions coupled to an 
$s$-wave superconducting condensate $\Phi$:
\begin{eqnarray}
{\cal H}({\bm k})=\left(
\begin{array}{cc}
s_x k_x+s_yk_y & \Phi^{*} \\
\Phi & -s_x k_x-s_y k_y
\end{array}
\right) .
\end{eqnarray}
This ${\cal H}({\bm k})$ has the following particle-hole symmetry, 
\begin{eqnarray}
{\cal C}{\cal H}({\bm k}){\cal C}^{-1}=-{\cal H}(-{\bm k}),
\,\,
{\cal C}=
\left(
\begin{array}{cc}
0 & is_y\\
-is_y & 0
\end{array}
\right) K.
\end{eqnarray}
In the ground state, the $s$-wave condensate $\Phi$ is uniform and can be taken as a
real constant, and thus the system preserves time-reversal symmetry as follows:
\begin{eqnarray}
{\cal T}{\cal H}({\bm k}){\cal T}^{-1}={\cal H}(-{\bm k}),
\,\, {\cal T}= 
\left(\begin{array}{cc}
is_y& 0\\
0 & is_y 
\end{array}\right)K.
\end{eqnarray}
Therefore, the Chern number is zero and it cannot be used for judging
the topology.  
However, for the Dirac system one can employ the so-called index theorem to see 
the existence of Majorana zero modes in vortices:
For the Dirac system, Majorana zero modes
are introduced as solutions of 
$
{\cal H}(-i{\bm \partial})|u_0^{(\pm)}\rangle=0,
$
where 
\begin{eqnarray}
H(-i{\bm \partial})=
\left(
\begin{array}{cc}
-is_x\partial_x-is_y\partial_y & \Phi^{*}({\bm x})\\
\Phi({\bm x}) & is_x\partial_x+is_y\partial_y
\end{array}
\right).
\label{eq4:Dirac_vortex}
\end{eqnarray}
This solution $|u_0^{(\pm)}\rangle$ also satisfies the equation
\begin{eqnarray}
\left(
\begin{array}{cc}
s_z & 0\\
0 & -s_z
\end{array}
\right)|u_0^{(\pm)}\rangle=\pm |u_0^{(\pm)}\rangle
\end{eqnarray}
at the same time.
The index theorem relates the number $N_{\pm}$ of such zero modes 
to the vorticity of the condensate $\Phi({\bm x})$ in
Eq. (\ref{eq4:Dirac_vortex}) as \cite{Weinberg1981}
\begin{eqnarray}
N_+-N_-=-\frac{1}{4\pi i}\oint_{C}dl_i 
\frac{(\Phi^*\partial_i \Phi-\Phi\partial_{i}\Phi^*)}{|\Phi|^2}.
\label{eq4:index_theorem}
\end{eqnarray} 
Note that 
the right hand side in Eq. (\ref{eq4:index_theorem})
counts the number of vortices in the system 
($C$ is a contour around vortices). 

The above theorem implies that a single vortex hosts an odd
number of Majorana zero modes, and an explicit analysis of the zero
modes indeed shows that the number of the zero modes is one
\cite{Jackiw-Rossi1981, Callan1985}.
Therefore, like a spinless chiral $p$-wave superconductor, 
a vortex obeys the non-Abelian anyon statistics.
The ground state of the present model keeps time-reversal symmetry, but the vortex state
breaks time-reversal symmetry. This local time-reversal breaking at vortices 
allows the $s$-wave pairing state to host non-Abelian anyons without contradiction. 

It is important to note that only a 2D system with an odd number of bulk Dirac
nodes can realize non-Abelian anyons in this manner. 
Otherwise, a vortex hosts an even number of
Majorana zero modes, which are equivalent to Dirac zero modes and the
non-Abelian nature is lost.
While no such condensed matter system was known in 2003, Fu and Kane 
explicitly demonstrated in 2008 that Majorana zero modes can be realized in the surface states 
of a topological insulator with proximity-induced $s$-wave pairing \cite{Fu-Kane2008}; 
this was actually a realization of the above scenario. 
The $s$-wave pairing state in the surface Dirac fermions hosting 
Majorana zero modes in vortices can be viewed as a non-Abelian topological
superconductor \cite{Fu-Kane2008, Fukui2010, Fukui2010b, Parente2014}. 

Note that this superconducting state in the 2D Dirac fermion system effectively realizes a 
spinless superconductor \cite{Fu-Kane2008}. This is because Dirac fermions
in 2D do not have spin degeneracy in the energy spectrum, unlike conventional
electrons obeying the Schr\"odinger equation. The absence of spin degeneracy in the 
spectrum is the essential prerequisite for obtaining a non-Abelian
Majorana zero mode in the $s$-wave pairing state.

\subsubsection{$s$-wave Rashba superconductor with Zeeman field}

Following the ground-breaking proposal by Fu and Kane in 2008 \cite{Fu-Kane2008}, 
it was shown in 2009 that even for ordinary Schr\"odinger electrons
having a parabolic energy dispersion, a clever use of Rashba spin splitting makes it 
possible to realize topological superconductivity hosting Majorana zero modes 
\cite{Sato-Takahashi-Fujimoto2009}. 

To understand the basic idea, we consider electrons with the Rashba spin-orbit coupling,
\begin{eqnarray}
\varepsilon({\bm k})=\frac{k_x^2+k_y^2}{2m}+
v(k_x s_y-k_y s_x)-\mu.
\label{eq4:Rashba}
\end{eqnarray} 
This is motivated by the success of the Dirac system, because the Rashba term 
$v(k_x s_y-k_ys_x)$ mimics the Dirac Hamiltonian. However, considering this term alone is 
not sufficient for realizing non-Abelian anyons: On one hand, the
index theorem [Eq. (\ref{eq4:index_theorem})] is not applicable in the
presence of the first term in Eq. (\ref{eq4:Rashba}); on the other hand, the Chern number 
to guarantee the Majorana zero mode vanishes due to time-reversal symmetry.
Therefore, to obtain a nonzero Chern number, one introduces the time-reversal-breaking Zeeman 
field, i.e., apply a magnetic field in the $z$-direction.
It can be shown that the $s$-wave superconducting state of the Rashba electrons in 
Eq. (\ref{eq4:Rashba}) becomes a topological superconductor with $|Ch|=1$ 
when the Zeeman field $H_z$ is larger than the critical value $H_z^{\rm c}$ 
\cite{Sato-Takahashi-Fujimoto2009, Sato-Takahashi-Fujimoto2010, Sau2010}, 
\begin{eqnarray}
H_z \ge
\sqrt{\mu^2+\Delta_{\rm s}^2}/\mu_{\rm B}
\equiv H_z^{\rm c},
\label{eq4:condition}
\end{eqnarray}
where $\Delta_{\rm s}$ is the $s$-wave pair potential and $\mu_{\rm B}$
is the magnetic moment of electron.
As a result, a vortex in this system becomes a non-Abelian anyon with a
Majorana zero mode in the core. 

Like the case of 2D Dirac fermions, the present system realizes an effectively 
spinless situation under the condition of Eq. (\ref{eq4:condition}). 
In the presence of the Zeeman term $\mu_{\rm B}H_z s_z$, 
the eigenvalues $\varepsilon_{\pm}({\bm k})$ of the normal-state Hamiltonian 
Eq. (\ref{eq4:Rashba}) are given by
\begin{eqnarray}
\varepsilon_{\pm}({\bm k})=
\frac{k_x^2+k_y^2}{2m}\pm \sqrt{v(k_x^2+k_y^2)+(\mu_{\rm B}H_z)^2}-\mu, 
\nonumber\\
\end{eqnarray}
and the upper band $\varepsilon_+({\bm k})$ is always positive under the
condition of Eq. (\ref{eq4:condition}). This means that the Fermi surface consists 
only of the spin-nondegenerate lower band $\varepsilon_-({\bm k})$ when 
Eq. (\ref{eq4:condition}) is satisfied, realizing an effectively spinless Fermi surface.
This property allows the Majorana zero mode to appear in the vortex core in the
$s$-wave pairing state, just like the case of a spinless chiral $p$-wave 
superconductor.

The system considered above is 2D, but the same idea can also be 
applied to a 1D Rashba system \cite{Lutchyn2010, Oreg2010}.
By putting $k_y=0$ in Eq. (\ref{eq4:Rashba}), one obtains 
\begin{eqnarray}
\varepsilon({\bm k})=\frac{k_x^2}{2m}+vk_x s_y-\mu,
\label{eq4:Rashba2}
\end{eqnarray} 
which gives a model of the Rashba electrons in a 1D nanowire. 
When a Zeeman field is applied in the $z$-direction, the spectrum becomes 
effectively spinless and an $s$-wave pairing state becomes a topological superconductor
under the condition of Eq. (\ref{eq4:condition}). In this case, the 1D 
${\bm Z}_2$ index is non-trivial, $(-1)^{\nu_{\rm 1d}}=-1$, which guarantees
the appearance of a Majorana zero mode at each end of the nanowire.

The two types of systems considered in this subsection, 2D Dirac fermions and Rashba electrons, 
are both characterized by a strong (or inherent) spin-orbit coupling. By now, the spin-orbit 
coupling is recognized to be the key to realizing topological superconductivity. 
There have been already many proposals for similar topological superconductors realized in an
$s$-wave pairings state \cite{Tanaka-Sato-Nagaosa2012, Qi11, Alicea2012, Beenakker2013, 
Sato-Fujimoto2016}.

\subsubsection{$d$-wave pairing with spin-orbit coupling}

Instead of $s$-wave pairing, one can consider $d_{x^2-y^2}$-wave or
$d_{xy}$-wave pairing for the realization of topological superconductivity.
It has been shown that the surface Dirac fermions in topological insulators host topological
superconductivity when proximity-coupled to a $d$-wave superconductor
\cite{Linder2010}.
Also, 2D $d$-wave superconductors with the Rashba
spin-orbit coupling can be topological in the presence of Zeeman fields
\cite{Sato-Fujimoto2010}. In this regard, it was recently shown that a nodal $d$-wave Rashba
superconductor can become a fully-gapped topological superconductor in the 
presence of the Zeeman field, because the breaking of inversion and time-reversal 
symmetries conspire to lift the node \cite{Daido2016}.
In 1D, $d_{x^2-y^2}$-wave superconductors with the Rashba spin-orbit
coupling can be topological even in the absence of Zeeman fields \cite{Wong2012}.

As for non-Abelian anyons, their appearance has been investigated in 2D 
chiral $d$-wave superconductors \cite{Sato-Takahashi-Fujimoto2010, Black-Schaffer2012,
Farrel2013,Sun2016}.

\subsection{Spin-rotation breaking and Majorana fermion}

\begin{table}[b]
\begin{center}
\begin{tabular}{|c|ccc|ccc|}
\hline
AZ class & TRS & PHS & SU(2)$_{\rm spin}$ & 1d & 2d & 3d \\
\hline
class D & - & \checkmark & - & ${\bm Z}_2$ & ${\bm Z}$ & 0 \\
class C & - & \checkmark & \checkmark & 0 & $2{\bm Z}$ & 0 \\
\hline
class DIII & \checkmark & \checkmark & - & ${\bm Z}_2$ & ${\bm Z}_2$ &
${\bm Z}$ \\
class CI & \checkmark & \checkmark & \checkmark & 0 & 0 & $2{\bm Z}$ \\
\hline
\end{tabular}
\caption{Topological periodic table for superconductors.}
\label{table}
\end{center} 
\end{table}

It is useful to understand why the spin-orbit coupling or the spin-triplet pairing
is important for realizing topological superconductivity.
To see the reason, examination of the topological periodic table is helpful
\cite{Schnyder2008, Kitaev2009}.
Such a table summarizes possible topological numbers in various dimensions with or without general
non-spatial symmetries, namely, time-reversal, particle-hole, and chiral symmetries 
\cite{Schnyder2008, Kitaev2009}.
Based on these non-spatial symmetries, Hamiltonians are classified into ten
Altland-Zirnbauer (AZ) classes \cite{Altland97}. 
Time-reversal-breaking superconductors having particle-hole symmetry 
belong to class D in the AZ scheme, and time-reversal-invariant
superconductors having additional particle-hole symmetry belong to
class DIII.
In Table \ref{table}, we show the topological periodic table
\cite{Schnyder2008, Kitaev2009} relevant to superconductors
(a part of this table was already shown in Table I). 
Corresponding to different topological numbers in Table \ref{table}, various
types of Majorana fermions are conceived. For instance, the ${\bm Z}_2$ index
in the 1D class D system corresponds to a Majorana zero mode at
the end of superconducting nanowires \cite{Kitaev2001}.

Now let us see what happens if there is no term to break the SU(2)
spin-rotation symmetry, such as the spin-orbit coupling or the spin-triple
pairing interaction. Such a situation corresponds to ordinary $s$-wave superconductors.
In this case, the additional SU(2) rotation symmetry doubles the possible
topological numbers (in up to three dimensions); namely, integer topological numbers become
even numbers, and ${\bm Z}_2$ indices become trivial.
(In terms of the AZ scheme, class D and DIII becomes class C and CI, respectively,
in the presence of the additional SU(2) spin-rotation symmetry.)
One can see these changes in the topological numbers in the topological periodic table 
shown in Table \ref{table}.

Importantly, in the presence of SU(2) spin-rotation symmetry, no topological number 
exists in 1D, which means that no Majorana fermions appears in superconducting nanowires
preserving spin-rotation symmetry.
Furthermore, only even numbers are possible for topological numbers in all the
topological cases in the presence of SU(2) spin-rotation symmetry.
This implies that only Dirac fermions (constructed from a pair of Majorana fermions)
are possible as topologically protected states, and hence the non-Abelian statistics is not realized.

From these consideration, it becomes evident why ordinary spin-singlet
superconductors, which preserve SU(2) spin-rotation symmetry, do not support Majorana boundary states. 
The physical origin of this situation lies in the fact that in ordinary superconductors,
the spin-orbit coupling is weak and can be neglected in the considerations of the topological nature.
Conversely, if the spin-orbit coupling cannot be neglected, even an $s$-wave pairing state can be a 
topological superconductor associated with Majorana fermions \footnote{This argument
does not imply that an explicit spin-rotation-breaking term is necessary for realizing 
Majorana fermions. Even when the microscopic Hamiltonian preserves full SU(2) spin-rotation 
symmetry, topological superconductivity with Majorana fermions can be realized if the 
SU(2) symmetry is spontaneously broken by the formation of spin-triplet Cooper pairs
\cite{Sun2014}}.

\section{Materials realizations}

There are broadly two categories of topological superconductors, intrinsic ones and artificially engineered ones. Intrinsic topological superconductors are those in which a topologically-nontrivial gap function naturally shows up. As already discussed, superconductors having odd-parity pairing are in most cases topological. Also, noncentrosymmetric superconductors, in which parity is not well-defined, can be topological when the spin-triplet component is strong. We will discuss concrete examples of such intrinsic topological insulators in this sections. One can also artificially engineer topological superconductivity in hybrid system consisting of a metal or semiconductor proximitized with a conventional $s$-wave superconductor. This category of topological superconductors is currently attracting significant attention due to its potential as a building block of Majorana-based qubit for topological quantum computation. In Sec. \ref{sec:hybrid}, we will discuss the essence of such hybrid topological superconductors.

\subsection{Candidates of intrinsic topological superconductors}

\subsubsection{{\rm Sr$_2$RuO$_4$}}

The superconductor Sr$_2$RuO$_4$, discovered in 1994 \cite{Maeno1994}, initially attracted attention because it is isostructural to the prototypical high-temperature cuprate superconductor La$_{2-x}$Sr$_x$CuO$_4$. This material is arguably the first superconductor in which the realization of a topological state is seriously discussed. Because of the tendency of ruthenium oxides to become ferromagnetic, spin-triplet pairing was theoretically proposed for this material \cite{Rice1995}. Nuclear magnetic resonance (NMR) experiments found that the Knight shift, which reflects the spin susceptibility, does not change across the superconducting transition temperature $T_c$, which supports spin-triplet pairing \cite{Ishida1998}. Furthermore, muon spin rotation ($\mu$SR) experiments detected the appearance of internal magnetic field below $T_c$, pointing to time-reversal-symmetry breaking \cite{Luke1998}. These experiments led to the expectation that a spin-triplet superconductivity with a definite chirality, often called chiral $p$-wave superconductivity \cite{Kallin2016}, is realized in Sr$_2$RuO$_4$. However, there are unresolved issues, such as the fact that the chiral domains have never been observed, which have been preventing the consensus on the chiral $p$-wave superconductivity. The situation is well summarized in recent review articles \cite{Maeno2012,Kallin2016}.

\begin{figure}[tb]
\centering
\includegraphics[width=0.55\columnwidth]{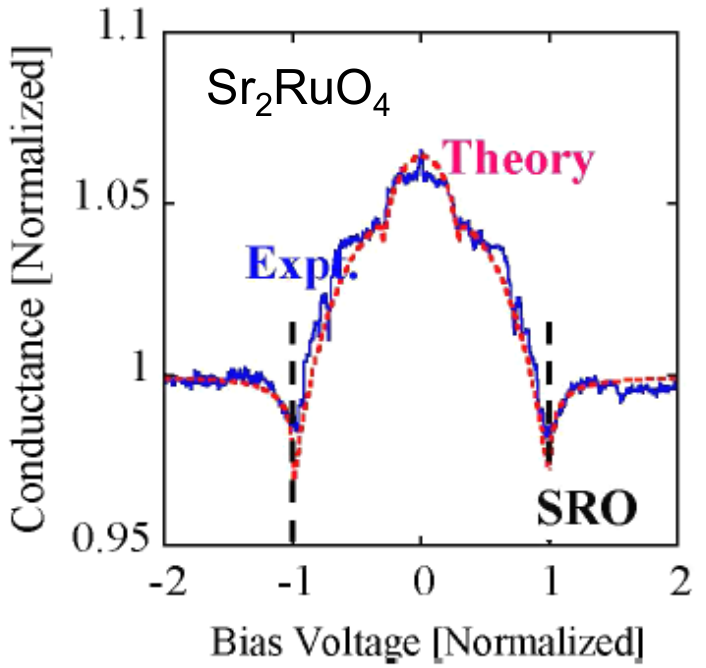}
\caption{Differential-conductance spectrum obtained from a tunnel junction made on a side surface of Sr$_2$RuO$_4$; theoretically calculated spectrum for the side surface containing Andreev bound states is also shown with red dotted line. Bias voltage is normalized by the superconducting gap. Reprinted with permission from Ref. \cite{Kashiwaya2011}; 
copyright (2011) by the American Physical Society.} 
\label{fig:Kashiwaya}
\end{figure}

Whereas the realization of chiral $p$-wave superconductivity in
Sr$_2$RuO$_4$ is still under debate, the spin-triplet pairing dictates
that the orbital part of the gap function must have odd parity, which
guarantees the topological nature of the superconductivity \cite{Yamashiro1997,Furusaki2001,Stone2004, Yada2014}. Indeed, surface Andreev bound state, which is naturally expected to accompany a topological bulk state, has been detected by a tunnel junction experiment (see Fig. \ref{fig:Kashiwaya}) \cite{Kashiwaya2011}. One should note, however, that in the classifications of possible topological states of matter \cite{Schnyder2008,Kitaev2009}, time-reversal-symmetry breaking superconductor can be topological in 1D and 2D, but not in 3D. Actually, Sr$_2$RuO$_4$ is a quasi-2D material with warped cylindrical Fermi surfaces \cite{Mackenzie2003}, and hence it harbors an essentially 2D topological superconductivity. This means that the surface Andreev bound states to be expected as topological boundary states will be present only on the side surfaces.

The expectation for Majorana fermions in Sr$_2$RuO$_4$ is a little complex. Since spin-orbit coupling is not strong in this material, the topological surface states preserve spin degeneracy and hence the Bogoliubov quasiparticle on the surface are spin degenerate. In such a case, the two spin subspaces are not independent from each other and, even if one defines a Majorana operator in each subspace, they can mix to form an ordinary fermion \cite{Ueno2013}. This means that Majorana fermions are not expected on the surface of Sr$_2$RuO$_4$. 

Nevertheless, if the Bogoliubov quasiparticles in the two spin subspaces cannot mix, they can be considered as two independent modes of Majorana fermions. It has been theoretically proposed \cite{Ueno2013} that, due to the mirror symmetry of the Sr$_2$RuO$_4$ crystal structure, there may be a magnetic-field-induced topological phase transition into a topological crystalline superconductor phase, in which two Majorana modes are each protected by symmetry.

In addition, it has been documented that half-quantized vortices show up
in Sr$_2$RuO$_4$ in oblique magnetic fields \cite{Budakian2011}. Such a
half-quantized vortex is expected to harbor a Majorana zero-mode in the
core, because the spin degeneracy is lifted there \cite{Tewari2007}. The resulting Majorana zero-mode is non-Abelian and is in principle useful for quantum computation.

\subsubsection{{\rm Cu$_x$Bi$_2$Se$_3$}}
\label{sec:CBS}

The superconductivity in Cu$_x$Bi$_2$Se$_3$ with $T_c$ up to $\sim$4 K was discovered in 2010 \cite{Hor2010}. This was a first material to show superconductivity upon doping charge carriers into a topological insulator. Such a superconductor is a promising ground to look for topological superconductivity, because of the following reasons: Since the topological surface states are present in topological insulators even when carriers are doped (as long the doping is not too heavy), one would expect that the superconductivity in the bulk would lead to proximity-induced superconductivity of the surface, where 2D topological superconductivity should be found. In this respect, ARPES experiments on Cu$_x$Bi$_2$Se$_3$ confirmed that at the doping level necessary for superconductivity ($\sim 2 \times 10^{20}$ cm$^{-3}$), the topological surface states are still well separated from the bulk states in the momentum space \cite{Wray2010}.

Already in 2010, a more exotic possibility was proposed for Cu$_x$Bi$_2$Se$_3$
by Fu and Berg \cite{Fu-Berg2010}. Due to the strong spin-orbit
coupling, the effective low-energy Hamiltonian of Cu$_x$Bi$_2$Se$_3$
becomes a massive Dirac Hamiltonian discussed in
Sec.\ref{sec:STM_TI}. The four-component basis set of this Hamiltonian
consists of two $p$-orbitals with opposite parities. Fu and Berg noticed
that, if pairing occurs between such orbitals with opposite parities,
the resulting gap function naturally obtains odd parity. In this case,
since Cu$_x$Bi$_2$Se$_3$ has only one bulk Fermi surface which encircles
the $\Gamma$ point \cite{Wray2010}, the Fermi surface criterion
\cite{Sato2009, Sato2010, Fu-Berg2010} in Sec.\ref{sec:odd_parity} is sufficient for concluding that the bulk superconducting state is topological. When the bulk of Cu$_x$Bi$_2$Se$_3$ is in such a topological superconducting state, helical Majorana fermions show up on the surface \cite{Fu-Berg2010}. They are gapless and dispersive Andreev bound states having the Majorana nature. Expected dispersion of the helical Majorana fermions depends on the symmetry of the gap function and the height of the chemical potential, as has been calculated in Refs. \cite{Hao-Lee2011, Sasaki2011,Hsieh2012,Yamakage2012}.

The crystal symmetry of Cu$_x$Bi$_2$Se$_3$ belongs to $D_{3d}$ point
group. According to the symmetry classifications, there are four
possible types of gap functions, $\Delta_1$ to $\Delta_4$ in Table \ref{table4:pair_potential}, that are classified based on the irreducible representations of the $D_{3d}$ point group, $A_{1g}$, $A_{1u}$, $A_{2u}$, and $E_u$, respectively \cite{Fu-Berg2010,Ando-Fu2015}. Among them, $\Delta_1$ is conventional even-parity type, while other three are all odd-parity and hence are topological. Note that due to the strong spin-orbit coupling, spin is no longer a good quantum number in Cu$_x$Bi$_2$Se$_3$ and the notions of spin singlet and triplet are only approximately valid. 

The synthesis of superconducting Cu$_x$Bi$_2$Se$_3$ is difficult. With a simple melt-growth method in which one melts all the constituent elements together and lets them crystallize by slowly cooling across the melting point, only a low volume fraction of the superconducting phase up to 30\% is available \cite{Hor2010}. This is because Cu atoms can occupy both the Bi sites as substitutional impurities and the intercalation sites in the van der Waals gap as intercalants; in the melt-growth method, one has little control over where the Cu atoms will go. To address this problem, a new synthesis method to employ electrochemical intercalation has been developed \cite{Kriener2011PRB}, and an improved volume fraction of the superconducting phase up to 80\% has been obtained; in this method, annealing the Cu-intercalated crystals at a high temperature just below the melting temperature is necessary for obtaining superconductivity. The exact of role of this annealing is not known. In fact, the precise positions of the Cu atoms in the superconducting phase of the crystals have not been elucidated. This is because the superconducting phase of Cu$_x$Bi$_2$Se$_3$ is fragile against mechanical stress, which makes it difficult to prepare samples for precise crystal structure analysis. The fragility of the superconducting phase is a serious problem in making devices of Cu$_x$Bi$_2$Se$_3$ for tunnel-junction or Josephson-junction experiments; for example, after exfoliation into thin flakes, it shows no superconductivity.

\begin{figure}[tb]
\centering
\includegraphics[width=0.55\columnwidth]{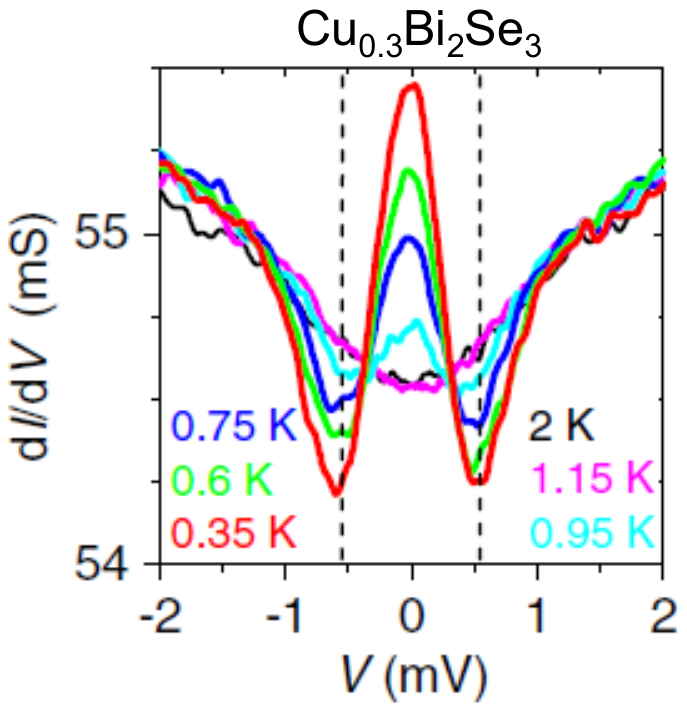}
\caption{Zero-bias conductance peak observed in Cu$_{0.3}$Bi$_2$Se$_3$ via point-contact spectroscopy. The data shown were taken in 0 T at various temperatures. Vertical dashed lines mark the superconducting gap energy. Adapted from Ref. \cite{Sasaki2011}; 
copyright (2011) by the American Physical Society.} 
\label{fig:Sasaki}
\end{figure}

Nevertheless, experiments made on best available samples have already given evidence for topological superconductivity. The first indication came in 2011 from conductance spectroscopy experiments made on the surface. By using a soft-point-contact technique to minimize mechanical stress, Sasaki {\it et al.} found a pronounced zero-bias peak in the differential conductance (Fig. \ref{fig:Sasaki}), whose origin was scrutinized in their experiment to be intrinsic \cite{Sasaki2011}. The shape of the spectra, which present clear minima at the gap energy, is in good qualitative agreement with theoretical calculations \cite{Yamakage2012} of point-contact spectra for the topological cases, all of which are accompanied by surface Majorana fermions.

\begin{figure}[tb]
\centering
\includegraphics[width=0.9\columnwidth]{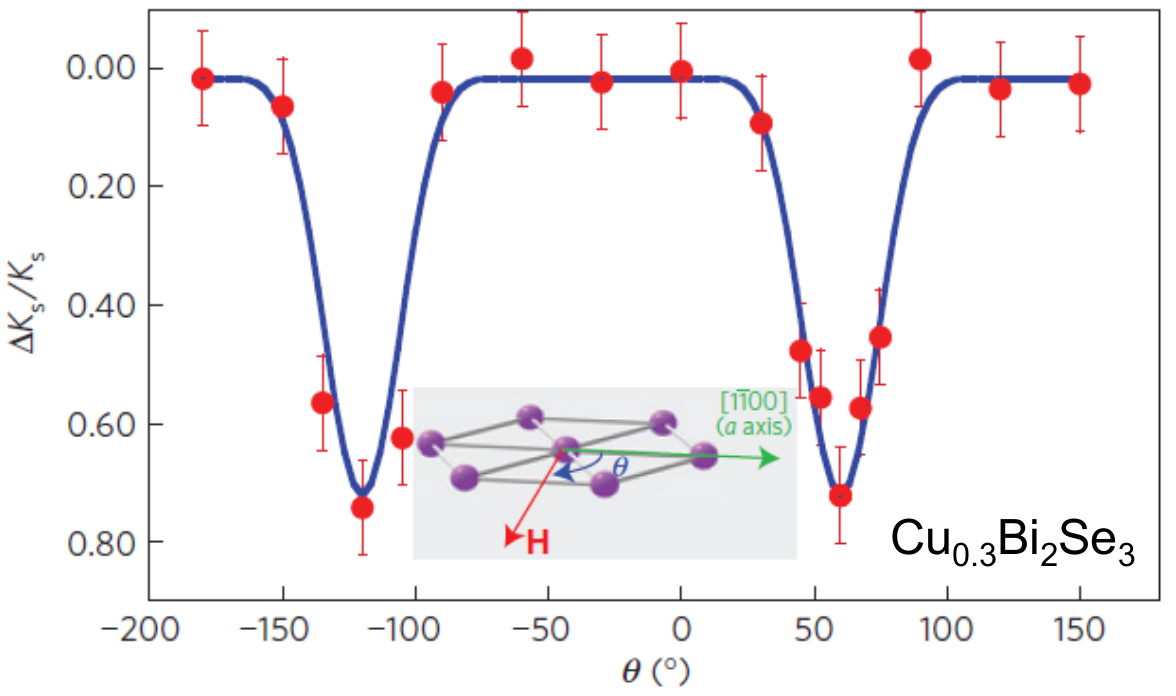}
\caption{Change in the NMR Knight shift between 3 K ($> T_c$ in 0.7 T) and 1.4 K ($< T_c$) measured on Cu$_{0.3}$Bi$_2$Se$_3$ in 0.7 T as a function of the magnetic-field angle which is rotated within the basal plane as depicted in the inset. These data give direct evidence that the spin susceptibility is anisotropic and hence the spin-rotation symmetry is spontaneously broken in the superconducting state. Adapted from Ref. \cite{Matano2016}.} 
\label{fig:NMR}
\end{figure}

Recently, NMR measurements of the Knight shift, which allows one to probe the spin susceptibility in the superconducting state, found that the spin-rotation symmetry is spontaneously broken in the superconducting state of Cu$_x$Bi$_2$Se$_3$ \cite{Matano2016}. Specifically, when the applied magnetic field is rotated in the basal plane, the Knight shift showed pronounced minima which are two-fold symmetric (Fig. \ref{fig:NMR}). Since the crystal structure is three-fold symmetric, the two-fold symmetry must be a result of spontaneously ordering of the electronic system to break spin-rotation symmetry \cite{Nagai2012,Hashimoto2013}. The most natural interpretation of this result is that the superconductivity occurs through pseudo-spin-triplet odd-parity pairing, and the spin angular momentum is pinned to a particular axis in the crystal, whose origin is not very clear at the moment; possible factors to play a role in the pinning are sample edges, some uniaxial strain, or some weak superstructure associated with intercalated Cu atoms, etc. In any case, the NMR result gave direct bulk evidence for topological superconductivity in Cu$_x$Bi$_2$Se$_3$. 

\begin{figure}[tb]
\centering
\includegraphics[width=0.65\columnwidth]{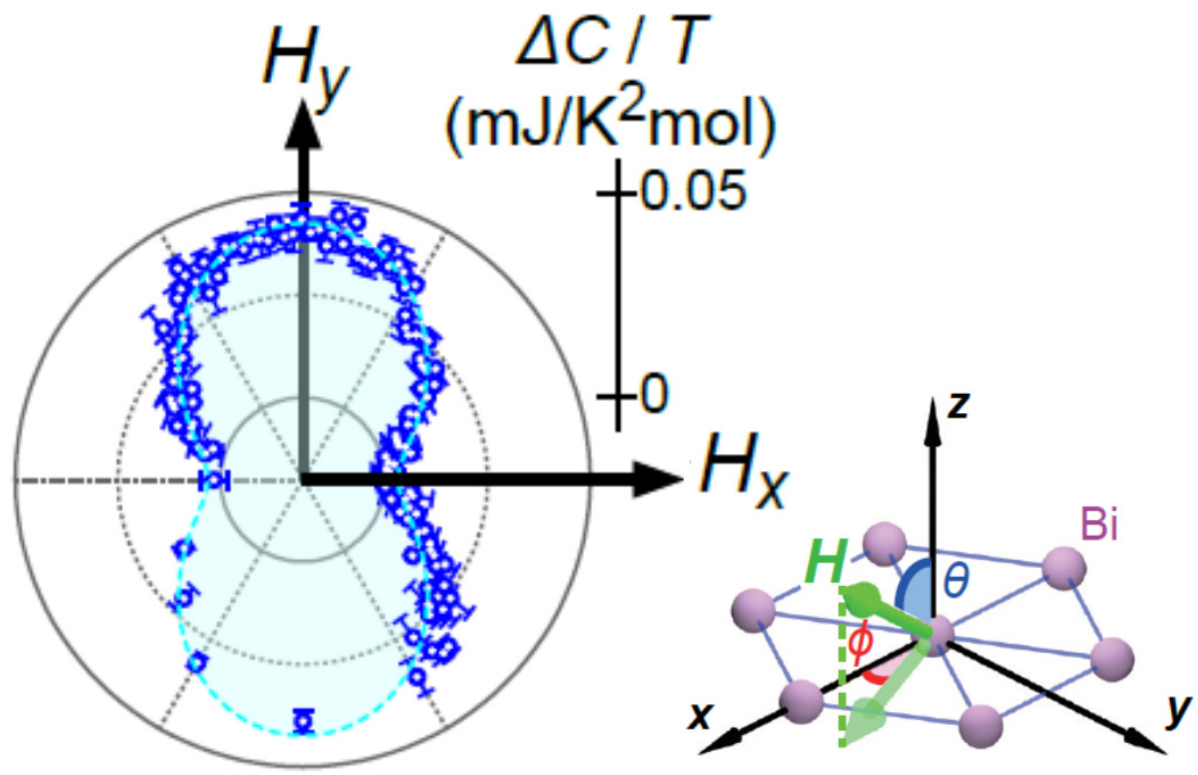}
\caption{Change in the specific heat of Cu$_{0.3}$Bi$_2$Se$_3$ upon rotating the magnetic-field orientation within the basal plane, measured in 0.3 T at 0.6 K; $x$ and $y$ axes are defined in the inset.
Adapted from Ref. \cite{Yonezawa2016}.} 
\label{fig:Yonezawa}
\end{figure}

More recently, specific heat measurements of Cu$_x$Bi$_2$Se$_3$ in applied magnetic fields have also found clear two-fold symmetry when the magnetic field is rotated in the basal plane (Fig. \ref{fig:Yonezawa}) \cite{Yonezawa2016}. Not only the magnetic-field-induced specific heat value in the superconducting state, but also the upper critical field $H_{c2}$ present the same two-fold symmetry. This result indicates that the superconducting gap {\it amplitude} has two-fold-symmetric minima in the momentum space, so that the number of quasiparticles induced by vortices becomes maximal when the magnetic-field orientation is perpendicular to the line connecting the two gap minima (Fig. \ref{fig:Yonezawa}). Such an in-plane anisotropy in the gap amplitude is only consistent with the $\Delta_4$ state among the four possibilities allowed for Cu$_x$Bi$_2$Se$_3$ \cite{Hashimoto2013}. Since this $\Delta_4$ state generates a subsidiary nematic order specified by a two-component order parameter, it has been called {\it nematic superconducting state} \cite{Fu2014}. Note that the $\Delta_4$ state has nodes in the simplest theory, but these nodes are not protected by any symmetry and hence they can be easily lifted by additional interactions \cite{Fu2014}. Therefore, the $\Delta_4$ state can be consistent with the specific-heat data \cite{Kriener2011PRL} which point to the absence of any node.

It is prudent to mention that the experimental situation on Cu$_x$Bi$_2$Se$_3$ was controversial until the NMR result came about. A prominent example was the STM experiment which claimed conventional $s$-wave superconductivity based on the shape of the spectra, which fitted well to a BCS-based model \cite{Levy2013}; however, there were puzzling features in the STM data, such as the extracted gap of 0.4 meV which is much smaller than the gap size of 0.7 meV obtained from the specific-heat jump \cite{Kriener2011PRL}. Since the $T_c$ and $H_{c2}$ were essentially the same between specific heat and STM experiments, the small gap suggested by the STM data was difficult to understand. Potential sources of the controversy have been theoretically discussed \cite{Mizushima2014}, and it could be due to the quasi-2D nature of the Fermi surface \cite{Lahoud2013} or simply the difficulty in preparing homogeneous and well-characterized samples. In any case, in view of the recent confirmations of the $\Delta_4$ state by bulk measurements, the zero-bias conductance peak found in Cu$_x$Bi$_2$Se$_3$ was likely to be the first experimental observation of Majorana fermions in a solid-state system.

Alongside the discovery of nematic superconductivity in Cu$_{0.3}$Bi$_2$Se$_3$ \cite{Matano2016, Yonezawa2016}, essentially the same two-fold symmetry has been observed in similar superconductors derived from Bi$_2$Se$_3$. In Sr$_{x}$Bi$_2$Se$_3$ \cite{Liu2015}, $H_{c2}$ presents clear two-fold in-plane anisotropy \cite{Pan2016, Nikitin2016, Du2017}. Also, Nb$_{x}$Bi$_2$Se$_3$ \cite{Qiu2015} presents two-fold in-plane anisotropy in the magnetization hysteresis loops \cite{Asaba2017}. Hence, it becomes increasingly clear that the topological $\Delta_4$ state generically shows up when Bi$_2$Se$_3$ is doped to become a superconductor.

\subsubsection{{\rm Sn$_{1-x}$In$_x$Te}}

\begin{figure}[tb]
\centering
\includegraphics[width=0.9\columnwidth]{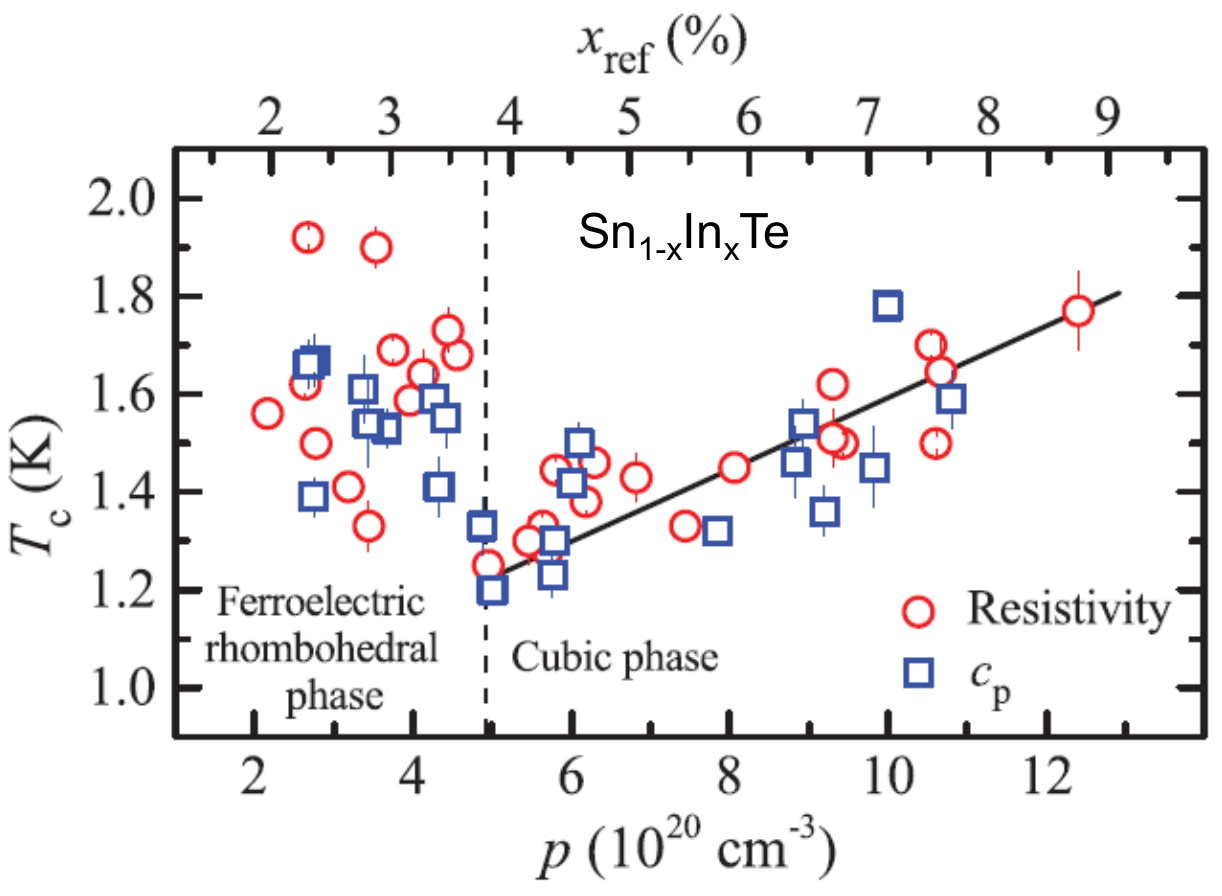}
\caption{Phase diagram of Sn$_{1-x}$In$_x$Te showing how $T_c$ changes with In doping $x_{\rm ref}$; the latter is calculated from the measured hole density $p$. Vertical dashed line divides the ferroelectric rhombohedral phase from the cubic phase; solid straight line is a guide to the eyes to mark the linear increase in $T_c$ with $x$ in the cubic phase. Adapted from Ref. \cite{Novak2013}; 
copyright (2012) by the American Physical Society.} 
\label{fig:Novak}
\end{figure}

SnTe is a topological crystalline insulator characterized by a nontrivial mirror Chern number \cite{HsiehTCI2012,Tanaka2012,Ando-Fu2015}, and it becomes a superconductor upon substituting 2\% or more of Sn with In, which introduces hole carriers \cite{Erickson2009,Novak2013}. While the hole density increases linearly with the In content $x$, the $T_c$ vs. $x$ phase diagram is complex (Fig. \ref{fig:Novak}), presenting a minimum in $T_c$ at $x \sim$ 0.04, where the structural transition from rhombohedral to cubic phase takes place; $T_c$ is a little higher ($\sim$1.6 K) in the rhombohedral phase with $x < 0.038$, and it is reduced to 1.2 K in the cubic phase at $x \simeq 0.04$ \cite{Novak2013}; increasing the $x$ value in the cubic phase leads to an almost linear increase in $T_c$ up to 4.6 K at $x$ = 0.45 \cite{Zhong2013}. Intriguingly, Sn$_{1-x}$In$_x$Te is the cleanest (i.e. presents the lowest residual resistivity) in samples showing the lowest $T_c$ in the cubic phase \cite{Novak2013}. It is exactly in those clean samples with $T_c \simeq$ 1.2 K where a pronounced zero-bias conductance peak similar to that in Cu$_x$Bi$_2$Se$_3$ has been observed by point-contact spectroscopy \cite{Sasaki2012}, which points to the existence of surface Andreev bound states. Note that in Sn$_{1-x}$In$_x$Te, nonmagnetic impurities {\it raise} the $T_c$ through an enhancement in the density of states at the Fermi energy \cite{Novak2013}, as discussed theoretically by Martin and Phillips \cite{Martin1997}, which is contrary to most other superconductors.

The low-energy effective Hamiltonian of SnTe is essentially the same as that for Cu$_x$Bi$_2$Se$_3$, besides the fact that the Fermi surfaces are located at the Brillouin-zone boundary, surrounding the four $L$ points \cite{Ando-Fu2015}. This reduces the possible irreducible representations of the gap functions to $A_{1g}$, $A_{1u}$, and $A_{2u}$, corresponding respectively to the $\Delta_1$, $\Delta_2$, and $\Delta_3$ states defined for Cu$_x$Bi$_2$Se$_3$. Among them, $\Delta_2$ and $\Delta_3$ states are odd-parity with nonzero topological invariants, even though the Fermi surfaces are now surrounding four time-reversal-invariant momenta \cite{Sasaki2012}. Hence, the zero-bias conductance peak observed in the point-contact spectroscopy is likely due to helical surface Majorana fermions associated with $\Delta_2$ or $\Delta_3$ states. 

High-quality single crystals of Sn$_{1-x}$In$_x$Te with 100\% superconducting volume fraction can be grown by a vapor transport method, but the In content is difficult to control in this method. Point-contact spectroscopy experiments performed on many Sn$_{1-x}$In$_x$Te samples found that the zero-bias conductance peak is observed only in the cleanest samples with $x \simeq$ 0.04 \cite{Novak2013}, suggesting that the conventional $\Delta_1$ state is competing with unconventional $\Delta_2$ or $\Delta_3$ states and disorder can easily flip the competition. This makes the studies of topological superconductivity in Sn$_{1-x}$In$_x$Te difficult. Nevertheless, availability of superconducting nanoplates \cite{Sasaki2015} suitable for device fabrications gives good prospects for detailed investigations of this material.

When the bulk of Sn$_{1-x}$In$_x$Te is in the conventional $\Delta_1$ state, it is expected that the topological surface states harbors a proximity-induced topological superconductivity. The ARPES experiments performed on Sn$_{1-x}$In$_x$Te at $x$ = 0.045 confirmed that the topological surface state remains intact after In doping \cite{Sato2013}. Since the parent material SnTe is a topological crystalline insulator, the topological surface states consist of four spin-nondegenerate Dirac cones and their locations in the surface Brillouin zone depends on the surface orientation \cite{Tanaka2013}. It has been theoretically proposed that the four Dirac cones give rise to four Majorana fermions at the end of the vortex core, but only two of them are symmetry protected to robustly remain at zero energy \cite{Fan-Gilbert-Bernevig2014}. The properties of such vortices binding a pair of Majorana zero-modes is an interesting subject to be pursued in Sn$_{1-x}$In$_x$Te in the conventional bulk superconducting state.

\subsubsection{Noncentrosymmetric superconductors}

In noncentrosymmetric superconductors where inversion symmetry is broken, parity is ill-defined and hence mixing of $s$-wave (spin-singlet) and $p$-wave (spin-triplet) states is allowed. Also, broken inversion symmetry allows Rashba and Dresselhaus spin-orbit coupling, which lifts the spin degeneracy and splits the Fermi surface. It has been shown \cite{Sato-Fujimoto2009,Tanaka-Yokoyama2009} that 2D noncentrosymmetric superconductors are topological when the $p$-wave gap is larger than the $s$-wave gap. In such a topological state, if it preserves time-reversal symmetry, helical Majorana fermions show up at the edge \cite{Sato-Fujimoto2009}. 

The first superconductor identified to be noncentrosymmetric was probably BiPd, which was discovered in 1952 \cite{Alekseevskii1952}; unfortunately, this material does not present any peculiar property that is ascribable to the noncentrosymmetric nature \cite{Peets2016}. In this regard, CePt$_3$Si \cite{Bauer2004} was the first noncentrosymmetric superconductor which presented a signature of unconventional superconductivity possibly related to the lack of inversion symmetry. 
This is a heavy-fermion system with strong
electron interactions, which means that there is a strong on-site
repulsion to favor unconventional superconductivity. Indeed, CePt$_3$Si
has a line node, which is a hallmark of unconventional superconductivity
\cite{Bauer2007}. The node could be a result of $s+p$-wave pairing,
because in such a case, the superconducting gap may be written as
$\Delta = \Delta_s - \Delta_p \sin \theta$ ($\Delta_s$ and $\Delta_p$
are the gap sizes of the $s$- and $p$-wave components, respectively),
which shows a sign change when $\Delta_s < \Delta_p$. However, the node
could also be due to the effect of coexisting antiferromagnetic order
\cite{Fujimoto2007}. To date, the origin of the node is not well
understood. The topological nature of 3D nodal noncentrosymmetric
superconductors has been analyzed \cite{Sato2006, Beri2010,
Schnyder2012} and it was predicted that surface flat bands of
topological origin should show up as a result of non-trivial bulk
topology \cite{Sato-Tanaka-Yada-Yokoyama2011,Schnyder-Ryu2011}. 

Li$_2$(Pd$_{1-x}$Pt$_x$)$_3$B \cite{Badica2005} is also a noncentrosymmetric superconductor presenting evidence for line nodes at Pt-rich compositions \cite{Nishiyama2007}. Contrary to CePt$_3$Si, this material has only weak electron correlations, which makes the $s+p$-wave pairing as likely origin of the line nodes. Intriguingly, nodes are present in Li$_2$Pt$_3$B but not in Li$_2$Pd$_3$B, which suggests that the source of the different pairing state lies in the difference in the strength of spin-orbit coupling. In fact, it was theoretically shown that strong spin-orbit coupling enhances $p$-wave component \cite{Fujimoto2007}. Unfortunately, the difficulty in growing single crystals of Li$_2$Pt$_3$B has been a hindrance for detailed studies of this interesting material. 

To search for a topological superconductor, 2D or quasi-2D superconductors with broken inversion symmetry would also be a fertile ground. In this regard, superconductivity induced at the surface of an insulator such as SrTiO$_3$ or KTaO$_3$ by ionic-liquid gating \cite{Ueno2014} is interesting. However, so far no evidence for $p$-wave components have been reported for these 2D superconductors.

\subsubsection{Nodal topological superconductors}

As mentioned in Sec. \ref{sec:general}, superconductors need not be
fully gapped in order to be topological. As long as a nontrivial
topological invariant is defined for the occupied states, nodal
superconductors can be called topological. For example, the $\Delta_3$
and $\Delta_4$ states of Cu$_x$Bi$_2$Se$_3$ have nodes, but an explicit
${\bm Z}_2$ topological invariant has been found to characterize their
topological nature \cite{Sasaki2011}. Similarly, concrete topological
invariants have been found for 3D noncentrosymmetric superconductors
\cite{Schnyder2012}.

\begin{figure}[tb]
\centering
\includegraphics[width=0.5\columnwidth]{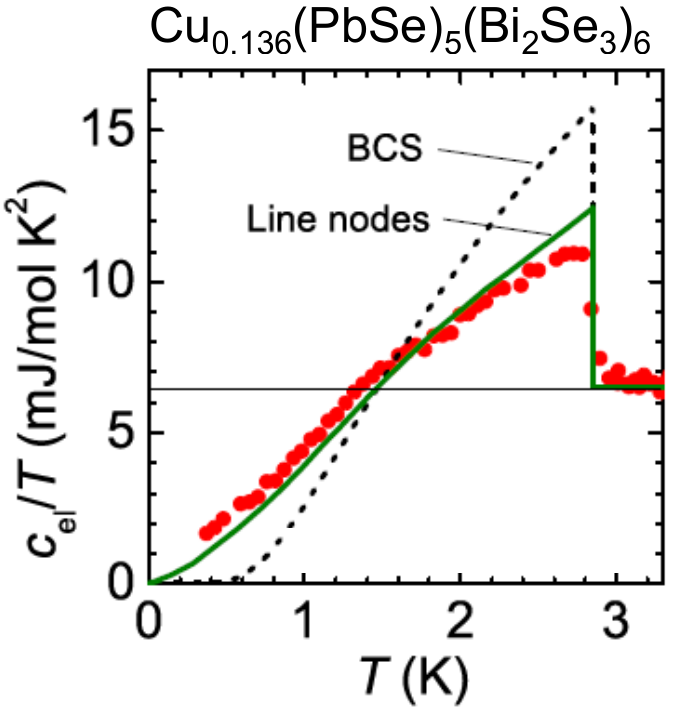}
\caption{Temperature dependence of the electronic specific heat measured in a nearly-100\%-superconducting Cu$_{0.136}$(PbSe)$_5$(Bi$_2$Se$_3$)$_6$ sample; the dashed line shows the weak-coupling BCS behavior, while the green solid line shows the theoretical curve for a quasi-2D superconducting state with line nodes. Adapted from Ref. \cite{Sasaki2014}; 
copyright (2014) by the American Physical Society.} 
\label{fig:CPSBS}
\end{figure}

An interesting superconductor related to Cu$_x$Bi$_2$Se$_3$ is Cu$_x$(PbSe)$_5$(Bi$_2$Se$_3$)$_6$. This superconductor was discovered in 2014 \cite{Sasaki2014} and its specific-heat behavior strongly suggests the existence of line nodes. A good news is that it is possible to synthesize 100\% superconducting sample of this material at $x \simeq$ 1.5 \cite{Sasaki2014}. The crystal structure of the parent material (PbSe)$_5$(Bi$_2$Se$_3$)$_6$ can be viewed as a naturally-formed heterostructure, in which a bilayer of PbSe and two quintuple-layers of Bi$_2$Se$_3$ alternate \cite{Kanatzidis2004,Nakayama2012,Segawa2015}. Since PbSe is an ordinary insulator, this natural heterostructure realizes a situation in which topological insulator units are separated by an ordinary insulator unit. Intriguingly, the ARPES study of (PbSe)$_5$(Bi$_2$Se$_3$)$_6$ found evidence that topological interface states are present at every interface between PbSe and Bi$_2$Se$_3$ units \cite{Nakayama2012}. 

(PbSe)$_5$(Bi$_2$Se$_3$)$_6$ has a van der Waals gap in the middle of the two-quintuple-layer Bi$_2$Se$_3$ unit, and when Cu atoms are intercalated into this van der Waals gap, the resulting Cu$_x$(PbSe)$_5$(Bi$_2$Se$_3$)$_6$ is a superconductor with $T_c$ = 2.85 K \cite{Sasaki2014}. Hence, the essential building blocks for superconductivity in Cu$_x$(PbSe)$_5$(Bi$_2$Se$_3$)$_6$ and Cu$_x$Bi$_2$Se$_3$ are basically the same, and the mechanism of superconductivity is also expected to be the same, although Cu$_x$(PbSe)$_5$(Bi$_2$Se$_3$)$_6$ is strongly quasi-2D due to its crystal structure. The Fu-Berg theory for Cu$_x$Bi$_2$Se$_3$ has been extended to the case of quasi-2D Fermi surface \cite{Hashimoto2014}, and it was shown that the $\Delta_4$ state has a pair of line nodes in the quasi-2D case, while other three possible states are all fully gapped. Therefore, it is most likely that the odd-parity $\Delta_4$ state is realized not only in Cu$_x$Bi$_2$Se$_3$ but also in Cu$_x$(PbSe)$_5$(Bi$_2$Se$_3$)$_6$. Probably the higher-order effect to lift the nodes in the $\Delta_4$ state is weak or absent in the latter, leading to different specific-heat behaviors (i.e. fully-gapped vs. nodal) in the two compounds. 

UPt$_3$ is an unconventional heavy-fermion superconductor discovered in
1984 \cite{Stewart1984}. It has a complex $B$ vs. $T$ phase diagram,
with three distinct superconducting phases A (high-$T$, low-$B$), B
(low-$T$, low-$B$), and C (low-$T$, high-$B$). Gap nodes are present in
all three phases and the pairing is most likely spin triplet; however,
the nature of each phase as well as the pairing mechanism are still
under debate \cite{Joynt-Taillefer2002,Izawa2012}. Recently, topological
crystalline superconductivity was proposed theoretically for the
B-phase, based on the assumption that UPt$_3$ is a spin-triplet $f$-wave
superconductor \cite{Tsutsumi2013}. In this putative phase, two linearly
dispersing Majorana modes appear at the edge of the $ab$ plane, and the
two modes do not mix because they are protected by mirror chiral
symmetry. 
Topological nodal structures in UPt$_3$ were also studied theoretically
\cite{Yanase2016, Kobayashi2016}.

YPtBi is a relatively new superconductor \cite{Butch2011} which is noncentrosymmetric and presents evidence for line nodes \cite{Paglione2016}. It has a very low carrier density of only 2 $\times$ 10$^{18}$ cm$^{-3}$, which cannot afford its $T_c$ of 0.78 K within the BCS theory \cite{Meinert2016}. (In this regard, the $T_c$ of up to 4 K with the low carrier density of 2 $\times$ 10$^{20}$ cm$^{-3}$ in Cu$_x$Bi$_2$Se$_3$ is also too high to be explained within the BCS theory.) It has been proposed that in YPtBi, pairing of $j$ =3/2 fermions may be taking place, which allows Cooper pairs to have quintet or septet total angular momentum \cite{Brydon2016}. The topological nature of such a novel superconductor should be investigated in future.

\subsection{Artificial topological superconductors}
\label{sec:hybrid}

Artificially engineered topological superconductivity in hybrid structures are currently attracting significant attention, because they offer exciting opportunities to create and manipulate non-Abelian Majorana zero modes \cite{Alicea2012,Beenakker2013}. The essential ingredients for such hybrid structures are spin-nondegenerate metal and proximity-induced $s$-wave superconductivity. The original idea came from Fu and Kane \cite{Fu-Kane2008}, who showed that if $s$-wave pairing is imposed on the topological surface states of a 3D topological insulator through superconducting proximity effect, due to the helical spin polarization of the surface states, the resulting superconducting state can be effectively viewed as a 2D $p$-wave superconductor harboring a Majorana zero mode in the vortex core.

To see this mechanism, let us consider the BdG equation 
${\cal H}=\sum_{\bm k}\Psi_{\bm k}^{\dagger}{\cal H}({\bm k})\Psi_{\bm k}/2$ 
with the BdG Hamiltonian
\begin{eqnarray}
{\cal H}({\bm k})=
\left(
\begin{array}{cc}
v{\bm s}\cdot{\bm k}-\mu & \Delta\\
\Delta & -v{\bm s}\cdot{\bm k}+\mu
\end{array}
\right), 
\label{eq7:s-waveDirac}
\end{eqnarray} 
where $\Delta$ is the proximity-induced $s$-wave pairing amplitude, ${\bm s}$ is the 
Pauli matrix of spin, and ${\bm s}\cdot {\bm k}=s_x k_x+s_yk_y$.
The normal-state Hamiltonian describes the spin-helical surface state. 
For the field operator $\Psi$, we take the Nambu representation $\Psi=(c_{\uparrow}, c_{\downarrow},
c^{\dagger}_{\downarrow}, -c^{\dagger}_{\uparrow})^t$. 

When one creates a vortex in the
condensate given by the 
above Hamiltonian, such a vortex hosts a Majorana zero mode and it obeys the
non-Abelian anyon statistics.
This result was originally derived \cite{Sato2003} for $\mu=0$ by using the index theorem
as discussed in Sec. \ref{sec:sato2003}, but a more intuitive argument can be done for $\mu\gg
|\Delta|$ \cite{Fu-Kane2008}: 
In the weak coupling case, the superconducting state is well-described
by the single band near the Fermi energy. 
The wavefunction of the single-band electron can be written as 
$(1, e^{i\theta})^t/\sqrt{2}$ with $e^{i\theta}=(k_x+ik_y)/k$, and in this basis the
single-band BdG Hamiltonian is transformed as
\begin{eqnarray}
&& \left[
\begin{array}{c}
(1, e^{-i\theta})\\
(e^{i\theta}, 1)
\end{array}
\right]
{\cal H}({\bm k})
\left[
\left(
\begin{array}{c}
1\\
e^{i\theta}
\end{array}
\right)
\left(
\begin{array}{c}
e^{-i\theta}\\
1
\end{array}
\right)
\right]
\nonumber\\
&&\rightarrow\left(
\begin{array}{cc}
v|{\bm k}|-\mu & \Delta e^{-i\theta}\\
\Delta^* e^{i\theta} & -v|{\bm k}|+\mu
\end{array}
\right).
\end{eqnarray}
It should be noted that this transformed BdG Hamiltonian describes a
spinless chiral $p$-wave pairing, although it is singular at ${\bm k}=0$. 
Because of the non-analyticity, this single-band Hamiltonian does not
have a well-defined Chern number, which is consistent with the time-reversal
invariance of the original Hamiltonian Eq. (\ref{eq7:s-waveDirac}); 
however, once a time-reversal-breaking vortex or a magnetic boundary
is introduced, the singularity is smeared out and
Majorana modes appear as in the case of a true spinless chiral $p$-wave superconductor
\cite{Fu-Kane2008}.



Later, it was recognized that a spin-nondegenerate 2D metal similar to the surface states of 3D topological insulators can be obtained by considering the combination of Rashba spin-orbit coupling and Zeeman effect in a 2D metal: Rashba spin-orbit coupling splits a spin degenerate band into a pair of spin nondegenerate bands, and Zeeman effect will open a gap at the crossing point of these two bands; when the chemical potential is tuned into this Zeeman gap, there is only one Fermi surface which has a helical spin polarization \cite{Sato-Fujimoto2009}. Therefore, if $s$-wave pairing is induced in such a state by using proximity effect, it also realizes an effective $p$-wave superconductor \cite{Sato-Fujimoto2009, Sau2010,Alicea2010,Sato-Takahashi-Fujimoto2010}.

A drawback of the above scheme to turn a Rashba-split 2D metal into a topological superconductor is that the size of the Zeeman splitting must be larger than the size of the induced superconducting gap. This condition is difficult to meet in ordinary 2D metals due to the orbital pair-breaking effect, and one needs to devise a way to go around this problem. Possible solutions are to use semiconductors with a large $g$ factor, to utilize exchange coupling for Zeeman effect, or to apply magnetic fields parallel to the 2D plane \cite{Sau2010,Alicea2010,Sato-Takahashi-Fujimoto2010}. 

With the same basic mechanism, it is possible to engineer a 1D topological superconductor. Inducing $s$-wave pairing in the 1D helical edge states of a 2D topological insulators gives rise to effective spinless 1D $p$-wave superconductivity, which was treated in the celebrated 1D Kitaev model \cite{Kitaev2001}. Nanowires of semiconductors having strong Rashba spin-orbit coupling (such as InAs or InSb) placed in magnetic fields can also serve as a platform to engineer 1D topological superconductivity \cite{Lutchyn2010,Oreg2010}. Those engineered 1D topological superconductors are accompanied by localized Majorana zero-mode at the edge. Since the edge of a 1D wire is much easier to access than a vortex in the middle of a 2D system, the engineered 1D topological superconductor is generally considered to be most practically useful.

Experimentally, proximity-induced superconductivity on the surface of 3D topological insulators has been studied by many groups and there is accumulating evidence that superconductivity can indeed be induced in the topological surface states \cite{Williams2012, Wang-XueSC2012, Fan-Lu2012, Wang-XueSC2013, Mason2013, Oostinga-Molenkamp2013, Harlingen2014, Snelder-Brinkman2014}. While it has been difficult to elucidate the topological nature of the induced 2D superconductivity, recent observation \cite{Wiedenmann-Molenkamp2016} of $4\pi$-periodic Josephson supercurrent in 3D HgTe topological insulator is encouraging. The experimental situation for induced 1D superconductivity in the topological edge states of the 2D topological insulator HgTe is probably more promising; for example, it has been shown that proximity-induced superconductivity is clearly established in the 1D helical edge states \cite{Hart2014} and the recent observation of half-frequency Josephson radiation gave reasonably strong evidence for the existence of Majorana fermions \cite{Deacon2016}.

Regarding the semiconductors with Rashba-split bands, there has been no experiment on a 2D system, and efforts are focused on 1D nanowires of InAs or InSb. Initial experiments based on point-contact spectroscopy on the edges of such nanowires proximitized by Al or Nb-based superconductor found zero-bias conductance peaks in finite magnetic fields \cite{Mourik2012, Heiblum2012, HQXu2012}. While this was an encouraging observation, zero-bias conductance peak in a junction made on a mesoscopic superconductor can be of various origins \cite{Franceschi2014}, and it was difficult to conclusively nail down the Majorana nature. This was particularly so when the junction shows a ``soft" gap, in which a lot of sub-gap excitations are present. However, recent experiment on InAs nanowire with epitaxially-grown Al superconductor add-layer, which realizes much cleaner proximity-induced superconductivity with a ``hard" gap, found convincing evidence for Majorana fermions at the edges of the nanowire through observation of the Majorana teleportation phenomenon \cite{Albrecht2016}. 

On a different route, by putting a 1D chain of ferromagnetically-ordered atoms on top of an $s$-wave superconductor having strong spin-orbit coupling, one can locally create 1D topological superconductivity in the portion of the superconductor beneath the atomic chains \cite{NadjPerge2013}. STM experiments have been perform on Pb(110) surface decorated with Fe atomic chains, and reasonably convincing evidence for a Majorana zero-mode localized at the edge has been reported \cite{NadjPerge2014}.

Finally, there is a very different avenue for artificially realizing a topological superconductor based on graphene. It has been proposed \cite{Nandkishore2012} that chiral $d$-wave superconductivity with a $d_{x^2-y^2} \pm id_{xy}$ gap structure may be realized if mono-layer graphene is heavily doped to the vicinity of van Hove singularity, which lies at the doping level of $\pm$1/8 from the pristine half-filled level. Such a state would be a time-reversal-symmetry breaking 2D topological superconductor with Majorana zero-modes in the vortex core.


\section{Properties of topological superconductors}

{\bf Zero-bias conductance peak --} 
Topological superconductors are without exception accompanied by gapless surface/edge modes because of the bulk-edge correspondence of topological systems. These boundary modes can be observed as in-gap states in STM or tunnel-junction experiments, which probe the surface density of states. Also, existence of boundary modes affects the Andreev reflection at the boundary, which leaves a signature in point-contact spectra; note that, depending on the transparency of the point contact, the spectral signature varies in point-contact spectroscopy \cite{BTK, Yamakage2012}. In both Cu$_x$Bi$_2$Se$_3$ \cite{Sasaki2011} and InSb nanowire \cite{Mourik2012}, a zero-bias conductance peak in the point-contact spectra was the first signature of Majorana fermions.

The tunneling conductance spectra of topological superconductors 
depend on their dimensions and symmetries.
In one-dimensional time-reversal-breaking topological superconductors,
there exist an isolated single Majorana zero mode at each end.
The tunneling conductance due to the isolated Majorana zero mode is
given by 
\begin{eqnarray}
\frac{dI}{dV}=\frac{2e^2}{h}\frac{1}{1+(eV/2\Gamma)^2}, 
\label{eq8:tunnel1}
\end{eqnarray}
where $I$ is the tunneling current, $V$ is the bias voltage, and
$\Gamma$ is the width of the spectrum.
It shows a zero-bias differential conductance peak
of height $2e^2/h$ \cite{Law2009,Flensberg2010, Ioselevich2013}.
If the single Majorana zero mode is coupled
with another Majorana zero mode at the other end, the tunneling
conductance is modified to \cite{Flensberg2010},
\begin{eqnarray}
\frac{dI}{dV}=\frac{2e^2}{h}
\frac{1}{1+(eV/2\Gamma-2t^2/eV\Gamma)^2}, 
\label{eq8:tunnel2}
\end{eqnarray}
where $t$ is the coupling between the Majorana zero modes at
different ends.
When $t/\Gamma$ is negligibly small, Eq. (\ref{eq8:tunnel2}) almost reproduces 
Eq. (\ref{eq8:tunnel1}) \cite{Ioselevich2013}, but when the mixing is
substantial, the tunneling conductance shows a dip rather than a peak.
In the latter case, the zero-bias conductance becomes zero. 

If the system has an additional symmetry, the quantized values of the
zero-bias conductance can be multiplied by an integer.
For example, in one-dimensional time-reversal-invariant superconductors, 
the zero-bias conductance peak is doubled \cite{Wang2012}, 
\begin{eqnarray}
\left.\frac{dI}{dV}\right|_{eV=0}=\frac{4e^2}{h},
\end{eqnarray}
due to the Kramers degeneracy.
Because of time-reversal symmetry, they host a Kramers pair of Majorana
end states, each of which induces resonant Andreev reflections of the
quantized zero-bias conductance $2e^2/h$.
Furthermore, if the system has chiral symmetry, the zero-bias peak can
be $2Ne^2/h$ with an integer $N$ \cite{Yamakage2014}.

In two dimensions, topological superconductors have 
gapless Majorana edge modes with linear dispersion
\begin{eqnarray}
E\sim v k_x 
\end{eqnarray}
in the time-reversal-breaking case, and 
\begin{eqnarray}
E\sim \pm v k_x 
\end{eqnarray} 
in the time-reversal-invariant case.
Such Majorana edge modes have a tendency to give rise to a broad
zero-bias peak, not a sharp one, in the tunneling conductance. 
Although the actual tunneling spectra can be complicated by multi-band
effects \cite{Yada2014}, 
in the case of simple chiral $p_x+ip_y$-wave superconductors
\cite{Yamashiro1997, Tanaka2009} and helical $p_x\pm ip_y$-wave
superconductors \cite{Tanaka2009b},
theoretical calculations found bell-sharped broad tunneling spectra. 

\begin{figure}[tb]
\centering
\includegraphics[width=0.8\columnwidth]{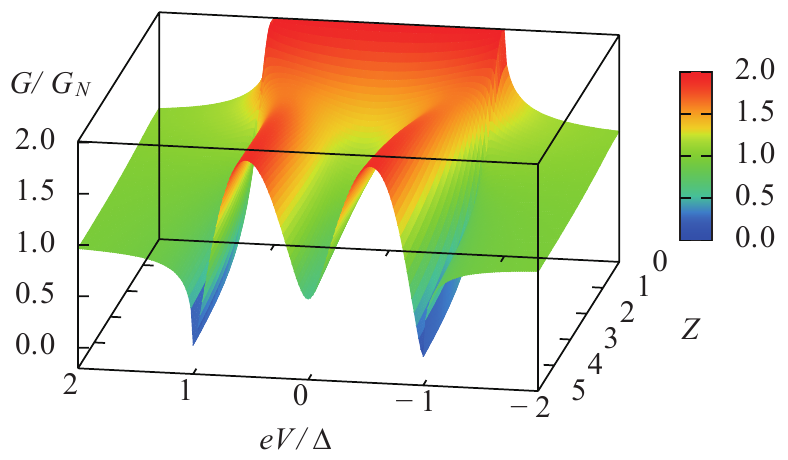}
\caption{Normalized tunneling conductance $G/G_N$ as a function of bias
voltage for the Balian-Werthamer (BW) phase, which is a fully-gapped 
three-dimensional time-reversal-invariant topological superconducting phase 
accompanied by surface helical Majorana fermions \cite{Schnyder2008, Sato2009}.
Here, the transmissivity (represented by $Z$) is continuously varied and the tunneling limit 
corresponds to $Z\rightarrow \infty$. Adapted from
Ref. \cite{Yamakage2012}; copyright (2012) by the American Physical Society.} 
\label{fig:BW2}
\end{figure}

In three-dimensional topological superconductors, the manifestations of 
surface Majorana fermions in the zero-bias conductance of tunneling 
spectroscopy would be subtle. For example, in the tunneling limit, 
the zero-bias conductance shows a zero-bias {\it dip} even in the presence of 
gapless surface helical Majorana fermion \cite{Asano2003, Yamakage2012}. 
The tunneling conductance through a two-dimensional interface between a
metal and a superconductor is given in the form
\begin{eqnarray}
\left.\frac{dI}{dV}\right|_{eV=0}=
\int_0^{2\pi} d\phi \int_0^{\pi/2} d\theta \sin\theta T(k_x,
k_y)|_{eV=0}, 
\nonumber\\
\label{eq8:3dconductance}
\end{eqnarray}
where 
$(k_x, k_y, k_z)=(k_{\rm F}\sin\theta\cos\phi, k_{\rm
F}\sin\theta\sin\phi, k_{\rm F}\cos\theta)$
is the momentum of an incident electron in the metal.
The interface
is taken to be normal to the $z$-axis, and the metal (superconductor)
is in the negative (positive) $z$ side.
Taking the momentum average of incident electrons results in the
volume factor $\sin\theta d\phi d\theta$ in the integral.
In the tunneling limit, there is no direct coupling between electrons in the
metal and Cooper pairs in the superconductor, and hence surface bound states
are needed to mediate the Andreev reflections. 
Therefore, the tunneling rate $T(k_x, k_y)$ at zero-bias is
proportional to the momentum-resolved surface density of state $N(k_x,
k_y, E)$ at zero energy,
\begin{eqnarray}
T(k_x, k_y)|_{eV=0} \propto N(k_x, k_y, E=0).
\end{eqnarray}
In the case of three-dimensional time-reversal-invariant topological 
superconductors, the surface states are
helical Majorana fermions with the following energy dispersion
\begin{eqnarray}
E\sim v\sqrt{k_x^2+k_y^2},
\end{eqnarray}
which give a non-zero value of $N(k_x,k_y,E=0)$ only when
$\sin\theta=0$.
Therefore, from the $\sin\theta$ factor in Eq. (\ref{eq8:3dconductance}), the
zero-bias conductance becomes zero in the tunneling limit.
We illustrate the behavior of the tunneling conductance of a typical 
three-dimensional topological superconductor in Fig. \ref{fig:BW2}.

\begin{figure}[bt]
\centering
\includegraphics[width=0.95\columnwidth]{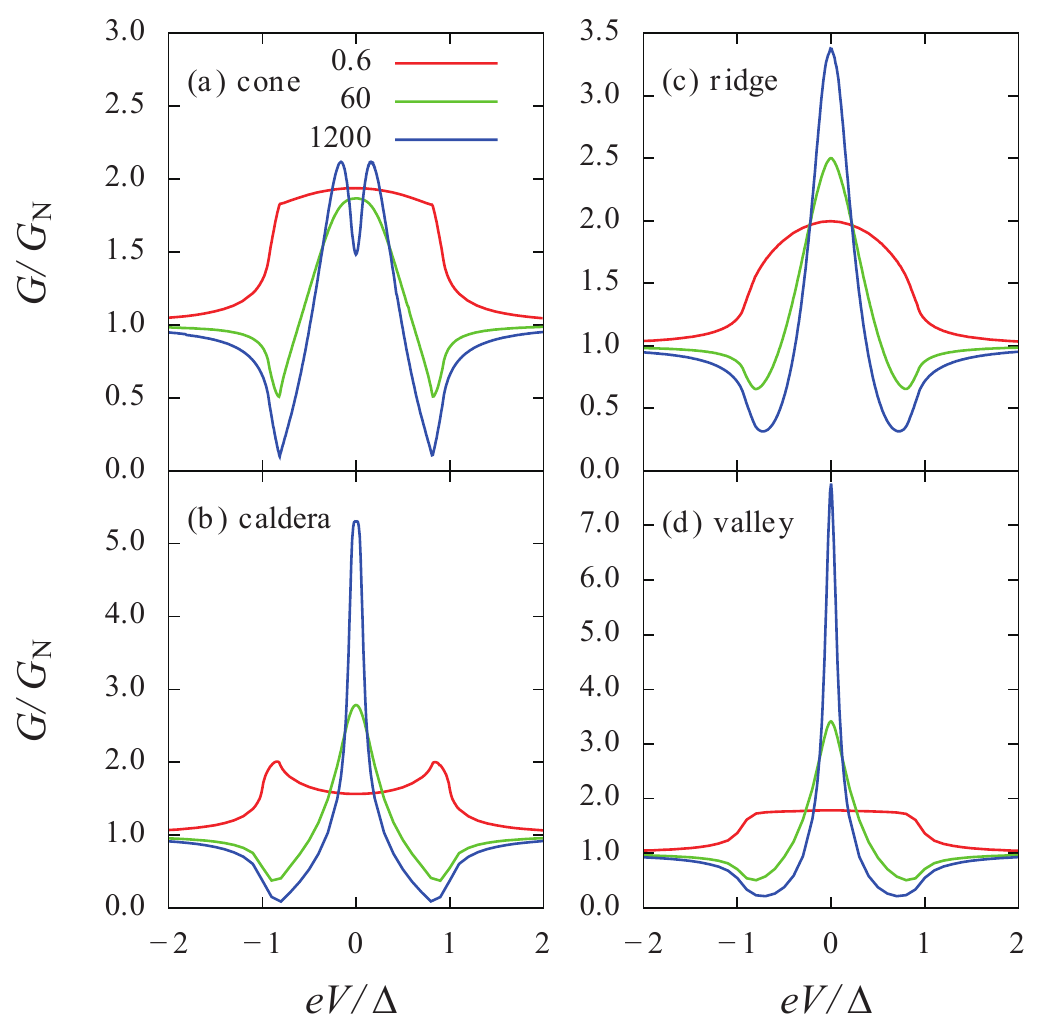}
\caption{Theoretically calculated tunneling spectra in three-dimensional
time-reversal-invariant topological superconductors for various types 
of dispersions of the helical Majorana fermions at the surface: 
(a) conventional X-shaped energy dispersion, 
(b) twisted energy dispersion, and (c, d) flat energy dispersions. 
Different colors indicate different transmissivities of the tunneling interface.
The transmissivity is high for red lines, intermediate for green lines,
and low for blue lines. Adapted from Ref. \cite{Yamakage2012}; 
copyright (2012) by the American Physical Society.} 
\label{fig:conductance}
\end{figure}

Despite the general behavior discussed above, 
the actual tunneling conductance of three-dimensional topological
superconductors may present zero-bias peaks for several reasons 
\cite{Yamakage2012,Hsieh2012}:
(i) When the tunneling barrier of the tunnel junction is low or the
transmissivity of the tunneling interface is high, the above
argument does not hold and a
zero-bias peak can appear; see Fig. \ref{fig:conductance}(a).
(ii) In the case of superconducting topological insulators, surface
Majorana fermions may have a twisted energy dispersion due to the existence
of surface Dirac fermions in the normal state.
Such a twisted dispersion enhances the surface density of states near zero
energy, leading to a zero-bias peak structure in the tunneling
conductance; see Fig. \ref{fig:conductance}(b).
(iii) If the bulk superconducting gap has nodes or deep minima, 
the energy dispersion 
of surface Majorana fermions can become almost flat along the path
connecting the projected bulk nodes in the surface Brillouin zone.
The flat dispersion enhances the surface density of states at zero
energy, resulting in a zero-bias peak in the tunneling conductance;
see Fig. \ref{fig:conductance}(c,d).
(iv) At finite temperature, the zero-bias dip may be smeared by thermal
broadening \cite{Hsieh2012}.

{\bf Thermal conductivity and spin current --}
The gapless surface/edge modes can also be probed by transport experiments. Although dc charge transport measurements are not useful because of the zero-resistivity nature of the superconducting state, one can employ thermal transport or spin transport for studying quasiparticles in the superconducting states \cite{Schnyder2008}. Note that Cooper pairs do not carry heat and phonons die away following the $T^3$ law, making gapless Majorana modes to be the only heat carriers at low temperature.
Therefore, a finite surface/edge thermal conductivity in a fully-gapped superconductor gives strong evidence for dispersive Majorana fermions. Also, when the surface/edge modes have a helical spin polarization as in the case of Cu$_x$Bi$_2$Se$_3$, a heat current is at the same time a spin current. Measurements of the heat-current-induced spin polarization on the surface or edge of a topological superconductor gives further evidence for Majorana fermions. 

{\bf Quantized thermal Hall conductivity --}
For 3D time-reversal-invariant topological superconductors, an interesting prediction for quantization of the thermal Hall effect has been made \cite{Wang-Qi-Zhang2011, Nomura2012, Shiozaki2013}. To observe this phenomenon, one needs to cover the whole surface with an $s$-wave superconductor to open a gap at the Dirac point of the helical Majorana-fermion dispersion. The predicted thermal Hall conductivity is $\kappa_{yx}/T = \pi^2 k_B^2/(12h)$. 

{\bf Nematicity --} 
As already mentioned in Sec. \ref{sec:CBS}, there is strong evidence that Cu$_x$Bi$_2$Se$_3$ realizes the $\Delta_4$ state which can be termed a nematic superconductor \cite{Fu2014}. The nematicity has so far been detected through uniaxial anisotropy in the Knight shift \cite{Matano2016}, specific heat \cite{Yonezawa2016}, and upper critical field \cite{Yonezawa2016}, in their dependencies on the orientation of the applied magnetic field. The occurrence of nematicity is also confirmed in similar superconductors, Sr$_x$Bi$_2$Se$_3$ \cite{Pan2016, Nikitin2016, Du2017} and Nb$_x$Bi$_2$Se$_3$ \cite{Asaba2017}. The nematicity should manifest itself in various other properties, such as thermal conductivity, penetration depth, elastic constants, sound velocity, etc. The nematicity would also give rise to a formation of nematic domains, and the domain boundaries may present unusual properties because the phase factor changes across the boundary. In fact, the physics of nematic domain boundaries would be a fertile ground for new discoveries.

{\bf Anomalous Josephson effects --}
Josephson junctions between a time-reversal-invariant topological superconductor and an $s$-wave superconductor contain surface/edge Majorana modes in the junction, which leads to an anomalous current-phase relationship. This is essentially because the Majorana mode remains gapless for the phase difference $\theta$ of 0 or $\pi$, but becomes gapped for $\theta \neq 0, \pi$, which leads to a half-period Fraunhofer pattern \cite{Chung-Horowitz-Qi2013}. For similar reasons, when one inserts a 3D time-reversal-invariant topological superconductor into the gap of a C-shaped $s$-wave superconductor to make a continuous ring, the flux quantization in the ring is predicted to occur in units of half the flux quantum, $\frac{h}{4e}$ \cite{Fu-Berg2010}. Intriguingly, if the C-shaped part of such a ring is replaced by a $d$-wave superconductor, the flux quantization will take the value $\frac{h}{2e}(n+\frac{1}{2})$ with integer $n$ \cite{Fu-Berg2010}.
Signatures of topological phase transitions in terms of a discontinuity in 
the Josephson current-phase relation was recently discussed \cite{Marra2016}.

The Josephson current-phase relation is also sensitive to the
symmetry of the system \cite{Yamakage2013}.
The Josephson current can be generally decomposed into a harmonic series, 
\begin{eqnarray}
J(\theta)=\sum_{n=1}(J_n\sin n\theta+I_n\cos n\theta),
\end{eqnarray}
where $J_n$ and $I_n$ diminish as $n$ increases.
Under time-reversal operation, $J(\theta)$ transforms to $-J(-\theta)$, which means
$I_n=0$ in the case of time-reversal-invariant junctions.
If the junction preserves mirror-reflection symmetry as well, the
current-phase relationship depends also on the mirror parity of the
gap function. When we consider a junction between a mirror-odd
superconductor and a conventional $s$-wave superconductor, mirror-reflection 
symmetry implies $J(\theta+\pi)=J(\theta)$, which leads to $J_{2n+1}=0$.
(Note also that $I_n=0$ as mentioned above.)
As a result, one obtains a Josephson current with an anomalous
current-phase relationship $J(\theta)\sim \sin 2\theta$.
Such an anomalous relation is useful for identifying the bulk pairing symmetry
of topological superconductors.

{\bf Odd-frequency Cooper pairs --}
The presence of surface Majorana fermions gives rise to
odd-frequency Cooper pairs at the interfaces \cite{Asano2013}.
Odd-frequency Cooper pairs are stable against disorders, which results in 
anomalous proximity effects in normal-metal/superconductor and
superconductor/normal-metal/superconductor junctions
\cite{Tanaka2004,Tanaka2005,Tanaka2005b,Tanaka2007,Ikegaya2015}.
For more details about the relation between topological superconductivity 
and odd-frequency
Cooper pairs, see a recent review \cite{Tanaka-Sato-Nagaosa2012}.

{\bf Majorana fermions in hybrid systems --}
For artificial topological superconductors engineered in hybrid systems, various predictions have been made for their peculiar properties to reflect the existence of Majorana fermions. Most prominent predictions are half-integer conductance quantization at a ballistic NS point-contact junction, $4\pi$-periodic Josephson effect, Majorana teleportation, interference effect around an island containing a Majorana fermion, and braiding phenomena associated with Majorana exchange. We refer interested readers to reviews dedicated to Majorana zero-modes for detailed discussions of these effects \cite{Alicea2012,Beenakker2013,Sato2016}

\section{Outlook}

While the concept of topological superconductivity has been prominent in theoretical physics since 2000 \cite{Read-Green}, their experimental investigations are still at its infancy. With the recent confirmations of the topological $\Delta_4$ state in Cu$_x$Bi$_2$Se$_3$ \cite{Matano2016,Yonezawa2016}, Majorana teleportation in InAs/Al nanowire \cite{Albrecht2016}, and the $4\pi$-periodic Josephson effect in HgTe/Nb junction \cite{Deacon2016}, concrete platforms for exploring the physics of topological superconductivity are being established. It is foreseeable that Majorana fermions in solid-state systems will soon become a well-grounded paradigm.

In intrinsic topological superconductors, investigations of novel phenomena associated with surface/edge Majorana fermions will be an important subject. The first step would be to study heat and spin transport properties, for which Majorana fermions will be responsible. Also, the quantization of thermal Hall conductivity would be intriguing as a novel manifestation of topology. In future, conceiving a Majorana qubit based on intrinsic topological superconductors for quantum computation would be an interesting and potentially important direction. Needless to say, discoveries of new types of intrinsic topological superconductors to widen our knowledge of topological materials remain important.

In hybrid systems, since the possibility to engineer topological superconductivity has already been confirmed, the interest is shifting toward controlling and reading the occupation of a Majorana zero-mode, which is the basis of qubit operation. The most exciting prospect is the demonstration of non-Abelian statistics associated with a Majorana zero-mode \cite{Read-Green}. Such an experiment requires braiding operation involving at least four Majorana fermions \cite{Ivanov2001} and confirmation of the flipping of a qubit. This is technically very demanding, but already concrete setups and protocols have been proposed \cite{Aasen2016}. Another exciting prospect of the hybrid systems is the potential to implement the ``surface code'' \cite{Bravyi-Kitaev2001,Freedman-Meyer2001}, a protocol to realize scalable quantum computation with built-in error corrections. The surface code can in principle be realized by using non-topological quantum states, but it has been argued that using Majorana fermions will make it much simpler to realize the surface code \cite{Vijay2015, Landau2016, Plugge2016, Manousakis2017}.

Regarding the theory of topological superconductors, one of the important remaining
problems is the complete understanding of topological crystalline superconductivity. 
Real materials have a variety of crystal symmetries, which could
enrich their topological structures. 
For instance, it was revealed recently that nonsymmorphic space group
symmetry such as glide reflection and screw rotation makes it possible to have 
novel ${\bm Z}_2$ and ${\bm Z}_4$ topological phases with
M\"{o}bius-twisted surface states \cite{Shiozaki-Sato-Gomi2016}.
To clarify if other exotic topological surfaces may exist or not,
more systematic studies are needed.
Furthermore, disorder and interaction effects on topological crystalline
superconductors have not been well understood yet. 
A naive anticipation that topological
crystalline phases are easily washed out by weak disorders is found to be
incorrect, and numerical studies have suggested that they could be
stable against disorders \cite{Ringel2012, Mong2012, Fu-Kane2012, Fulga2014}.
Also, in the presence of a magnetic point group symmetry,
symmetry-protected Majorana modes
present the so-called Majorana Ising property, by which 
the coupling between the Majorana modes and disorders is strongly
suppressed \cite{Chung09, Shindou10, Mizushima2012, Shiozaki2014}.
Another interesting problem in topological crystalline
superconductors is to study their topological quantum phase transitions; 
intriguingly, it has been predicted that some space supersymmetry
appears at such a topological phase transition point \cite{Grover2014,Jian2015}.

To facilitate the hunt for new topological superconductors, further understanding of
topological nodal superconductors is desirable.
Most discovered unconventional superconductors including high-$T_{\rm c}$ cuprates are nodal, 
and they also have topologically non-trivial properties. Some of them have globally-defined bulk topological
numbers, which implies that the nodal topological phase can be stable in the presence of
disorder. 
Moreover, like Weyl semimetals, gapless bulk nodes could be a source of
anomaly-related quantum phenomena.

Regarding the mechanism of topological superconductivity, a promising
direction is to study orbital fluctuations in the presence of strong
spin-orbit coupling. As was shown in Sec.\ref{sec:route}, inter-orbital pairing
interactions can lead to spin-triplet topological
superconductivity even without spin-dependent pairing interactions. 
In this regard, identifying the microscopic mechanism of odd-parity Cooper pairs 
in Sr$_2$RuO$_4$ and Cu$_x$Bi$_2$Se$_3$ would be important.

Finally, revealing hitherto-unknown topological phenomena is of particular
interest. In this regard, a systematic search based on general frameworks such as topological
quantum field theories might give universal and tractable predictions for novel
topological phenomena.


\section{Acknowledgments}

The authors acknowledges Kouji Segawa, Alexey Taskin, Satoshi Sasaki, Zhi Ren,
Markus Kriener, Mario Novak, Fan Yang, Kazuma Eto, Takafumi Sato, Seigo
Souma, Takashi Takahashi, Keiji Yada, Ai Yamakage, Shingo Kobayashi,
Tatsuki Hashimoto, Satoshi Fujimoto, Takeshi Mizushima, Ken
Shiozaki, Kiyonori Gomi, Yoichi Yanase, Takehito Yokoyama and
Yukio Tanaka for collaborations. 
This work was supported by a Grant-in-Aid for Scientific Research on
Innovative Areas ``Topological Materials Science'' (KAKENHI Grant
Nos. JP15H05855 and JP15H05853) from JSPS of Japan.
The work of MS was performed in part at the Aspen
Center for Physics, which is supported by National Science Foundation grant
PHY-1066293.
Y.A. is supported by DFG through CRC1238 ``Control and Dynamics of
Quantum Materials'', Project A04.
\bibliography{library}

\begin{thebibliography}{100}
\expandafter\ifx\csname url\endcsname\relax
  \def\url#1{\texttt{#1}}\fi
\expandafter\ifx\csname urlprefix\endcsname\relax\def\urlprefix{URL }\fi
\providecommand{\bibinfo}[2]{#2}
\providecommand{\eprint}[2][]{\url{#2}}

\bibitem{TKNN}
\bibinfo{author}{Thouless, D.~J.}, \bibinfo{author}{Kohmoto, M.},
  \bibinfo{author}{Nightingale, M.~P.} \& \bibinfo{author}{Dennijs, M.}
\newblock \bibinfo{title}{Quantized {H}all conductance in a two-dimensional
  periodic potential}.
\newblock \emph{\bibinfo{journal}{Phys. Rev. Lett.}}
  \textbf{\bibinfo{volume}{49}}, \bibinfo{pages}{405--408}
  (\bibinfo{year}{1982}).

\bibitem{Kohmoto85}
\bibinfo{author}{Kohmoto, M.}
\newblock \bibinfo{title}{Topological invariant and the quantization of the
  {H}all conductance}.
\newblock \emph{\bibinfo{journal}{Ann. Phys.}} \textbf{\bibinfo{volume}{160}},
  \bibinfo{pages}{343} (\bibinfo{year}{1985}).

\bibitem{Kane-Mele1}
\bibinfo{author}{Kane, C.~L.} \& \bibinfo{author}{Mele, E.~J.}
\newblock \bibinfo{title}{Quantum spin {H}all effect in graphene}.
\newblock \emph{\bibinfo{journal}{Phys. Rev. Lett.}}
  \textbf{\bibinfo{volume}{95}}, \bibinfo{pages}{226801}
  (\bibinfo{year}{2005}).

\bibitem{Kane-Mele2}
\bibinfo{author}{Kane, C.~L.} \& \bibinfo{author}{Mele, E.~J.}
\newblock \bibinfo{title}{Z$_2$ topological order and the quantum spin {H}all
  effect}.
\newblock \emph{\bibinfo{journal}{Phys. Rev. Lett.}}
  \textbf{\bibinfo{volume}{95}}, \bibinfo{pages}{146802}
  (\bibinfo{year}{2005}).

\bibitem{Moore-Balents}
\bibinfo{author}{Moore, J.~E.} \& \bibinfo{author}{Balents, L.}
\newblock \bibinfo{title}{Topological invariants of time-reversal-invariant
  band structures}.
\newblock \emph{\bibinfo{journal}{Phys. Rev. B}} \textbf{\bibinfo{volume}{75}},
  \bibinfo{pages}{121306} (\bibinfo{year}{2007}).

\bibitem{Fu-Kane-Mele}
\bibinfo{author}{Fu, L.}, \bibinfo{author}{Kane, C.~L.} \&
  \bibinfo{author}{Mele, E.~J.}
\newblock \bibinfo{title}{Topological insulators in three dimensions}.
\newblock \emph{\bibinfo{journal}{Phys. Rev. Lett.}}
  \textbf{\bibinfo{volume}{98}}, \bibinfo{pages}{106803}
  (\bibinfo{year}{2007}).

\bibitem{Roy2009}
\bibinfo{author}{Roy, R.}
\newblock \bibinfo{title}{Topological phases and the quantum spin {H}all effect
  in three dimensions}.
\newblock \emph{\bibinfo{journal}{Phys. Rev. B}} \textbf{\bibinfo{volume}{79}},
  \bibinfo{pages}{195322} (\bibinfo{year}{2009}).

\bibitem{AndoReview}
\bibinfo{author}{Ando, Y.}
\newblock \bibinfo{title}{Topological insulator materials}.
\newblock \emph{\bibinfo{journal}{J. Phys. Soc. Jpn.}}
  \textbf{\bibinfo{volume}{82}}, \bibinfo{pages}{102001}
  (\bibinfo{year}{2013}).

\bibitem{Schnyder2008}
\bibinfo{author}{Schnyder, A.~P.}, \bibinfo{author}{Ryu, S.},
  \bibinfo{author}{Furusaki, A.} \& \bibinfo{author}{Ludwig, A. W.~W.}
\newblock \bibinfo{title}{Classification of topological insulators and
  superconductors in three spatial dimensions}.
\newblock \emph{\bibinfo{journal}{Phys. Rev. B}} \textbf{\bibinfo{volume}{78}},
  \bibinfo{pages}{195125} (\bibinfo{year}{2008}).

\bibitem{Kitaev2009}
\bibinfo{author}{Kitaev, A.}
\newblock \bibinfo{title}{Periodic table for topological insulators and
  superconductors}.
\newblock \emph{\bibinfo{journal}{AIP Conf. Proc.}}
  \textbf{\bibinfo{volume}{1134}}, \bibinfo{pages}{22--30}
  (\bibinfo{year}{2009}).

\bibitem{VolovikHe3}
\bibinfo{author}{Volovik, G.~E.}
\newblock \emph{\bibinfo{title}{The Universe in a Helium Droplet}}
  (\bibinfo{publisher}{Oxford University Press}, \bibinfo{year}{2003}).

\bibitem{Read-Green}
\bibinfo{author}{Read, N.} \& \bibinfo{author}{Green, D.}
\newblock \bibinfo{title}{Paired states of fermions in two dimensions with
  breaking of parity and time-reversal symmetries and the fractional quantum
  {H}all effect}.
\newblock \emph{\bibinfo{journal}{Phys. Rev. B}} \textbf{\bibinfo{volume}{61}},
  \bibinfo{pages}{10267--10297} (\bibinfo{year}{2000}).

\bibitem{Kitaev2001}
\bibinfo{author}{Kitaev, A.~Y.}
\newblock \bibinfo{title}{Unpaired {M}ajorana fermions in quantum wires}.
\newblock \emph{\bibinfo{journal}{Physics-Uspekhi}}
  \textbf{\bibinfo{volume}{44}}, \bibinfo{pages}{131} (\bibinfo{year}{2001}).

\bibitem{Sato2003}
\bibinfo{author}{Sato, M.}
\newblock \bibinfo{title}{Non-abelian statistics of axion strings}.
\newblock \emph{\bibinfo{journal}{Phys. Lett. B}}
  \textbf{\bibinfo{volume}{575}}, \bibinfo{pages}{126--130}
  (\bibinfo{year}{2003}).

\bibitem{Alicea2012}
\bibinfo{author}{Alicea, J.}
\newblock \bibinfo{title}{New directions in the pursuit of {M}ajorana fermions
  in solid state systems}.
\newblock \emph{\bibinfo{journal}{Rep. Prog. Phys.}}
  \textbf{\bibinfo{volume}{75}}, \bibinfo{pages}{076501}
  (\bibinfo{year}{2012}).

\bibitem{Beenakker2013}
\bibinfo{author}{Beenakker, C. W.~J.}
\newblock \bibinfo{title}{Search for {M}ajorana fermions in superconductors}.
\newblock \emph{\bibinfo{journal}{Annu. Rev. Condens. Matter Phys.}}
  \textbf{\bibinfo{volume}{4}}, \bibinfo{pages}{113--136}
  (\bibinfo{year}{2013}).

\bibitem{Sato2016}
\bibinfo{author}{{Sato}, M.} \& \bibinfo{author}{{Fujimoto}, S.}
\newblock \bibinfo{title}{Majorana fermions and topology in superconductors}.
\newblock \emph{\bibinfo{journal}{J. Phys. Soc. Jpn.}}
  \textbf{\bibinfo{volume}{85}}, \bibinfo{pages}{072001}
  (\bibinfo{year}{2016}).

\bibitem{Berry84}
\bibinfo{author}{Berry, M.}
\newblock \bibinfo{title}{Quantum phase factors accompanying adiabatic
  changes}.
\newblock \emph{\bibinfo{journal}{Proc. R. Soc. Lond. A}}
  \textbf{\bibinfo{volume}{392}}, \bibinfo{pages}{45} (\bibinfo{year}{1984}).

\bibitem{Simon83}
\bibinfo{author}{Simon, B.}
\newblock \bibinfo{title}{Holonomy, the quantum adiabatic theorem, and
  {B}erry's phase}.
\newblock \emph{\bibinfo{journal}{Phys. Rev. Lett.}}
  \textbf{\bibinfo{volume}{51}}, \bibinfo{pages}{2167--2170}
  (\bibinfo{year}{1983}).

\bibitem{Avron1983}
\bibinfo{author}{Avron, J.~E.}, \bibinfo{author}{Seiler, R.} \&
  \bibinfo{author}{Simon, B.}
\newblock \bibinfo{title}{Homotopy and quantization in condensed matter
  physics}.
\newblock \emph{\bibinfo{journal}{Phys. Rev. Lett.}}
  \textbf{\bibinfo{volume}{51}}, \bibinfo{pages}{51--53}
  (\bibinfo{year}{1983}).

\bibitem{Bernevig2006}
\bibinfo{author}{Bernevig, B.~A.} \& \bibinfo{author}{Zhang, S.-C.}
\newblock \bibinfo{title}{Quantum spin {H}all effect}.
\newblock \emph{\bibinfo{journal}{Phys. Rev. Lett.}}
  \textbf{\bibinfo{volume}{96}}, \bibinfo{pages}{106802}
  (\bibinfo{year}{2006}).

\bibitem{Qi2008}
\bibinfo{author}{Qi, X.-L.}, \bibinfo{author}{Hughes, T.~L.} \&
  \bibinfo{author}{Zhang, S.-C.}
\newblock \bibinfo{title}{Topological field theory of time-reversal invariant
  insulators}.
\newblock \emph{\bibinfo{journal}{Phys. Rev. B}} \textbf{\bibinfo{volume}{78}},
  \bibinfo{pages}{195424} (\bibinfo{year}{2008}).

\bibitem{Sheng2006}
\bibinfo{author}{Sheng, D.~N.}, \bibinfo{author}{Weng, Z.~Y.},
  \bibinfo{author}{Sheng, L.} \& \bibinfo{author}{Haldane, F. D.~M.}
\newblock \bibinfo{title}{Quantum spin {H}all effect and topologically
  invariant {C}hern numbers}.
\newblock \emph{\bibinfo{journal}{Phys. Rev. Lett.}}
  \textbf{\bibinfo{volume}{97}}, \bibinfo{pages}{036808}
  (\bibinfo{year}{2006}).

\bibitem{Fu-Kane2007}
\bibinfo{author}{Fu, L.} \& \bibinfo{author}{Kane, C.~L.}
\newblock \bibinfo{title}{Topological insulators with inversion symmetry}.
\newblock \emph{\bibinfo{journal}{Phys. Rev. B}} \textbf{\bibinfo{volume}{76}},
  \bibinfo{pages}{045302} (\bibinfo{year}{2007}).

\bibitem{Hasan10}
\bibinfo{author}{Hasan, M.~Z.} \& \bibinfo{author}{Kane, C.~L.}
\newblock \bibinfo{title}{\textit{Colloquium} : Topological insulators}.
\newblock \emph{\bibinfo{journal}{Rev. Mod. Phys.}}
  \textbf{\bibinfo{volume}{82}}, \bibinfo{pages}{3045--3067}
  (\bibinfo{year}{2010}).

\bibitem{Qi11}
\bibinfo{author}{Qi, X.-L.} \& \bibinfo{author}{Zhang, S.-C.}
\newblock \bibinfo{title}{Topological insulators and superconductors}.
\newblock \emph{\bibinfo{journal}{Rev. Mod. Phys.}}
  \textbf{\bibinfo{volume}{83}}, \bibinfo{pages}{1057--1110}
  (\bibinfo{year}{2011}).

\bibitem{Zhang2009}
\bibinfo{author}{Zhang, H.} \emph{et~al.}
\newblock \bibinfo{title}{Topological insulators in {Bi}$_2${Se}$_3$,
  {Bi}$_2${Te}$_3$ and {Sb}$_2${Te}$_3$ with a single {Dirac} cone on the
  surface}.
\newblock \emph{\bibinfo{journal}{Nat. Phys.}} \textbf{\bibinfo{volume}{5}},
  \bibinfo{pages}{438} (\bibinfo{year}{2009}).

\bibitem{Nielsen-Ninomiya1983}
\bibinfo{author}{{Nielsen}, H.~B.} \& \bibinfo{author}{{Ninomiya}, M.}
\newblock \bibinfo{title}{{The Adler-Bell-Jackiw anomaly and {Weyl} fermions in
  a crystal}}.
\newblock \emph{\bibinfo{journal}{Phys. Lett. B}}
  \textbf{\bibinfo{volume}{130}}, \bibinfo{pages}{389--396}
  (\bibinfo{year}{1983}).

\bibitem{Murakami2007}
\bibinfo{author}{Murakami, S.}
\newblock \bibinfo{title}{Phase transition between the quantum spin {H}all and
  insulator phases in 3d: emergence of a topological gapless phase}.
\newblock \emph{\bibinfo{journal}{New Journal of Physics}}
  \textbf{\bibinfo{volume}{9}}, \bibinfo{pages}{356} (\bibinfo{year}{2007}).

\bibitem{Wan2011}
\bibinfo{author}{Wan, X.}, \bibinfo{author}{Turner, A.~M.},
  \bibinfo{author}{Vishwanath, A.} \& \bibinfo{author}{Savrasov, S.~Y.}
\newblock \bibinfo{title}{Topological semimetal and fermi-arc surface states in
  the electronic structure of pyrochlore iridates}.
\newblock \emph{\bibinfo{journal}{Phys. Rev. B}} \textbf{\bibinfo{volume}{83}},
  \bibinfo{pages}{205101} (\bibinfo{year}{2011}).

\bibitem{Burkov-Balents2011}
\bibinfo{author}{Burkov, A.~A.} \& \bibinfo{author}{Balents, L.}
\newblock \bibinfo{title}{Weyl semimetal in a topological insulator
  multilayer}.
\newblock \emph{\bibinfo{journal}{Phys. Rev. Lett.}}
  \textbf{\bibinfo{volume}{107}}, \bibinfo{pages}{127205}
  (\bibinfo{year}{2011}).

\bibitem{Nielsen-Ninomiya1981}
\bibinfo{author}{Nielsen, H.~B.} \& \bibinfo{author}{Ninomiya, M.}
\newblock \bibinfo{title}{{Absence of neutrinos on a lattice. (I). Proof by
  homotopy theory}}.
\newblock \emph{\bibinfo{journal}{Nuclear Phys. B}}
  \textbf{\bibinfo{volume}{185}}, \bibinfo{pages}{20} (\bibinfo{year}{1981}).

\bibitem{Nielsen-Ninomiya1981b}
\bibinfo{author}{Nielsen, H.~B.} \& \bibinfo{author}{Ninomiya, M.}
\newblock \bibinfo{title}{{Absence of neutrinos on a lattice. (II). Intuitive
  topological proof}}.
\newblock \emph{\bibinfo{journal}{Nuclear Phys. B}}
  \textbf{\bibinfo{volume}{193}}, \bibinfo{pages}{173} (\bibinfo{year}{1981}).

\bibitem{Young2012}
\bibinfo{author}{Young, S.~M.} \emph{et~al.}
\newblock \bibinfo{title}{Dirac semimetal in three dimensions}.
\newblock \emph{\bibinfo{journal}{Phys. Rev. Lett.}}
  \textbf{\bibinfo{volume}{108}}, \bibinfo{pages}{140405}
  (\bibinfo{year}{2012}).

\bibitem{Wang2012}
\bibinfo{author}{Wang, Z.} \emph{et~al.}
\newblock \bibinfo{title}{Dirac semimetal and topological phase transitions in
  {A}$_{3}${B}i ({A}$=${N}a, {K}, {R}b)}.
\newblock \emph{\bibinfo{journal}{Phys. Rev. B}} \textbf{\bibinfo{volume}{85}},
  \bibinfo{pages}{195320} (\bibinfo{year}{2012}).

\bibitem{Wang2013}
\bibinfo{author}{Wang, Z.}, \bibinfo{author}{Weng, H.}, \bibinfo{author}{Wu,
  Q.}, \bibinfo{author}{Dai, X.} \& \bibinfo{author}{Fang, Z.}
\newblock \bibinfo{title}{Three-dimensional {Dirac} semimetal and quantum
  transport in {C}d${}_{3}${A}s${}_{2}$}.
\newblock \emph{\bibinfo{journal}{Phys. Rev. B}} \textbf{\bibinfo{volume}{88}},
  \bibinfo{pages}{125427} (\bibinfo{year}{2013}).

\bibitem{Yang2014}
\bibinfo{author}{{Yang}, B.-J.} \& \bibinfo{author}{{Nagaosa}, N.}
\newblock \bibinfo{title}{{Classification of stable three-dimensional {Dirac}
  semimetals with nontrivial topology}}.
\newblock \emph{\bibinfo{journal}{Nat. Commun.}} \textbf{\bibinfo{volume}{5}},
  \bibinfo{pages}{4898} (\bibinfo{year}{2014}).

\bibitem{Novak2015}
\bibinfo{author}{Novak, M.}, \bibinfo{author}{Sasaki, S.},
  \bibinfo{author}{Segawa, K.} \& \bibinfo{author}{Ando, Y.}
\newblock \bibinfo{title}{Large linear magnetoresistance in the dirac semimetal
  {TlBiSSe}}.
\newblock \emph{\bibinfo{journal}{Phys. Rev. B}} \textbf{\bibinfo{volume}{91}},
  \bibinfo{pages}{041203} (\bibinfo{year}{2015}).

\bibitem{Neupane2014}
\bibinfo{author}{{Neupane}, M.} \emph{et~al.}
\newblock \bibinfo{title}{{Observation of a three-dimensional topological
  {Dirac} semimetal phase in high-mobility Cd$_{3}$As$_{2}$}}.
\newblock \emph{\bibinfo{journal}{Nat. Commun.}} \textbf{\bibinfo{volume}{5}},
  \bibinfo{pages}{3786} (\bibinfo{year}{2014}).

\bibitem{Borisenko2014}
\bibinfo{author}{Borisenko, S.} \emph{et~al.}
\newblock \bibinfo{title}{Experimental realization of a three-dimensional
  {Dirac} semimetal}.
\newblock \emph{\bibinfo{journal}{Phys. Rev. Lett.}}
  \textbf{\bibinfo{volume}{113}}, \bibinfo{pages}{027603}
  (\bibinfo{year}{2014}).

\bibitem{Liu2014}
\bibinfo{author}{{Liu}, Z.~K.} \emph{et~al.}
\newblock \bibinfo{title}{{A stable three-dimensional topological {Dirac}
  semimetal Cd$_{3}$As$_{2}$}}.
\newblock \emph{\bibinfo{journal}{Nat. Mater.}} \textbf{\bibinfo{volume}{13}},
  \bibinfo{pages}{677--681} (\bibinfo{year}{2014}).

\bibitem{Yi2014}
\bibinfo{author}{{Yi}, H.} \emph{et~al.}
\newblock \bibinfo{title}{{Evidence of Topological Surface State in
  Three-Dimensional {Dirac} Semimetal Cd$_{3}$As$_{2}$}}.
\newblock \emph{\bibinfo{journal}{Sci. Rep.}} \textbf{\bibinfo{volume}{4}},
  \bibinfo{pages}{6106} (\bibinfo{year}{2014}).

\bibitem{Sigrist-Ueda1991}
\bibinfo{author}{Sigrist, M.} \& \bibinfo{author}{Ueda, K.}
\newblock \bibinfo{title}{Phenomenological theory of unconventional
  superconductivity}.
\newblock \emph{\bibinfo{journal}{Rev. Mod. Phys.}}
  \textbf{\bibinfo{volume}{63}}, \bibinfo{pages}{239--311}
  (\bibinfo{year}{1991}).

\bibitem{Frigeri2004}
\bibinfo{author}{Frigeri, P.~A.}, \bibinfo{author}{Agterberg, D.~F.},
  \bibinfo{author}{Koga, A.} \& \bibinfo{author}{Sigrist, M.}
\newblock \bibinfo{title}{Superconductivity without inversion symmetry: Mnsi
  versus
  {${\mathrm{C}\mathrm{e}\mathrm{P}\mathrm{t}}_{3}\mathrm{S}\mathrm{i}$}}.
\newblock \emph{\bibinfo{journal}{Phys. Rev. Lett.}}
  \textbf{\bibinfo{volume}{92}}, \bibinfo{pages}{097001}
  (\bibinfo{year}{2004}).

\bibitem{Noncentro}
\bibinfo{author}{Bauer, E.} \& \bibinfo{author}{Ed., M.~S.}
\newblock \emph{\bibinfo{title}{Non-centrosymmetric Superconductors
  Introduction and Overview}} (\bibinfo{publisher}{Springer},
  \bibinfo{year}{2012}).

\bibitem{Caroli1964}
\bibinfo{author}{{Caroli}, C.}, \bibinfo{author}{{De Gennes}, P.~G.} \&
  \bibinfo{author}{{Matricon}, J.}
\newblock \bibinfo{title}{{Bound Fermion states on a vortex line in a type II
  superconductor}}.
\newblock \emph{\bibinfo{journal}{Physics Letters}}
  \textbf{\bibinfo{volume}{9}}, \bibinfo{pages}{307--309}
  (\bibinfo{year}{1964}).

\bibitem{Kashiwaya00}
\bibinfo{author}{Kashiwaya, S.} \& \bibinfo{author}{Tanaka, Y.}
\newblock \bibinfo{title}{Tunnelling effects on surface bound states in
  unconventional superconductors}.
\newblock \emph{\bibinfo{journal}{Rep. Prog. Phys.}}
  \textbf{\bibinfo{volume}{63}}, \bibinfo{pages}{1641} (\bibinfo{year}{2000}).

\bibitem{Lofwander2001}
\bibinfo{author}{Löfwander, T.}, \bibinfo{author}{Shumeiko, V.~S.} \&
  \bibinfo{author}{Wendin, G.}
\newblock \bibinfo{title}{Andreev bound states in high- t c superconducting
  junctions}.
\newblock \emph{\bibinfo{journal}{Superconductor Science and Technology}}
  \textbf{\bibinfo{volume}{14}}, \bibinfo{pages}{R53} (\bibinfo{year}{2001}).

\bibitem{Hara1986}
\bibinfo{author}{Hara, J.} \& \bibinfo{author}{Nagai, K.}
\newblock \bibinfo{title}{A polar state in a slab as a soluble model of
  $p$-wave fermi superfluid in finite geometry}.
\newblock \emph{\bibinfo{journal}{Progress of Theoretical Physics}}
  \textbf{\bibinfo{volume}{76}}, \bibinfo{pages}{1237--1249}
  (\bibinfo{year}{1986}).

\bibitem{Hu1994}
\bibinfo{author}{Hu, C.-R.}
\newblock \bibinfo{title}{Midgap surface states as a novel signature for
  $d_{x_a^2-x_b^2}$-wave superconductivity}.
\newblock \emph{\bibinfo{journal}{Phys. Rev. Lett.}}
  \textbf{\bibinfo{volume}{72}}, \bibinfo{pages}{1526--1529}
  (\bibinfo{year}{1994}).

\bibitem{Kopnin-Salomaa1991}
\bibinfo{author}{Kopnin, N.~B.} \& \bibinfo{author}{Salomaa, M.~M.}
\newblock \bibinfo{title}{Mutual friction in superfluid $^{3}\mathrm{He}$:
  Effects of bound states in the vortex core}.
\newblock \emph{\bibinfo{journal}{Phys. Rev. B}} \textbf{\bibinfo{volume}{44}},
  \bibinfo{pages}{9667--9677} (\bibinfo{year}{1991}).

\bibitem{Tanaka-Kashiwaya1995}
\bibinfo{author}{Tanaka, Y.} \& \bibinfo{author}{Kashiwaya, S.}
\newblock \bibinfo{title}{Theory of tunneling spectroscopy of $\mathit{d}$-wave
  superconductors}.
\newblock \emph{\bibinfo{journal}{Phys. Rev. Lett.}}
  \textbf{\bibinfo{volume}{74}}, \bibinfo{pages}{3451--3454}
  (\bibinfo{year}{1995}).

\bibitem{Soliton}
\bibinfo{author}{Manton, N.} \& \bibinfo{author}{Sutcliffe, P.}
\newblock \emph{\bibinfo{title}{Topological Solitons}}
  (\bibinfo{publisher}{Cambridge University Press}, \bibinfo{year}{2004}).

\bibitem{Sato-Fujimoto2009}
\bibinfo{author}{Sato, M.} \& \bibinfo{author}{Fujimoto, S.}
\newblock \bibinfo{title}{Topological phases of noncentrosymmetric
  superconductors: Edge states, {Majorana} fermions, and non-abelian
  statistics}.
\newblock \emph{\bibinfo{journal}{Phys. Rev. B}} \textbf{\bibinfo{volume}{79}},
  \bibinfo{pages}{094504} (\bibinfo{year}{2009}).

\bibitem{Tanaka-Yokoyama2009}
\bibinfo{author}{Tanaka, Y.}, \bibinfo{author}{Yokoyama, T.},
  \bibinfo{author}{Balatsky, A.~V.} \& \bibinfo{author}{Nagaosa, N.}
\newblock \bibinfo{title}{Theory of topological spin current in
  noncentrosymmetric superconductors}.
\newblock \emph{\bibinfo{journal}{Phys. Rev. B}} \textbf{\bibinfo{volume}{79}},
  \bibinfo{pages}{060505} (\bibinfo{year}{2009}).

\bibitem{Grinevich1988}
\bibinfo{author}{Grinevich, P.~G.} \& \bibinfo{author}{Volovik, G.~E.}
\newblock \bibinfo{title}{Topology of gap nodes in superfluid3he: $\pi_4$
  homotopy group for {${}^3$He-B} disclination}.
\newblock \emph{\bibinfo{journal}{J. Low Temp. Phys.}}
  \textbf{\bibinfo{volume}{72}}, \bibinfo{pages}{371--380}
  (\bibinfo{year}{1988}).

\bibitem{Tanaka-Sato-Nagaosa2012}
\bibinfo{author}{Tanaka, Y.}, \bibinfo{author}{Sato, M.} \&
  \bibinfo{author}{Nagaosa, N.}
\newblock \bibinfo{title}{Symmetry and topology in superconductors
  –odd-frequency pairing and edge states–}.
\newblock \emph{\bibinfo{journal}{J. Phys. Soc. Jpn.}}
  \textbf{\bibinfo{volume}{81}}, \bibinfo{pages}{011013}
  (\bibinfo{year}{2012}).

\bibitem{Hatsugai1993}
\bibinfo{author}{Hatsugai, Y.}
\newblock \bibinfo{title}{Chern number and edge states in the integer quantum
  {H}all effect}.
\newblock \emph{\bibinfo{journal}{Phys. Rev. Lett.}}
  \textbf{\bibinfo{volume}{71}}, \bibinfo{pages}{3697--3700}
  (\bibinfo{year}{1993}).

\bibitem{Teo-Kane2010}
\bibinfo{author}{Teo, J. C.~Y.} \& \bibinfo{author}{Kane, C.~L.}
\newblock \bibinfo{title}{Topological defects and gapless modes in insulators
  and superconductors}.
\newblock \emph{\bibinfo{journal}{Phys. Rev. B}} \textbf{\bibinfo{volume}{82}},
  \bibinfo{pages}{115120} (\bibinfo{year}{2010}).

\bibitem{Sato-Tanaka-Yada-Yokoyama2011}
\bibinfo{author}{Sato, M.}, \bibinfo{author}{Tanaka, Y.},
  \bibinfo{author}{Yada, K.} \& \bibinfo{author}{Yokoyama, T.}
\newblock \bibinfo{title}{Topology of {Andreev} bound states with flat
  dispersion}.
\newblock \emph{\bibinfo{journal}{Phys. Rev. B}} \textbf{\bibinfo{volume}{83}},
  \bibinfo{pages}{224511} (\bibinfo{year}{2011}).

\bibitem{Essin2011}
\bibinfo{author}{Essin, A.~M.} \& \bibinfo{author}{Gurarie, V.}
\newblock \bibinfo{title}{Bulk-boundary correspondence of topological
  insulators from their respective green's functions}.
\newblock \emph{\bibinfo{journal}{Phys. Rev. B}} \textbf{\bibinfo{volume}{84}},
  \bibinfo{pages}{125132} (\bibinfo{year}{2011}).

\bibitem{Teo2013}
\bibinfo{author}{Teo, J. C.~Y.} \& \bibinfo{author}{Hughes, T.~L.}
\newblock \bibinfo{title}{Existence of {Majorana}-fermion bound states on
  disclinations and the classification of topological crystalline
  superconductors in two dimensions}.
\newblock \emph{\bibinfo{journal}{Phys. Rev. Lett.}}
  \textbf{\bibinfo{volume}{111}}, \bibinfo{pages}{047006}
  (\bibinfo{year}{2013}).

\bibitem{Ueno2013}
\bibinfo{author}{Ueno, Y.}, \bibinfo{author}{Yamakage, A.},
  \bibinfo{author}{Tanaka, Y.} \& \bibinfo{author}{Sato, M.}
\newblock \bibinfo{title}{Symmetry-protected {Majorana} fermions in topological
  crystalline superconductors: Theory and application to
  {${\mathrm{Sr}}_{2}{\mathrm{RuO}}_{4}$}}.
\newblock \emph{\bibinfo{journal}{Phys. Rev. Lett.}}
  \textbf{\bibinfo{volume}{111}}, \bibinfo{pages}{087002}
  (\bibinfo{year}{2013}).

\bibitem{Benalcazar2014}
\bibinfo{author}{Benalcazar, W.~A.}, \bibinfo{author}{Teo, J. C.~Y.} \&
  \bibinfo{author}{Hughes, T.~L.}
\newblock \bibinfo{title}{Classification of two-dimensional topological
  crystalline superconductors and {Majorana} bound states at disclinations}.
\newblock \emph{\bibinfo{journal}{Phys. Rev. B}} \textbf{\bibinfo{volume}{89}},
  \bibinfo{pages}{224503} (\bibinfo{year}{2014}).

\bibitem{Shiozaki2014}
\bibinfo{author}{Shiozaki, K.} \& \bibinfo{author}{Sato, M.}
\newblock \bibinfo{title}{Topology of crystalline insulators and
  superconductors}.
\newblock \emph{\bibinfo{journal}{Phys. Rev. B}} \textbf{\bibinfo{volume}{90}},
  \bibinfo{pages}{165114} (\bibinfo{year}{2014}).

\bibitem{Tsutsumi2013}
\bibinfo{author}{Tsutsumi, Y.} \emph{et~al.}
\newblock \bibinfo{title}{{UPt$_3$} as a topological crystalline
  superconductor}.
\newblock \emph{\bibinfo{journal}{J. Phys. Soc. Jpn.}}
  \textbf{\bibinfo{volume}{82}}, \bibinfo{pages}{113707}
  (\bibinfo{year}{2013}).

\bibitem{Chang2014}
\bibinfo{author}{Chang, P.-Y.}, \bibinfo{author}{Matsuura, S.},
  \bibinfo{author}{Schnyder, A.~P.} \& \bibinfo{author}{Ryu, S.}
\newblock \bibinfo{title}{Majorana vortex-bound states in three-dimensional
  nodal noncentrosymmetric superconductors}.
\newblock \emph{\bibinfo{journal}{Phys. Rev. B}} \textbf{\bibinfo{volume}{90}},
  \bibinfo{pages}{174504} (\bibinfo{year}{2014}).

\bibitem{Lee2016}
\bibinfo{author}{Lee, D.} \& \bibinfo{author}{Schnyder, A.~P.}
\newblock \bibinfo{title}{Structure of vortex-bound states in spin-singlet
  chiral superconductors}.
\newblock \emph{\bibinfo{journal}{Phys. Rev. B}} \textbf{\bibinfo{volume}{93}},
  \bibinfo{pages}{064522} (\bibinfo{year}{2016}).

\bibitem{Tsutsumi2015}
\bibinfo{author}{Tsutsumi, Y.}, \bibinfo{author}{Kawakami, T.},
  \bibinfo{author}{Shiozaki, K.}, \bibinfo{author}{Sato, M.} \&
  \bibinfo{author}{Machida, K.}
\newblock \bibinfo{title}{Symmetry-protected vortex bound state in superfluid
  {$^{3}\mathrm{He}\text{-}B$ phase}}.
\newblock \emph{\bibinfo{journal}{Phys. Rev. B}} \textbf{\bibinfo{volume}{91}},
  \bibinfo{pages}{144504} (\bibinfo{year}{2015}).

\bibitem{Fu2011}
\bibinfo{author}{Fu, L.}
\newblock \bibinfo{title}{Topological crystalline insulators}.
\newblock \emph{\bibinfo{journal}{Phys. Rev. Lett.}}
  \textbf{\bibinfo{volume}{106}}, \bibinfo{pages}{106802}
  (\bibinfo{year}{2011}).

\bibitem{Mizushima2012}
\bibinfo{author}{Mizushima, T.}, \bibinfo{author}{Sato, M.} \&
  \bibinfo{author}{Machida, K.}
\newblock \bibinfo{title}{Symmetry protected topological order and spin
  susceptibility in superfluid {$^{3}$He-B}}.
\newblock \emph{\bibinfo{journal}{Phys. Rev. Lett.}}
  \textbf{\bibinfo{volume}{109}}, \bibinfo{pages}{165301}
  (\bibinfo{year}{2012}).

\bibitem{Zhang-Kane-Mele2013}
\bibinfo{author}{Zhang, F.}, \bibinfo{author}{Kane, C.~L.} \&
  \bibinfo{author}{Mele, E.~J.}
\newblock \bibinfo{title}{Topological mirror superconductivity}.
\newblock \emph{\bibinfo{journal}{Phys. Rev. Lett.}}
  \textbf{\bibinfo{volume}{111}}, \bibinfo{pages}{056403}
  (\bibinfo{year}{2013}).

\bibitem{Chiu2013}
\bibinfo{author}{Chiu, C.-K.}, \bibinfo{author}{Yao, H.} \&
  \bibinfo{author}{Ryu, S.}
\newblock \bibinfo{title}{Classification of topological insulators and
  superconductors in the presence of reflection symmetry}.
\newblock \emph{\bibinfo{journal}{Phys. Rev. B}} \textbf{\bibinfo{volume}{88}},
  \bibinfo{pages}{075142} (\bibinfo{year}{2013}).

\bibitem{Morimoto2013}
\bibinfo{author}{Morimoto, T.} \& \bibinfo{author}{Furusaki, A.}
\newblock \bibinfo{title}{Topological classification with additional symmetries
  from clifford algebras}.
\newblock \emph{\bibinfo{journal}{Phys. Rev. B}} \textbf{\bibinfo{volume}{88}},
  \bibinfo{pages}{125129} (\bibinfo{year}{2013}).

\bibitem{Shiozaki-Sato-Gomi2016}
\bibinfo{author}{Shiozaki, K.}, \bibinfo{author}{Sato, M.} \&
  \bibinfo{author}{Gomi, K.}
\newblock \bibinfo{title}{Topology of nonsymmorphic crystalline insulators and
  superconductors}.
\newblock \emph{\bibinfo{journal}{Phys. Rev. B}} \textbf{\bibinfo{volume}{93}},
  \bibinfo{pages}{195413} (\bibinfo{year}{2016}).

\bibitem{Teo-Fu-Kane2008}
\bibinfo{author}{Teo, J. C.~Y.}, \bibinfo{author}{Fu, L.} \&
  \bibinfo{author}{Kane, C.~L.}
\newblock \bibinfo{title}{Surface states and topological invariants in
  three-dimensional topological insulators: Application to
  {${\text{Bi}}_{1\ensuremath{-}x}{\text{Sb}}_{x}$}}.
\newblock \emph{\bibinfo{journal}{Phys. Rev. B}} \textbf{\bibinfo{volume}{78}},
  \bibinfo{pages}{045426} (\bibinfo{year}{2008}).

\bibitem{HsiehTCI2012}
\bibinfo{author}{Hsieh, T.~H.} \emph{et~al.}
\newblock \bibinfo{title}{Topological crystalline insulators in the snte
  material class}.
\newblock \emph{\bibinfo{journal}{Nat. Commun.}} \textbf{\bibinfo{volume}{3}},
  \bibinfo{pages}{982} (\bibinfo{year}{2012}).

\bibitem{Tanaka2012}
\bibinfo{author}{Tanaka, Y.} \emph{et~al.}
\newblock \bibinfo{title}{Experimental realization of a topological crystalline
  insulator in snte}.
\newblock \emph{\bibinfo{journal}{Nat. Phys.}} \textbf{\bibinfo{volume}{8}},
  \bibinfo{pages}{800--803} (\bibinfo{year}{2012}).

\bibitem{Meng2012}
\bibinfo{author}{Meng, T.} \& \bibinfo{author}{Balents, L.}
\newblock \bibinfo{title}{Weyl superconductors}.
\newblock \emph{\bibinfo{journal}{Phys. Rev. B}} \textbf{\bibinfo{volume}{86}},
  \bibinfo{pages}{054504} (\bibinfo{year}{2012}).

\bibitem{Sato2006}
\bibinfo{author}{Sato, M.}
\newblock \bibinfo{title}{Nodal structure of superconductors with time-reversal
  invariance and {${\bf Z}_{2}$} topological number}.
\newblock \emph{\bibinfo{journal}{Phys. Rev. B}} \textbf{\bibinfo{volume}{73}},
  \bibinfo{pages}{214502} (\bibinfo{year}{2006}).

\bibitem{Beri2010}
\bibinfo{author}{B\'eri, B.}
\newblock \bibinfo{title}{Topologically stable gapless phases of
  time-reversal-invariant superconductors}.
\newblock \emph{\bibinfo{journal}{Phys. Rev. B}} \textbf{\bibinfo{volume}{81}},
  \bibinfo{pages}{134515} (\bibinfo{year}{2010}).

\bibitem{Schnyder-Ryu2011}
\bibinfo{author}{Schnyder, A.~P.} \& \bibinfo{author}{Ryu, S.}
\newblock \bibinfo{title}{Topological phases and surface flat bands in
  superconductors without inversion symmetry}.
\newblock \emph{\bibinfo{journal}{Phys. Rev. B}} \textbf{\bibinfo{volume}{84}},
  \bibinfo{pages}{060504} (\bibinfo{year}{2011}).

\bibitem{Mizuno2010}
\bibinfo{author}{Tanaka, Y.}, \bibinfo{author}{Mizuno, Y.},
  \bibinfo{author}{Yokoyama, T.}, \bibinfo{author}{Yada, K.} \&
  \bibinfo{author}{Sato, M.}
\newblock \bibinfo{title}{Anomalous {Andreev} bound state in noncentrosymmetric
  superconductors}.
\newblock \emph{\bibinfo{journal}{Phys. Rev. Lett.}}
  \textbf{\bibinfo{volume}{105}}, \bibinfo{pages}{097002}
  (\bibinfo{year}{2010}).

\bibitem{Yada2011}
\bibinfo{author}{Yada, K.}, \bibinfo{author}{Sato, M.},
  \bibinfo{author}{Tanaka, Y.} \& \bibinfo{author}{Yokoyama, T.}
\newblock \bibinfo{title}{Surface density of states and topological edge states
  in noncentrosymmetric superconductors}.
\newblock \emph{\bibinfo{journal}{Phys. Rev. B}} \textbf{\bibinfo{volume}{83}},
  \bibinfo{pages}{064505} (\bibinfo{year}{2011}).

\bibitem{Schnyder2012}
\bibinfo{author}{Schnyder, A.~P.}, \bibinfo{author}{Brydon, P. M.~R.} \&
  \bibinfo{author}{Timm, C.}
\newblock \bibinfo{title}{Types of topological surface states in nodal
  noncentrosymmetric superconductors}.
\newblock \emph{\bibinfo{journal}{Phys. Rev. B}} \textbf{\bibinfo{volume}{85}},
  \bibinfo{pages}{024522} (\bibinfo{year}{2012}).

\bibitem{Ikegaya2016}
\bibinfo{author}{Ikegaya, S.}, \bibinfo{author}{Suzuki, S.-I.},
  \bibinfo{author}{Tanaka, Y.} \& \bibinfo{author}{Asano, Y.}
\newblock \bibinfo{title}{Quantization of conductance minimum and index
  theorem}.
\newblock \emph{\bibinfo{journal}{Phys. Rev. B}} \textbf{\bibinfo{volume}{94}},
  \bibinfo{pages}{054512} (\bibinfo{year}{2016}).

\bibitem{Kobayashi2014}
\bibinfo{author}{Kobayashi, S.}, \bibinfo{author}{Shiozaki, K.},
  \bibinfo{author}{Tanaka, Y.} \& \bibinfo{author}{Sato, M.}
\newblock \bibinfo{title}{Topological {B}lount's theorem of odd-parity
  superconductors}.
\newblock \emph{\bibinfo{journal}{Phys. Rev. B}} \textbf{\bibinfo{volume}{90}},
  \bibinfo{pages}{024516} (\bibinfo{year}{2014}).

\bibitem{Zhao2016}
\bibinfo{author}{Zhao, Y.~X.}, \bibinfo{author}{Schnyder, A.~P.} \&
  \bibinfo{author}{Wang, Z.~D.}
\newblock \bibinfo{title}{Unified theory of {$PT$} and {$CP$} invariant
  topological metals and nodal superconductors}.
\newblock \emph{\bibinfo{journal}{Phys. Rev. Lett.}}
  \textbf{\bibinfo{volume}{116}}, \bibinfo{pages}{156402}
  (\bibinfo{year}{2016}).

\bibitem{Blount}
\bibinfo{author}{Blount, E.~I.}
\newblock \bibinfo{title}{Symmetry properties of triplet superconductors}.
\newblock \emph{\bibinfo{journal}{Phys. Rev. B}} \textbf{\bibinfo{volume}{32}},
  \bibinfo{pages}{2935--2944} (\bibinfo{year}{1985}).

\bibitem{Agterberg2016}
\bibinfo{author}{{Agterberg}, D.~F.}, \bibinfo{author}{{Brydon}, P.~M.~R.} \&
  \bibinfo{author}{{Timm}, C.}
\newblock \bibinfo{title}{{Inflated nodes in superconductors with broken
  time-reversal symmetry}}.
\newblock \emph{\bibinfo{journal}{ArXiv e-prints}}  (\bibinfo{year}{2016}).
\newblock arXiv:\eprint{1608.06461}.

\bibitem{Norman1995}
\bibinfo{author}{Norman, M.~R.}
\newblock \bibinfo{title}{Odd parity and line nodes in heavy-fermion
  superconductors}.
\newblock \emph{\bibinfo{journal}{Phys. Rev. B}} \textbf{\bibinfo{volume}{52}},
  \bibinfo{pages}{15093--15094} (\bibinfo{year}{1995}).

\bibitem{Micklitz2009}
\bibinfo{author}{Micklitz, T.} \& \bibinfo{author}{Norman, M.~R.}
\newblock \bibinfo{title}{Odd parity and line nodes in nonsymmorphic
  superconductors}.
\newblock \emph{\bibinfo{journal}{Phys. Rev. B}} \textbf{\bibinfo{volume}{80}},
  \bibinfo{pages}{100506} (\bibinfo{year}{2009}).

\bibitem{Nomoto2016}
\bibinfo{author}{{Nomoto}, T.} \& \bibinfo{author}{{Ikeda}, H.}
\newblock \bibinfo{title}{Symmetry protected line nodes in non-symmorphic
  magnetic space groups: Applications to {UCoGe} and {UPd$_2$Al$_3$}}.
\newblock \emph{\bibinfo{journal}{ArXiv e-prints}}  (\bibinfo{year}{2016}).
\newblock arXiv:\eprint{1610.04679}.

\bibitem{Yang-Pan2014}
\bibinfo{author}{Yang, S.~A.}, \bibinfo{author}{Pan, H.} \&
  \bibinfo{author}{Zhang, F.}
\newblock \bibinfo{title}{Dirac and {Weyl} superconductors in three
  dimensions}.
\newblock \emph{\bibinfo{journal}{Phys. Rev. Lett.}}
  \textbf{\bibinfo{volume}{113}}, \bibinfo{pages}{046401}
  (\bibinfo{year}{2014}).

\bibitem{Chiu2014}
\bibinfo{author}{Chiu, C.-K.} \& \bibinfo{author}{Schnyder, A.~P.}
\newblock \bibinfo{title}{Classification of reflection-symmetry-protected
  topological semimetals and nodal superconductors}.
\newblock \emph{\bibinfo{journal}{Phys. Rev. B}} \textbf{\bibinfo{volume}{90}},
  \bibinfo{pages}{205136} (\bibinfo{year}{2014}).

\bibitem{Kobayashi2015}
\bibinfo{author}{Kobayashi, S.}, \bibinfo{author}{Tanaka, Y.} \&
  \bibinfo{author}{Sato, M.}
\newblock \bibinfo{title}{Fragile surface zero-energy flat bands in
  three-dimensional chiral superconductors}.
\newblock \emph{\bibinfo{journal}{Phys. Rev. B}} \textbf{\bibinfo{volume}{92}},
  \bibinfo{pages}{214514} (\bibinfo{year}{2015}).

\bibitem{Kobayashi2016}
\bibinfo{author}{Kobayashi, S.}, \bibinfo{author}{Yanase, Y.} \&
  \bibinfo{author}{Sato, M.}
\newblock \bibinfo{title}{Topologically stable gapless phases in nonsymmorphic
  superconductors}.
\newblock \emph{\bibinfo{journal}{Phys. Rev. B}} \textbf{\bibinfo{volume}{94}},
  \bibinfo{pages}{134512} (\bibinfo{year}{2016}).

\bibitem{Micklitz2016}
\bibinfo{author}{{Micklitz}, T.} \& \bibinfo{author}{{Norman}, M.~R.}
\newblock \bibinfo{title}{{Nodal-line superconductors and band-sticking}}.
\newblock \emph{\bibinfo{journal}{ArXiv e-prints}}  (\bibinfo{year}{2016}).
\newblock arXiv:\eprint{1611.07590}.

\bibitem{Mizushima2016}
\bibinfo{author}{{Mizushima}, T.} \emph{et~al.}
\newblock \bibinfo{title}{{Symmetry-Protected Topological Superfluids and
  Superconductors --From the Basics to {$^{3}$He}--}}.
\newblock \emph{\bibinfo{journal}{J. Phys. Soc. Jpn.}}
  \textbf{\bibinfo{volume}{85}}, \bibinfo{pages}{022001}
  (\bibinfo{year}{2016}).

\bibitem{Silaev2012}
\bibinfo{author}{Silaev, M.~A.} \& \bibinfo{author}{Volovik, G.~E.}
\newblock \bibinfo{title}{Topological fermi arcs in superfluid {${}^{3}$He}}.
\newblock \emph{\bibinfo{journal}{Phys. Rev. B}} \textbf{\bibinfo{volume}{86}},
  \bibinfo{pages}{214511} (\bibinfo{year}{2012}).

\bibitem{Goswami2013}
\bibinfo{author}{{Goswami}, P.} \& \bibinfo{author}{{Balicas}, L.}
\newblock \bibinfo{title}{Topological properties of possible {Weyl}
  superconducting states of {URu$_{2}$Si$_{2}$}}.
\newblock \emph{\bibinfo{journal}{ArXiv e-prints}}  (\bibinfo{year}{2013}).
\newblock arXiv:\eprint{1312.3632}.

\bibitem{Fischer2014}
\bibinfo{author}{Fischer, M.~H.} \emph{et~al.}
\newblock \bibinfo{title}{Chiral $d$-wave superconductivity in srptas}.
\newblock \emph{\bibinfo{journal}{Phys. Rev. B}} \textbf{\bibinfo{volume}{89}},
  \bibinfo{pages}{020509} (\bibinfo{year}{2014}).

\bibitem{Goswami2015}
\bibinfo{author}{Goswami, P.} \& \bibinfo{author}{Nevidomskyy, A.~H.}
\newblock \bibinfo{title}{Topological {Weyl} superconductor to diffusive
  thermal {H}all metal crossover in the {$B$} phase of {${\mathrm{UPt}}_{3}$}}.
\newblock \emph{\bibinfo{journal}{Phys. Rev. B}} \textbf{\bibinfo{volume}{92}},
  \bibinfo{pages}{214504} (\bibinfo{year}{2015}).

\bibitem{Bo2015}
\bibinfo{author}{Lu, B.}, \bibinfo{author}{Yada, K.}, \bibinfo{author}{Sato,
  M.} \& \bibinfo{author}{Tanaka, Y.}
\newblock \bibinfo{title}{Crossed surface flat bands of {Weyl} semimetal
  superconductors}.
\newblock \emph{\bibinfo{journal}{Phys. Rev. Lett.}}
  \textbf{\bibinfo{volume}{114}}, \bibinfo{pages}{096804}
  (\bibinfo{year}{2015}).

\bibitem{Ryu2002}
\bibinfo{author}{Ryu, S.} \& \bibinfo{author}{Hatsugai, Y.}
\newblock \bibinfo{title}{Topological origin of zero-energy edge states in
  particle-hole symmetric systems}.
\newblock \emph{\bibinfo{journal}{Phys. Rev. Lett.}}
  \textbf{\bibinfo{volume}{89}}, \bibinfo{pages}{077002}
  (\bibinfo{year}{2002}).

\bibitem{Brydon2011}
\bibinfo{author}{Brydon, P. M.~R.}, \bibinfo{author}{Schnyder, A.~P.} \&
  \bibinfo{author}{Timm, C.}
\newblock \bibinfo{title}{Topologically protected flat zero-energy surface
  bands in noncentrosymmetric superconductors}.
\newblock \emph{\bibinfo{journal}{Phys. Rev. B}} \textbf{\bibinfo{volume}{84}},
  \bibinfo{pages}{020501} (\bibinfo{year}{2011}).

\bibitem{Sato-Fujimoto2010}
\bibinfo{author}{Sato, M.} \& \bibinfo{author}{Fujimoto, S.}
\newblock \bibinfo{title}{Existence of {Majorana} fermions and topological
  order in nodal superconductors with spin-orbit interactions in external
  magnetic fields}.
\newblock \emph{\bibinfo{journal}{Phys. Rev. Lett.}}
  \textbf{\bibinfo{volume}{105}}, \bibinfo{pages}{217001}
  (\bibinfo{year}{2010}).

\bibitem{Sasaki2011}
\bibinfo{author}{Sasaki, S.} \emph{et~al.}
\newblock \bibinfo{title}{Topological superconductivity in
  {Cu$_{x}$Bi$_{2}$Se$_{3}$}}.
\newblock \emph{\bibinfo{journal}{Phys. Rev. Lett.}}
  \textbf{\bibinfo{volume}{107}}, \bibinfo{pages}{217001}
  (\bibinfo{year}{2011}).

\bibitem{Yuval2015}
\bibinfo{author}{Baum, Y.}, \bibinfo{author}{Posske, T.},
  \bibinfo{author}{Fulga, I.~C.}, \bibinfo{author}{Trauzettel, B.} \&
  \bibinfo{author}{Stern, A.}
\newblock \bibinfo{title}{Coexisting edge states and gapless bulk in
  topological states of matter}.
\newblock \emph{\bibinfo{journal}{Phys. Rev. Lett.}}
  \textbf{\bibinfo{volume}{114}}, \bibinfo{pages}{136801}
  (\bibinfo{year}{2015}).

\bibitem{Yuval2015b}
\bibinfo{author}{Baum, Y.}, \bibinfo{author}{Posske, T.},
  \bibinfo{author}{Fulga, I.~C.}, \bibinfo{author}{Trauzettel, B.} \&
  \bibinfo{author}{Stern, A.}
\newblock \bibinfo{title}{Gapless topological superconductors: Model
  hamiltonian and realization}.
\newblock \emph{\bibinfo{journal}{Phys. Rev. B}} \textbf{\bibinfo{volume}{92}},
  \bibinfo{pages}{045128} (\bibinfo{year}{2015}).

\bibitem{Majorana37}
\bibinfo{author}{Majorana, E.}
\newblock \bibinfo{title}{Teoria simmetrica dell'elettrone e del positrone}.
\newblock \emph{\bibinfo{journal}{Nuovo Cimento}}
  \textbf{\bibinfo{volume}{14}}, \bibinfo{pages}{171} (\bibinfo{year}{1937}).

\bibitem{Wilczek2009}
\bibinfo{author}{{Wilczek}, F.}
\newblock \bibinfo{title}{{Majorana returns}}.
\newblock \emph{\bibinfo{journal}{Nature Physics}}
  \textbf{\bibinfo{volume}{5}}, \bibinfo{pages}{614--618}
  (\bibinfo{year}{2009}).

\bibitem{Ivanov2001}
\bibinfo{author}{Ivanov, D.~A.}
\newblock \bibinfo{title}{Non-abelian statistics of half-quantum vortices in
  $\mathit{p}$-wave superconductors}.
\newblock \emph{\bibinfo{journal}{Phys. Rev. Lett.}}
  \textbf{\bibinfo{volume}{86}}, \bibinfo{pages}{268--271}
  (\bibinfo{year}{2001}).

\bibitem{Alicea2011}
\bibinfo{author}{{Alicea}, J.}, \bibinfo{author}{{Oreg}, Y.},
  \bibinfo{author}{{Refael}, G.}, \bibinfo{author}{{von Oppen}, F.} \&
  \bibinfo{author}{{Fisher}, M.~P.~A.}
\newblock \bibinfo{title}{{Non-Abelian statistics and topological quantum
  information processing in 1D wire networks}}.
\newblock \emph{\bibinfo{journal}{Nature Physics}}
  \textbf{\bibinfo{volume}{7}}, \bibinfo{pages}{412--417}
  (\bibinfo{year}{2011}).
\newblock arXiv:\eprint{1006.4395}.

\bibitem{Sau2011}
\bibinfo{author}{Sau, J.~D.}, \bibinfo{author}{Clarke, D.~J.} \&
  \bibinfo{author}{Tewari, S.}
\newblock \bibinfo{title}{Controlling non-abelian statistics of {Majorana}
  fermions in semiconductor nanowires}.
\newblock \emph{\bibinfo{journal}{Phys. Rev. B}} \textbf{\bibinfo{volume}{84}},
  \bibinfo{pages}{094505} (\bibinfo{year}{2011}).

\bibitem{Kotetes2013}
\bibinfo{author}{Kotetes, P.}, \bibinfo{author}{Sch\"{o}n, G.} \&
  \bibinfo{author}{Shnirman, A.}
\newblock \bibinfo{title}{Engineering and manipulating topological qubits in 1d
  quantum wires}.
\newblock \emph{\bibinfo{journal}{Journal of the Korean Physical Society}}
  \textbf{\bibinfo{volume}{62}}, \bibinfo{pages}{1558--1563}
  (\bibinfo{year}{2013}).

\bibitem{Sau2010b}
\bibinfo{author}{Sau, J.~D.}, \bibinfo{author}{Tewari, S.} \&
  \bibinfo{author}{Das~Sarma, S.}
\newblock \bibinfo{title}{Universal quantum computation in a semiconductor
  quantum wire network}.
\newblock \emph{\bibinfo{journal}{Phys. Rev. A}} \textbf{\bibinfo{volume}{82}},
  \bibinfo{pages}{052322} (\bibinfo{year}{2010}).

\bibitem{Halperin2012}
\bibinfo{author}{Halperin, B.~I.} \emph{et~al.}
\newblock \bibinfo{title}{Adiabatic manipulations of {Majorana} fermions in a
  three-dimensional network of quantum wires}.
\newblock \emph{\bibinfo{journal}{Phys. Rev. B}} \textbf{\bibinfo{volume}{85}},
  \bibinfo{pages}{144501} (\bibinfo{year}{2012}).

\bibitem{Hyart2013}
\bibinfo{author}{Hyart, T.} \emph{et~al.}
\newblock \bibinfo{title}{Flux-controlled quantum computation with {Majorana}
  fermions}.
\newblock \emph{\bibinfo{journal}{Phys. Rev. B}} \textbf{\bibinfo{volume}{88}},
  \bibinfo{pages}{035121} (\bibinfo{year}{2013}).

\bibitem{Kraus2013}
\bibinfo{author}{Kraus, C.~V.}, \bibinfo{author}{Zoller, P.} \&
  \bibinfo{author}{Baranov, M.~A.}
\newblock \bibinfo{title}{Braiding of atomic {Majorana} fermions in wire
  networks and implementation of the {Deutsch-Jozsa} algorithm}.
\newblock \emph{\bibinfo{journal}{Phys. Rev. Lett.}}
  \textbf{\bibinfo{volume}{111}}, \bibinfo{pages}{203001}
  (\bibinfo{year}{2013}).

\bibitem{Liu2013}
\bibinfo{author}{Liu, X.-J.} \& \bibinfo{author}{Lobos, A.~M.}
\newblock \bibinfo{title}{Manipulating {Majorana} fermions in quantum nanowires
  with broken inversion symmetry}.
\newblock \emph{\bibinfo{journal}{Phys. Rev. B}} \textbf{\bibinfo{volume}{87}},
  \bibinfo{pages}{060504} (\bibinfo{year}{2013}).

\bibitem{Chiu2015}
\bibinfo{author}{Chiu, C.-K.}, \bibinfo{author}{Vazifeh, M.~M.} \&
  \bibinfo{author}{Franz, M.}
\newblock \bibinfo{title}{Majorana fermion exchange in strictly one-dimensional
  structures}.
\newblock \emph{\bibinfo{journal}{EPL (Europhysics Letters)}}
  \textbf{\bibinfo{volume}{110}}, \bibinfo{pages}{10001}
  (\bibinfo{year}{2015}).

\bibitem{Amorim2015}
\bibinfo{author}{Amorim, C.~S.}, \bibinfo{author}{Ebihara, K.},
  \bibinfo{author}{Yamakage, A.}, \bibinfo{author}{Tanaka, Y.} \&
  \bibinfo{author}{Sato, M.}
\newblock \bibinfo{title}{Majorana braiding dynamics in nanowires}.
\newblock \emph{\bibinfo{journal}{Phys. Rev. B}} \textbf{\bibinfo{volume}{91}},
  \bibinfo{pages}{174305} (\bibinfo{year}{2015}).

\bibitem{Kitaev2003}
\bibinfo{author}{Kitaev, A.}
\newblock \bibinfo{title}{Fault-tolerant quantum computation by anyons}.
\newblock \emph{\bibinfo{journal}{Annals of Physics}}
  \textbf{\bibinfo{volume}{303}}, \bibinfo{pages}{2 -- 30}
  (\bibinfo{year}{2003}).

\bibitem{Nayak2008}
\bibinfo{author}{Nayak, C.}, \bibinfo{author}{Simon, S.~H.},
  \bibinfo{author}{Stern, A.}, \bibinfo{author}{Freedman, M.} \&
  \bibinfo{author}{Das~Sarma, S.}
\newblock \bibinfo{title}{Non-abelian anyons and topological quantum
  computation}.
\newblock \emph{\bibinfo{journal}{Rev. Mod. Phys.}}
  \textbf{\bibinfo{volume}{80}}, \bibinfo{pages}{1083--1159}
  (\bibinfo{year}{2008}).

\bibitem{Chung09}
\bibinfo{author}{Chung, S.~B.} \& \bibinfo{author}{Zhang, S.-C.}
\newblock \bibinfo{title}{Detecting the {Majorana} fermion surface state of
  {$^{3}\mathrm{He}\mathrm{\text{-}}B$} through spin relaxation}.
\newblock \emph{\bibinfo{journal}{Phys. Rev. Lett.}}
  \textbf{\bibinfo{volume}{103}}, \bibinfo{pages}{235301}
  (\bibinfo{year}{2009}).

\bibitem{Nagato2009}
\bibinfo{author}{Nagato, Y.}, \bibinfo{author}{Higashitani, S.} \&
  \bibinfo{author}{Nagai, K.}
\newblock \bibinfo{title}{Strong anisotropy in spin susceptibility of
  superfluid {${}^3$He-B} film caused by surface bound states}.
\newblock \emph{\bibinfo{journal}{J. Phys. Soc. Jpn.}}
  \textbf{\bibinfo{volume}{78}}, \bibinfo{pages}{123603}
  (\bibinfo{year}{2009}).

\bibitem{Shindou10}
\bibinfo{author}{Shindou, R.}, \bibinfo{author}{Furusaki, A.} \&
  \bibinfo{author}{Nagaosa, N.}
\newblock \bibinfo{title}{Quantum impurity spin in {Majorana} edge fermions}.
\newblock \emph{\bibinfo{journal}{Phys. Rev. B}} \textbf{\bibinfo{volume}{82}},
  \bibinfo{pages}{180505} (\bibinfo{year}{2010}).

\bibitem{Sato2009}
\bibinfo{author}{Sato, M.}
\newblock \bibinfo{title}{Topological properties of spin-triplet
  superconductors and fermi surface topology in the normal state}.
\newblock \emph{\bibinfo{journal}{Phys. Rev. B}} \textbf{\bibinfo{volume}{79}},
  \bibinfo{pages}{214526} (\bibinfo{year}{2009}).

\bibitem{Sato2010}
\bibinfo{author}{Sato, M.}
\newblock \bibinfo{title}{Topological odd-parity superconductors}.
\newblock \emph{\bibinfo{journal}{Phys. Rev. B}} \textbf{\bibinfo{volume}{81}},
  \bibinfo{pages}{220504} (\bibinfo{year}{2010}).

\bibitem{Fu-Berg2010}
\bibinfo{author}{Fu, L.} \& \bibinfo{author}{Berg, E.}
\newblock \bibinfo{title}{Odd-parity topological superconductors: Theory and
  application to {Cu$_x$Bi$_2$Se$_3$}}.
\newblock \emph{\bibinfo{journal}{Phys. Rev. Lett.}}
  \textbf{\bibinfo{volume}{105}}, \bibinfo{pages}{097001}
  (\bibinfo{year}{2010}).

\bibitem{Brydon2014}
\bibinfo{author}{Brydon, P. M.~R.}, \bibinfo{author}{Das~Sarma, S.},
  \bibinfo{author}{Hui, H.-Y.} \& \bibinfo{author}{Sau, J.~D.}
\newblock \bibinfo{title}{Odd-parity superconductivity from phonon-mediated
  pairing: Application to
  {${\mathrm{Cu}}_{x}{\mathrm{Bi}}_{2}{\mathrm{Se}}_{3}$}}.
\newblock \emph{\bibinfo{journal}{Phys. Rev. B}} \textbf{\bibinfo{volume}{90}},
  \bibinfo{pages}{184512} (\bibinfo{year}{2014}).

\bibitem{Fu2014}
\bibinfo{author}{Fu, L.}
\newblock \bibinfo{title}{Odd-parity topological superconductor with nematic
  order: Application to {Cu$_{x}$Bi$_2$Se$_3$}}.
\newblock \emph{\bibinfo{journal}{Phys. Rev. B}} \textbf{\bibinfo{volume}{90}},
  \bibinfo{pages}{100509} (\bibinfo{year}{2014}).

\bibitem{Hashimoto2013}
\bibinfo{author}{Hashimoto, T.}, \bibinfo{author}{Yada, K.},
  \bibinfo{author}{Yamakage, A.}, \bibinfo{author}{Sato, M.} \&
  \bibinfo{author}{Tanaka, Y.}
\newblock \bibinfo{title}{Bulk electronic state of superconducting topological
  insulator}.
\newblock \emph{\bibinfo{journal}{J. Phys. Soc. Jpn.}}
  \textbf{\bibinfo{volume}{82}}, \bibinfo{pages}{044704}
  (\bibinfo{year}{2013}).

\bibitem{Hashimoto2014}
\bibinfo{author}{Hashimoto, T.}, \bibinfo{author}{Yada, K.},
  \bibinfo{author}{Yamakage, A.}, \bibinfo{author}{Sato, M.} \&
  \bibinfo{author}{Tanaka, Y.}
\newblock \bibinfo{title}{Effect of fermi surface evolution on superconducting
  gap in superconducting topological insulator}.
\newblock \emph{\bibinfo{journal}{Supercond. Sci. Technol.}}
  \textbf{\bibinfo{volume}{27}}, \bibinfo{pages}{104002}
  (\bibinfo{year}{2014}).

\bibitem{Nagai2012}
\bibinfo{author}{Nagai, Y.}, \bibinfo{author}{Nakamura, H.} \&
  \bibinfo{author}{Machida, M.}
\newblock \bibinfo{title}{Rotational isotropy breaking as proof for
  spin-polarized {Cooper} pairs in the topological superconductor
  {Cu${}_{x}$Bi${}_{2}$Se${}_{3}$}}.
\newblock \emph{\bibinfo{journal}{Phys. Rev. B}} \textbf{\bibinfo{volume}{86}},
  \bibinfo{pages}{094507} (\bibinfo{year}{2012}).

\bibitem{Hao-Lee2011}
\bibinfo{author}{Hao, L.} \& \bibinfo{author}{Lee, T.~K.}
\newblock \bibinfo{title}{Surface spectral function in the superconducting
  state of a topological insulator}.
\newblock \emph{\bibinfo{journal}{Phys. Rev. B}} \textbf{\bibinfo{volume}{83}},
  \bibinfo{pages}{134516} (\bibinfo{year}{2011}).

\bibitem{Yamakage2012}
\bibinfo{author}{Yamakage, A.}, \bibinfo{author}{Yada, K.},
  \bibinfo{author}{Sato, M.} \& \bibinfo{author}{Tanaka, Y.}
\newblock \bibinfo{title}{Theory of tunneling conductance and surface-state
  transition in superconducting topological insulators}.
\newblock \emph{\bibinfo{journal}{Phys. Rev. B}} \textbf{\bibinfo{volume}{85}},
  \bibinfo{pages}{180509} (\bibinfo{year}{2012}).

\bibitem{Hsieh2012}
\bibinfo{author}{Hsieh, T.~H.} \& \bibinfo{author}{Fu, L.}
\newblock \bibinfo{title}{Majorana fermions and exotic surface {Andreev} bound
  states in topological superconductors: Application to {Cu$_x$Bi$_2$Se$_3$}}.
\newblock \emph{\bibinfo{journal}{Phys. Rev. Lett.}}
  \textbf{\bibinfo{volume}{108}}, \bibinfo{pages}{107005}
  (\bibinfo{year}{2012}).

\bibitem{Takami2014}
\bibinfo{author}{Takami, S.}, \bibinfo{author}{Yada, K.},
  \bibinfo{author}{Yamakage, A.}, \bibinfo{author}{Sato, M.} \&
  \bibinfo{author}{Tanaka, Y.}
\newblock \bibinfo{title}{Quasi-classical theory of tunneling spectroscopy in
  superconducting topological insulator}.
\newblock \emph{\bibinfo{journal}{J. Phys. Soc. Jpn.}}
  \textbf{\bibinfo{volume}{83}}, \bibinfo{pages}{064705}
  (\bibinfo{year}{2014}).

\bibitem{Mizushima2014}
\bibinfo{author}{Mizushima, T.}, \bibinfo{author}{Yamakage, A.},
  \bibinfo{author}{Sato, M.} \& \bibinfo{author}{Tanaka, Y.}
\newblock \bibinfo{title}{Dirac-fermion-induced parity mixing in
  superconducting topological insulators}.
\newblock \emph{\bibinfo{journal}{Phys. Rev. B}} \textbf{\bibinfo{volume}{90}},
  \bibinfo{pages}{184516} (\bibinfo{year}{2014}).

\bibitem{Nakosai2012}
\bibinfo{author}{Nakosai, S.}, \bibinfo{author}{Tanaka, Y.} \&
  \bibinfo{author}{Nagaosa, N.}
\newblock \bibinfo{title}{Topological superconductivity in bilayer {Rashba}
  system}.
\newblock \emph{\bibinfo{journal}{Phys. Rev. Lett.}}
  \textbf{\bibinfo{volume}{108}}, \bibinfo{pages}{147003}
  (\bibinfo{year}{2012}).

\bibitem{Cho2012}
\bibinfo{author}{Cho, G.~Y.}, \bibinfo{author}{Bardarson, J.~H.},
  \bibinfo{author}{Lu, Y.-M.} \& \bibinfo{author}{Moore, J.~E.}
\newblock \bibinfo{title}{Superconductivity of doped {Weyl} semimetals:
  Finite-momentum pairing and electronic analog of the {${}^{3}$He-$A$} phase}.
\newblock \emph{\bibinfo{journal}{Phys. Rev. B}} \textbf{\bibinfo{volume}{86}},
  \bibinfo{pages}{214514} (\bibinfo{year}{2012}).

\bibitem{Bednik2015}
\bibinfo{author}{Bednik, G.}, \bibinfo{author}{Zyuzin, A.~A.} \&
  \bibinfo{author}{Burkov, A.~A.}
\newblock \bibinfo{title}{Superconductivity in {Weyl} metals}.
\newblock \emph{\bibinfo{journal}{Phys. Rev. B}} \textbf{\bibinfo{volume}{92}},
  \bibinfo{pages}{035153} (\bibinfo{year}{2015}).

\bibitem{Zhou2016}
\bibinfo{author}{Zhou, T.}, \bibinfo{author}{Gao, Y.} \& \bibinfo{author}{Wang,
  Z.~D.}
\newblock \bibinfo{title}{Superconductivity in doped inversion-symmetric {Weyl}
  semimetals}.
\newblock \emph{\bibinfo{journal}{Phys. Rev. B}} \textbf{\bibinfo{volume}{93}},
  \bibinfo{pages}{094517} (\bibinfo{year}{2016}).

\bibitem{Murakami-Nagaosa2003}
\bibinfo{author}{Murakami, S.} \& \bibinfo{author}{Nagaosa, N.}
\newblock \bibinfo{title}{Berry phase in magnetic superconductors}.
\newblock \emph{\bibinfo{journal}{Phys. Rev. Lett.}}
  \textbf{\bibinfo{volume}{90}}, \bibinfo{pages}{057002}
  (\bibinfo{year}{2003}).

\bibitem{Li-Haldane2015}
\bibinfo{author}{Li, Y.} \& \bibinfo{author}{Haldane, D.~M.}
\newblock \bibinfo{title}{Topological nodal {Cooper} pairing in doped {Weyl}
  metals}.
\newblock \emph{\bibinfo{journal}{ArXiv e-prints}}  (\bibinfo{year}{2015}).
\newblock arXiv:\eprint{1510.01730}.

\bibitem{Kobayashi-Sato2015}
\bibinfo{author}{Kobayashi, S.} \& \bibinfo{author}{Sato, M.}
\newblock \bibinfo{title}{Topological superconductivity in {Dirac} semimetals}.
\newblock \emph{\bibinfo{journal}{Phys. Rev. Lett.}}
  \textbf{\bibinfo{volume}{115}}, \bibinfo{pages}{187001}
  (\bibinfo{year}{2015}).

\bibitem{Hashimoto2016}
\bibinfo{author}{Hashimoto, T.}, \bibinfo{author}{Kobayashi, S.},
  \bibinfo{author}{Tanaka, Y.} \& \bibinfo{author}{Sato, M.}
\newblock \bibinfo{title}{Superconductivity in doped {Dirac} semimetals}.
\newblock \emph{\bibinfo{journal}{Phys. Rev. B}} \textbf{\bibinfo{volume}{94}},
  \bibinfo{pages}{014510} (\bibinfo{year}{2016}).

\bibitem{He2015}
\bibinfo{author}{{He}, L.~P.} \emph{et~al.}
\newblock \bibinfo{title}{{Pressure-induced superconductivity in the
  three-dimensional {Dirac} semimetal Cd$_3$As$_2$}}.
\newblock \emph{\bibinfo{journal}{npj Quantum Materials}}
  \textbf{\bibinfo{volume}{1}}, \bibinfo{pages}{16014} (\bibinfo{year}{2016}).

\bibitem{Aggarwal2016}
\bibinfo{author}{{Aggarwal}, L.} \emph{et~al.}
\newblock \bibinfo{title}{{Unconventional superconductivity at mesoscopic point
  contacts on the 3D {Dirac} semimetal Cd$_{3}$As$_{2}$}}.
\newblock \emph{\bibinfo{journal}{Nat. Mater.}} \textbf{\bibinfo{volume}{15}},
  \bibinfo{pages}{32--37} (\bibinfo{year}{2016}).

\bibitem{Wang2016}
\bibinfo{author}{{Wang}, H.} \emph{et~al.}
\newblock \bibinfo{title}{{Observation of superconductivity induced by a point
  contact on 3D {Dirac} semimetal Cd$_{3}$As$_{2}$ crystals}}.
\newblock \emph{\bibinfo{journal}{Nat. Mater.}} \textbf{\bibinfo{volume}{15}},
  \bibinfo{pages}{38--42} (\bibinfo{year}{2016}).

\bibitem{Moore-Read1991}
\bibinfo{author}{Moore, G.~W.} \& \bibinfo{author}{Read, N.}
\newblock \bibinfo{title}{{Nonabelions in the fractional quantum {H}all
  effect}}.
\newblock \emph{\bibinfo{journal}{Nucl. Phys.}}
  \textbf{\bibinfo{volume}{B360}}, \bibinfo{pages}{362--396}
  (\bibinfo{year}{1991}).

\bibitem{Weinberg1981}
\bibinfo{author}{Weinberg, E.~J.}
\newblock \bibinfo{title}{{Index Calculations for the Fermion-Vortex System}}.
\newblock \emph{\bibinfo{journal}{Phys. Rev.}} \textbf{\bibinfo{volume}{D24}},
  \bibinfo{pages}{2669} (\bibinfo{year}{1981}).

\bibitem{Jackiw-Rossi1981}
\bibinfo{author}{{Jackiw}, R.} \& \bibinfo{author}{{Rossi}, P.}
\newblock \bibinfo{title}{{Zero modes of the vortex-fermion system}}.
\newblock \emph{\bibinfo{journal}{Nuclear Physics B}}
  \textbf{\bibinfo{volume}{190}}, \bibinfo{pages}{681--691}
  (\bibinfo{year}{1981}).

\bibitem{Callan1985}
\bibinfo{author}{Callan, C.~G., Jr.} \& \bibinfo{author}{Harvey, J.~A.}
\newblock \bibinfo{title}{{Anomalies and Fermion Zero Modes on Strings and
  Domain Walls}}.
\newblock \emph{\bibinfo{journal}{Nucl. Phys.}}
  \textbf{\bibinfo{volume}{B250}}, \bibinfo{pages}{427--436}
  (\bibinfo{year}{1985}).

\bibitem{Fu-Kane2008}
\bibinfo{author}{Fu, L.} \& \bibinfo{author}{Kane, C.~L.}
\newblock \bibinfo{title}{Superconducting proximity effect and {Majorana}
  fermions at the surface of a topological insulator}.
\newblock \emph{\bibinfo{journal}{Phys. Rev. Lett.}}
  \textbf{\bibinfo{volume}{100}}, \bibinfo{pages}{096407}
  (\bibinfo{year}{2008}).

\bibitem{Fukui2010}
\bibinfo{author}{Fukui, T.} \& \bibinfo{author}{Fujiwara, T.}
\newblock \bibinfo{title}{Topological stability of {Majorana} zero modes in
  superconductor–topological insulator systems}.
\newblock \emph{\bibinfo{journal}{J. Phys. Soc. Jpn.}}
  \textbf{\bibinfo{volume}{79}}, \bibinfo{pages}{033701}
  (\bibinfo{year}{2010}).

\bibitem{Fukui2010b}
\bibinfo{author}{Fukui, T.}
\newblock \bibinfo{title}{Majorana zero modes bound to a vortex line in a
  topological superconductor}.
\newblock \emph{\bibinfo{journal}{Phys. Rev. B}} \textbf{\bibinfo{volume}{81}},
  \bibinfo{pages}{214516} (\bibinfo{year}{2010}).

\bibitem{Parente2014}
\bibinfo{author}{Parente, V.}, \bibinfo{author}{Campagnano, G.},
  \bibinfo{author}{Giuliano, D.}, \bibinfo{author}{Tagliacozzo, A.} \&
  \bibinfo{author}{Guinea, F.}
\newblock \bibinfo{title}{Topological defects in topological insulators and
  bound states at topological superconductor vortices}.
\newblock \emph{\bibinfo{journal}{Materials}} \textbf{\bibinfo{volume}{7}},
  \bibinfo{pages}{1652–1686} (\bibinfo{year}{2014}).

\bibitem{Sato-Takahashi-Fujimoto2009}
\bibinfo{author}{Sato, M.}, \bibinfo{author}{Takahashi, Y.} \&
  \bibinfo{author}{Fujimoto, S.}
\newblock \bibinfo{title}{Non-abelian topological order in $s$-wave superfluids
  of ultracold fermionic atoms}.
\newblock \emph{\bibinfo{journal}{Phys. Rev. Lett.}}
  \textbf{\bibinfo{volume}{103}}, \bibinfo{pages}{020401}
  (\bibinfo{year}{2009}).

\bibitem{Sato-Takahashi-Fujimoto2010}
\bibinfo{author}{Sato, M.}, \bibinfo{author}{Takahashi, Y.} \&
  \bibinfo{author}{Fujimoto, S.}
\newblock \bibinfo{title}{Non-abelian topological orders and {Majorana}
  fermions in spin-singlet superconductors}.
\newblock \emph{\bibinfo{journal}{Phys. Rev. B}} \textbf{\bibinfo{volume}{82}},
  \bibinfo{pages}{134521} (\bibinfo{year}{2010}).

\bibitem{Sau2010}
\bibinfo{author}{Sau, J.~D.}, \bibinfo{author}{Lutchyn, R.~M.},
  \bibinfo{author}{Tewari, S.} \& \bibinfo{author}{Das~Sarma, S.}
\newblock \bibinfo{title}{Generic new platform for topological quantum
  computation using semiconductor heterostructures}.
\newblock \emph{\bibinfo{journal}{Phys. Rev. Lett.}}
  \textbf{\bibinfo{volume}{104}}, \bibinfo{pages}{040502}
  (\bibinfo{year}{2010}).

\bibitem{Lutchyn2010}
\bibinfo{author}{Lutchyn, R.~M.}, \bibinfo{author}{Sau, J.~D.} \&
  \bibinfo{author}{Das~Sarma, S.}
\newblock \bibinfo{title}{Majorana fermions and a topological phase transition
  in semiconductor-superconductor heterostructures}.
\newblock \emph{\bibinfo{journal}{Phys. Rev. Lett.}}
  \textbf{\bibinfo{volume}{105}}, \bibinfo{pages}{077001}
  (\bibinfo{year}{2010}).

\bibitem{Oreg2010}
\bibinfo{author}{Oreg, Y.}, \bibinfo{author}{Refael, G.} \&
  \bibinfo{author}{von Oppen, F.}
\newblock \bibinfo{title}{Helical liquids and {Majorana} bound states in
  quantum wires}.
\newblock \emph{\bibinfo{journal}{Phys. Rev. Lett.}}
  \textbf{\bibinfo{volume}{105}}, \bibinfo{pages}{177002}
  (\bibinfo{year}{2010}).

\bibitem{Sato-Fujimoto2016}
\bibinfo{author}{Sato, M.} \& \bibinfo{author}{Fujimoto, S.}
\newblock \bibinfo{title}{Majorana fermions and topology in superconductors}.
\newblock \emph{\bibinfo{journal}{J. Phys. Soc. Jpn.}}
  \textbf{\bibinfo{volume}{85}}, \bibinfo{pages}{072001}
  (\bibinfo{year}{2016}).

\bibitem{Linder2010}
\bibinfo{author}{Linder, J.}, \bibinfo{author}{Tanaka, Y.},
  \bibinfo{author}{Yokoyama, T.}, \bibinfo{author}{Sudb\o{}, A.} \&
  \bibinfo{author}{Nagaosa, N.}
\newblock \bibinfo{title}{Unconventional superconductivity on a topological
  insulator}.
\newblock \emph{\bibinfo{journal}{Phys. Rev. Lett.}}
  \textbf{\bibinfo{volume}{104}}, \bibinfo{pages}{067001}
  (\bibinfo{year}{2010}).

\bibitem{Daido2016}
\bibinfo{author}{Daido, A.} \& \bibinfo{author}{Yanase, Y.}
\newblock \bibinfo{title}{Paramagnetically induced gapful topological
  superconductors}.
\newblock \emph{\bibinfo{journal}{Phys. Rev. B}} \textbf{\bibinfo{volume}{94}},
  \bibinfo{pages}{054519} (\bibinfo{year}{2016}).

\bibitem{Wong2012}
\bibinfo{author}{Wong, C. L.~M.} \& \bibinfo{author}{Law, K.~T.}
\newblock \bibinfo{title}{Majorana kramers doublets in
  ${d}_{{x}^{2}\ensuremath{-}{y}^{2}}$-wave superconductors with {Rashba}
  spin-orbit coupling}.
\newblock \emph{\bibinfo{journal}{Phys. Rev. B}} \textbf{\bibinfo{volume}{86}},
  \bibinfo{pages}{184516} (\bibinfo{year}{2012}).

\bibitem{Black-Schaffer2012}
\bibinfo{author}{Black-Schaffer, A.~M.}
\newblock \bibinfo{title}{Edge properties and {Majorana} fermions in the
  proposed chiral $d$-wave superconducting state of doped graphene}.
\newblock \emph{\bibinfo{journal}{Phys. Rev. Lett.}}
  \textbf{\bibinfo{volume}{109}}, \bibinfo{pages}{197001}
  (\bibinfo{year}{2012}).

\bibitem{Farrel2013}
\bibinfo{author}{Farrell, A.} \& \bibinfo{author}{Pereg-Barnea, T.}
\newblock \bibinfo{title}{Topological superconductivity without proximity
  effect}.
\newblock \emph{\bibinfo{journal}{Phys. Rev. B}} \textbf{\bibinfo{volume}{87}},
  \bibinfo{pages}{214517} (\bibinfo{year}{2013}).

\bibitem{Sun2016}
\bibinfo{author}{{Sun}, S.-J.}, \bibinfo{author}{{Chung}, C.-H.},
  \bibinfo{author}{{Chang}, Y.-Y.}, \bibinfo{author}{{Tsai}, W.-F.} \&
  \bibinfo{author}{{Zhang}, F.-C.}
\newblock \bibinfo{title}{{Helical {Majorana} fermions in
  $d_{x^2-y^2}+id_{xy}$-wave topological superconductivity of doped correlated
  quantum spin {H}all insulators}}.
\newblock \emph{\bibinfo{journal}{Sci. Rep.}} \textbf{\bibinfo{volume}{6}},
  \bibinfo{pages}{24102} (\bibinfo{year}{2016}).

\bibitem{Altland97}
\bibinfo{author}{Altland, A.} \& \bibinfo{author}{Zirnbauer, M.~R.}
\newblock \bibinfo{title}{Nonstandard symmetry classes in mesoscopic
  normal-superconducting hybrid structures}.
\newblock \emph{\bibinfo{journal}{Phys. Rev. B}} \textbf{\bibinfo{volume}{55}},
  \bibinfo{pages}{1142--1161} (\bibinfo{year}{1997}).

\bibitem{Note1}
\bibinfo{note}{This argument does not imply that an explicit
  spin-rotation-breaking term is necessary for realizing Majorana fermions.
  Even when the microscopic Hamiltonian preserves full SU(2) spin-rotation
  symmetry, topological superconductivity with Majorana fermions can be
  realized if the SU(2) symmetry is spontaneously broken by the formation of
  spin-triplet Cooper pairs \cite {Sun2014}}.

\bibitem{Maeno1994}
\bibinfo{author}{Maeno, Y.} \emph{et~al.}
\newblock \bibinfo{title}{Superconductivity in a layered perovskite without
  copper}.
\newblock \emph{\bibinfo{journal}{Nature}} \textbf{\bibinfo{volume}{372}},
  \bibinfo{pages}{532--534} (\bibinfo{year}{1994}).

\bibitem{Rice1995}
\bibinfo{author}{Rice, T.~M.} \& \bibinfo{author}{Sigrist, M.}
\newblock \bibinfo{title}{{Sr$_2$ RuO$_4$} : an electronic analogue of
  {${}^3$He}?}
\newblock \emph{\bibinfo{journal}{J. Phys.: Condens. Matter.}}
  \textbf{\bibinfo{volume}{7}}, \bibinfo{pages}{L643} (\bibinfo{year}{1995}).

\bibitem{Ishida1998}
\bibinfo{author}{Ishida, K.} \emph{et~al.}
\newblock \bibinfo{title}{Spin-triplet superconductivity in {Sr$_2$RuO$_4$}
  identified by {${}^{17}$O} knight shift}.
\newblock \emph{\bibinfo{journal}{Nature}} \textbf{\bibinfo{volume}{396}},
  \bibinfo{pages}{658--660} (\bibinfo{year}{1998}).

\bibitem{Luke1998}
\bibinfo{author}{Luke, G.~M.} \emph{et~al.}
\newblock \bibinfo{title}{Time-reversal symmetry-breaking superconductivity in
  {Sr$_2$RuO$_4$}}.
\newblock \emph{\bibinfo{journal}{Nature}} \textbf{\bibinfo{volume}{394}},
  \bibinfo{pages}{558--561} (\bibinfo{year}{1998}).

\bibitem{Kallin2016}
\bibinfo{author}{Catherine, K.} \& \bibinfo{author}{John, B.}
\newblock \bibinfo{title}{Chiral superconductors}.
\newblock \emph{\bibinfo{journal}{Rep. Prog. Phys.}}
  \textbf{\bibinfo{volume}{79}}, \bibinfo{pages}{054502}
  (\bibinfo{year}{2016}).

\bibitem{Maeno2012}
\bibinfo{author}{Maeno, Y.}, \bibinfo{author}{Kittaka, S.},
  \bibinfo{author}{Nomura, T.}, \bibinfo{author}{Yonezawa, S.} \&
  \bibinfo{author}{Ishida, K.}
\newblock \bibinfo{title}{Evaluation of spin-triplet superconductivity in
  {Sr$_{2}$RuO$_{4}$}}.
\newblock \emph{\bibinfo{journal}{J. Phys. Soc. Jpn.}}
  \textbf{\bibinfo{volume}{81}}, \bibinfo{pages}{011009}
  (\bibinfo{year}{2012}).

\bibitem{Kashiwaya2011}
\bibinfo{author}{Kashiwaya, S.} \emph{et~al.}
\newblock \bibinfo{title}{Edge states of {Sr$_2$RuO$_4$} detected by in-plane
  tunneling spectroscopy}.
\newblock \emph{\bibinfo{journal}{Phys. Rev. Lett.}}
  \textbf{\bibinfo{volume}{107}}, \bibinfo{pages}{077003}
  (\bibinfo{year}{2011}).

\bibitem{Yamashiro1997}
\bibinfo{author}{Yamashiro, M.}, \bibinfo{author}{Tanaka, Y.} \&
  \bibinfo{author}{Kashiwaya, S.}
\newblock \bibinfo{title}{Theory of tunneling spectroscopy in superconducting
  {${\mathrm{Sr}}_{2}{\mathrm{RuO}}_{4}$}}.
\newblock \emph{\bibinfo{journal}{Phys. Rev. B}} \textbf{\bibinfo{volume}{56}},
  \bibinfo{pages}{7847--7850} (\bibinfo{year}{1997}).

\bibitem{Furusaki2001}
\bibinfo{author}{Furusaki, A.}, \bibinfo{author}{Matsumoto, M.} \&
  \bibinfo{author}{Sigrist, M.}
\newblock \bibinfo{title}{Spontaneous {H}all effect in a chiral \textit{p}
  -wave superconductor}.
\newblock \emph{\bibinfo{journal}{Phys. Rev. B}} \textbf{\bibinfo{volume}{64}},
  \bibinfo{pages}{054514} (\bibinfo{year}{2001}).

\bibitem{Stone2004}
\bibinfo{author}{Stone, M.} \& \bibinfo{author}{Roy, R.}
\newblock \bibinfo{title}{Edge modes, edge currents, and gauge invariance in
  ${p}_{x}{+ip}_{y}$ superfluids and superconductors}.
\newblock \emph{\bibinfo{journal}{Phys. Rev. B}} \textbf{\bibinfo{volume}{69}},
  \bibinfo{pages}{184511} (\bibinfo{year}{2004}).

\bibitem{Yada2014}
\bibinfo{author}{Yada, K.}, \bibinfo{author}{Golubov, A.~A.},
  \bibinfo{author}{Tanaka, Y.} \& \bibinfo{author}{Kashiwaya, S.}
\newblock \bibinfo{title}{Microscopic theory of tunneling spectroscopy in
  {Sr$_2$RuO$_4$}}.
\newblock \emph{\bibinfo{journal}{J. Phys. Soc. Jpn.}}
  \textbf{\bibinfo{volume}{83}}, \bibinfo{pages}{074706}
  (\bibinfo{year}{2014}).

\bibitem{Mackenzie2003}
\bibinfo{author}{Mackenzie, A.~P.} \& \bibinfo{author}{Maeno, Y.}
\newblock \bibinfo{title}{The superconductivity of
  {${\mathrm{Sr}}_{2}{\mathrm{RuO}}_{4}$} and the physics of spin-triplet
  pairing}.
\newblock \emph{\bibinfo{journal}{Rev. Mod. Phys.}}
  \textbf{\bibinfo{volume}{75}}, \bibinfo{pages}{657--712}
  (\bibinfo{year}{2003}).

\bibitem{Budakian2011}
\bibinfo{author}{Jang, J.} \emph{et~al.}
\newblock \bibinfo{title}{Observation of half-height magnetization steps in
  {Sr$_2$RuO$_4$}}.
\newblock \emph{\bibinfo{journal}{Science}} \textbf{\bibinfo{volume}{331}},
  \bibinfo{pages}{186--188} (\bibinfo{year}{2011}).

\bibitem{Tewari2007}
\bibinfo{author}{Tewari, S.}, \bibinfo{author}{Das~Sarma, S.},
  \bibinfo{author}{Nayak, C.}, \bibinfo{author}{Zhang, C.} \&
  \bibinfo{author}{Zoller, P.}
\newblock \bibinfo{title}{Quantum computation using vortices and {Majorana}
  zero modes of a ${p}_{x}+i{p}_{y}$ superfluid of fermionic cold atoms}.
\newblock \emph{\bibinfo{journal}{Phys. Rev. Lett.}}
  \textbf{\bibinfo{volume}{98}}, \bibinfo{pages}{010506}
  (\bibinfo{year}{2007}).

\bibitem{Hor2010}
\bibinfo{author}{Hor, Y.~S.} \emph{et~al.}
\newblock \bibinfo{title}{Superconductivity in {Cu$_x$Bi$_2$Se$_3$} and its
  implications for pairing in the undoped topological insulator}.
\newblock \emph{\bibinfo{journal}{Phys. Rev. Lett.}}
  \textbf{\bibinfo{volume}{104}}, \bibinfo{pages}{057001}
  (\bibinfo{year}{2010}).

\bibitem{Wray2010}
\bibinfo{author}{Wray, L.~A.} \emph{et~al.}
\newblock \bibinfo{title}{Observation of topological order in a superconducting
  doped topological insulator}.
\newblock \emph{\bibinfo{journal}{Nat. Phys.}} \textbf{\bibinfo{volume}{6}},
  \bibinfo{pages}{855--859} (\bibinfo{year}{2010}).

\bibitem{Ando-Fu2015}
\bibinfo{author}{Ando, Y.} \& \bibinfo{author}{Fu, L.}
\newblock \bibinfo{title}{Topological crystalline insulators and topological
  superconductors: From concepts to materials}.
\newblock \emph{\bibinfo{journal}{Annu. Rev. Condens. Matter Phys.}}
  \textbf{\bibinfo{volume}{6}}, \bibinfo{pages}{361--381}
  (\bibinfo{year}{2015}).

\bibitem{Kriener2011PRB}
\bibinfo{author}{Kriener, M.} \emph{et~al.}
\newblock \bibinfo{title}{Electrochemical synthesis and superconducting phase
  diagram of {Cu$_{x}$Bi$_{2}$Se$_{3}$}}.
\newblock \emph{\bibinfo{journal}{Phys. Rev. B}} \textbf{\bibinfo{volume}{84}},
  \bibinfo{pages}{054513} (\bibinfo{year}{2011}).

\bibitem{Matano2016}
\bibinfo{author}{Matano, K.}, \bibinfo{author}{Kriener, M.},
  \bibinfo{author}{Segawa, K.}, \bibinfo{author}{Ando, Y.} \&
  \bibinfo{author}{Zheng, G.-q.}
\newblock \bibinfo{title}{Spin-rotation symmetry breaking in the
  superconducting state of {Cu$_x$Bi$_2$Se$_3$}}.
\newblock \emph{\bibinfo{journal}{Nat Phys}} \textbf{\bibinfo{volume}{12}},
  \bibinfo{pages}{852--854} (\bibinfo{year}{2016}).

\bibitem{Yonezawa2016}
\bibinfo{author}{{Yonezawa}, S.} \emph{et~al.}
\newblock \bibinfo{title}{Thermodynamic evidence for nematic superconductivity
  in {Cu$_x$Bi$_2$Se$_3$}}.
\newblock \emph{\bibinfo{journal}{Nat. Phys.}} \textbf{\bibinfo{volume}{13}},
  \bibinfo{pages}{123--126} (\bibinfo{year}{2017}).

\bibitem{Kriener2011PRL}
\bibinfo{author}{Kriener, M.}, \bibinfo{author}{Segawa, K.},
  \bibinfo{author}{Ren, Z.}, \bibinfo{author}{Sasaki, S.} \&
  \bibinfo{author}{Ando, Y.}
\newblock \bibinfo{title}{Bulk superconducting phase with a full energy gap in
  the doped topological insulator {Cu$_{x}$Bi$_{2}$Se$_{3}$}}.
\newblock \emph{\bibinfo{journal}{Phys. Rev. Lett.}}
  \textbf{\bibinfo{volume}{106}}, \bibinfo{pages}{127004}
  (\bibinfo{year}{2011}).

\bibitem{Levy2013}
\bibinfo{author}{Levy, N.} \emph{et~al.}
\newblock \bibinfo{title}{Experimental evidence for $s$-wave pairing symmetry
  in superconducting {Cu$_x$Bi$_2$Se$_3$} single crystals using a scanning
  tunneling microscope}.
\newblock \emph{\bibinfo{journal}{Phys. Rev. Lett.}}
  \textbf{\bibinfo{volume}{110}}, \bibinfo{pages}{117001}
  (\bibinfo{year}{2013}).

\bibitem{Lahoud2013}
\bibinfo{author}{Lahoud, E.} \emph{et~al.}
\newblock \bibinfo{title}{Evolution of the fermi surface of a doped topological
  insulator with carrier concentration}.
\newblock \emph{\bibinfo{journal}{Phys. Rev. B}} \textbf{\bibinfo{volume}{88}},
  \bibinfo{pages}{195107} (\bibinfo{year}{2013}).

\bibitem{Liu2015}
\bibinfo{author}{Liu, Z.} \emph{et~al.}
\newblock \bibinfo{title}{Superconductivity with topological surface state in
  {S}r$_x${B}i$_2${S}e$_3$}.
\newblock \emph{\bibinfo{journal}{J. Am. Chem. Soc.}}
  \textbf{\bibinfo{volume}{137}}, \bibinfo{pages}{10512--10515}
  (\bibinfo{year}{2015}).

\bibitem{Pan2016}
\bibinfo{author}{Pan, Y.} \emph{et~al.}
\newblock \bibinfo{title}{Rotational symmetry breaking in the topological
  superconductor {S}r$_x${B}i$_2${S}e$_3$ probed by upper-critical field
  experiments}.
\newblock \emph{\bibinfo{journal}{Sci. Rep.}} \textbf{\bibinfo{volume}{6}},
  \bibinfo{pages}{28632} (\bibinfo{year}{2016}).

\bibitem{Nikitin2016}
\bibinfo{author}{Nikitin, A.~M.}, \bibinfo{author}{Pan, Y.},
  \bibinfo{author}{Huang, Y.~K.}, \bibinfo{author}{Naka, T.} \&
  \bibinfo{author}{de~Visser, A.}
\newblock \bibinfo{title}{High-pressure study of the basal-plane anisotropy of
  the upper critical field of the topological superconductor
  {S}r$_x${B}i$_2${S}e$_3$}.
\newblock \emph{\bibinfo{journal}{Phys. Rev. B}} \textbf{\bibinfo{volume}{94}},
  \bibinfo{pages}{144516} (\bibinfo{year}{2016}).

\bibitem{Du2017}
\bibinfo{author}{Du, G.} \emph{et~al.}
\newblock \bibinfo{title}{Superconductivity with two-fold symmetry in
  topological superconductor {S}r$_x${B}i$_2${S}e$_3$}.
\newblock \emph{\bibinfo{journal}{Sci. China-Phys. Mech. Astron.}}
  \textbf{\bibinfo{volume}{60}}, \bibinfo{pages}{037411}
  (\bibinfo{year}{2017}).

\bibitem{Qiu2015}
\bibinfo{author}{{Qiu}, Y.} \emph{et~al.}
\newblock \bibinfo{title}{{Time reversal symmetry breaking superconductivity in
  topological materials}}.
\newblock \emph{\bibinfo{journal}{ArXiv e-prints}}  (\bibinfo{year}{2015}).
\newblock arXiv:\eprint{1512.03519}.

\bibitem{Asaba2017}
\bibinfo{author}{Asaba, T.} \emph{et~al.}
\newblock \bibinfo{title}{Rotational symmetry breaking in a trigonal
  superconductor nb-doped {B}i$_2${S}e$_3$}.
\newblock \emph{\bibinfo{journal}{Phys. Rev. X}} \textbf{\bibinfo{volume}{7}},
  \bibinfo{pages}{011009} (\bibinfo{year}{2017}).

\bibitem{Novak2013}
\bibinfo{author}{Novak, M.}, \bibinfo{author}{Sasaki, S.},
  \bibinfo{author}{Kriener, M.}, \bibinfo{author}{Segawa, K.} \&
  \bibinfo{author}{Ando, Y.}
\newblock \bibinfo{title}{Unusual nature of fully gapped superconductivity in
  {In-doped SnTe}}.
\newblock \emph{\bibinfo{journal}{Phys. Rev. B}} \textbf{\bibinfo{volume}{88}},
  \bibinfo{pages}{140502} (\bibinfo{year}{2013}).

\bibitem{Erickson2009}
\bibinfo{author}{Erickson, A.~S.}, \bibinfo{author}{Chu, J.~H.},
  \bibinfo{author}{Toney, M.~F.}, \bibinfo{author}{Geballe, T.~H.} \&
  \bibinfo{author}{Fisher, I.~R.}
\newblock \bibinfo{title}{Enhanced superconducting pairing interaction in
  indium-doped tin telluride}.
\newblock \emph{\bibinfo{journal}{Phys. Rev. B}} \textbf{\bibinfo{volume}{79}},
  \bibinfo{pages}{024520} (\bibinfo{year}{2009}).

\bibitem{Zhong2013}
\bibinfo{author}{Zhong, R.~D.} \emph{et~al.}
\newblock \bibinfo{title}{Optimizing the superconducting transition temperature
  and upper critical field of {Sn${}_{1-x}$In${}_{x}$Te}}.
\newblock \emph{\bibinfo{journal}{Phys. Rev. B}} \textbf{\bibinfo{volume}{88}},
  \bibinfo{pages}{020505} (\bibinfo{year}{2013}).

\bibitem{Sasaki2012}
\bibinfo{author}{Sasaki, S.} \emph{et~al.}
\newblock \bibinfo{title}{Odd-parity pairing and topological superconductivity
  in a strongly spin-orbit coupled semiconductor}.
\newblock \emph{\bibinfo{journal}{Phys. Rev. Lett.}}
  \textbf{\bibinfo{volume}{109}}, \bibinfo{pages}{217004}
  (\bibinfo{year}{2012}).

\bibitem{Martin1997}
\bibinfo{author}{Martin, I.} \& \bibinfo{author}{Phillips, P.}
\newblock \bibinfo{title}{Local pairing at $u$ impurities in bcs
  superconductors can enhance ${T}_{c}$}.
\newblock \emph{\bibinfo{journal}{Phys. Rev. B}} \textbf{\bibinfo{volume}{56}},
  \bibinfo{pages}{14650--14654} (\bibinfo{year}{1997}).

\bibitem{Sasaki2015}
\bibinfo{author}{Sasaki, S.} \& \bibinfo{author}{Ando, Y.}
\newblock \bibinfo{title}{Superconducting {Sn$_{1-x}$In$_x$Te} nanoplates}.
\newblock \emph{\bibinfo{journal}{Cryst. Growth Des.}}
  \textbf{\bibinfo{volume}{15}}, \bibinfo{pages}{2748--2752}
  (\bibinfo{year}{2015}).

\bibitem{Sato2013}
\bibinfo{author}{Sato, T.} \emph{et~al.}
\newblock \bibinfo{title}{Fermiology of the strongly spin-orbit coupled
  superconductor {Sn$_{1-x}$In$_{x}$Te}: Implications for topological
  superconductivity}.
\newblock \emph{\bibinfo{journal}{Phys. Rev. Lett.}}
  \textbf{\bibinfo{volume}{110}}, \bibinfo{pages}{206804}
  (\bibinfo{year}{2013}).

\bibitem{Tanaka2013}
\bibinfo{author}{Tanaka, Y.} \emph{et~al.}
\newblock \bibinfo{title}{Two types of {Dirac}-cone surface states on the (111)
  surface of the topological crystalline insulator {SnTe}}.
\newblock \emph{\bibinfo{journal}{Phys. Rev. B}} \textbf{\bibinfo{volume}{88}},
  \bibinfo{pages}{235126} (\bibinfo{year}{2013}).

\bibitem{Fan-Gilbert-Bernevig2014}
\bibinfo{author}{Fang, C.}, \bibinfo{author}{Gilbert, M.~J.} \&
  \bibinfo{author}{Bernevig, B.~A.}
\newblock \bibinfo{title}{New class of topological superconductors protected by
  magnetic group symmetries}.
\newblock \emph{\bibinfo{journal}{Phys. Rev. Lett.}}
  \textbf{\bibinfo{volume}{112}}, \bibinfo{pages}{106401}
  (\bibinfo{year}{2014}).

\bibitem{Alekseevskii1952}
\bibinfo{author}{Alekseevskii, N.~E.}
\newblock \bibinfo{title}{Sverkhprovodimost soedinenii sistemy {B}i-{P}d}.
\newblock \emph{\bibinfo{journal}{Zh. Eksp. Teor. Fiz.}}
  \textbf{\bibinfo{volume}{23}}, \bibinfo{pages}{484} (\bibinfo{year}{1952}).

\bibitem{Peets2016}
\bibinfo{author}{Peets, D.~C.} \emph{et~al.}
\newblock \bibinfo{title}{Upper critical field of the noncentrosymmetric
  superconductor {BiPd}}.
\newblock \emph{\bibinfo{journal}{Phys. Rev. B}} \textbf{\bibinfo{volume}{93}},
  \bibinfo{pages}{174504} (\bibinfo{year}{2016}).

\bibitem{Bauer2004}
\bibinfo{author}{Bauer, E.} \emph{et~al.}
\newblock \bibinfo{title}{Heavy fermion superconductivity and magnetic order in
  noncentrosymmetric
  {${\mathrm{C}\mathrm{e}\mathrm{P}\mathrm{t}}_{3}\mathrm{S}\mathrm{i}$}}.
\newblock \emph{\bibinfo{journal}{Phys. Rev. Lett.}}
  \textbf{\bibinfo{volume}{92}}, \bibinfo{pages}{027003}
  (\bibinfo{year}{2004}).

\bibitem{Bauer2007}
\bibinfo{author}{Bauer, E.} \emph{et~al.}
\newblock \bibinfo{title}{Heavy fermion superconductivity and antiferromagnetic
  ordering in {CePt$_3$Si} without inversion symmetry}.
\newblock \emph{\bibinfo{journal}{J. Phys. Soc. Jpn.}}
  \textbf{\bibinfo{volume}{76}}, \bibinfo{pages}{051009}
  (\bibinfo{year}{2007}).

\bibitem{Fujimoto2007}
\bibinfo{author}{Fujimoto, S.}
\newblock \bibinfo{title}{Electron correlation and pairing states in
  superconductors without inversion symmetry}.
\newblock \emph{\bibinfo{journal}{J. Phys. Soc. Jpn.}}
  \textbf{\bibinfo{volume}{76}}, \bibinfo{pages}{051008}
  (\bibinfo{year}{2007}).

\bibitem{Badica2005}
\bibinfo{author}{Badica, P.}, \bibinfo{author}{Kondo, T.} \&
  \bibinfo{author}{Togano, K.}
\newblock \bibinfo{title}{Superconductivity in a new pseudo-binary
  {Li$_2$B(Pd$_{1-x}$Pt$_x$)3} ($x=0-1$) boride system}.
\newblock \emph{\bibinfo{journal}{J. Phys. Soc. Jpn.}}
  \textbf{\bibinfo{volume}{74}}, \bibinfo{pages}{1014--1019}
  (\bibinfo{year}{2005}).

\bibitem{Nishiyama2007}
\bibinfo{author}{Nishiyama, M.}, \bibinfo{author}{Inada, Y.} \&
  \bibinfo{author}{Zheng, G.-q.}
\newblock \bibinfo{title}{Spin triplet superconducting state due to broken
  inversion symmetry in ${\mathrm{li}}_{2}{\mathrm{pt}}_{3}\mathrm{B}$}.
\newblock \emph{\bibinfo{journal}{Phys. Rev. Lett.}}
  \textbf{\bibinfo{volume}{98}}, \bibinfo{pages}{047002}
  (\bibinfo{year}{2007}).

\bibitem{Ueno2014}
\bibinfo{author}{Ueno, K.} \emph{et~al.}
\newblock \bibinfo{title}{Field-induced superconductivity in electric double
  layer transistors}.
\newblock \emph{\bibinfo{journal}{J. Phys. Soc. Jpn.}}
  \textbf{\bibinfo{volume}{83}}, \bibinfo{pages}{032001}
  (\bibinfo{year}{2014}).

\bibitem{Sasaki2014}
\bibinfo{author}{Sasaki, S.}, \bibinfo{author}{Segawa, K.} \&
  \bibinfo{author}{Ando, Y.}
\newblock \bibinfo{title}{Superconductor derived from a topological insulator
  heterostructure}.
\newblock \emph{\bibinfo{journal}{Phys. Rev. B}} \textbf{\bibinfo{volume}{90}},
  \bibinfo{pages}{220504} (\bibinfo{year}{2014}).

\bibitem{Kanatzidis2004}
\bibinfo{author}{Kanatzidis, M.~G.}
\newblock \bibinfo{title}{Structural evolution and phase homologies for
  "design" and prediction of solid-state compounds}.
\newblock \emph{\bibinfo{journal}{Accounts Chem. Res.}}
  \textbf{\bibinfo{volume}{38}}, \bibinfo{pages}{359--368}
  (\bibinfo{year}{2004}).

\bibitem{Nakayama2012}
\bibinfo{author}{Nakayama, K.} \emph{et~al.}
\newblock \bibinfo{title}{Manipulation of topological states and the bulk band
  gap using natural heterostructures of a topological insulator}.
\newblock \emph{\bibinfo{journal}{Phys. Rev. Lett.}}
  \textbf{\bibinfo{volume}{109}}, \bibinfo{pages}{236804}
  (\bibinfo{year}{2012}).

\bibitem{Segawa2015}
\bibinfo{author}{Segawa, K.}, \bibinfo{author}{Taskin, A.~A.} \&
  \bibinfo{author}{Ando, Y.}
\newblock \bibinfo{title}{Pb5bi24se41: A new member of the homologous series
  forming topological insulator heterostructures}.
\newblock \emph{\bibinfo{journal}{J. Solid State Chem.}}
  \textbf{\bibinfo{volume}{221}}, \bibinfo{pages}{196--201}
  (\bibinfo{year}{2015}).

\bibitem{Stewart1984}
\bibinfo{author}{Stewart, G.~R.} \& \bibinfo{author}{Brandt, B.~L.}
\newblock \bibinfo{title}{High-field specific heats of {$A15$
  ${\mathrm{V}}_{3}$Si and ${\mathrm{Nb}}_{3}$Sn}}.
\newblock \emph{\bibinfo{journal}{Phys. Rev. B}} \textbf{\bibinfo{volume}{29}},
  \bibinfo{pages}{3908--3912} (\bibinfo{year}{1984}).

\bibitem{Joynt-Taillefer2002}
\bibinfo{author}{Joynt, R.} \& \bibinfo{author}{Taillefer, L.}
\newblock \bibinfo{title}{The superconducting phases of {UPt$_3$}}.
\newblock \emph{\bibinfo{journal}{Rev. Mod. Phys.}}
  \textbf{\bibinfo{volume}{74}}, \bibinfo{pages}{235--294}
  (\bibinfo{year}{2002}).

\bibitem{Izawa2012}
\bibinfo{author}{Machida, Y.} \emph{et~al.}
\newblock \bibinfo{title}{Twofold spontaneous symmetry breaking in the
  heavy-fermion superconductor ${\mathrm{upt}}_{3}$}.
\newblock \emph{\bibinfo{journal}{Phys. Rev. Lett.}}
  \textbf{\bibinfo{volume}{108}}, \bibinfo{pages}{157002}
  (\bibinfo{year}{2012}).

\bibitem{Yanase2016}
\bibinfo{author}{Yanase, Y.}
\newblock \bibinfo{title}{Nonsymmorphic {Weyl} superconductivity in
  ${\mathrm{upt}}_{3}$ based on ${E}_{2u}$ representation}.
\newblock \emph{\bibinfo{journal}{Phys. Rev. B}} \textbf{\bibinfo{volume}{94}},
  \bibinfo{pages}{174502} (\bibinfo{year}{2016}).

\bibitem{Butch2011}
\bibinfo{author}{Butch, N.~P.}, \bibinfo{author}{Syers, P.},
  \bibinfo{author}{Kirshenbaum, K.}, \bibinfo{author}{Hope, A.~P.} \&
  \bibinfo{author}{Paglione, J.}
\newblock \bibinfo{title}{Superconductivity in the topological semimetal
  yptbi}.
\newblock \emph{\bibinfo{journal}{Phys. Rev. B}} \textbf{\bibinfo{volume}{84}},
  \bibinfo{pages}{220504} (\bibinfo{year}{2011}).

\bibitem{Paglione2016}
\bibinfo{author}{{Kim}, H.} \emph{et~al.}
\newblock \bibinfo{title}{{Beyond Spin-Triplet: Nodal Topological
  Superconductivity in a Noncentrosymmetric Semimetal}}.
\newblock \emph{\bibinfo{journal}{ArXiv e-prints}}  (\bibinfo{year}{2016}).
\newblock arXiv:\eprint{1603.03375}.

\bibitem{Meinert2016}
\bibinfo{author}{Meinert, M.}
\newblock \bibinfo{title}{Unconventional superconductivity in {YPtBi} and
  related topological semimetals}.
\newblock \emph{\bibinfo{journal}{Phys. Rev. Lett.}}
  \textbf{\bibinfo{volume}{116}}, \bibinfo{pages}{137001}
  (\bibinfo{year}{2016}).

\bibitem{Brydon2016}
\bibinfo{author}{Brydon, P. M.~R.}, \bibinfo{author}{Wang, L.},
  \bibinfo{author}{Weinert, M.} \& \bibinfo{author}{Agterberg, D.~F.}
\newblock \bibinfo{title}{Pairing of $j=3/2$ fermions in half-heusler
  superconductors}.
\newblock \emph{\bibinfo{journal}{Phys. Rev. Lett.}}
  \textbf{\bibinfo{volume}{116}}, \bibinfo{pages}{177001}
  (\bibinfo{year}{2016}).

\bibitem{Alicea2010}
\bibinfo{author}{Alicea, J.}
\newblock \bibinfo{title}{Majorana fermions in a tunable semiconductor device}.
\newblock \emph{\bibinfo{journal}{Phys. Rev. B}} \textbf{\bibinfo{volume}{81}},
  \bibinfo{pages}{125318} (\bibinfo{year}{2010}).

\bibitem{Williams2012}
\bibinfo{author}{Williams, J.~R.} \emph{et~al.}
\newblock \bibinfo{title}{Unconventional {Josephson} effect in hybrid
  superconductor-topological insulator devices}.
\newblock \emph{\bibinfo{journal}{Phys. Rev. Lett.}}
  \textbf{\bibinfo{volume}{109}}, \bibinfo{pages}{056803}
  (\bibinfo{year}{2012}).

\bibitem{Wang-XueSC2012}
\bibinfo{author}{Wang, M.-X.} \emph{et~al.}
\newblock \bibinfo{title}{The coexistence of superconductivity and topological
  order in the {Bi$_2$Se$_3$} thin films}.
\newblock \emph{\bibinfo{journal}{Science}} \textbf{\bibinfo{volume}{336}},
  \bibinfo{pages}{52--55} (\bibinfo{year}{2012}).

\bibitem{Fan-Lu2012}
\bibinfo{author}{Yang, F.} \emph{et~al.}
\newblock \bibinfo{title}{Proximity-effect-induced superconducting phase in the
  topological insulator {Bi${}_{2}$Se${}_{3}$}}.
\newblock \emph{\bibinfo{journal}{Phys. Rev. B}} \textbf{\bibinfo{volume}{86}},
  \bibinfo{pages}{134504} (\bibinfo{year}{2012}).

\bibitem{Wang-XueSC2013}
\bibinfo{author}{Wang, E.} \emph{et~al.}
\newblock \bibinfo{title}{Fully gapped topological surface states in
  {Bi$_2$Se$_3$} films induced by a $d$-wave high-temperature superconductor}.
\newblock \emph{\bibinfo{journal}{Nat. Phys.}} \textbf{\bibinfo{volume}{9}},
  \bibinfo{pages}{621--625} (\bibinfo{year}{2013}).

\bibitem{Mason2013}
\bibinfo{author}{Cho, S.} \emph{et~al.}
\newblock \bibinfo{title}{Symmetry protected {Josephson} supercurrents in
  three-dimensional topological insulators}.
\newblock \emph{\bibinfo{journal}{Nat. Commun.}} \textbf{\bibinfo{volume}{4}},
  \bibinfo{pages}{1689} (\bibinfo{year}{2013}).

\bibitem{Oostinga-Molenkamp2013}
\bibinfo{author}{Oostinga, J.~B.} \emph{et~al.}
\newblock \bibinfo{title}{Josephson supercurrent through the topological
  surface states of strained bulk hgte}.
\newblock \emph{\bibinfo{journal}{Phys. Rev. X}} \textbf{\bibinfo{volume}{3}},
  \bibinfo{pages}{021007} (\bibinfo{year}{2013}).

\bibitem{Harlingen2014}
\bibinfo{author}{Finck, A. â.~â.}, \bibinfo{author}{Kurter, C.},
  \bibinfo{author}{Hor, Y.~â.} \& \bibinfo{author}{Van~Harlingen, D.~â.}
\newblock \bibinfo{title}{Phase coherence and {Andreev} reflection in
  topological insulator devices}.
\newblock \emph{\bibinfo{journal}{Phys. Rev. X}} \textbf{\bibinfo{volume}{4}},
  \bibinfo{pages}{041022} (\bibinfo{year}{2014}).

\bibitem{Snelder-Brinkman2014}
\bibinfo{author}{Snelder, M.} \emph{et~al.}
\newblock \bibinfo{title}{Josephson supercurrent in a topological insulator
  without a bulk shunt}.
\newblock \emph{\bibinfo{journal}{Supercond. Sci. Technol.}}
  \textbf{\bibinfo{volume}{27}}, \bibinfo{pages}{104001}
  (\bibinfo{year}{2014}).

\bibitem{Wiedenmann-Molenkamp2016}
\bibinfo{author}{Wiedenmann, J.} \emph{et~al.}
\newblock \bibinfo{title}{4$\pi$-periodic {Josephson} supercurrent in
  hgte-based topological {Josephson} junctions}.
\newblock \emph{\bibinfo{journal}{Nat Commun}} \textbf{\bibinfo{volume}{7}},
  \bibinfo{pages}{10303} (\bibinfo{year}{2016}).

\bibitem{Hart2014}
\bibinfo{author}{Hart, S.} \emph{et~al.}
\newblock \bibinfo{title}{Induced superconductivity in the quantum spin {H}all
  edge}.
\newblock \emph{\bibinfo{journal}{Nat. Phys.}} \textbf{\bibinfo{volume}{10}},
  \bibinfo{pages}{638--643} (\bibinfo{year}{2014}).

\bibitem{Deacon2016}
\bibinfo{author}{{Deacon}, R.~S.} \emph{et~al.}
\newblock \bibinfo{title}{{Josephson radiation from gapless {Andreev} bound
  states in HgTe-based topological junctions}}.
\newblock \emph{\bibinfo{journal}{ArXiv e-prints}}  (\bibinfo{year}{2016}).
\newblock arXiv:\eprint{1603.09611}.

\bibitem{Mourik2012}
\bibinfo{author}{Mourik, V.} \emph{et~al.}
\newblock \bibinfo{title}{Signatures of {Majorana} fermions in hybrid
  superconductor-semiconductor nanowire devices}.
\newblock \emph{\bibinfo{journal}{Science}} \textbf{\bibinfo{volume}{336}},
  \bibinfo{pages}{1003--7} (\bibinfo{year}{2012}).

\bibitem{Heiblum2012}
\bibinfo{author}{Das, A.} \emph{et~al.}
\newblock \bibinfo{title}{Zero-bias peaks and splitting in an al-inas nanowire
  topological superconductor as a signature of {Majorana} fermions}.
\newblock \emph{\bibinfo{journal}{Nat. Phys.}} \textbf{\bibinfo{volume}{8}},
  \bibinfo{pages}{887--895} (\bibinfo{year}{2012}).

\bibitem{HQXu2012}
\bibinfo{author}{Deng, M.~T.} \emph{et~al.}
\newblock \bibinfo{title}{Anomalous zero-bias conductance peak in a nb–insb
  nanowire–nb hybrid device}.
\newblock \emph{\bibinfo{journal}{Nano Lett.}} \textbf{\bibinfo{volume}{12}},
  \bibinfo{pages}{6414--6419} (\bibinfo{year}{2012}).

\bibitem{Franceschi2014}
\bibinfo{author}{Lee, E. J.~H.} \emph{et~al.}
\newblock \bibinfo{title}{Spin-resolved {Andreev} levels and parity crossings
  in hybrid superconductor-semiconductor nanostructures}.
\newblock \emph{\bibinfo{journal}{Nat. Nano.}} \textbf{\bibinfo{volume}{9}},
  \bibinfo{pages}{79--84} (\bibinfo{year}{2014}).

\bibitem{Albrecht2016}
\bibinfo{author}{Albrecht, S.~M.} \emph{et~al.}
\newblock \bibinfo{title}{Exponential protection of zero modes in {Majorana}
  islands}.
\newblock \emph{\bibinfo{journal}{Nature}} \textbf{\bibinfo{volume}{531}},
  \bibinfo{pages}{206--209} (\bibinfo{year}{2016}).

\bibitem{NadjPerge2013}
\bibinfo{author}{Nadj-Perge, S.}, \bibinfo{author}{Drozdov, I.~K.},
  \bibinfo{author}{Bernevig, B.~A.} \& \bibinfo{author}{Yazdani, A.}
\newblock \bibinfo{title}{Proposal for realizing {Majorana} fermions in chains
  of magnetic atoms on a superconductor}.
\newblock \emph{\bibinfo{journal}{Phys. Rev. B}} \textbf{\bibinfo{volume}{88}},
  \bibinfo{pages}{020407} (\bibinfo{year}{2013}).

\bibitem{NadjPerge2014}
\bibinfo{author}{Nadj-Perge, S.} \emph{et~al.}
\newblock \bibinfo{title}{Observation of {Majorana} fermions in ferromagnetic
  atomic chains on a superconductor}.
\newblock \emph{\bibinfo{journal}{Science}} \textbf{\bibinfo{volume}{346}},
  \bibinfo{pages}{602--607} (\bibinfo{year}{2014}).

\bibitem{Nandkishore2012}
\bibinfo{author}{Nandkishore, R.}, \bibinfo{author}{Levitov, L.~S.} \&
  \bibinfo{author}{Chubukov, A.~V.}
\newblock \bibinfo{title}{Chiral superconductivity from repulsive interactions
  in doped graphene}.
\newblock \emph{\bibinfo{journal}{Nat. Phys.}} \textbf{\bibinfo{volume}{8}},
  \bibinfo{pages}{158--163} (\bibinfo{year}{2012}).

\bibitem{BTK}
\bibinfo{author}{Blonder, G.~E.}, \bibinfo{author}{Tinkham, M.} \&
  \bibinfo{author}{Klapwijk, T.~M.}
\newblock \bibinfo{title}{Transition from metallic to tunneling regimes in
  superconducting microconstrictions: Excess current, charge imbalance, and
  supercurrent conversion}.
\newblock \emph{\bibinfo{journal}{Phys. Rev. B}} \textbf{\bibinfo{volume}{25}},
  \bibinfo{pages}{4515--4532} (\bibinfo{year}{1982}).

\bibitem{Law2009}
\bibinfo{author}{Law, K.~T.}, \bibinfo{author}{Lee, P.~A.} \&
  \bibinfo{author}{Ng, T.~K.}
\newblock \bibinfo{title}{Majorana fermion induced resonant {Andreev}
  reflection}.
\newblock \emph{\bibinfo{journal}{Phys. Rev. Lett.}}
  \textbf{\bibinfo{volume}{103}}, \bibinfo{pages}{237001}
  (\bibinfo{year}{2009}).

\bibitem{Flensberg2010}
\bibinfo{author}{Flensberg, K.}
\newblock \bibinfo{title}{Tunneling characteristics of a chain of {Majorana}
  bound states}.
\newblock \emph{\bibinfo{journal}{Phys. Rev. B}} \textbf{\bibinfo{volume}{82}},
  \bibinfo{pages}{180516} (\bibinfo{year}{2010}).

\bibitem{Ioselevich2013}
\bibinfo{author}{Ioselevich, P.~A.} \& \bibinfo{author}{Feigel'man, M.~V.}
\newblock \bibinfo{title}{Tunneling conductance due to a discrete spectrum of
  {Andreev} states}.
\newblock \emph{\bibinfo{journal}{New J. Phys.}} \textbf{\bibinfo{volume}{15}},
  \bibinfo{pages}{055011} (\bibinfo{year}{2013}).

\bibitem{Yamakage2014}
\bibinfo{author}{{Yamakage}, A.} \& \bibinfo{author}{{Sato}, M.}
\newblock \bibinfo{title}{{Interference of {Majorana} fermions in NS
  junctions}}.
\newblock \emph{\bibinfo{journal}{Physica E Low-Dimensional Systems and
  Nanostructures}} \textbf{\bibinfo{volume}{55}}, \bibinfo{pages}{13--19}
  (\bibinfo{year}{2014}).

\bibitem{Tanaka2009}
\bibinfo{author}{Tanaka, Y.}, \bibinfo{author}{Yokoyama, T.} \&
  \bibinfo{author}{Nagaosa, N.}
\newblock \bibinfo{title}{Manipulation of the {Majorana} fermion, {Andreev}
  reflection, and {Josephson} current on topological insulators}.
\newblock \emph{\bibinfo{journal}{Phys. Rev. Lett.}}
  \textbf{\bibinfo{volume}{103}}, \bibinfo{pages}{107002}
  (\bibinfo{year}{2009}).

\bibitem{Tanaka2009b}
\bibinfo{author}{Tanaka, Y.}, \bibinfo{author}{Yokoyama, T.},
  \bibinfo{author}{Balatsky, A.~V.} \& \bibinfo{author}{Nagaosa, N.}
\newblock \bibinfo{title}{Theory of topological spin current in
  noncentrosymmetric superconductors}.
\newblock \emph{\bibinfo{journal}{Phys. Rev. B}} \textbf{\bibinfo{volume}{79}},
  \bibinfo{pages}{060505} (\bibinfo{year}{2009}).

\bibitem{Asano2003}
\bibinfo{author}{Asano, Y.}, \bibinfo{author}{Tanaka, Y.},
  \bibinfo{author}{Matsuda, Y.} \& \bibinfo{author}{Kashiwaya, S.}
\newblock \bibinfo{title}{A theoretical study of tunneling conductance in
  {${\mathrm{PrOs}}_{4}{\mathrm{Sb}}_{12}$} superconducting junctions}.
\newblock \emph{\bibinfo{journal}{Phys. Rev. B}} \textbf{\bibinfo{volume}{68}},
  \bibinfo{pages}{184506} (\bibinfo{year}{2003}).

\bibitem{Wang-Qi-Zhang2011}
\bibinfo{author}{Wang, Z.}, \bibinfo{author}{Qi, X.-L.} \&
  \bibinfo{author}{Zhang, S.-C.}
\newblock \bibinfo{title}{Topological field theory and thermal responses of
  interacting topological superconductors}.
\newblock \emph{\bibinfo{journal}{Phys. Rev. B}} \textbf{\bibinfo{volume}{84}},
  \bibinfo{pages}{014527} (\bibinfo{year}{2011}).

\bibitem{Nomura2012}
\bibinfo{author}{Nomura, K.}, \bibinfo{author}{Ryu, S.},
  \bibinfo{author}{Furusaki, A.} \& \bibinfo{author}{Nagaosa, N.}
\newblock \bibinfo{title}{Cross-correlated responses of topological
  superconductors and superfluids}.
\newblock \emph{\bibinfo{journal}{Phys. Rev. Lett.}}
  \textbf{\bibinfo{volume}{108}}, \bibinfo{pages}{026802}
  (\bibinfo{year}{2012}).

\bibitem{Shiozaki2013}
\bibinfo{author}{Shiozaki, K.} \& \bibinfo{author}{Fujimoto, S.}
\newblock \bibinfo{title}{Electromagnetic and thermal responses of {$Z$}
  topological insulators and superconductors in odd spatial dimensions}.
\newblock \emph{\bibinfo{journal}{Phys. Rev. Lett.}}
  \textbf{\bibinfo{volume}{110}}, \bibinfo{pages}{076804}
  (\bibinfo{year}{2013}).

\bibitem{Chung-Horowitz-Qi2013}
\bibinfo{author}{Chung, S.~B.}, \bibinfo{author}{Horowitz, J.} \&
  \bibinfo{author}{Qi, X.-L.}
\newblock \bibinfo{title}{Time-reversal anomaly and {Josephson} effect in
  time-reversal-invariant topological superconductors}.
\newblock \emph{\bibinfo{journal}{Phys. Rev. B}} \textbf{\bibinfo{volume}{88}},
  \bibinfo{pages}{214514} (\bibinfo{year}{2013}).

\bibitem{Marra2016}
\bibinfo{author}{Marra, P.}, \bibinfo{author}{Citro, R.} \&
  \bibinfo{author}{Braggio, A.}
\newblock \bibinfo{title}{Signatures of topological phase transitions in
  {Josephson} current-phase discontinuities}.
\newblock \emph{\bibinfo{journal}{Phys. Rev. B}} \textbf{\bibinfo{volume}{93}},
  \bibinfo{pages}{220507} (\bibinfo{year}{2016}).

\bibitem{Yamakage2013}
\bibinfo{author}{Yamakage, A.}, \bibinfo{author}{Sato, M.},
  \bibinfo{author}{Yada, K.}, \bibinfo{author}{Kashiwaya, S.} \&
  \bibinfo{author}{Tanaka, Y.}
\newblock \bibinfo{title}{Anomalous {Josephson} current in superconducting
  topological insulator}.
\newblock \emph{\bibinfo{journal}{Phys. Rev. B}} \textbf{\bibinfo{volume}{87}},
  \bibinfo{pages}{100510} (\bibinfo{year}{2013}).

\bibitem{Asano2013}
\bibinfo{author}{Asano, Y.} \& \bibinfo{author}{Tanaka, Y.}
\newblock \bibinfo{title}{Majorana fermions and odd-frequency {Cooper} pairs in
  a normal-metal nanowire proximity-coupled to a topological superconductor}.
\newblock \emph{\bibinfo{journal}{Phys. Rev. B}} \textbf{\bibinfo{volume}{87}},
  \bibinfo{pages}{104513} (\bibinfo{year}{2013}).

\bibitem{Tanaka2004}
\bibinfo{author}{Tanaka, Y.} \& \bibinfo{author}{Kashiwaya, S.}
\newblock \bibinfo{title}{Anomalous charge transport in triplet superconductor
  junctions}.
\newblock \emph{\bibinfo{journal}{Phys. Rev. B}} \textbf{\bibinfo{volume}{70}},
  \bibinfo{pages}{012507} (\bibinfo{year}{2004}).

\bibitem{Tanaka2005}
\bibinfo{author}{Tanaka, Y.}, \bibinfo{author}{Kashiwaya, S.} \&
  \bibinfo{author}{Yokoyama, T.}
\newblock \bibinfo{title}{Theory of enhanced proximity effect by midgap
  {Andreev} resonant state in diffusive normal-metal/triplet superconductor
  junctions}.
\newblock \emph{\bibinfo{journal}{Phys. Rev. B}} \textbf{\bibinfo{volume}{71}},
  \bibinfo{pages}{094513} (\bibinfo{year}{2005}).

\bibitem{Tanaka2005b}
\bibinfo{author}{Tanaka, Y.}, \bibinfo{author}{Asano, Y.},
  \bibinfo{author}{Golubov, A.~A.} \& \bibinfo{author}{Kashiwaya, S.}
\newblock \bibinfo{title}{Anomalous features of the proximity effect in triplet
  superconductors}.
\newblock \emph{\bibinfo{journal}{Phys. Rev. B}} \textbf{\bibinfo{volume}{72}},
  \bibinfo{pages}{140503} (\bibinfo{year}{2005}).

\bibitem{Tanaka2007}
\bibinfo{author}{Tanaka, Y.}, \bibinfo{author}{Golubov, A.~A.},
  \bibinfo{author}{Kashiwaya, S.} \& \bibinfo{author}{Ueda, M.}
\newblock \bibinfo{title}{Anomalous {Josephson} effect between even- and
  odd-frequency superconductors}.
\newblock \emph{\bibinfo{journal}{Phys. Rev. Lett.}}
  \textbf{\bibinfo{volume}{99}}, \bibinfo{pages}{037005}
  (\bibinfo{year}{2007}).

\bibitem{Ikegaya2015}
\bibinfo{author}{Ikegaya, S.}, \bibinfo{author}{Asano, Y.} \&
  \bibinfo{author}{Tanaka, Y.}
\newblock \bibinfo{title}{Anomalous proximity effect and theoretical design for
  its realization}.
\newblock \emph{\bibinfo{journal}{Phys. Rev. B}} \textbf{\bibinfo{volume}{91}},
  \bibinfo{pages}{174511} (\bibinfo{year}{2015}).

\bibitem{Aasen2016}
\bibinfo{author}{{Aasen}, D.} \emph{et~al.}
\newblock \bibinfo{title}{{Milestones toward {Majorana}-based quantum
  computing}}.
\newblock \emph{\bibinfo{journal}{ArXiv e-prints}}  (\bibinfo{year}{2015}).
\newblock arXiv:\eprint{1511.05153}.

\bibitem{Bravyi-Kitaev2001}
\bibinfo{author}{Bravyi, A., S.~Kitaev}.
\newblock \bibinfo{title}{Quantum codes on a lattice with boundary}.
\newblock \emph{\bibinfo{journal}{Quantum Computers and Computing}}
  \textbf{\bibinfo{volume}{2}}, \bibinfo{pages}{43} (\bibinfo{year}{2001}).

\bibitem{Freedman-Meyer2001}
\bibinfo{author}{Freedman, H.~M.} \& \bibinfo{author}{Meyer, A.~D.}
\newblock \bibinfo{title}{Projective plane and planar quantum codes}.
\newblock \emph{\bibinfo{journal}{Foundations of Computational Mathematics}}
  \textbf{\bibinfo{volume}{1}}, \bibinfo{pages}{325--332}
  (\bibinfo{year}{2001}).

\bibitem{Vijay2015}
\bibinfo{author}{Vijay, S.}, \bibinfo{author}{Hsieh, T.~H.} \&
  \bibinfo{author}{Fu, L.}
\newblock \bibinfo{title}{Majorana fermion surface code for universal quantum
  computation}.
\newblock \emph{\bibinfo{journal}{Phys. Rev. X}} \textbf{\bibinfo{volume}{5}},
  \bibinfo{pages}{041038} (\bibinfo{year}{2015}).

\bibitem{Landau2016}
\bibinfo{author}{Landau, L.~A.} \emph{et~al.}
\newblock \bibinfo{title}{Towards realistic implementations of a {Majorana}
  surface code}.
\newblock \emph{\bibinfo{journal}{Phys Rev Lett}}
  \textbf{\bibinfo{volume}{116}}, \bibinfo{pages}{050501}
  (\bibinfo{year}{2016}).

\bibitem{Plugge2016}
\bibinfo{author}{Plugge, S.} \emph{et~al.}
\newblock \bibinfo{title}{Roadmap to {M}ajorana surface codes}.
\newblock \emph{\bibinfo{journal}{Phys. Rev. B}} \textbf{\bibinfo{volume}{94}},
  \bibinfo{pages}{174514} (\bibinfo{year}{2016}).

\bibitem{Manousakis2017}
\bibinfo{author}{{Manousakis}, J.}, \bibinfo{author}{{Altland}, A.},
  \bibinfo{author}{{Bagrets}, D.}, \bibinfo{author}{{Egger}, R.} \&
  \bibinfo{author}{{Ando}, Y.}
\newblock \bibinfo{title}{{Majorana qubits in topological insulator nanoribbon
  architecture}}.
\newblock \emph{\bibinfo{journal}{ArXiv e-prints}}  (\bibinfo{year}{2017}).
\newblock arXiv:\eprint{1702.02845}.

\bibitem{Ringel2012}
\bibinfo{author}{Ringel, Z.}, \bibinfo{author}{Kraus, Y.~E.} \&
  \bibinfo{author}{Stern, A.}
\newblock \bibinfo{title}{Strong side of weak topological insulators}.
\newblock \emph{\bibinfo{journal}{Phys. Rev. B}} \textbf{\bibinfo{volume}{86}},
  \bibinfo{pages}{045102} (\bibinfo{year}{2012}).

\bibitem{Mong2012}
\bibinfo{author}{Mong, R. S.~K.}, \bibinfo{author}{Bardarson, J.~H.} \&
  \bibinfo{author}{Moore, J.~E.}
\newblock \bibinfo{title}{Quantum transport and two-parameter scaling at the
  surface of a weak topological insulator}.
\newblock \emph{\bibinfo{journal}{Phys. Rev. Lett.}}
  \textbf{\bibinfo{volume}{108}}, \bibinfo{pages}{076804}
  (\bibinfo{year}{2012}).

\bibitem{Fu-Kane2012}
\bibinfo{author}{Fu, L.} \& \bibinfo{author}{Kane, C.~L.}
\newblock \bibinfo{title}{Topology, delocalization via average symmetry and the
  symplectic anderson transition}.
\newblock \emph{\bibinfo{journal}{Phys. Rev. Lett.}}
  \textbf{\bibinfo{volume}{109}}, \bibinfo{pages}{246605}
  (\bibinfo{year}{2012}).

\bibitem{Fulga2014}
\bibinfo{author}{Fulga, I.~C.}, \bibinfo{author}{van Heck, B.},
  \bibinfo{author}{Edge, J.~M.} \& \bibinfo{author}{Akhmerov, A.~R.}
\newblock \bibinfo{title}{Statistical topological insulators}.
\newblock \emph{\bibinfo{journal}{Phys. Rev. B}} \textbf{\bibinfo{volume}{89}},
  \bibinfo{pages}{155424} (\bibinfo{year}{2014}).

\bibitem{Grover2014}
\bibinfo{author}{{Grover}, T.}, \bibinfo{author}{{Sheng}, D.~N.} \&
  \bibinfo{author}{{Vishwanath}, A.}
\newblock \bibinfo{title}{{Emergent Space-Time Supersymmetry at the Boundary of
  a Topological Phase}}.
\newblock \emph{\bibinfo{journal}{Science}} \textbf{\bibinfo{volume}{344}},
  \bibinfo{pages}{280--283} (\bibinfo{year}{2014}).

\bibitem{Jian2015}
\bibinfo{author}{Jian, S.-K.}, \bibinfo{author}{Jiang, Y.-F.} \&
  \bibinfo{author}{Yao, H.}
\newblock \bibinfo{title}{Emergent spacetime supersymmetry in 3d {Weyl}
  semimetals and 2d {Dirac} semimetals}.
\newblock \emph{\bibinfo{journal}{Phys. Rev. Lett.}}
  \textbf{\bibinfo{volume}{114}}, \bibinfo{pages}{237001}
  (\bibinfo{year}{2015}).

\bibitem{Sun2014}
\bibinfo{author}{Sun, K.}, \bibinfo{author}{Chiu, C.-K.},
  \bibinfo{author}{Hung, H.-H.} \& \bibinfo{author}{Wu, J.}
\newblock \bibinfo{title}{Tuning between singlet, triplet, and mixed pairing
  states in an extended hubbard chain}.
\newblock \emph{\bibinfo{journal}{Phys. Rev. B}} \textbf{\bibinfo{volume}{89}},
  \bibinfo{pages}{104519} (\bibinfo{year}{2014}).

\end{thebibliography}

\end{document}